# DRAFT : Task System and Item Architecture (TSIA)

Burkhard D. Burow

*Postfach 1163, 73241 Wernau, Germany*
*burow@tsia.org*

March 1999

During its execution, a task is independent of all other tasks. For an application which executes in terms of tasks, the application definition can be free of the details of the execution. Many projects have demonstrated that a task system (TS) can provide such an application with a parallel, distributed, heterogeneous, adaptive, dynamic, real-time, interactive, reliable, secure or other execution. A task consists of items and thus the application is defined in terms of items. An item architecture (IA) can support arrays, routines and other structures of items, thus allowing for a structured application definition. Taking properties from many projects, the support can extend through to currying, application defined types, conditional items, streams and other definition elements. A task system and item architecture (TSIA) thus promises unprecedented levels of support for application execution and definition.

This publication may be referenced as:

Burkhard D. Burow. *DRAFT : Task System and Item Architecture (TSIA)*, March 1999.
(Available at http://www.tsia.org and elsewhere).

**Table of Contents**

# 1 Introduction

## *1.1 The Thesis*

An application is given by its *definition*. Since an application *execution* merely acts out the definition, the details of the execution are irrelevant to the definition. The definition may be freed of from these details by separating the application definition from its execution.

The *environment* consists of all the resources required to execute an application. For a computing application, the environment includes at least a computer. For a given application, the environment may include many other resources. For example, for an interactive application, the resources include the human interacting with the application. The environment also may include operating systems and other systems that control and facilitate the use of the resources.

In the beginning of computing, an application was defined in terms of an environment. Since the environment consisted of little more than a computer, the application thus was defined in the machine code of the computer. This original relationship between the application definition, the environment and the application execution is illustrated in Figure 1a). Defining an application in terms of an environment is ineffective since the application definition thus includes the irrelevant details of the execution. In addition, such a definition generally is ineffective since an environment is unlikely to provide good support for a structured application definition.

The arrival of programming languages in the 1950’s allowed for an application definition independent of the environment [Fortran]. As illustrated in Figure 1b), an application can be defined using a programming language which in turn uses the environment to execute the application. A programming language thus achieves the desired separation between the application definition and the application execution. In addition, such a definition can be effective since a programming language can support a structured application definition [Structured Programming : Data].

To date, programming languages generally do not provide *execution elements*. These elements include parallel, distributed, heterogeneous, adaptive, dynamic, real-time, interactive, reliable, secure or other execution. The separation between the application definition and execution achieved by current programming languages, as illustrated in Figure 1b), thus generally is suitable only for an application which requires no execution elements.

Other applications, requiring one or more execution elements, thus generally cannot be defined solely in terms of current programming languages. Instead, such an application is defined in terms of a language and the environment. As illustrated in Figure 1c), the desired separation between the application definition and execution thus is lost. For example, parallel execution currently often is achieved by including message passing communication or shared memory synchronization in the application definition. Since these and other execution details are irrelevant to the application definition, an application definition free of execution details is desired.

During its execution, a *task* is independent of all other tasks. For an application which executes in terms of tasks, many projects have demonstrated that a *task system* (TS, pronounced “tee ess”) can provide such an application with execution elements. Tasks, task systems and other details of application execution are presented in chapter 2.

As illustrated in Figure 1d), a TS allows the application definition to be defined in terms of a programming language and the TS. The application is not defined in terms of the environment. A TS thus separates the application definition from the application execution. The application definition thus is free of the execution details.

To date, TS support only a limited variety of applications. Defining an application in terms of a TS implies that the application is defined in terms of tasks. Relatively few applications are suitable for a definition in terms of tasks.

This thesis recognizes that a TS does not require an application to be defined in terms of tasks. Instead, a TS only requires that an application execute in terms of tasks.

A task consists of *items* and thus an application can be defined in terms of items. A large variety of applications can be defined in terms of items. Combining properties from various projects, an *item architecture* (IA, pronounced “ya”) allows an application defined in terms of items to execute in terms of tasks. Items, item architectures and other details of application definition are presented in chapter 3.

Figure 1e) illustrates the introduction of a IA to a TS. An application then is defined using a programming language which uses a IA to feed tasks to a TS which executes the tasks in the environment, thus executing the application.

A IA can support arrays, routines and other structures of items, thus allowing for a structured application definition. Taking properties from many projects, the support can extend through to currying, application defined types, conditional items, streams and other definition elements. A IA thus can allow a programming language to better support a structured application definition.

A *task system and item architecture* (TSIA, pronounced “tsia” with “ts” as in “tsunami”) thus may be summarized as follows. A TS promises that an application which executes in terms of tasks may be provided with execution elements. A IA promises that an application with a structured definition can execute in terms of tasks. Thus TSIA promises that an application with a structured definition can be provided with execution elements. A few general comments on this thesis follow.

TSIA is very compatible with current computing and as such augments current computing. Thus while TSIA raises areas of computing to new heights, it is not a radical departure from current computing.

For example, a IA does not introduce new structures for items, at least not in this presentation. Instead a IA introduces new possibilities for implementing structures. In other words, a TSIA does not so much introduce new programming languages as it does introduce new possibilities for implementing programming languages. As illustrated in Figure 1e), a programming language is implemented in terms of TSIA and the environment.

Similarly, a TSIA can use the resources and systems of a current environment. While TSIA ultimately may lead to changes in the typical environment, TSIA does not require such changes.

The compatibility between TSIA and current computing allows an application to have a part which involves TSIA and a part which does not. Thus for example, a library of routines using TSIA may be part of an application which otherwise does not involve TSIA. As illustrated in Figure 1f), an application may be defined in terms of a programming language, a IA, a TS and the environment. Thus throughout this thesis, mention of an application using TSIA also implicitly refers to just part of an application.

As a natural progression for current computing, TSIA incorporates properties from very many projects. The initial TSIA presented in this thesis thus requires few new properties. Since a generic TSIA has not yet been implemented, the feasibility of TSIA only can be argued. By requiring few new properties, the argument is greatly simplified.

A variety of TSIA may be implemented, each emphasizing different features of the TS or of the IA. A TSIA thus may be more suitable than another for a given application. For example, the TSIA supporting a home appliance may or may not be the same TSIA supporting the transactions of a bank.

In short, allowing an application to have a structured definition and providing it with execution elements no longer is an unsolved problem. For a large variety of applications, the solution is just a matter of implementing TSIA.

### *1.2 The Presentation*

This presentation is an initial introduction to TSIA. Before anything else, interest in TSIA first must be established. The emphasis of the presentation thus is a demonstration that a TSIA indeed can provide a structured application definition with execution elements.

If in this presentation a name initially is *emphasized*, then the name either is introduced by this presentation or is not yet in common use. Unless noted otherwise, a name is introduced only when no previous name is in common use. In general, only the name is introduced; the notion usually already exists.

The large power and scope of TSIA are difficult to contain within an initial introduction and thus requires the presentation to be accompanied by the following caveats.

TSIA spans the large computing areas of execution and definition. This presentation explicitly introduces TSIA within an implicit description of execution and definition. As an initial introduction, this presentation is not a comprehensive description of TSIA. Instead, presented is an outline of TSIA, within a rough description of execution and definition. For example, programming languages, input data, libraries of routines, execution requirements and more may define an application but this presentation examines only programming languages.

Few properties of TSIA are mentioned in this presentation; even fewer are described. This includes properties known elsewhere in computing. For example, this presentation mentions few of the details required to implement TSIA. Similarly, properly motivating TSIA would require a description of current computing and its limitations. Similarly, this presentation addresses few of the possibilities for TSIA and few of the consequences of TSIA. For example, like other models [Interdisciplinary Coordination], the task model may have uses outside of computing. In general, such properties are beyond the scope of this presentation.

For each of the few properties of TSIA that are addressed in this introduction, it is well beyond the scope of this presentation to present the origin and evolution of that property in computing. Instead, only a fraction of relevant research and experience is mentioned and even this generally is just a citation.

In large part, this presentation demonstrates TSIA without defining it. For example, many different execution elements are mentioned, but no attempt is made to define the term execution element. The preference of demonstrations over definitions has two justifications. Firstly, with only limited experience to date with TSIA, definitions are premature. Secondly, demonstrations allow for a simpler and more direct introduction to TSIA. How-ever, demonstrations without definitions do have a negative property. An understanding of TSIA based on this presentation must take care that it addresses the essence of TSIA and not the irrelevant detail of a demonstration.

## 2 Application Execution

This chapter demonstrates that a TSIA can provide an application with a large variety of execution elements. Chapter 3 demonstrates that TSIA can do so for a large variety of applications. This chapter thus focuses on application execution and on the task system (TS) of TSIA. Chapter 3 focuses on application definition and on the item architecture (IA) of TSIA.

### *2.1 The Precursors of the Name Task*

The name task has had many precursors. These include the task [Fx][Jade], the independent task [Spawn], the thread (not to be confused with the thread as known in this thesis and elsewhere in computing) [Cilk-NOW][Coarse Grain Dataflow], the regular Mentat object [Mentat], the upcall [RTU][SUMO], the filter predicate [Packet Filter], the event [168/E][Funnel], the soft-instruction and others [SISA].

Precursors of the task are cited throughout this presentation. By mentioning a few original names, the above list of precursors aims to help relate this presentation to the presentations of precursors.

### *2.2 A Task is Not a Process*

For an application which executes in terms of tasks, many projects have demonstrated that a task system (TS) can provide such an application with execution elements [168/E][Cilk-NOW][DBC][DNA][Funnel][GLU][Internet Computing][Jade][Linda-Piranha] [Mentat] [Nimrod] [Packet Filter] [RTU] [SCHEDULE] [SETI@home][Spawn][SUMO][Supervisor-Worker].

From these previous TS, TSIA adopts and unifies the notion of tasks and the role of the TS. As described in section 2.11, the combined experiences of previous TS demonstrate that a TS can provide an application with a large variety of execution elements.

The task model is an alternative to the familiar process model. During its execution, a task is independent of all other tasks. In contrast, during its execution, a process may depend on other processes. In other words, during its execution, a task does not communicate with other tasks. In contrast, during its execution, a process may communicate with other processes.

Other names for the process model include concurrent system or concurrency. In the process model, an executing application consists of one or more processes [CSP]. A process executes arbitrarily many actions. In order to coordinate the application execution, processes communicate, typically via shared memory [Linda][OpenMP][Threads] or message passing [Active Messages][MPI]. In comparison to a task, a thread [Threads] is a minor variation on a process and thus is a process as far as this thesis is concerned. There exist many elaborations of the process model. Active objects is an example [Active Objects].

Unfortunately, in some previous projects the process or thread has been named the task. Vice versa and as already mentioned in section 2.1, the task sometimes has been named the thread. Thus in this presentation the behavior, not the name, identifies whether a previous project is based on the task model or the process model.

Despite a vast amount of accumulated research and experience, the process model presently does not allow an application definition to be provided with execution elements.

An elegant example of an argument for such an application definition and execution within the process model may be found elsewhere [Coordination]. This wish for the process model may be used to gauge the achievements by the task model of this thesis.

In the process model, each process determines what actions it executes. In other words, an application controls its own execution. A system supporting the process model thus is subordinate to the application. As a slave to the application, such a system has the difficult chore of trying to coerce the environment to ideally obey the execution commands of the application. As a result, a system based on the process model has great difficulty providing an application with execution elements [Cilk-5][Linda].

The task model does not require an application to control its own execution. This is one of the major differences between the task model and the process model. The TS thus takes control of the application execution. This is based on the recognition that the execution requirements of an application do not require the application to command its own execution. This approach is similar to that of a windowing system based on events dispatched to callbacks [X], as discussed in subsection 2.11.16. By commanding the execution, the TS gains degrees of freedom which make it relatively easy to satisfy the execution requirements of the application.

The process model, especially with execution elements, often provides a fragile application execution, with indeterminate results, priority inversion, deadlock or other ills. In contrast, the task model, even with execution elements, provides a robust application execution, without the ills common to the process model. This is a second major difference between the task model and the process model.

As illustrated in Figure 2a), the process model pervades throughout the application execution and definition. If the environment uses the process model, as is very common in current computing practice, the task model of course can co-exist with the process model. As illustrated in Figure 2b), the TS may be implemented using the process model. The ills of the process model then are overcome within the TS. With the process model thus restricted to the application execution, the task model can be used for the application definition. As far as the task model is concerned, the process model is a low level execution detail of the environment just like any other such detail.

A task is a single action to be executed. Compared to the many actions of a process, this is a third major difference between the task model and the process model. Furthermore, a task is constructed such that it may be simply managed by the TS. The action of a task is often a computation and these are most of the actions examined in this presentation. Examples of other actions include data storage or interaction with the real world, such as moving the arm of a robot.

In the task model, the execution of an application thus is reduced to the simple situation of managing the execution of tasks. It is the simple management of each task which allows the TS to provide an application with execution elements. Section 2.4 begins the introduction to such an application execution in terms of tasks. Many of the issues raised in each section are further described and discussed in subsequent sections.

In short, because a task excludes execution, a task system can encapsulate execution. A task system thus allows an application definition to be free of execution details. In contrast, because a process includes execution, a process system cannot encapsulate execution. It would seem that by definition, the process model embeds execution details into the application definition.

### 2.3 The Goto Precedent

A communication between processes is very similar to the use of the goto statement within a process. A communication is like a more powerful goto. A goto statement then is like a process communicating with itself. The precedent set by the goto statement thus may be a suitable analogy in the argument for the task model over the process model [SISA].

The undisciplined use of the goto statement long has been considered harmful for an application definition [GOTO]. The argument against such use implicitly is an argument for higher level constructs. These include the `if then` conditional construct, the `case` choice construct and the `while` loop construct. These higher level constructs support structured programming for the application definition [Structured Programming]. The higher level constructs can be imitated with a disciplined use of goto.

This thesis argues for the task model. As described in section 2.5, an executing task does not communicate with any other task. The task model thus is free of the goto-like communication of the process model. In general, the task model allows an application definition to be separate from its execution. This elimination of irrelevant execution details is the first step to a structured application definition. The relationship between structured programming and a structured application definition is discussed in subsection 3.17.4.

By arguing for the task model, this chapter implicitly argues against using the process model for an application definition. Of course in the application execution, processes and communication may be present as low level details, just as a goto may be present as a jump in the machine code. As for the argument against goto, the essence of this thesis is the identification and elimination of low level details, including processes, communication and everything else irrelevant to the application definition, and the promotion of suitable high level alternatives.

The arguments against goto and those for tasks rest on the same basis:

> **“.. our intellectual powers are rather geared to master static relations
> .. our powers to visualize processes evolving in time are relatively poorly developed.” [GOTO]**

Though presumably not originally intended, this thesis encourages taking literally the word processes in the above quote. By executing an application in terms of tasks, TSIA eliminates processes from the application definition. As demonstrated in chapter 3, the application definition thus is reduced to static relations.

### 2.4 Introducing an Application Execution in terms of Tasks

As illustrated in Figure 3, the tasks of an application execute via the task pool. Each task in the pool is unique. The pool, including each task in the pool, is managed by the TS. From the application definition, a new task enters the pool via the item architecture (IA). Once the task system (TS) has executed a task, the task exits the pool. A task thus normally executes once.

Chapter 3 describes how a IA allows an application definition to execute in terms of tasks. The description there includes examples of application execution via the task pool. The origin of the task in the pool thus is not described in this chapter. Instead, this chapter considers a task once it has entered the task pool.

The task pool thus is the main interface between the TS and the IA. The task pool thus also is the main link between this chapter and chapter 3.

At any moment during the execution of an application, an arbitrary number of tasks may be in the pool. If the entire application executes in terms of tasks, then the application execution begins once it places a task into the pool and ends once there are no more tasks in the pool.

The TS serves the application execution. The TS thus is perhaps best imagined as a run-time service for an application. For each task placed by the application into the pool, the TS manages the task through to its completed execution. Any application at any moment may place one or more tasks into the pool. Thus as described in section 1.1, an application may have a part which involves TSIA and a part which does not. The part which does not may execute outside of the task model. For example, if the environment uses the process model then that part of the application may use the process model.

### *2.5 The Essence of a Task*

A task is a single action to be executed, constructed such that it may be simply managed by the TS. Though it is a most general definition, this operational definition from section 2.2 is not a practical definition of a task. Arriving at a practical definition requires identifying the essential nature which allows a task to satisfy the operational definition. Analyzing systems based on tasks, including previous TS, yields the following observation:

**"During its execution, a task is independent of all other tasks."**

Known as *task autonomy*, this observation determines the constraints and the freedoms for a task.

The expression task autonomy is preferred over the expression task independence. The latter is too easily confused with the dependence or independence of a task, outside of its execution, on other tasks. Because they do not occur during the execution, such dependencies do not violate task autonomy.

Achieving task autonomy is described in the next two sections. The remainder of this section outlines the benefits of this achievement for the TS.

Task autonomy implies that during execution a task does not communicate with any other task. In addition to those differences identified in section 2.2, this is a major difference between the task model and the process model.

Task autonomy thus implies that, once started, a task simply executes to completion. In other words, the execution of a task requires no further attention or assistance from the TS. Since there are no details to concern the TS, an executing task may be treated by the TS as a black box.

Since nothing of interest to the TS occurs during the execution of a task, a task execution may be considered to occur indivisibly. For an application execution, the task thus is the fundamental unit of execution or action.

### *2.6 The Definition of a Task in terms of Items*

A task consists of arbitrarily many *items*. An item is either an *in*, an *out* or an *inout*.

If an executing task is treated as a black box, then the ins of a task include everything required for the execution of the task. Likewise, the outs include everything resulting from the execution of the task. In other words, the ins include everything that affects the task, while the outs include everything affected by the task. A task thus is an act which creates outs from ins. The ins and outs may be arbitrarily large and sophisticated.

An inout behaves both as in and as an out. Instead of discussing inouts explicitly, this presentation generally implicitly includes inouts in the discussions both of ins and of outs.

Since the ins and outs include everything involved with the task, the ins and outs define the task. As illustrated in Figure 4, the *task definition* thus is given by the ins, outs and inouts of the task. In the execution of the task, the ins creates the outs. The task definition allows the details of its execution to be ignored, as required for a black box.

Because it allows the details of the task execution to be ignored, it usually is simplest to think of a task in terms of its definition. Section 2.8 outlines the few actions of the TS required by the task execution. This and the next section continue presenting the task definition.

As described above, task autonomy requires the task definition to include everything that affects or is affected by the task. Task autonomy thus implies that a task is free of side-effects. A task thus enjoys referential transparency [Monads], as it is known in functional computing. With referential transparency, the same outcome results from two executions of the same task. Similarly, effectively the same task yields effectively the same outcome. In functional computing, if the outcome of a function is known, referential transparency often is used to replace the function by its outcome, instead of executing the function. As described in subsection 2.11.7 and in section 3.14, a TS can make the same use of referential transparency. In addition, the TS also makes essentially the opposite use of referential transparency. As described in subsection 2.11.3, referential transparency allows a task to execute more than once.

The task definition includes all the items of the task. Examples of such ins and outs follow. Obviously the ins and outs respectively may include the data required by and produced by the task. An in may be the instruction which encodes the action to be performed by the task. An in may be the software licence or other authorization required in order to execute the task. An in may be the computer processor or processors required to execute the task. An in may be a question and another in may be a human operator whose answer to the question is the out of the task. An inout may be the arm of a robot. A task of a distributed application may involve a particular item, not just a copy of that item.

In addition to compatibility with present computing techniques, TSIA also seems to be suitable for promising future techniques. For example, instead of using a traditional instruction and a computer processor, a configured field programmable gate array (FPGA) or similar device may encode and execute the action of a task [Reconfigurable Architectures]. The configured FPGA is then the out of a previous task and the in of a subsequent task or tasks. Another example of a promising computing technique arrives at a solution of a large problem through the evolution of a population of candidate solutions [Evolutionary Computing]. Each candidate solution, as data or as an instruction, is the out of a task and the in of a subsequent task or tasks.

Because it includes all the items of a task, the task definition introduces a very powerful symmetry to a task. Although an item may be vastly different in many respects from another item, each is simply an item, as far as the task and the TS are concerned. For example, the instruction of a task usually is considered to be quite distinct from an argument. However, for a task the two are very similar since they are both items. Likewise, the exclusive use of a computer processor or of a human user is very similar for a task to the exclusive use of an area of memory for the data of an out. The symmetry between human items and software items has been pursued elsewhere [DSL].

Because of the symmetry introduced by the task definition, the management of a task is reduced to the coordination of the items of the task. The simple management of each task thus is achieved. As already mentioned in section 2.2, it is the simple management of each task which allows the TS to provide an application with execution elements.

There is an important difference between management and coordination. Management is coordination combined with control. The TS manages a task. The TS thus controls and is responsible for a task. In contrast, the TS coordinates the items of a task. In addition, the TS may manage an item, but this control is not required by the TS. For example, the TS does not control the human user of an interactive application. Instead, the TS merely coordinates the role of the user in the application execution. Similarly, the TS need not control computing resources. Instead, the TS simply may use computing resources, as allocated by a tool controlling the resources [LSF]. Similarly, the TS may coordinate shared memory which manages that an out of data from one task is available as an in of data to a subsequent task. Alternatively, the TS may use message passing to directly manage the data.

### *2.6.1 The Task Model versus the Process Model*

As defined in section 2.5, a task is the unit of execution. Thus by definition, every application execution implicitly involves tasks. In the process model, tasks are only implicit in the application. For the execution, the application effectively assembles the items required for each task. Message passing is an example of such assembly [Active Messages][MPI]. In the task model, tasks are explicit in the application. The application thus only defines the items required for each task. The TS, not the application, then assembles the items required for each task. The responsibility for assembling the items of each task, by the application or by the TS, respectively, is a way of expressing perhaps the fundamental difference between the process model and the task model.

## *2.7 The Origin of a Task Definition*

As illustrated in Figure 5, a task definition is divided among the IA, the TS and the TE. The task executor (TE) is introduced below. The boundaries in the task definition between the IA, the TS and the TE are not fixed. Instead the boundaries depend on the *execution situation*. These flexible boundaries are a result of the symmetry introduced by the task definition and are a further property of the simple management of a task.

Throughout this presentation, the term *execution situation*, as used in the previous paragraph for example, includes all factors associated with an application execution. The execution situation thus includes the application, the environment, the required execution elements, the state of the execution and other factors.

As described in chapter 3, the item architecture (IA) is used to define an application such that it executes in terms of tasks. For any given task, the application defines, via programming languages and the IA, only the items which are significant to the application definition. For example, for a task performing a computation, the items in the application definition generally only include the instruction and the data. For any given task, the IA thus defines the items significant to the application definition. The remaining items are defined by the TS or the TE.

The TS manages the execution of the application. For any given task, the TS defines the items necessary for the execution elements required by the application. For example, the items of interest to the TS may include scarce computing resources such as processors and memory. For any given task, the TS thus defines the items significant to the application execution. The remaining items, those defined by neither the IA nor the TS, are defined by the TE.

In principle and often in practice, it is convenient to introduce a *task executor* (TE). The TE manages the execution of an individual task. The TS thus is spared these details. The TE is an explicit or an implicit part of the TS. If the TE is not explicitly discussed, then the TE is implicitly part of the TS.

By construction, the TE thus defines the items which are not defined by the IA or the TS. The items defined by the TE thus are neither significant to the application definition nor to the application execution. Since the items defined by the TE are insignificant, their definition often is only implicit and not explicit.

The processor type used to execute a task can provide an example of the flexible boundaries in the task definition between the IA, the TS and the TE. Usually the processor type is irrelevant to the application definition and execution. The processor type thus usually is an in defined by the TE. Even within the TE, the processor type usually is not explicitly noted.

Nevertheless, there are situations where the processor type is significant to the application definition or execution [Heterogeneous]. The processor type is then an in defined by the IA or the TS, respectively. For example, the IA defines the processor type if the task is a calculation which depends on some feature of the processor, such as the particular floating point implementation. Similarly, the TS may choose to define the processor type of a task after noticing from previous tasks that the instruction of the task performs markedly better on that processor type.

### *2.8 The Execution of a Task*

In order to achieve task autonomy, a task must be provided before execution with its items. During its execution, a task thus is independent of all other tasks. For example and as described in chapter 3, an out of a task is very often an in of a subsequent task. Such a dependence between tasks does not violate task autonomy. As for any other item, such a dependence simply must be satisfied before the task executes.

As described in the section 2.6, the task definition reduces the management of a task to the simple coordination of the items of the task. The details of each item may be ignored. Task autonomy merely requires a simple management which ensures that a task is provided before execution with its items.

The above simple management is performed by the TS in cooperation with the TE. At least conceptually, the TS and the TE thus play two roles for a task. In addition to the above management role, the TS and the TE provide part of the task definition, as described in the previous section. The two roles thus follow the separation between the task definition and the task execution.

The complete life-cycle of a task may be summarized as follows:

- The application definition, via the IA, enters a new task into the task pool of the TS. At this point, the task definition includes only the items significant to the application definition.

- The TS adds to the task definition the items significant to the task execution.
- The TS, in cooperation with the TE, assembles the significant items of the task.
  The TS then leaves the task with the TE.
  Since the items may include out of previous tasks, ensuring the previous execution of such tasks is among the chores of the TS. The management of such dependencies between tasks is discussed further in subsection 2.11.2.
- The TE explicitly or implicitly defines any remaining items required for the execution.
- The TE assembles the remaining items.
- The TE executes the task.
- The task leaves the task pool of the TS.
- The outs produced by the task are coordinated by the TS and TE, typically as ins to a subsequent task or tasks.

At least in concept, it is convenient to identify the TE as the place where a task actually executes. This holds even when all the items of a task are defined by the IA and TS and none are defined by the TE. The TS passes a task to the TE for execution. Once the task has executed, the TS passes another task to the TE for execution.

For a variety of reasons, as described in section 2.11, the execution of an application may involve multiple TE. In contrast, at least in concept, the execution of an application involves only a single TS. The multiple TE may or may not be identical copies of one another. For example, the multiple TE are different if each is specialized to perform a particular set of instructions. A TS with multiple TE is illustrated in Figure 6. Each task is passed by the TS to a TE for execution.

### *2.9 The Classic Application*

A particular type of application long has enjoyed a definition free of execution details and has been provided with execution elements [168/E][DBC][DNA][Funnel] [Internet Computing] [Linda-Piranha] [Nimrod] [Packet Filter] [RTU] [SETI@home] [Spawn][SUMO][Supervisor-Worker]. This presentation refers to this particular type of application as the *classic application*. In return for a minimum of effort, a simple system can provide a classic application with execution elements.

A pseudocode program for the classic application is shown in Figure 7a). It has the following simple definition. The input data consists of independent pieces of data. Each input piece is independently used to produce an independent piece of output data. Each such production obviously can correspond to a task. A classic application thus may execute in terms of tasks. A simple system which provides a such classic application with execution elements thus is identified in this thesis as a rudimentary TS.

An example of a classic application is the simulation of some system, where the simulation consists of many independent trials [Funnel][Nimrod][Spawn]. Each trial corresponds to a task for which an in provides a unique initial condition and the corresponding out contains the result. Another example arises when many independent measurements are made by some system, where each measurement subsequently requires some form of processing [168/E][DBC][SETI@home]. The identification of the independent packets in networking provides another example [Packet Filter]. Similarly, distributed and real-time support for multimedia may perform an independent processing of each independent frame or other unit of the continuous media [RTU][SUMO]. Yet another example is given by a problem whose solution requires the evaluation of a large number of candidate solutions [DNA][Evolutionary Computing][Internet Computing].

Each task of the classic application has as an in an independent piece of input data. Similarly, each task has as an out an independent piece of output data. This in and this out may be the only items of each task significant to the application definition. The remaining items are defined by the TS and the TE. For example, if the same instruction is used by each task of the classic application, then the instruction is an implicit in which need not be explicitly defined for each task by the application definition. If a classic application consists of `N` ins, then the entire application is defined by `N` tasks, as illustrated in Figure 7b). At the start of the execution for the classic application, the `N` tasks may be imagined to be in the task pool illustrated in Figure 3. Once a task completes, it leaves the pool. Therefore, the execution of a classic application completes once each of its tasks has completed and there thus are no more tasks in the pool.

A classic application may be provided with execution elements. As already stated above, this property of a classic application has long been popularly recognized and exploited. In contrast, the origin of this property is not yet popularly known. This thesis identifies and promotes the origin.

The origin of this property of a classic application has two parts. Firstly, a classic application has a straightforward execution in terms of tasks, as described above. Secondly, any application which executes in terms of tasks may be provided with execution elements, as demonstrated in the next two sections.

For an application which executes in terms of tasks, the ability to provide it with execution elements also has been recognized for applications other than the classic application. For example, in order execute efficiently, some large computational fluid dynamics applications and similar applications essentially execute in terms of tasks [CFD][Guard].

Chapter 1 introduced TSIA as a derivation from previous TS. In fact, this thesis originates from extensive experience with the rudimentary and heavily used TS of a classic application [Funnel].

The classic application provides two types of support for this thesis. Firstly, the classic application is an excellent introductory example application, since it has a straightforward execution in terms of tasks. The second type of support is given by the many existing rudimentary TS, each providing a classic application with execution elements. Just like a general TS, a rudimentary TS executes an application in terms of tasks. In many ways a rudimentary TS thus demonstrates the feasibility of a TS for TSIA just as well as is demonstrated by a general TS. Therefore, previous rudimentary TS in addition to more general TS are referenced in the section 2.11 in order to demonstrate the feasibility of a TS for TSIA.

### *2.10 The Feasibility of the TS*

A TS is feasible because TSIA requires an application to execute in terms of tasks. The previous sections of this chapter outline what it means for an application to execute in terms of tasks. This and the next section demonstrate that a TS is indeed feasible.

The feasibility of a TS essentially has two parts. The external feasibility must demonstrate that the TS can provide an application with execution elements. The internal feasibility must demonstrate that the TS can be built.

Because an application executes in terms of tasks, a TS can provide it with execution elements. This external feasibility of a TS may be summarized in three points. Firstly, the TS views each application in terms of tasks. Thus decoupled from each application definition, the TS is suitable for a large variety of applications. Secondly and as introduced in section 2.2, the TS commands the application execution, not vice versa. The TS thus performs the 'global' management of the application execution. Thirdly and as described in section 2.8, the management of any individual task merely is a coordination of its items. Therefore, the 'local' management of each task is simple and the TS thus may focus on the 'global' management of the application execution. A TS thus is in an excellent position to satisfy the execution requirements of the application.

As already implicitly introduced in subsection 1.1 and illustrated in Figure 1, a TS is defined in terms of the environment. A TS thus is an application like many others. Defined in terms of the environment, the definition of such an application is full of execution details. The internal feasibility of a TS thus is due to the very large number of similar existing applications. As introduced in subsection 1.1 and as illustrated in Figure 1c), an application to date requiring execution elements generally has had to be defined in terms of the environment. For example, an application to date with a parallel execution generally is defined in terms of the environment. Such existing applications include previous TS. Admittedly, a TS is an arbitrarily large and sophisticated application, but it is not fundamentally different from other applications defined in terms of the environment. No fundamental difference between a TS and other applications is introduced by the fact that a TS manages the execution of another application. Like other applications, the TS thus merely is an assembly of existing tools and techniques to define an application in terms of the environment. Of course, new tools and techniques eventually may be introduced. The internal feasibility of a TS thus is due to the fact that the implementation of the TS requires the same pain and effort as that required for any other application full of execution details.

Other than the execution details, much of the internal feasibility of a TS concerns the simple accounting associated with managing the task pool, including its task and items.

The external and internal feasibility of a TS thus may be summarized as follows. The external feasibility allows an application managed by the TS to have a definition free of execution details. Instead, the execution details are a part of the TS. For the internal feasibility, the execution details inside a TS are not different than those inside any other application defined in terms of the environment.

A TS thus does not magically eliminate execution details. A TS is feasible since it instead merely moves the execution details of an application to the TS. The sequence of illustrations in Figure 8 concisely presents the feasibility of a TS. Figure 8a) is identical to Figure 1c) and illustrates the situation of an application before the introduction of a TS. In Figure 8b), tasks allow the execution details to be separated from the remaining application definition. This separation corresponds to the external feasibility of a TS. In Figure 8c), the execution details may be performed by a TS. This performance corresponds to the internal feasibility of a TS. Re-arranging Figure 8c) arrives at Figure 8d), which is identical to Figure 1d) and illustrates the introduction of a TS.

The feasibility of a TS may be equivalently described using a simple difference between the process model and the task model. As described in subsection 2.6.1, in the process model the application definition implicitly is responsible for assembling the items of a task. In the task model, this responsibility still exists, but is instead an explicit part of the TS.

### *2.11 Examples of Execution Elements*

The general arguments presented in section 2.10 for the feasibility of TSIA are made more concrete by a list of examples in this section. Each example illustrates an execution element, as provided by the TS to an application.

Only a few of the examples are presented explicitly in terms of the of the internal and external feasibility described in section 2.10. Nevertheless, it should be obvious in each example that both parts of the feasibility are satisfied. For example for the internal feasibility, in each of the examples, the tools and techniques used by the TS have been used by previous applications defined in terms of the environment.

Though some execution elements may depend on another, elements are largely orthogonal to each other. For the most part, the examples thus may be examined independently. For example, reliable computing is largely independent of whether or not the execution involves parallel computing. In arguing that the TS is feasible, it thus is valid to examine individual execution elements.

The above orthogonality also results in tremendous power when execution elements are combined. For example, adaptive and reliable real-time computing allows a mission-critical real-time application to execute in an environment where a redundant copy of a resource can be introduced or removed at any time. As another example, the TS achievement of execution elements may be gauged using the yardstick of metacomputing [Legion][Metacomputing][Globus].

Since a large number of existing applications defined in terms of the environment have implemented any given execution element, this thesis describes little of the details underlying each element. Any particular method given in an example is not meant to indicate the best or only method of providing a particular execution element. Instead, the method only is given in order to concretely illustrate that the TS can provide an application with the execution element. The reader is free to substitute any given method by personal favorites.

Many of the examples are supported by references to previous TS which have satisfied the execution element. As described in the section 2.9, the previous TS include rudimentary TS used for a classic application as well as more general TS.

As described in section 2.9, a classic application has a straightforward execution in terms of tasks. A rudimentary TS thus may satisfy the execution elements discussed in the examples. By keeping the simple classic application in mind, the reader should be able to easily envision and understand each example. Basically the reader only has to remember that for the classic application each task corresponds to the independent production of an out from an in. The reader also should keep in mind the following summary of the TS. Execution elements may be provided to any application which executes in terms of tasks. Thus any execution element provided to a classic application also can be provided to any other application which executes in terms of tasks.

#### *2.11.1 The TS and the Environment*

Before discussing the execution elements and other execution issues, a few general words about the implementation of a TS are in order. This subsection thus addresses the relation-ship between the TS and the environment, while the following subsections primarily address the relationship between the TS and the application.

As described in section 2.4, a TS is perhaps best imagined as a run-time service provided to the application. As introduced in section 1.1, a TS is built on top of the underlying environment. The TS thus is perhaps best imagined as a run-time service which is only loosely coupled to the environment. Loosely coupled means that the implementation of the TS requires little to no modification of an existing environment [Cilk-NOW][Funnel][Mentat]. The possibility of such a loosely coupled TS supports the feasibility of a TS. In addition, imagining a TS as a loosely coupled system helps maintain the conceptual distinction between the TS and the environment.

Of course a TS need not be only loosely coupled to the environment. While remaining as the application interface to the environment, a TS could be tightly coupled to the environment. For example, the TS could be absorbed into the operating system of the computing environment. Alternatively, the TS could take over from the operating system some of the responsibility for the computing environment [Exokernel].

### *2.11.2 The Dependencies between Tasks*

Each task of an application consists of an arbitrarily large and sophisticated set of items. Since each of its tasks has a very simple set of items, the classic application may be served by a rudimentary TS. Other applications require a more general TS. The number and sophistication of the items of each task essentially are irrelevant to the arguments of this chapter. The next chapter addresses the items of each task resulting from the application definition. In this chapter, all that really matters is that an application executes in terms of tasks.

As described in chapter 3, the ins and outs of a task usually are the ins and outs of other tasks. The items of each task thus define the dependencies between tasks.

The dependencies between tasks must be known to the TS, since the tasks must execute in a sequence which satisfies the dependencies. For example, the dependencies arising from the application definition may be specified explicitly to the TS [Cilk-NOW][Jade][SCHEDULE] or be automatically recognized by the TS [GLU][Mentat].

Of course, independent tasks may execute in parallel. For example, two tasks are independent if the items of one task are independent of the items of the other task. Of course, identical in effectively may be independent if the in may be copied or otherwise simultaneously accessed.

Dependence analysis determines and describes the dependencies between tasks. In the past, dependence analysis often has addressed particular types of items. An example is the analysis of items associated with the equivalent of a single or a few machine instructions of a computer processor [Coarse Grain Dataflow][Compiler Transformations][Dataflow][Dataflow Architectures]. In essence, dependence analysis is concerned with the connection of items between tasks. Since it is not concerned with the internal details of any in or out, dependence analysis is well suited to the all inclusive items of TSIA [GLU][Mentat].

In summary, dependencies between tasks are no impediment to the feasibility of a TS.

### *2.11.3 Executing a Task More than Once*

Certain execution situations may require a task to execute more than once. For example, a task may execute more than once if recreating an out of the task is more efficient than stor-ing or communicating that out. Another example is described in subsection 2.11.13. If an application execution is to continue after a computer crash, those tasks executed after the most recent checkpoint are tasks which must execute again.

As introduced in section 2.4, a task normally executes once. If required, a task may execute more than once since all items of the task are identified to the TS. Thus the outs of each execution are identical. As described in section 2.6, this referential transparency is the result of the items of a task defining everything that affects and is affected by the task. In order for a task to execute again, the TS merely has to reassemble the items of the task.

Occasionally a task cannot execute again because one or more of its items are unavailable. For example, a task observing the real world at a given instant cannot execute again since it is impossible to return to that instant in time. Other items also may be beyond the coordination of the TS, but with effort the corresponding tasks may be executed again. For example, some items involve humans, databases or other large or complex systems and thus may require transaction-style tasks.

Despite exceptions such as those mentioned above, for most tasks it generally is simple for a task to execute more than once if required by the execution.

### *2.11.4 Performant Computing*

Performance may be defined as the capacity to produce desired results with a minimum expenditure of energy, space, time or other resources. Efficient computing thus might be another name for performant computing. The performance of an application execution may refer to any one or combination of the required execution elements.

A TS facilitates improvements in the performance of an application execution; in particular in the performance of the individual components of the execution: application, TS, environment. Because a TS uncouples these three components, a component can be changed while the other two components are kept constant. Thus with little regard for the other two components, the performance of each individual component can be studied and improved. Two examples of such improvement are given below.

Firstly, the uncoupling of the application, TS and environment facilitates for each component any increased sophistication and complexity required for increased performance. For example, since the environment is hidden behind a TS, there is no effect on an application if increased performance requires a more complex computing environment.

Secondly, a TS serves many applications, so it is worthwhile for the developers of a TS to investigate, evaluate and incorporate experiences and research results on the theory and techniques of execution. The effort invested into implementing these experiences and results may be amortized over the execution of many applications. These experiences and results may pertain to any of the many topics involved in an execution. Examples pertaining to scheduling, communication and to the execution of a task are described in the next three subsections, respectively.

An analogy may be drawn between a TS and a compiler for a programming language. A compiler not only hides from the application definition the low level details of machine code, but generally also generates very performant machine code. Similarly, a TS not only hides from the application definition the low level details of execution, but also can provide a very performant execution.

For the performance of an application execution, one may ask how an optimal performance is to be achieved. This question applies for each of the many topics, such as com-munication and scheduling, which may contribute to the performance of an application. Even with a fixed set of resources and a constant application definition throughout the execution, it often is difficult to statically determine how the TS can provide an optimal performance. Such a static determination is even more difficult if the resources or application definition may change during the course of the execution. Nonetheless, the TS is of course free to use a static determination as an initial approximation to optimal performance. The following paragraph ignores applications and parts of applications for which such a static determination is feasible and sufficient.

Instead of a static determination of the application execution, optimal performance often requires a dynamic determination. In other words, feedback may achieve an optimal performance [Adapt][Paradyn][Slim Binaries][Synthesis]. By monitoring and adapting to the performance, the TS can move the execution towards an optimal performance. In order to better examine the performance possibilities, the TS even could deliberately vary the application execution and measuring the resulting performance. The dynamic attempt at an optimal performance is part of the adaptive execution described in subsection 2.11.14. Task autonomy greatly allows the TS to adapt to and experiment with the execution. For example, an otherwise identical task may be run on two different computers in order to measure their performance. The results can be used to adjust and improve the execution of similar subsequent tasks.

Despite the best efforts of the TS, achieving the optimal performance of an application execution may require instructions or at least hints from a human operator or some other system outside the TS. As well as being available to the TS, the performance measurements thus also must be available outside the TS. For example, such information is required in order for a human operator to understand, debug and improve the performance of an application execution [Paradyn]. Similar issues arise for the application developer, for example when evaluating the performance of alternative algorithms for a part of the application definition.

### *2.11.5 Scheduling includes Throttling and Mapping*

Scheduling determines how to best use the resources of the environment to execute an application. A key to such a best use is the flexibility of the application execution. In TSIA, scheduling has two sources of flexibility. As illustrated in Figure 9, scheduling includes throttling and mapping. In essence, throttling controls the entry of tasks into the task pool from the application definition, while mapping controls the exit of tasks from the task pool onto the resources of the environment. The boundary between mapping and throttling is not always firm. Mapping is a part of the TS and is described in subsection 2.11.5.1 below.

As introduced in section 3.6, throttling is the part of IA which controls the expression of the application definition in terms of tasks. The classic application allows for a very simple example of throttling. Nominally each in and out pair of the classic application corresponds to a task. Throttling allows more than one in and out pair to correspond to a task. Since there often is a fixed overhead associated with each task, such throttling can improve the efficiency of the execution.

Throttling thus allows an application definition to be expressed in one of a variety of sets of tasks. Between the sets, the individual tasks can vary as can the moment in the execution when a given task enters the task pool.

Since it is constrained by the resources of the environment, the mapping generally prefers certain sets of tasks over others. A part of scheduling is the communication of these preferences from the mapping to the throttling.

#### *2.11.5.1 Mapping*

As described in subsection 2.11.2, the tasks of an application must execute in a sequence which satisfies the dependencies between the tasks. For any given application, there generally are a large variety of valid sequences, usually with a wide variety of performances. A performant execution thus usually requires identifying a performant sequence. Mapping is this evaluation of sequences and the identification of a performant sequence [Mapping] [Spawn]. For some TS, a simple mapping can deliver efficient and predictable performance [Cilk-NOW].

Since its tasks are independent of one another, a classic application has a particularly simple mapping [Mapping]. If each task corresponds to a job, then the mapping of a classic application is essentially the same as that of a traditional batch system. It thus is not surprising that the rudimentary TS of some classic applications essentially are batch systems [DBC][DNA][Nimrod].

Vice versa, with each job corresponding to a task, a traditional batch system is essentially a rudimentary TS. For example, all items required by a job, such as input and output files, can be declared to the batch system. Thus automatically restarting job after a power outage is very familiar in a batch system and demonstrates that a job, like a task, can obey task autonomy and referential transparency.

The characteristics which identify a performant sequence depend on the execution situation. An example of such a characteristic is locality. There usually is a cost associated with assembling the items of a task. A typical cost is communication. There usually is a lesser cost if a fraction of the assembly can be taken over from a previous task. The degree to which an assembly of items can be reused is referred to as locality. For example, if a task on a computer produces data as an out, it often is efficient for that data to be an in of a subsequent task on that computer [CFD]. Similarly, if the 50 Mbyte code of an instruction occupies the entire memory of a computer, it is often efficient for that computer to execute several tasks involving that instruction before executing a different large instruction.

Task autonomy considerably simplifies achieving good locality since it guarantees that the execution of a task never requires an item other than those defining the task. Thus the execution of a task cannot spoil locality. In addition, since all the items of a task are known, the cost of a particular assembly of items can be accurately determined in advance. Mapping obviously requires such advance information in order to choose between alternate assemblies for a task.

### *2.11.6 Communication*

Assembling the items of a task may involve the communication of an item. Effective and efficient communication thus is an important ingredient to a performant application execution. The ability of the TS to ensure such communication has two poles which may be introduced as follows. The TS does not just determine how to best communicate between points A and B. The TS first determines whether A and B should communicate at all.

At the macroscopic pole, the TS controls the application execution. The TS thus is able to incorporate communication in scheduling the execution of the application. The communication requirements of the application execution thus can be adapted to the communication available in the environment.

For a simple example, consider a classic application executing with a single TS master and multiple TE slaves, as illustrated in Figure 6. Pretend that the TS and each TE is running on an individual machine. The machine running the TS thus should be chosen to have sufficient aggregate bandwidth to serve all the TE with items.

Also at the macroscopic pole, the TS can recognize when it is cheaper to recreate an item, rather than to communicate that item.

At the microscopic pole, the TS assembles the items of each task. The TS thus is able to provide the best possible communication to any individual item. This is demonstrated by the following examples.

For the communication of any item, the TS is free to use any mechanism available. A single application execution thus may involve a mixture of message passing and shared memory communication.

The TS can obey the usual efficiencies. For example, the TS can avoid needlessly copying a large item of data. The communication of such an item within a computer then involves only a pointer to the item, not a copy of the item. Vice versa, for a small item the TS can avoid the overhead of indirection by passing a copy of the item, not a pointer to the item.

The communication performance required by the application execution also may involve additional features. For example, an application execution may be distributed over trusted computers, which are connected by an untrusted network. The TS can help the security of the execution by encrypting items during communication.

Latencies in communication often are overcome using buffers and similar devices. Such techniques are relatively easily implemented by the TS since an executing task requires no communication. The TS thus has considerable freedom in planning communication. For example, latency tolerance on a computer processor can be achieved as follows. In addition to the presently executing task, there always is at least one task with all items available and thus ready to execute [ADAM]. In other words, the TS can overlap computation and communication.

The conditions which allow the TS to provide an application execution with performant communication also seem suitable for data persistence. Regardless of the data's lifetime within an application execution or across separate executions of different applications, data persistence allows all data to be created and manipulated in a uniform manner [Grasshopper]. The TS coordinates all items, including all data items, of the application execution. The TS thus would seem to be a well-suited gateway between the application and a system for data persistence. At least conceptually, such a system could be independent of the TS and of the operating system or other resources of the environment.

### *2.11.7 Variations from a Normal Task Execution*

Performance and other execution elements may require or encourage slight variations from the normal execution of a task. Normally, the TS assembles for each task the items and the task then executes to completion. Seven variations from the normal execution are mentioned here.

A task, which previously has executed successfully to completion, may execute again. Subsection 2.11.3 mentions several execution situations requiring this variation.

In the second variation, one popular in functional computing, a task never executes. Instead, its outs are taken from another task. As described in section 2.6, if the ins of two tasks are effectively identical, then referential transparency assures that the outs also would be effectively identical. An example of this second variation is described in section 3.14.

In the third variation, a task may begin execution before all its items are available. Once such a missing item is required by the execution, the task must block until the item is available [Jade]. The fourth variation is very similar to the third. Long before a task finishes its execution, the task may indicate that an item no longer is required and thus already is available for any subsequent task [Cilk-NOW][Jade][Mentat].

The remaining three variations force a task's items to be available before the usual completion of the task. For example, another task thus may gain quick access to the computer processor. In the fifth variation, the task execution is suspended. Such a suspension is transparent to the task, but can involve high costs. Each item used by another task during the suspension eventually has to be restored in order for the suspended task to continue. In the sixth variation the task simply is killed and thus any progress made by the task is ignored [Funnel]. As described in subsection 2.11.3, the killed task can execute again. This sixth variation also is transparent to the task, but can have similarly high costs to those of the fifth variation. Because it frees resource quickly, the sixth variation is often suitable when making opportunistic use of otherwise idle resources and the time comes to return a resource to its rightful owner [Linda-Piranha].

Unlike the other variations, the seventh variation is not transparent to the task. Fortunately, this variation can be made transparent to the application definition by the programming language or IA. In this variation a task can respond to a yield request by the TS. The TS may signal the yield request using a variable periodically examined by the task. Such a variable introduces no race conditions or other ills, since the task only reads the variable. The task abandons the normal execution by making alternate arrangements for its outs and then exiting prematurely [Linda-Piranha]. For example and as described in chapter 3, the task could delegate the responsibility for its outs to one or more new tasks. Alternatively, a premature out of the task could be acceptable as is for the subsequent tasks expecting it as in [RTU].

The seventh variation also would seem to be well-suited for a solution to a so-called time-dependent problem. In the solution a task returns outs, even if a yield request by the TS occurs. Once resources are again available, a subsequent task may improve the outs [Time-Dependent Problems]. For example, the control of a robot may include many such time-dependent problems. If the robot has no time or other resources for the ultimate outs, the original outs allows for a best-effort control of the robot.

A variation from a normal task execution may not be used to allow a task to communicate during execution and thus to violate task autonomy. As long as such abuse of the variations does not occur, the variations do not change the fundamental nature of an execution in terms of tasks. Thus the variations used for application performance generally are not pursued beyond this subsection. Any variation pursued is one required for another execution element.

### *2.11.8 Secure and Accountable Computing*

A TS naturally accommodates security and accounting mechanisms. These allow an application to use resources far beyond what is presently possible.

An application defined in terms of the environment examines and controls resources. Such an application thus risks the security of the resources and the environment. Therefore such an application only can be executed by a trusted user of the environment. In this subsection, the environment of the trusted user is named the local environment. The trusted user is named a local user and has a local account. The resources of the local environment are the local resources. Such local users, accounts, environments and resources are common computing practice. This practice remains in this discussion.

Local environments may be combined to create a global environment with global resources [Legion]. In addition to being a local user of a local environment, a user may have a global account and be a global user of a global environment. A global user is **not** a trusted user of the environments making up the global environment. Such global environments are not yet part of computing practice. This subsection argues that a TS allows global environments to become a part of computing practice.

An application defined in terms of the environment has an execution restricted to the local environment. Although the execution might benefit greatly from using global resources, security allows such an application to use only local resources.

In contrast, an application definition not defined in terms of the environment is free to execute using global resources. As described below, since the application definition is free of execution details, its use of global resources is secure and accountable.

In general, an application definition can be separated into two parts. One part is defined in terms of the environment and is full of execution details. The other part is free of execution details. An execution of such an application is illustrated in Figure 10. The part of the application definition full of execution details only can use local resources. In contrast, the part free of execution details can use local or global resources.

The part of an application definition full of execution details and similarly the local environment is common in present computing practice and thus is not pursued further in this presentation. For simplicity, in the remainder of this subsection the entire application definition thus is assumed to be free of execution details.

The execution of an application uses resources, but an application definition free of execution details neither examines nor controls the resources. Regardless of the type and number of resources used, such an application thus cannot harm the environment. For example, such an application definition does not access computing resources, via system calls or otherwise. The absence of such system calls in the application definition easily can be verified and enforced. In addition, the application execution is restricted to its own address space by the memory protection of the computer, by the programming language or by other means. With no system calls and access to only its own address space, such an application can neither examine nor control computing resources [Slim Binaries].

An application not defined in terms of the environment thus poses no security risk to the environment. For example, such an application thus cannot determine any information about other users or applications. In fact, such an application even cannot determine if there are other users or applications in the environment. The global environment is secure.

Just as such an application poses no security risk to an environment, it also poses no other risks. For example, if an application is to opportunistically use otherwise idle resources, a TS can ensure that this use truly does not in any way affect other users of the environment. Such opportunism is further described in subsection 2.11.14.

The TS management of the execution thus allows an application to use global resources. In order to do so, the TS is a local user of each local environment which makes up the global environment. The TS thus is just an agent acting for the application; the TS is not part of the application. Instead, the TS is a trusted user of each local environment, similar to any other local user. Admittedly, the TS is an extraordinary local user in that it safely allows the local resources to be used by any application in the global environment.

Since the TS is a trusted local user and the application is not, the application must not be able to corrupt the TS and thus gain intimate access to the local environment. For example, achieving this security may be helped by the division of the TS into a TS and a TE, as described in section 2.7. The local environment then may be secured at the boundary between the TS and the TE. Unlike the TS, but like the application, the TE is not a trusted local user. The TE then can work closely with the application, as may be required for an efficient execution. The details of security thus are within the TS and TE, away from the application. Security thus is no different than any other execution element provided to an application by the TS.

Since a TS fully controls an application execution, it can account for all the resources used by the execution. For a computing application this would include the use of computer processors, memory, communication and software. In addition, when a variety of resources are available, the accounting machinery allows a TS to choose between resources in order to minimize the cost of an application execution [Spawn]. Such a minimization of cost is part of the execution performance optimization described in subsection 2.11.5.

In short, a present-day application executes using only local resources. The alternative provided by the TS allows an application also to use global resources.

The use of local and global resources requires local and global accounts respectively. A local account serves two purposes. It allows intimate access to the local resources and it can account for the resources used. A global account serves only the latter purpose. Thus a global account allows resources to be used and accounted for, without allowing intimate access to those resources.

The global environment allowed by the TS bears similarities to the world's present-day telecommunication system. There are also similarities to an electricity grid which allows a user to supply or receive electricity. Instead of electrical power, the TS allows a user to supply or receive computing power, for example.

On the supply side, TS security and accounting mechanisms allow an environment to make its resources available to any application in the global environment. Unfortunately there exists a security problem on the demand side, which is not implicitly solved by a TS: How to ensure that the owner or a local user of a local environment neither spy on nor corrupt an application? For example, while a TS allows the Red automobile company to sell computing resources to the Blue automobile company, how can Blue be assured that Red does not spy upon Blue applications? Just as encryption can secure static data, is there a mechanism to secure execution?

Since the execution cannot harm the environment and since all resources used can be accounted for, a TS allows an application execution to use an immense amount of resources. Take for example a very computation-intense application, for which the above demand side security problem is solved or irrelevant. Bandwidth willing and with a TS on each computer, such an application could execute on a million or more of the computers on the Internet [Internet Computing][SETI@home]. The market forces of demand and supply will determine the cost of such an execution [Spawn].

Additional arguments motivating the above computing model are described elsewhere [Client-Utility].

### *2.11.9 Parallel Computing*

As already described in subsection 2.11.2, at any given moment during its execution an application may have independent tasks in the pool. Independent tasks may execute in parallel.

Since a TE manages the execution of an individual task, it is simplest, in concept and often in practice, if the parallel execution of tasks uses multiple copies of TE. As illustrated in Figure 6, each task is thus passed by the TS to a copy of the TE execution. With the TS as the supervisor and each TE as a worker, Figure 6 thus is simply an illustration of the age-old supervisor-worker model for parallel computing [168/E][Supervisor-Worker].

As an example, consider the execution of a classic application using a network of computers. Then the TE essentially may be a process running a copy of the classic application. Such a TE process then runs on each computer. A classic application easily can act effectively as a TE. For example, the `READ(A)` and `WRITE(B)` of a classic application, as illustrated in Figure 7a), may communicate with the TS [Funnel]. The actual execution of a task thus occurs within a TE process. As is generally the case, there thus is very little overhead associated with a task. A task thus is lightweight. As a process, the TE is heavyweight, but its initial overhead is incurred infrequently.

The above example also demonstrates that the age-old process model can exist within TSIA. As a low level detail, the process model is hidden from the application definition, as is required for an application definition free of the details of the execution.

Having a TS provide parallel computing is relatively well understood. Probably more TS have addressed parallel computing than any other execution element. This is true both for general TS [Cilk-NOW][GLU][Jade][Linda-Piranha][Mentat][SCHEDULE], as well as for the rudimentary TS of the classic application [DBC][DNA][Funnel][Supervisor-Worker].

The characteristics of the execution situation determine the maximum parallelism of an application. The TS may continually determine and adapt to this maximum throughout the execution. This is part of the dynamic performance optimization described in subsection 2.11.5.

### *2.11.10 Distributed Computing*

Distributed computing refers to an application execution which is distributed across resources.

An application may require distributed computing for a variety of reasons. Three such reasons follow. Firstly, an application may require a variety of resources. For example, resources such as graphics devices, sensors and robot arms may be required for the inter-action between the application and its physical environment. Secondly, the application may be geographically distributed. Thirdly, an execution element such as parallel computing or reliability may require the application to be distributed. For example, consider a classic application running on a network of computers, as described in subsection 2.11.9.

At least in concept, if not always in practice, a TE may be associated with each of the resources. Then a distributed execution uses multiple TE, as illustrated in Figure 6.

For a distributed application, the TS typically also is distributed across the resources used by the application. On each resource, in addition to the TE, there may be a local part of the TS, managing the execution on that resource. It is assumed here for simplicity that one of the local TS also is the master TS, coordinating the entire application execution. Obviously if required, also this coordination can be distributed. For example, an application executing across a wide area network (WAN) might require a hierarchy of TS masters [DNA]. Though no-frills distributed computing can be performed with only a TS master and without the local TS on the other computing resources, many of the execution elements, as described below, require the local TS. In short, the TS shown in Figure 6 is not restricted to a single computing resource. Instead the TS may be active across all computing resources involved in the execution of the application.

As for other execution elements, TSIA transparently provides an application with a distributed execution [Cilk-NOW][DNA][Funnel][GLU][Linda-Piranha][Mentat]. The word transparent is just another way of saying the application definition remains free of the execution details.

In apparent contradiction to the above statement of this thesis, it has been argued in a report that it is not possible to transparently provide an application with a distributed execution [Distributed Computing]. This apparent contradiction between the report and this thesis has a very simple resolution.

The essence of the report convincingly and clearly argues that the system controlling the execution of the application must deal with the details of the distributed execution and thus cannot enjoy a transparent distributed execution. This thesis is in full agreement with this result of the report.

The report and this thesis only diverge because the report assumes that the system controlling the execution is the application itself. This assumption thus leads the report to conclude that an application cannot enjoy a transparent distributed execution. In contrast to the assumption, a TS is the system controlling the application execution in TSIA, as introduced in section 2.2. Thus the TS can deal with the details of the distributed execution, in a fashion transparent to the application. The report and this thesis thus are not in contradiction.

The resolution of the contradiction also can be explained in other words. If the report is to be accepted as is, then within the context of the report, the TS should be assumed to be part of the application, since the TS controls the execution of the application.

According to the report, a transparent distributed execution does not allow an application to control its own execution. Any system in which an application controls its own execution thus is unable to provide a transparent distributed execution. Research and experience corroborate this result [Linda].

One may speculate that the result of the report and the above discussion applies not only to distributed computing, but also to other execution elements. In other words, in addition to distributed computing, there may be other execution elements which only can be transparently provided to an application if the application does not control its own execution. Such speculation is corroborated by this thesis.

### *2.11.11 Heterogeneous Computing*

As mentioned in subsection 2.11.10, an application execution may require a variety of resources. This variety is a part of heterogeneous computing. In addition to any intrinsic variety of resources required by an application execution, extrinsic heterogeneous computing also may be introduced to an execution when the underlying environment consists of a variety of resources. For example, intrinsic and extrinsic heterogeneity is frequent when a network of computers is used to execute an application [Heterogeneous]. In general, distributed computing frequently implies extrinsic heterogeneity.

Extrinsic heterogeneity also is common for an interactive application. For example, in an international environment, an individual user may choose the language for the interaction. Another example arises when the interaction may occur via a keyboard and screen, via a microphone and speaker or via other mechanisms. These are examples of extrinsic heterogeneity since an application generally has no intrinsic interest in a user's particular choice of language or of interaction mechanism.

Only some of the variety introduced by intrinsic heterogeneity is relevant to the application definition. The remaining variety is irrelevant, as is the variety introduced by extrinsic heterogeneity. A transparent execution element is one whose details are hidden from the application definition. Transparency emphasizes the relevant by hiding the irrelevant. Thus, heterogeneous computing is transparent when the irrelevant variety is hidden from the application definition.

As described below, the feasibility of transparent heterogeneous computing by a TS essentially has two parts: an internal feasibility and an external feasibility, as introduced in section 2.10.

The internal feasibility of transparent heterogeneous computing is implicit to a TS. As described at the very beginning of this section, TSIA moves the execution details of an application into a TS. Thus a TS implicitly hides from the application any irrelevant variety of execution details introduced by heterogeneity. For example, heterogeneity may introduce a variety of tools controlling access to the resources [LSF]. Similarly, heterogeneity may result in a variety of communication mechanisms between resources [Active Messages][MPI]. Since such details are buried within the TS; any variety among these details is implicitly hidden from the application. Porting a TS to use a new variety of details is of course as difficult and painful as porting any other application full of execution details, but this is of no concern for an application using the TS.

The external feasibility of transparent heterogeneous computing must ensure the compatibility of the items of any given task. For example, the instruction of the task has to be suitable for the computer processor used to execute the task [Slim Binaries]. Similarly, the format of data has to be compatible with that expected by the instruction. Such data may be as simple as a four byte floating point number or as complex as an arbitrarily large and sophisticated data structure. For example, a computer routine unable to answer a question may defer to a human. The format of the question for the computer routine may or may not be a format suitable for a human [DSL].

If the items of a task are incompatible, compatibility may be achieved by converting some of the items. Such a conversion changes the format of an item, but maintains its essence. A conversion thus generally is reversible, though some conversions are not perfectly reversible. For example, imperfection can be introduced when converting between floating point number formats or when converting between human languages.

In general, it is simple for a TS to recognize that the items of a tasks are incompatible and to coordinate the required conversions. Depending on factors such as the type of the item and on the description of the item available to the TS, the TS may be able to automatically convert the item. Otherwise the TS may use a conversion supplied by the application.

As part of the performance optimization described in subsection 2.11.5, a TS generally minimizes the conversions required by an application execution. For example, the total amount of conversion can be adjusted by choosing the default format of each item. Minimizing conversions also is important when the conversions are imperfect.

#### *2.11.12 The State of the Application Execution*

Because a TS requires an application to execute in terms of tasks, this thesis encourages and assumes a very simple definition of the state of the application execution. As described in section 2.4, the task is the unit of execution. The state of an application execution thus is entirely given by that moment's collection of tasks to be completed. As also described in section 2.4, the moment's collection of tasks is the pool of tasks managed by the TS. Among the tasks to be completed are those executing at that moment.

Each task consist of items. Thus the state of the application execution is simply a set of items. Each item of the state belongs to a task to be completed. The state may include outs of completed tasks, but only if they are ins for tasks to be completed. The state otherwise does not involve tasks which have completed nor does it involve their items.

A recording of the state of the application execution is known as a checkpoint of that particular moment in the execution. Transparent to the application definition, a TS can record a checkpoint. As described below, the feasibility of a transparent checkpoint by a TS essentially has two parts: an internal feasibility and an external feasibility, as introduced in section 2.10.

The internal feasibility refers to the description of the task pool. This describes which item belong to which task and includes all the tasks belonging to the application execution at that moment. Since it manages the task pool, the TS can provide such a description.

The external feasibility refers to the description of the individual items associated with the task pool. For example, the description of an item of data is the value of that data. Not every item may allow a description or require a description in the checkpoint. For example, the description of an instruction generally is available elsewhere. While for simple items such as data, the TS may be able to record a description, other items may require a procedure supplied by the application.

In a TS, recording a checkpoint is orthogonal to other execution elements. For example, the checkpoint is portable across different types of computers, because of the arguments for heterogeneous computing presented in subsection 2.11.11.

In this presentation, a checkpoint simply is recorded to a so-called journal. If the history of checkpoints is kept in the journal, then the history of the execution may be followed. If sufficient changes of state are recorded in the journal, then the exact progress of the execution can be followed directly or can be recreated: task by task and item by item. For example, with a sufficient frequency of checkpoints, all the tasks which executed on a particular computer processor are known. In practice, the frequency of checkpoints and the number that are kept in the journal will depend on the execution situation.

Checkpoints and the journal are important for a number of execution elements. For example, the journal allows for the analysis of the performance of an application execution. As described in subsections 2.11.13 and 2.11.15, respectively, the journal is important for the reliability of the application execution and for debugging the application.

In the simple definition of execution state, the TS has no interest in any state inside any task. There thus is no need for a traditional checkpoint of any task. The TS could use facilities to checkpoint a task, but at considerable cost and complexity for relatively little return. For example, since a checkpoint of a task is just like that for a process, it generally is not portable to a different computer. Such a checkpoint uses a machine dependent encoding of the data, includes the state of the processor registers and involves other low level details irrelevant to the application definition [Checkpointing]. Nevertheless, a task with an extremely long execution might desire a checkpoint. Rather than checkpoint the task, this thesis encourages breaking up such a long task into a series of shorter tasks.

In summary, since a TS controls the application execution, since the application executes in terms of tasks and since a task consists of items, the TS can efficiently and completely record the state and progress of the execution.

#### *2.11.13 Reliable Computing*

A reliable application rarely fails. Reliability is the ability of an application execution to meet its performance requirements despite the fault of any part involved in the execution. A reliable application thus is said to be fault tolerant.

The ability of TSIA to provide an application with transparent reliability does not necessarily require new tools and techniques for reliability. TSIA simply may use existing tools and techniques for reliability [Reliability]. For TSIA, reliability thus is no different from any other execution element. As described in section 2.10, TSIA makes an execution element transparent to an application definition by moving the associated execution details from the definition into the TS. Existing techniques and technology for reliability thus may be used by a TS.

Since the two topics are related, the discussion of reliability in this section is similar to the discussion in subsection 2.11.12 of checkpoints and the state of the application execution. For example, following a computer crash, the reliability of some applications is sufficient if on the rebooted or on another computer the application immediately continues from the last checkpoint. Obviously if an application uses multiple computers, this continuation applies only to the tasks of the application on the crashed computer, since the other tasks of the application are unaffected. The TS provides this reliability at no cost to the application execution.

While the TS can provide the simple reliability described above by continuing the application execution, this subsection proceeds to applications which require higher reliability, albeit at a cost to the application execution. Redundancy allows an application execution to survive the fault of any part involved in the execution [Reliability].

As described below, the feasibility of transparent reliability by a TS essentially has two parts: an internal feasibility and an external feasibility as introduced in section 2.10.

The internal feasibility refers to the reliability of the task pool. As for checkpoints, this concerns which item belong to which task and includes all the tasks belonging to the application execution at that moment. Since it manages the task pool, the TS can provide such reliability. For example, in a distributed application, the TS typically also is distributed across the resources used by the application, as described in subsection 2.11.10. A distributed TS can incorporate redundancy, thus allowing it to ensure the reliability of the task pool. Such reliability may use existing tools and techniques [ISIS].

The external feasibility refers to the reliability of each item of each task in the task pool. The TS is in an excellent position to provide or to help provide redundancy and thus reliability to any item of the application. This ability of the TS is very much a result of task autonomy and the resulting referential transparency of the task. Redundancy of an item implies the redundancy of its task. Referential transparency allows the redundancy of a task [Functional Fault Tolerance].

For example, computer hardware as an item can be made reliable using modular redundancy [Reliability]. Task autonomy allows a task to execute on multiple replicas of a hardware module. A voter mechanism can compare the outs from the replicas and determine the correct out using, for example, majority vote. The multiple execution of a task on the replicas and the voter mechanism can be managed by the TS with more or less guidance from the application.

Similar to hardware, software as an item can be made reliable using N-version programming [Reliability]. Each task executes using multiple differing versions of the instruction. Again a voter mechanism can determine the correct out.

The symmetry of items within a task, as seen by a TS, allows reliability through redundancy to easily extend to many types of items. For example, a nuclear power plant application could use a TS to gather confirmation from at least three human operators before a safety procedure is overridden.

High reliability via redundancy adds cost to the execution of an application. A TS can minimize this cost. As part of the performance optimization described in subsection 2.11.5, the TS can determine and use a minimal redundancy which achieves the required level of reliability.

#### *2.11.14 Adaptive Computing and Dynamic Computing*

Adaptive computing allows a rapid change in the resources used for an application execution [Cilk-NOW][GLU][Linda-Piranha]. Dynamic computing allows a rapid change in the application definition used for an application execution [Dynamic Computing]. A rapid change implies that the effort required for the change is proportional to the size of the change and thus not proportional to the size of the application. Such a change in the resources or in the application definition even may occur during the application execution. An example of adaptive computing is described in subsection 2.11.13. When a computer crashes, adaptation provides reliability by continuing the application on another computer.

The TS can provide adaptive computing and dynamic computing since each simply corresponds to changing the items of tasks. Because of the symmetry of the items of a task, as seen by the TS, adaptive computing and dynamic computing are very similar. The TS does not care much if an item has its origins in the application definition or in the resources of the environment. Since the TS coordinates the assembly of items for a task to execute, the TS can allow the items to be changed. Task autonomy assures that changing the item of a task has no hidden impact on any other task and thus on the remainder of the application. The change of an item may rely on existing technology. For example, the TS may use dynamic linking to access new instructions for tasks [Slim Binaries].

A number of different situations result in adaptive computing. Three examples follow. Firstly, the owner of an application may inform the TS of new performance or cost boundaries for the execution. The TS then would adjust the use of resources appropriately. Secondly and as described in subsection 2.11.4, the TS may vary the use of resources in order to dynamically determine and achieve the optimal performance of the application execution. Thirdly, there may be a change in the environment supplying the resources for the application execution. For example, such changes in the environment are frequent if the application execution is making opportunistic use of otherwise idle resources [Cilk-NOW][Funnel][Internet Computing][Linda-Piranha][SETI@home][Spawn]. Thus if such an application executes using a number of computers in parallel, that number may vary.

Dynamic computing is valuable in a variety of situations. Four examples follow. Firstly, the ability to change the application definition during execution can help satisfy the reliability requirements of an application [Dynamic Computing]. Such a change to the application definition may correct an error or introduce a new feature. Secondly, dynamic computing can be very valuable for debugging the application definition, as described in subsection 2.11.15. For example, dynamic computing allows a developer to easily introduce to the execution a variety of scenarios to be survived by the application. Thirdly, dynamic computing is a great method to configure or extend an application at execution [Dynamic Computing][Slim Binaries]. The resulting versions of the application, individually customized for each situation, can have great advantages over an alternative monolithic version which would attempt to satisfy all situations. Fourthly and as described in subsection 2.11.4, the ability to periodically recompile and relink an instruction of the application, driven by the measured performance of the execution, can move the execution towards an optimal execution [Paradyn][Slim Binaries].

Though it primarily addresses the application execution, TS also benefits the application definition. As introduced in section 1.1, the separation of the application definition from the execution probably is the greatest benefit provided by TS to an application definition. In addition to this indirect benefit, the TS can directly benefit the application definition. Dynamic computing as described above is one example. Another example is the debugging environment possible in the TS, as described in the next subsection.

### *2.11.15 Debugging*

TSIA allows for a debugger which incorporates many sophisticated techniques including those of a variety of powerful debuggers. These techniques allow the debugging of a TSIA application to go far beyond that allowed by a traditional debugger. Such debugging traditionally involves repeatedly stopping the application, examining its state, and then either continuing or restarting the application in order to stop at an earlier point in the execution. The TSIA debugger can incorporate productive alternatives to such traditional debugging. For example, examining the state of an application execution can often be automated [Guard]. As another example, returning to an earlier point in the execution does not require restarting the application [Immediacy]. Before describing how the TSIA debugger goes beyond traditional debuggers, this subsection first shows that a TSIA debugger is at least as good as a traditional debugger.

This subsection addresses debugging the application definition, not the application execution. The application execution is provided by the TS and in this subsection is assumed to be correct. Debugging the application execution implies debugging the TS and this is orthogonal to debugging the application definition. For example, debugging the performance of the application execution is mentioned in subsection 2.11.5. In general, since the application execution and its execution elements are transparent to the application definition, the application execution also is transparent to debugging the application definition and thus also to the TSIA debugger.

With the execution elements out of the way, the TSIA debugger meets the first requirement to act like a traditional debugger. Execution elements are of no concern to such a traditional debugger since it helps debug applications which require no execution elements. For the purposes of this subsection, such an application requiring no execution elements is known as a sequential application.

Avoiding the execution elements thus is the first property of TSIA which allow its debugger to act like a traditional debugger. As explained below, other primary properties include the journal, task autonomy and the incorporation of a traditional debugger.

The traditional execution of the routines of a sequential application generally uses a stack [Recursive Techniques][Stack]. The traditional debugger thus obtains a history of the execution from the stack. The execution by TSIA of the tasks of an application does not use a stack. Thus the TSIA debugger must obtain a history of the execution from elsewhere. As described in subsection 2.11.12, the history of the execution may be recorded as checkpoints to a journal. It is assumed here that a very detailed history is available.

Each routine of a sequential application executes in isolation and thus may be debugged in isolation. By the definition of a sequential application, during the execution of a routine, no other routine executes. Thus the executing routine is alone in modifying the state of the sequential application. In TSIA, task autonomy allows each task to execute in isolation and thus to be debugged in isolation.

Debugging a task in a TSIA application thus is very similar to the traditional debugging of a routine in a sequential application. When debugging a task or routine there are only two possibilities. Either the bug is or is not recognized from the information within the task or routine. If it is recognized, debugging thus is successfully completed. An example is when the bug is in the code of the routine or in the instruction of the task. If the bug is not recognized within the task or routine, then information is required from a previously executed task or routine. For a sequential application, the stack is used to travel the history of the execution. For a TSIA application, the journal is used. Thus in an iterative fashion, tasks or routines are examined until the bug is found.

The fact that the TSIA debugger at least can behave like a traditional debugger can be demonstrated by having the TSIA execute an originally sequential application. It is irrelevant to the debugging whether the new execution of the application introduces parallel computing or other execution elements. In this demonstration a bug eventually causes a task to fail. It is unimportant here whether additional tasks fail or whether the task fails by crashing, by failing an assertion or by entering an infinite loop. The TSIA debugger essentially can use a traditional debugger to present the failed task. For example, if the instruction of the task is written in Fortran, then a Fortran debugger may be used. A traditional debugger is sufficient since the execution within a task is essentially the same as the exe-cution of a routine in a a sequential application. The state of the task at the failure thus may be examined.

Having shown above that the TSIA debugger can be at least as good as a traditional debugger, the TSIA debugger then can go beyond the traditional debugger. Since the items of the tasks are available from the journal, the TSIA debugger can launch the traditional debugger with the task at the start of its execution. The TSIA debugger thus effectively can move backward through the application execution [Immediacy]. The user then can determine if the combination of items defining the task is correct and can follow the execution of the task. If the task definition is incorrect, then the journal allows the TSIA debugger to return the user to the 'parent' routine which defined the failed task. The 'parent' routine then can be debugged in the same fashion. Similarly, if one of the items of the failed task was incorrect and the item was an out of a 'previous' task, then the user may return to this 'previous' task.

The classic application described in section 2.9 can provide a simple example of the merits of the TSIA debugger. The instruction of a classic application is assumed to fail for a particular in. Since a task has failed, the identity of the in is known to the TSIA. The TSIA debugger thus may launch a traditional debugger with the instruction and the particular in. The user thus immediately may begin to determine the cause of the bug. For comparison, consider achieving the same effect in the usual process model. The user first would have to identify the particular in and then to halt the execution of the application in the debugger at the beginning of the processing of the in.

Dynamic computing, as described in subsection 2.11.14, introduces considerable power to debugging an application in TSIA. For example, assume that the TSIA is configured to suspend an application if any of its tasks fail. In fact, the TSIA could continue executing any tasks which do not depend on any failed task. If a task fails, the debugger is used to identify the bug, as described above. If the nature of the bug is such that the failure is constrained to the failed task, then dynamic computing allows the application to be repaired during the execution. For example, if the bug is in the instruction of the failed task then the old version of the instruction can be thrown out and a new version installed. Recall that task autonomy allows an item of task easily to be replaced since no hidden dependencies are involved. Using the information from the journal, the previously failed task may re-execute and the suspended application may continue its execution.

Much of software development is not the creation of new applications, but instead is the maintenance and improvement of existing applications. Relative debugging automates the comparison between the result of a new and the result of an old version of an application [Guard]. A change in an application thus easily can be followed in terms of the change in the results. Relative debugging thus allows for the rapid change of an application. As for adaptive computing and dynamic computing, a rapid change implies that the effort required for the change is proportional to the size of the change and thus not proportional to the size of the application nor to the size of its results.

The TS can provide excellent relative debugging. For example, imagine that a new version of an instruction is to be introduced to an application. Other than being faster, the new instruction is to provide the same results as the old instruction. Task autonomy allows the new instruction to be thoroughly tested using relative debugging. A task using the new instruction runs alongside a task using the old instruction. The results of the old instruction are used by the application. The results of the new instruction are not used by the application and instead just are compared to those of the old instruction. Any differences are reported. Once relative debugging has thoroughly validated it, the new instruction can replace the old instruction.

### *2.11.16 Reactive Computing*

After parallel computing, reactive computing probably has been to date the most popular execution element for the task model and similar models. Reactive computing broadly includes real-time computing, concurrent computing and other related areas [SISA]. Reactive computing thus ranges from the hard real-time computing of a robot, through to the concurrent computing of an operating system.

The soft-instruction software architecture (SISA) is an implementation of the task model for reactive computing. Since SISA is well described elsewhere [SISA], this presentation does not repeat demonstrating the use of the task model for reactive computing. The description of SISA also presents the long history of the soft-instruction and thus of the task model.

In order to help relate the description [SISA] of SISA to this presentation, the matching names follow. The task and the TS in TSIA are known as the soft-instruction and the dispatcher in SISA, respectively. An application in TSIA roughly corresponds to a unique soft-instruction stream. The dispatcher generally deals with multiple such streams, each executing a common reactive control table logic.

Not all applications require general purpose reactive computing such as that provided by SISA. For example, a classic application is satisfied by much simpler support for reactive computing, as for other execution elements. An excellent introduction to the task model for reactive computing thus is given by the classic applications [Packet Filter][RTU][SUMO], already mentioned in section 2.9.

As introduced in section 2.1, the task model is similar to that of a windowing system based on events dispatched to callbacks [X]. In fact there are many previous event-driven systems for reactive computing. In such a system, an event is dispatched to an entity known by many different names including callback [X], event handler [Threads], upcall [Upcall] and passive input or output [Passive]. In such a system, the handling of an event is similar to a task. It is no more than similar since the handling of an event generally does not obey what would correspond to task autonomy. Exceptions are systems for classic applications, as mentioned above. There the definition of the classic application directly leads to task autonomy. Because of their similarity to the task model, previous event-driven systems are valuable precursors to using the task model for reactive computing. The techniques of the precursors can be implemented with task autonomy, thus achieving reactive computing within the task model. In fact, some event-driven systems already seem to approach the task model [Escaping the Event Loop].

## *2.12 Summary*

In the task model, an application execution explicitly consists of the execution of an arbitrary number and variety of tasks. The execution is managed by a task system (TS). During its execution, a task is independent of all other tasks. This task autonomy fundamentally simplifies the management of the execution. A task is defined by the items making up the task. An item of one task also may be an item of another task and thus may introduce a dependency between tasks. The items of a task include everything required for its execution and thus range from data through to computer processors. For each task, the application definition defines only those items relevant to the application definition; the TS defines the remaining items. A task executes once the TS has assembled all its items.

The task model thus results in a simple application execution. As demonstrated by many projects, the simplicity allows a TS to provide the application execution with execution elements including performant, secure, accountable, parallel, distributed, heterogeneous, reliable, adaptive, dynamic and reactive computing.

Any application which executes in terms of tasks thus may be provided with a variety of execution elements. Not the application, but the TS exerts the effort required for the execution elements. The application definition thus can be free of the details of the execution.

The next chapter demonstrates that a structured application definition can execute in terms of tasks.

# 3 Application Definition

This chapter demonstrates that an *item architecture* (IA) allows any one of a large variety of applications to be defined such that the application executes in terms of tasks. Furthermore, the IA allows for a structured application definition. As demonstrated in the previous chapter, an application which executes in terms of tasks can have a definition free of execution details and can be provided by a task system (TS) with a variety of execution elements. In combination, the TS and the IA of TSIA thus support the execution and definition requirements of a large variety of applications.

This thesis thus first solves the execution requirements of an application. The price of the TS solution is the constraint that an application execute in terms of tasks. Then, within this constraint, the IA solves the definition requirements of an application. In hindsight, the resulting application definition can be so convenient that the constraint of tasks is not a curse, but a blessing.

Tasks allow for a clean interface between the IA and the TS and thus between this and the previous chapter. The IA submits tasks to the task pool; the TS executes tasks from the task pool. The description of the TS in the previous chapter thus assumes the origin of tasks via the IA from the application definition. Similarly, the description of the IA in this chapter assumes the execution of tasks via the TS for the application execution.

This chapter introduces the IA in a bottom-up style. In other words, the description of each part of the IA generally relies only on parts described previously in the chapter. The bottom-up style thus helps argue the feasibility of the IA. Unfortunately, the ultimate aim of the IA - support for a structured application definition - thus is first addressed in section 3.17.

## *3.1 An Item Architecture*

### *3.1.1 The Item*

As shown implicitly in this chapter, the item is the fundamental unit of application definition. In contrast and as identified in section 2.5, the task is the fundamental unit of application execution. The task is fundamental since there is no unit of execution simpler than a task. Similarly, the item is fundamental since there is no unit of definition simpler than an item.

It is vital to not mistakenly identify the task as the fundamental unit of definition. Only a small variety of applications, such as the classic applications, conveniently may be defined if the task is the unit of definition.

As described in section 2.6, a task is made up of items. With the item as the unit of definition, any one of a large variety of applications may be defined such that the application executes in terms of tasks. If the task is a constraint for the application definition, then the item is the degree of freedom allowing the constraint to be satisfied. The focus on the item identifies techniques which allow an application to execute in terms of tasks. Delegation, introduced in subsection 3.2.2, is a prime example of such a technique.

Indirectly, the TS thus demands that an application be defined in terms of items. This is satisfying, since the aims of an application are expressed in terms of items: using ins, an application produces outs. Tasks are only the means to accomplish the aims. The IA and the application definition thus pursue items, not tasks.

### 3.1.2 *Introducing an Item Architecture*

*Definition elements* are the tools and techniques used to define an application. A definition element may be directly provided by a programming language. Among other alternatives, a definition element may be constructed using other definition elements. For example, iteration is a definition element used in many application definitions. Some programming languages directly provide for iteration via loop constructs such as the `do` of Fortran or the `for` of C. Alternatively and as demonstrated in subsection 3.3.2, iteration may be achieved by using recursion.

An *item architecture* (IA) implements definition elements.

**"In an item architecture, a definition element is defined in terms of items and executes in terms of tasks."**

An application definition using such definition elements thus is defined in terms of items and executes in terms of tasks.

This thesis primarily concerns the execution of an application, not the definition of an application. Thus this chapter has little intrinsic interest in the application definition. Instead, the interest in the application definition primarily is due to its role in the application execution.

This thesis does not introduce any new definition elements. Instead, this chapter merely demonstrates that a variety of definition elements may be defined in terms of items and execute in terms of tasks. Via these definition elements, a IA thus allows any one of a large variety of application definitions to execute in terms of tasks.

This presentation makes no claim that a IA can implement all possible definition elements. As explained below, a IA does not have to; the definition elements implemented by a IA may be used in conjunction with traditional implementations of these and other definition elements. Nonetheless, a substantial variety of definition elements is implemented by the IA of this presentation. The definition elements range from iteration, through to the hierarchical use of routines, through to the strict or non-strict evaluation of arguments to a routine, through to easy and effective computing using arbitrarily large arrays.

One of the most important features of a IA is that it extends, not replaces, definition elements. In other words, it provides an alternative, not a replacement, implementation of a definition element. All implementations of all definition elements may be used to define an application. For example, an application may define some instances of iteration in terms of a loop of C or Fortran; other instances may use recursion. Thus an application using a IA in part may be defined in Fortran and have those parts compiled using a traditional Fortran compiler.

Since a IA provides an alternative implementation of definition elements, this chapter frequently compares a IA implementation to a traditional implementation. Equivalently, a IA execution is compared to a traditional execution. For example, a IA execution often is compared to a Fortran execution. The traditional implementation of a definition element is assumed to be familiar and the primary purpose of the comparison is to explain the IA implementation. Though not mentioned explicitly in the comparisons, a IA execution implicitly can include any of the execution elements presented in chapter 2. The straightforward availability of the execution elements of course is the motivation for the IA execution.

Since a IA implements of a definition element, this thesis concerns the semantics or meaning of a definition element, not its syntax. As explained in section 3.5, all syntax largely is equivalent for the purposes of this thesis. For example, seen from this perspective, there is little difference between the C and the Fortran programming languages. Similarly, latter sections of this chapter compare a Fortran execution to a IA execution for a routine written in Fortran. The fact that the routine happens to be written in Fortran, instead of in another language, is by and large irrelevant to the concerns of this thesis.

### *3.1.3 Introducing a Ia Language*

For a number and variety of reasons, this presentation introduces an *item architecture language*. Three major reasons follow.

Firstly, a ia language can be restricted to only the definition elements which are implemented by a IA. The ia language then serves solely to express an application definition in terms of items and tasks. The ia language thus is simple since it serves no other purpose. For example, the ia language need not support computation. Pure and simple ia code is the result. The ia routines of an application thus easily may be identified with the IA of TSIA.

Secondly, a ia language can make explicit all information required by IA in order to implement a definition element in terms of items and tasks. For example, a ia language can distinguish between the ins, inouts and outs of each task.

Thirdly, the ia language allows much of an application to be defined using existing programming languages, compilers, and other implementations of definition elements. Thus for example, computation may be defined using existing languages such as C++ and Fortran.

An example alternative to the ia language would be to introduce the abilities of IA to an existing programming language [Cilk-NOW][Jade][Mentat].

As introduced above, the ia language cleanly extends current computing practice. As an extension, the ia language allows for simple demonstrations of how TSIA can be suitable for a large variety of applications. Thus in the examples of this presentation, an application definition often consists of two parts: a part using the ia language and a part using current computing practice. By taking current computing practice for granted, this presentation can focus on the ia language and thus on the IA.

In this chapter, the comparison of a IA implementation to a traditional implementation of a definition element often simply refers to ia code and traditional code. The latter usually is imperative code such as Fortran code. In the comparison, the IA implementation of ia code and the traditional imperative implementation of imperative code is implicitly assumed.

### *3.1.4 The Classic Application*

The classic application, introduced in section 2.8, allows for a simple introduction to the ia language of this thesis. A ia program for the classic application is presented in Figure 11a). The ad hoc syntax of the ia language is introduced in Figure 11b). The syntax is similar to that of the C programming language [C]. Since the classic application has a straightforward execution in terms of tasks, it is satisfying that its ia program almost is as simple as its pseudocode program in Figure 7a). In addition to the classic application, a variety of other applications are used in this chapter to describe the IA and the ia language of this thesis.

### *3.1.5 The Abilities and the Feasibility of a IA*

The purpose of this chapter is demonstrate some of the abilities of a IA. Furthermore, the abilities are chosen to be easily feasible. Therefore, the reader is asked to focus on the abilities and the feasibility of the example applications presented in this chapter. This requires understanding the semantics of the ia language and its implementation by a IA. In other words, the meaning of the ia code of each example is to be understood.

In contrast to the semantics, the syntax of the ia language presented here is unimportant and thus is best ignored to the greatest extent possible.

Many different ia languages are possible. The differences are not just in syntax, but more importantly in abilities and in semantics. The possibility of different ia languages is like the existence of many different imperative programming languages. The ia language presented in this thesis is just one of the many different possible ia languages. The ia language presented here is neither definitive nor complete.

A part of the ability of the IA is its suitability for an application with a large and complex definition. While the IA is presented here using applications with a small and simple definition, the abilities of the IA scale to applications with a large and complex definition. For example, an application execution in TSIA does not require a global analysis of the application definition. Such a global analysis does not scale well as application definitions increase in size or complexity. Instead, the IA allows each routine of the application definition to be compiled separately, as in a traditional C or Fortran environment. Such separate compilation scales well since the compilation of each routine essentially is independent of the size or complexity of the entire application definition.

### *3.1.6 The Purpose of a IA*

The purpose of a IA and of a ia language is to define a task in terms of items. A IA places no restrictions on the items. For example, an instruction may correspond to a single machine instruction or to a multi-million line program in an arbitrary programming language.

For each item of each task, if the item is an in then the item must have an origin, if the item is an out then the item must have a destination.

In the ia program for the classic application presented in Figure 11a), there are three origins for items. Firstly, the item `produce` is declared to be available as an instruction. Secondly, the file `"a.dat"` is declared to be available as an array of items. Each element of the array is an element of type `bytes`, which is simply a sequence of arbitrarily many bytes. Thirdly, the item `b[k]` is made available by the execution of the task `produce(a[k];;b[k])`.

There are two destinations for items in the ia program for the classic application. Firstly, the file `"b.dat"` is declared as a destination for an array of items, with each element of type `bytes`. Secondly, the item `a[k]` is a destination provided by the execution of the task `produce(a[k];;b[k])`.

A task thus is the origin of each of its out, each of which requires a destination. Similarly, a task thus is the destination of each of its in, each of which requires an origin. Obviously, an out of one task thus may be an in of another task. A task thus may depend on another task.

The above use of the file `"a.dat"` and similarly `"b.dat"` is feasible. For example, such a file of items can be accompanied by a file called an item directory. The item directory contains `N` entries, sequentially ordered from the first to the last item. Each entry contains two numbers. The first number is the position of the first byte of the item in the file of items. The second is the position of the last byte. Thus known is the number `N` of items in the item file, the length of each item as well as the storage order of the items in the item file.

### *3.1.7 The Application Execution Differs from its Definition*

The motivation of a IA and a ia language is to allow for the difference between the application definition and its execution. As introduced already in section 1.1, an application is given by its definition, while the execution merely puts the definition into action. In this thesis, satisfying both the execution and the definition of an application requires the freedom that the application execution can differ from its definition.

The difference between the application definition and its execution has many properties. Three examples follow. Firstly, execution elements are part of the execution, but not the definition. Secondly, in the definition a task occurs exactly once; in the execution the task may execute an arbitrary number of times. Thirdly, because the definition can be independent of the execution, the definition can be determinate, even if the execution is not. The third property is discussed extensively in subsection 3.1.8.

A ia language allows for the difference between the application definition and its execution because ia code only specifies what is to be achieved by the application execution, not how it is to be achieved.

For example, the IA implementation of the ia language may be contrasted with the implementation an imperative language such as C or Fortran. Imperative code specifies what is to be achieved by specifying how it is to be achieved. In other words, imperative code defines an application by describing its execution. Modulo minor differences introduced by optimization, for imperative code there thus is little difference between the definition and its execution. For example, if a routine is called once in imperative code, the routine is called exactly once in the execution of the code. The execution of a routine in imperative code thus is unlike the execution of a task in ia code.

The above comparison between a ia language and an imperative language may be described as follows. In a ia language, an application explicitly defines the dependencies between items. In exchange for obeying the dependencies, TSIA controls the execution. In contrast, in an imperative language, the application definition controls the execution. The dependencies between items thus implicitly are buried within the definition. A ia language thus is dependence-driven, making dependencies explicit and control implicit. Vice versa, an imperative language is control-driven, making control explicit and dependencies implicit.

### *3.1.8 Determinate and Indeterminate Definitions and Executions*

The results of an application execution are defined by the application definition. Thus if the results of the application execution are determinate, then the application definition is said to be determinate. Because an application definition can be independent of its execution, the definition can be determinate, even if the execution is not. This independence is very valuable since a determinate application definition has many advantages over an indeterminate one. Vice versa, an indeterminate application execution has many advantages over an determinate one. For example, adaptive computing in TSIA allows a determinate application definition to execute on an indeterminate number of computers. Having thus introduced determinate and indeterminate application definitions and executions, the remainder of this section provides a more detailed discussion.

In IA an application definition is determinate if the outs of each of its tasks are determinate. Two alternative conditions allow an out of a task to be determinate. In the first condition, an out of a task is determinate if the ins of the task are determinate. Alternatively, in the second condition, an out of a task is determinate if the out is independent of any indeterminate nature of an in.

An example of the first condition is an application execution using an indeterminate number of identical computers. As described in subsection 2.11.14, such indeterminate or varying amounts of resources are a part of adaptive computing. Since the computers are identical, this item of each task is determinate and thus the outs of each task are determinate. Thus the application definition is determinate even though the application execution is indeterminate.

An example of the second condition is an application execution using an indeterminate type of computer. As described in subsection 2.11.11, such indeterminate or varying types of resources are a part of heterogeneous computing. The out of a task thus is determinate only if it is independent of the type of computer executing the task. If this independence holds for each out then the application definition is determinate even though the application execution is indeterminate.

In contrast, if the out of a task depends on the type of computer executing the task then the application definition is indeterminate due to the indeterminate application execution. The indeterminate type of computer for each such dependent task is a part of the indeterminate application definition. The determinate application definition only may be restored by making determinate the type of computer for each such dependent task. The particular type of computer for each such dependent task then is a part of the determinate application definition.

The above discussion may be roughly summarized as follows. The application definition includes everything it depends on. An application definition thus is determinate if everything it depends on is determinate. A determinate application definition thus allows everything that it does not depend on to be indeterminate. Thus as demonstrated above, because the definition can be independent of the execution, the definition can be determinate, even if the execution is not.

For example, since an application definition depends on the code within its instructions, the code is part of the application definition. Thus a determinate application definition requires the code to be determinate. More generally, if an application definition depends on data then the data is part of the application definition. Only if the data is determinate can the application definition be determinate.

A more refined example is an application definition which includes data input by a human user. Such data is indeterminate and thus so is the application definition. If the data is recorded it is made determinate. For example, by once recording and often replaying human input, an interactive application may be made determinate. A determinate application is much easier to develop, debug and test than an indeterminate one.

Similarly, a determinate application definition may be achieved for an application definition originally indeterminate due to its execution on indeterminate types of computers. As described above, if the out of some task depends on the type of computer executing the task then the application definition is indeterminate. For each of these tasks, the type of computer can be included in checkpoints recorded by TSIA and described in subsection 2.11.12. If for each task the recorded type replaces the original indeterminate type of computer, then a determinate application definition can result. Thus if the original execution crashes, the now determinate application definition can be executed again and can be very valuable to a developer trying to debug the crash.

An application definition may well be entirely determinate. In contrast, an indeterminate application definition is unlikely to be entirely indeterminate. Instead, for an indeterminate application definition probably only part is indeterminate and the remainder is determinate. Nonetheless, for the simplicity of this presentation, an application definition is deemed either determinate or indeterminate. Similarly, for a given indeterminate out, it is beyond the scope of this presentation to pursue how this indeterminate property propagates through the application definition. For example, an indeterminate out might not even ultimately cause indeterminate results and thus an indeterminate application definition.

A determinate application definition has tremendous advantages over an indeterminate one. An advantage for application development is that the result of changing a determinate application definition is clear. In contrast, the result of changing an indeterminate application definition is not clear since the results are indeterminate. An advantage for IA is that a determinate application definition never requires the efforts described above to make determinate an initially indeterminate application definition.

An indeterminate application definition thus generally is avoided. Unfortunately, due to the nature of the application definition or execution, an indeterminate application definition may be unavoidable. An example of such an application definition is one which includes data input by a human. An example of such an application execution is one for which the resources or the performance require the execution flexibility gained by an indeterminate application definition. Thus even if an application definition depends on the type of computer used for the execution, the execution may use indeterminate types of computers if determinate types are unavailable.

In contrast to the desired determinate application definition, an indeterminate application execution has tremendous advantages over a determinate application execution. As described in subsection 3.1.9, the indeterminate application execution of TSIA may be compared to the determinate application execution of an imperative language such as C or Fortran. The indeterminate application execution allows TSIA to satisfy the execution elements required by the application.

Making determinate a previously indeterminate part of the execution thus generally is avoided. Unfortunately this may be unavoidable. An example is described above. There an application definition depends on the type of computer used for the execution. While the original execution may use indeterminate types, a second execution may use the recorded and now determinate types in order to assist debugging. By making determinate this part of the execution, the second execution is not as flexible as the first. This trade-off is acceptable if the aim of the first execution is high performance, while the aim of the second execution is to debug the application definition.

In short, TSIA attempts to achieve as an application definition as determinate as possible, while preserving a sufficiently indeterminate application execution. At the boundary between the application definition and execution, a determinate part of the application definition occasionally may have to be sacrificed in order to satisfy the execution. Vice versa, an indeterminate part of the application execution may have to be sacrificed in order to satisfy the definition. In addition to those of this subsection, subsection 3.7.5.2 and section 3.23 describe further examples of this trade-off.

### *3.1.9 The Application Definition*

In order to understand the application definition, ia code may be treated as if the ia language were an imperative programming language such as Fortran or C. As described in the next subsection, such treatment of ia code is not completely appropriate in order to understand the application execution.

A major aspect of imperative languages shared by the ia language of this presentation is that a top to bottom order of tasks is used to describe the dependencies between tasks. For example, `a(;x;);b(;x;)` thus must execute in the order shown.

The imperative treatment of ia code for the application definition is both accurate and convenient. An algorithm may be understood, coded and tested using an imperative programming language, before being implemented using a IA. For example, the ia program for the classic application presented in Figure 11a) easily is understood if treated as an imperative program. Essentially only changes in the syntax of the program would be required in order to compile the ia program for the classic application as a C or Fortran program.

As introduced in subsection 3.1.7, imperative code defines an application by describing its execution. Such a definition can be compact and exact. When ia code is treated as imperative code, it is to benefit from the good ability of imperative code to define an application. In an imperative implementation, the execution implicit in the application definition is the only possible execution. In contrast, in a IA implementation other executions easily are possible. For ia code, the execution implicit in the imperative application definition thus is just one of many possible executions. The IA thus allows the application execution to differ from the application definition.

### *3.1.10 The Application Execution*

A IA execution submits tasks to the task pool. The IA thus satisfies for the application definition the constraint of tasks made by the TS. A simple example is the execution of the ia program for the classic application presented in Figure 11a). The execution submits `N` tasks, illustrated in Figure 7b), to the task pool.

As a task, a ia routine has no communication with any of the *child tasks* it submits to the task pool. The requirements of task autonomy thus are met. After submitting its child tasks into the task pool, the *parent task* thus can complete execution and exit before any of its child tasks even begin execution. In fact, to understand the execution, the following usually is advisable:

**“Pretend that the parent task completes execution<br>before any of its child tasks begin execution.”**

This order is not required by TSIA, but may help to initially understand the execution.

The execution of ia code as a task may be compared to the synchronous execution of imperative code. When a child routine is called, the parent imperative code blocks until the child routine completes and returns. In contrast, the execution of a child task is beyond synchronous and even beyond asynchronous [Active Objects]; it is autonomous. As dictated by task autonomy, a task has no communication, and thus no synchronization, not even between a parent task and a child task.

Ia code can be compiled. Thus the dependencies between tasks can be recognized at compile time. This and other information can be forwarded to the task pool for the TS. Thus a compiler already can provide the TS with initial information on the variety of executions allowed by the definition.

To demonstrate the feasibility of compiling ia code, the ia compiler for the ia language could translate ia code to C code which accesses the task pool via TS library calls or shared memory. Like any other imperative code, the C code is compiled by a C compiler.

The feasibility of translating ia code to C code may be demonstrated using the ia program for the classic application presented in Figure 11a). A translation is shown in Figure 12. In each iteration of the C `for` loop, a unique child task `produce(a[k] ;;b[k])` is submitted to the task pool. This probably is an inefficient way to submit `N` tasks to the task pool, since it presumably is more efficient for the task to submit and for the TS to receive and manage all `N` tasks at once. In any case, this example demonstrates that a translation of ia code to C code is feasible. Because it submits a single task at a time to the task pool, the translated application code of this example is very similar to application source code in Cilk-NOW [Cilk-NOW].

In addition to demonstrating its feasibility, the above example translation into C code highlights an important difference between the application definition and execution. As described in subsection 3.1.9, the application definition may be treated imperatively. As described above in this subsection, this imperative treatment can extend to the execution, except for the execution of each child task. In an imperative execution, the equivalent child routine executes synchronously. In TSIA, each child task is submitted to the task pool and then executes autonomously.

### *3.1.11 Items Connect Origins and Destinations*

An item connects an origin to a destination or destinations. The implementation of the connection between origin and destinations is left up to the TS.

Several items are in the ia program for the classic application presented in Figure 11a). For example, the ia program uses two arrays of items. Each array element `a[k]` originating in the input file `"a.dat"` is connected to its destination in the task `produce(a[k];;b[k])`. Similarly, each array element `b[k]` is connected to its destination in the output file `"b.dat"`.

An item requires an origin and a destination. Furthermore, an item has no existence outside of its origin and destination. These qualities of an item lead to a simple transparent application definition, as desired for a flexible application execution. Two examples of this transparency follow. Firstly, the dependencies between tasks are obvious. For each task, the origin of each in is known, as is the destination of each out. Thus in the ia program of the classic application, the tasks obviously are independent of one another. Secondly, the state of the application execution is very simple. At any moment in the execution, the origin and the destination of any item is known.

An item in a ia language is not the same as a variable in an imperative language. For example, a variable exists independent of its origin and destination. In contrast, an item is given by its origin and destination. Every item requires an origin and destination and this can be ensured by a ia compiler.

As in IA, functional computing also avoids variables. In contrast to IA, much functional computing also avoids items. For example, large expressions can result from the lack of items to describe intermediate states [STATE]. Similarly, without an item for each out, a routine in functional computing thus easily only can have a single out -- the value of the routine or function.

### *3.1.12 The Type of an Item*

Type checking, described below, heterogeneous computing, described in subsection 2.11.11, as well as mixed definition computing, described in subsection 3.5.3, are on a wide spectrum of issues involving the type of an item. Of the many type issues, only a few possibilities for TSIA are included in this presentation. For example, no mention is made of the type issues concerning structures.

Type checking is an example of the convenience that a ia language can provide to the application definition. Type checking is not required by a IA. Instead it is an example of a definition element that can be provided by a IA.

Type checking ensures that the type of an item at its origin is compatible with the type expected at the destination. For example, in the ia program for the classic application presented in Figure 11a), each item `a[k]` originating in the file `"a.dat"`, is of type `bytes`, which matches the type of item expected in the destination `produce(a[k];;b[k])`.

In the ia language, type checking is reduced to its essence since the type of an item, as given by the origin, is directly checked against the type expected at the destination. For example, this differs from the indirect type checking of an imperative language. There the type of the origin is checked against a variable and the variable is checked against the destination. In an imperative language, type checking thus generally requires the type of the variable to be declared. As described in subsection 3.1.11, an item in ia code has no independent existence, never mind a declared type.

Type checking requires knowing the type of an item at its origin and at its destination. For example, in the ia program for the classic application presented in Figure 11a), `produce(bytes;;bytes)` declares the type of items expected by a task using the instruction `produce`. This *task prototype* is similar to a function prototype as declared for example in the C programming language [C].

If the type of the origin and of the destination of an item are incompatible, conversion may achieve compatibility. This assumes that the incompatibility is not due to an error in the application definition. Converted may be the origin, the destination or the item between the origin and the destination.

### *3.1.13 TSIA and Functional Computing*

Functional computing in many ways precedes IA. The precedences of functional computing would seem to originate in the similarity between task autonomy and a function free of side-effects. The precedences cross many different issues, including the use of referential transparency and of non-strict evaluation, the role of dependencies, the distinction between definition and execution, as well as implicit execution elements and a variety of executions for any one definition.

Both the IA and functional computing aim to support an application definition and execution, but from different starting points. Crudely speaking, functional computing focused first on the application definition. Focus on the application execution came second. In contrast, the starting point of a IA is the TS and its task model. In TSIA, application definition thus came second to application execution.

It long has been argued that due to referential transparency, functional computing is a natural candidate to support both the application definition and execution [Functional Fault Tolerance][Parallel Functional][Future Order][STATE]. Previous efforts have had limited success in corroborating this argument, in part due to the functional computing problem of state [STATE]. If TSIA has more success than previous efforts, perhaps it is because TSIA starts with the application execution, not the application definition.

An application transforms the state of some part of the world [Towards]. The state and the state transformers defined by the application definition are put into action by the application execution. Definition elements include the tools and techniques used to define states and state transformers. For example, imperative languages such as Fortran provide arrays, global variables and other elements for defining state.

Since it manages the execution, the TS thus manages state and the transformation of state. The IA thus must provide the TS with definitions of state and of state transformers. For example, in the ia program for the classic application presented in Figure 11a), the declaration of the instruction `produce` and of the files `"a.dat"` and `"b.dat"` is a description of state. The task `produce(a[k];;b[k])` is a state transformer.

While lousy for dealing with state itself, functional computing is wonderful for state transformers [STATE]. In other words, functional computing excels in composing functions or equivalently tasks. As demonstrated in some of the following sections of this chapter, functional computing and its implementation thus are a rich source of ideas and techniques for definition elements for transformation and for their implementation in IA.

## *3.2 Delegation*

### *3.2.1 Evaluation*

A task consists of items. Within a task, an item is *evaluated* if the task involves the value of the item. Consider for example the Fortran routine in Figure 13. In the execution of the routine `f` as a task, the instruction `f`, the in `a` and the out `b` each are evaluated.

Evaluation is a familiar part of current computing practice. By and large, current computing practice assumes that each item of each task is evaluated. This section breaks this assumption.

### *3.2.2 Delegation*

*Delegation* is the ability of a task to delegate responsibility for any of its items to one or more new tasks. In this context, the task is known as the parent task and the new tasks are its child tasks. For example, in Figure 14a), the parent task `a(x;;y)` delegates the responsibility for its in `x` and for its out `y` to the task `b(x;;y)`. A slightly more involved example is given in Figure 14b). There the parent task `c(x;;y,z)` delegates its outs `y` and `z` to the tasks `d(x;;y,q)` and `e(x,y,q;;z)`, respectively. In addition, the item `q` is an out of the first child and an in of the second.

The *responsibility* of a task thus includes all of its items and corresponds to the work of the task and of all of its children and further descendants. In contrast, the *work* of a task does not include the work of its children and further descendants.

As described in subsection 3.1.10, it usually is best to pretend that the parent task completes execution before any of its child tasks begin execution. This suggestion has a simple interpretation:

**"Pretend that in the task pool the parent task is replaced by its child tasks."**

As an example, the execution of the task shown in Figure 14a) results in the changes to the task pool illustrated in Figure 14c). Similarly, Figure 14b) corresponds to Figure 14d). As explained in Figure 14e), the arrow and the line are used to describe the execution as seen from the task pool.

In the illustrations of this presentation of the task pool, tasks are ordered from top to bottom, as in ia code, such that the dependencies of items between tasks are satisfied from top to bottom.

Delegation thus is a very powerful technique since it allows for both task autonomy and for a structured application definition. For example, delegation allows for the parent and child tasks of a hierarchy of routines, but without communication between tasks -- as required by task autonomy.

#### *3.2.2.1 Implementing Delegation*

In contrast to evaluation, the delegation of an item does not involve the value of the item. Instead, only a unique reference to the item must be available to the task. Consider for example the ia routine in Figure 14a). In the execution from the task pool of the routine `a(x;;y)` as a task, neither the in `x` nor the out `y` are evaluated. Instead, `x` and `y` are delegated to the task `b(x;;y)` and reappear in the task pool. When task `a(x;;y)` is replaced by task `b(x;;y)`, the unique references ensure that the `x` and `y` of the two tasks are the same.

The exact nature of the reference of an item is unimportant at this point. For example, for some items the reference may conveniently correspond to a storage location. For other items, such as a human user of an application, such a correspondence may be nonsense.

#### *3.2.2.2 Evaluate or Delegate or Ignore*

A task may either evaluate, delegate or ignore an item. By definition, an item is ignored if it is neither evaluated nor delegated by the task.

For simplicity, this presentation assumes that the value of an item includes the reference to the item. Then if an item is both evaluated and delegated by a task, the item simply can be treated as if evaluated.

Current computing practice largely assumes that each item of each task is evaluated. The ability to delegate or ignore an item thus introduces new degrees of freedom. Since evaluated items already are very familiar, this presentation focuses on delegated items and on ignored items.

#### *3.2.2.3 A Variation of Tail Recursion*

Delegation is a variation of tail recursion, including its implementation using continuations. Tail recursion and its implementation using continuations are techniques from functional computing [IMPERATIVE]. In this presentation, the use of the name tail recursion implicitly refers to the experience in functional computing. Tail recursion also is known by the names proper tail recursion, tail calling and by other names. As an example of a tail recursive routine in a functional language, the routine in Figure 14a) may be rewritten in Scheme as `(define (a x) (b x))`. In rewriting the example, the out `y` is assumed to be returned as the function value.

A comparison of tail recursion and delegation, including the use of continuation in the implementation of each, is beyond the scope of this presentation. Nonetheless, an initial comparison is roughly outlined here. In functional computing, a continuation represents the entire remaining application execution. Given a continuation as an explicit argument, a function can jump to the remaining application execution. Since there is no return from a jump, tail recursion thus is achieved. In the task pool of TSIA, the tasks and their items represent the entire remaining application execution. The equivalent of the continuation thus implicitly is part of each item and the whole is managed by the TS. When a task delegates an item to another task, the continuation identifies the source or the destination of that item. Thus continuation allows for delegation in TSIA, as it does in functional computing. In short, a continuation relates the items of a task or the result of a function to the remaining application execution. Given this relationship, delegation or equivalently tail recursion is possible.

The implicit continuation of TSIA is an aspect of a dependence-driven application definition. Such a definition explicitly defines the dependencies between items. In exchange for obeying the dependencies, TSIA controls the execution. In contrast, the explicit continuation in functional computing is an aspect of a control-driven application definition. Such a definition controls the execution. The dependencies between items thus implicitly are buried within the definition.

Just as delegation is a variation of tail recursion, the implementation of delegation using the task pool may be seen as a variation of an implementation of tail recursion. The UUO handler [RABBIT], and similarly the apply-like procedure [sml2c], may be seen as a task pool for one task. For example, the apply-like procedure is very similar to an equally small task pool imitated in subsection 3.3.3.

#### *3.2.2.4 Delegation in Previous Task Systems*

Cilk-NOW and Mentat are each a previous TS which provides delegation [Cilk-NOW] [Mentat]. Though their delegation is not as general as that of TSIA, Cilk-NOW and Mentat nevertheless strongly corroborate the feasibility of delegation for a TS. In addition, it was Cilk-NOW and Mentat which introduced to this thesis the notion of delegation for a TS.

As an example to demonstrate delegation in Cilk-NOW and in Mentat, the simple ia routine of Figure 14a) is rewritten in Figure 15a) and b) in their respective languages. In rewriting the example for Mentat, the out `y` is assumed to be returned as the function value.

In Mentat, delegation is provided by the return-to-future function, `rtf()`. The technique is identified as a form of tail recursion or continuation passing. Cilk-NOW identifies the technique of delegation as a form of continuation passing.

### *3.2.3 The Delegation Style*

Delegation obviously is similar to the usual call to a routine or function, but with a crucial characteristic. The delegation of responsibility for an item is true delegation in the sense that the item is not returned by a child task to its parent. Responsibility for the item truly is transferred from the parent to the child.

#### *3.2.3.1 Delegation versus Subordination*

In contrast to delegation, the usual function call may not truly transfer responsibility for an item from a parent to a child. Then the responsibility ultimately is retained by the parent since the item is returned by the child to its parent.

The return of the item is a form of communication between the parent and child. Then the parent is not autonomous. For example, executed as imperative code, the parent `a(x;;y)` of Figure 14a) cannot exit until it receives item `y` upon the return of the child `b(x;;y)`.

When the responsibility for an item is retained by the parent, the child essentially is supervised by the parent. In order to distinguish it from delegation, this presentation thus refers to such use as *subordination*.

Through delegation, the execution in TSIA of each child, that is of each task, is supervised by the TS. As explained in chapter 2, the TS thus is able to provide the application execution with execution elements.

#### *3.2.3.2 The Delegation Style versus the Subordinate Style*

Delegation or subordination, as discussed in the previous subsection, is a property of the parent task.

Subordination occurs when the parent task requires that the child task evaluate its items. This occurs when the code of the parent task is in the *subordinate style*. For example in Figure 16a), the parent task `sg(x;;y)` cannot delegate to the child `h(x;;z)` since the out `z` is required by subsequent code in the parent. Thus `h(x;;z)` must evaluate `z` and must be subordinate to `sg(x;;y)`.

In contrast, in the *delegation style* the parent allows the child task to evaluate or delegate its items. For example in Figure 16a), the parent task `dg(x;;y)` can delegate to the child `h(x;;z)` and the child `k(z;;y)`.

#### *3.2.3.3 The Universality of the Delegation Style*

**"Any application definition may be in the delegation style."**
This is the universality of the delegation style. In other words, the use of any child by any parent may be in the delegation style.

The universality of the delegation style seems to be the equivalent of compiling any application definition to the continuation passing style [RABBIT].

A two step argument validates the universality of the delegation style. The first step is the assumption that the only alternative to the delegation style is the subordinate style. The second step, as given in the next paragraph, shows that any definition in the subordinate style may be converted to the delegation style. As an example of the second step, consider the code of Figure 16. The routine `k(z;;y)` is defined such that the subordinate style `sg(x;;y)` and the delegation style `dg(x;;y)` effectively have the same application definition.

The second step holds that the use of any child by any parent, originally in the subordinate style, may be converted to the delegation style. This is argued by examining any child used by any parent. Either the child is the last thing done by the parent or it is not.

If the use of the child **is** the last thing done by the parent, then the parent obviously does not use any item returned by the child. The use of the child by the parent already is in the delegation style. For example, in Figure 14a) the use of the routine `b(x;;y)` is the last thing done by the parent `a(x;;y)`. The use thus is in the delegation style.

The use of a child as the last thing done by a parent is known as a tail call. Moreover, if the child uses the same instruction as the parent, the recursion is known as a tail recursion. The name tail recursion often is used instead of tail call, even when recursion is not involved. The name tail transfer also has been suggested [IMPERATIVE].

If the use of the child **is not** the last thing done by the parent, then there is parent code subsequent to the use of the child. This subsequent code may use items returned by the child. Then the use of the child by the parent is in the subordinate style. For example, in Figure 16a), the subsequent code `y=2*z` uses the item `z` returned by the child `h(x;;z)`.

Figure 17 illustrates the conversion of a generic parent from the subordination style to the delegation style. Every parent in the subordinate style consists of three parts. The first part is the code prior to the use of the child. The second part is the use of the child. The third part is the code subsequent to the use of the child. In the parent `p()` of Figure 17a), these three parts are labelled `//prior code`, `child(..;..;..)` and `// subsequent code`, respectively. The ellipsis `..` indicate arbitrary arguments.

In the conversion from the subordinate style to the delegation style, the parent task is split into 2 tasks, as illustrated in Figure 17b). Hence the delegation style also is known as the split-phase style [ADAM][Charm]. The first of these tasks, `pp()`, is the new parent task and replaces the original parent task, `p()`. From the original parent task, the new parent task contains the code prior to the use of the child, the use of the child and the use of the second task, `ps()`. This latter task is known as the *subsequent task* and from the original parent task contains the code subsequent to the use of the child.

The arguments of the subsequent task, `ps()`, includes all items required from the prior code, as well as any items required from the child. The resulting number of arguments may well be very large, but this is not a problem.

In order to delegate to a child, the delegation style thus requires that a parent also delegate to any code subsequent to the child. In order to do so, the subsequent code is placed into the so-called subsequent task. The new parent task obviously is in the delegation style, since there is no code subsequent to the child and the subsequent task. The conversion from the subordinate style to the delegation style thus is complete.

Obviously a tail call is a delegation style parent, but as the special case where there is no subsequent task or equivalently where the subsequent task is null.

In the above conversion from the subordinate style to the delegation style, some simplifications are made. For example, all the subsequent code was moved to the subsequent task. Of course, it is not necessary to move all the subsequent code. Instead, it is sufficient to move only the subsequent code which depends on items returned by the child. For example, the routine `a(b;;c,d) {x(b;;c); d=2*b;}` is in the delegation style. There is subsequent code, but it is independent of the items returned by the child `x(b;;c)`.

Another simplification in the above conversion is the omission of the possibility that the child is used from within an iterative loop of the parent. A solution is to implement iteration using recursion, as discussed in section 3.3. In a sense, the delegation style thus can lead to recursion. Vice versa, recursion makes heavy use of the delegation style.

#### *3.2.3.4 Illustrating the Delegation Style*

The delegation style allows for the parent and child tasks of a hierarchy of routines. Before illustrating this property of the delegation style, this section reviews the hierarchy of routines as allowed in the traditional subordinate style.

A routine A in the subordinate style is illustrated in Figure 18a). The subordinate style allows for a hierarchy of routines since encapsulated within any routine can be the use of another routine. This is illustrated in Figure 18b), where routine A subordinates to routine B. The use of a child routine by a parent routine is said to be encapsulated since the use of the child is not visible to a routine using the parent.

Because of encapsulation, the child of one routine in turn may be the parent of another. For example in Figure 18c), routine A uses routine B which in turn uses routine C.

Encapsulation greatly benefits the hierarchy of routines, helping it scale to arbitrary size. Because of encapsulation, each parent-child relationship is local. Thus the complexity of the hierarchy is proportional to the number of routines in the hierarchy. In contrast, if each routine in the hierarchy were aware of every other routine, then the complexity of the hierarchy would be proportional to the square of the number of routines. In other words, encapsulation hides the entire complexity of a hierarchy from any routine in and any routine using the hierarchy.

A routine of course can use arbitrarily many other routines. An example of a small hierarchy of routines is shown in Figure 18d).

In a fashion very similar to that of the subordinate style, the delegation style also allows for a hierarchy of routines. A task A in the delegation style is illustrated in Figure 18e). The delegation style allows for a hierarchy of routines since encapsulated within any task can be the use of another task. This is illustrated in Figure 18f), where task AP delegates to task B. As explained in subsection 3.2.3.3, the delegation divides the original subordinate-style routine A into a prior task AP and a subsequent task AS. For simplicity, this subsection uses the subsequent task, even if it is null. Usually, a null task is removed.

In the subordinate style, the subordination couples the call of a routine with its return. For example, in Figure 18b), B is called by A and returns to A. In the delegation style, the task pool or equivalently a continuation [IMPERATIVE], decouples the use of a task from its return. For example, in Figure 18f), B is called by AP and returns to AS.

The encapsulation and its benefits are essentially the same in the subordinate style and in the delegation style. For example in Figure 18g), task AP delegates to task BP which in turn delegates to task C. A task of course can delegate to arbitrarily many tasks. An example of a small hierarchy of tasks is shown in Figure 18h).

For a hierarchy of routines, Figure 18 nicely illustrates the correspondence between the subordinate style and the delegation style. In Figure 18, a) corresponds to e), b) to f) and so on.

In contrast to the subordinate style, the delegation style decouples the call and the return of each routine or task. In the subordinate style, the calling hierarchy and the return hierarchy are not only similar, they are the same one hierarchy. In the delegation style, the calling hierarchy and the return hierarchy are similar, but they are two separate hierarchies. The similarity of the two hierarchies is illustrated in the symmetry of each of Figure 18f), g) and h). The similarity or symmetry between the two hierarchies is broken if any null subsequent tasks are removed or if there are additional dependencies between the descendants.

As explained in subsection 3.2.5, encapsulated within the delegation style may be the subordinate style. Vice versa, encapsulated within the subordinate style may be the delegation style. Thus a hierarchy of routines can be a mixture of the delegation style and the subordinate style. This is an excellent example of how TSIA extends, not replaces, current computing practice.

An application definition written in an imperative language such as Fortran or C is an example of a hierarchy of routines in the subordinate style. This subsection thus illustrates how a Fortran or C program can be converted to the delegation style. More probably, the conversion is to a mixture of the delegation style and the subordinate style. The conversion implies that TSIA can execute a large variety of applications, but does not necessarily imply an efficient and effective execution. Such an execution may well require a change of algorithms or other changes to the application definition.

The delegation style generally is not explicitly discussed outside of this subsection. Instead it is used implicitly and used heavily. The delegation style is vital to a IA; it is the requirement on the application definition which allows for delegation in the application execution.

### *3.2.4 Ancestor Delegation*

Delegation may be used with a large variety of existing languages, including C++ and Fortran, without changing the language or its compilers.

A simple example using delegation with Fortran is shown in Figure 19a) and b). The task starts with the Fortran routine `a(x,y)`, which is declared in ia as `a(x;;y)`. The Fortran routine calls the ia routine `b(x;;y)`, which delegates the responsibility for the out `y` to the task `c(x;;y)`. As illustrated in Figure 19c), the execution of the task `a(x1;;y1)` from the task pool, thus leaves `c(x1;;y1)` in the task pool. The hierarchy of routines is illustrated in Figure 19d).

In the above example, the ia routine `b(x;;y)` completes without having evaluated its out `y`. Thus when `b(x;;y)` returns to the Fortran routine `a(x;;y)`, the Fortran routine can not use `y`, since `y` has not been evaluated. In other words, the Fortran routine must be in the delegation style. Thus, as for the ia routine `b(x;;y)`, the Fortran routine `a(x;;y)` completes without having evaluated its out `y`. In other words, because it is written in the delegation style, the Fortran routine `a(x;;y)` does not care whether or not `y` is evaluated by its child `b(x;;y)`.

*Ancestor delegation* is the name given here to this ability for the execution of a ia routine to delegate an item originally given to an ancestor of the ia routine. The ancestor is the initial routine of the task. In the above example, the ancestor is the simple Fortran routine `a(x;;y)` and is the immediate parent of the ia routine. Ancestor delegation accommodates an arbitrarily large and complex chain of ancestors between the initial routine of the task and the ia routine. For example, instead of a static hierarchy of routines, a task may have a chain of ancestors which vary at run-time and are implemented using pointers to functions.

The assumption for ancestor delegation by a ia routine that all ancestors are in the delegation style is a reasonable assumption. As described in subsection 3.2.3.3, any routine can be in the delegation style. Furthermore, apt delegation can be used to override ancestor delegation, as described in subsection 3.4.3. The override allows an ancestor to be in the subordinate style.

Ancestor delegation is a property of the implementation of delegation within a task. Outside a task, for example as seen from the task pool, ancestor delegation is not visible.

Implementing ancestor delegation can be straightforward and feasible, as demonstrated below using the above example. The demonstration assumes a Fortran implementation which passes an argument by its address. Though not required by the Fortran standard, this argument passing mechanism is used by many current Fortran compilers. The address of an argument thus may be used as the unique reference to an item. The need for such a unique reference was introduced in subsection 3.2.2.1.

The ia routine `b(x;;y)` may be just one of several routines making up a task. Using the unique reference for the out `y`, the ia routine `b(x;;y)` can recognize that its out `y` is an out expected from the task. By assuming that the out `y` is not used within the task, `b(x;;y)` may delegate the out `y` to the task `c(x;;y)`. In other words, the delegation style is assumed for out `y` throughout the task. Thus the assumption of the delegation style allows for ancestor delegation.

### *3.2.5 Apt Delegation*

*Apt delegation* is the ability of a ia routine implementation to recognize if an item may be delegated or if the parent of the ia routine requires the item to be evaluated. In other words, apt delegation recognizes if the parent is in the delegation style or in the subordination style, respectively.

Like ancestor delegation, apt delegation is a property of the implementation of delegation within a task. Also apt delegation is not visible outside of a task.

A simple example of apt delegation using Fortran is shown in Figure 20a) and b). The hierarchy of routines is illustrated in Figure 20d). The application begins by executing `program evordel` which calls `a(3;;y)`. In this case, the ia routine `a(3;;y)` is neither a task nor part of a task. Thus the out `y` is not an out expected from a task. As a first result, the ia routine `a(3;;y)` thus cannot delegate the item `y` to another task. That is, the responsibility for an item only can be delegated from one task to another task. Without an original task, there can be no delegation. As a second result, consistent with the first, the ia routine `a(3;;y)` assumes that the item `y` must be evaluated before the routine returns to its parent. The assumption of the second result is correct. Once `a(3;;y)` returns, `program evordel` will execute `print*,y`. The need for `a(3;;y)` to evaluate `y` thus is correctly recognized, as promised by apt delegation.

The ia routine a(3;;y) places the task b(3;;y) into the task pool. If y were delegated, then a(3;;y) would exit immediately. Instead, the evaluation of y requires a(3;;y) to block until y is evaluated.

From the task pool, the task b(3;;y) then may execute. Unlike for the routine a(3;;y), the out y is an out expected from the task b(3;;y). Apt delegation thus allows the task b(3;;y) to delegate the item y. As illustrated in Figure 20c), b(3;;y) thus may replace itself in the task pool by c(3;;y).

From the task pool, the Fortran task c(3;;y) then may evaluate y. Implicitly, for any out which is not delegated to another task, the out is assumed to have been evaluated by the execution of the task. The routine a(3;;y) then can unblock and return to program evordel, which executes print*,y and terminates normally.

The above example thus demonstrates apt delegation, including its feasibility. Though not explicit in the example, it is emphasized that, just like delegation in general, apt delegation is independent for each item of a task. Thus for example, a ia routine may delegate some items while having to evaluate others.

In addition, apt delegation also works if the ia routine is preceded in the task by an arbitrary hierarchy of routines. In other words, apt delegation can work with ancestor delegation.

### *3.2.6 An Application in Whole or in Part may use TSIA*

The application of Figure 20a) and b) demonstrates that it is possible for part of an application to use TSIA. For example, the a(x;;y), b(x;;y) and c(x;;y) could be routines from a library available to program evordel. Due to apt delegation, program evordel may use the library as a usual Fortran library, even though the library is implemented using TSIA. Transparent to the application, the library thus may involve any number of execution elements.

Alternatively, instead of only part of an application using TSIA, the entire application can use TSIA. An example is given by the classic application of Figure 11a). The routine main() is a task which loads N tasks into the task pool. Thus the entire classic application uses TSIA.

As another example, the routine program evordel of Figure 20a), written in the subordinate style, may be rewritten in the delegation style, as illustrated in Figure 21a) and b). The routine c(x;;y) of Figure 20a) and the routines a(x;;y) and b(x;;y) of Figure 20b) remain unchanged. The hierarchy of routines is illustrated in Figure 21c). The application execution is begun by entering the task begin_evordel(;;) into the task pool. Via ia_evordel(3;;), the task begin_evordel(;;) is replaced by the tasks a(3;;y) and end_evordel(y;;). Thus in contrast to an imperative execution, the routine program begin_evordel returns before the application execution has completed! Continuing with the application execution, the task a(3;;y) replaces itself by b(3;;y), which replaces itself by c(3;;y), which evaluates the item y. The execution of the task end_evordel(3;;) places no tasks into the task pool. Thus with no tasks remaining in the task pool, the execution of the application is finished.

Reliability is an obvious example of an advantage when the whole, instead of only part, of an application uses TSIA. The disadvantage of having only part of an application use TSIA can be demonstrated using program evordel of Figure 20a). None of the resources used for the execution of the routine program evordel may fail, if the application execution is to complete successfully. Of course, resources used by the TSIA part of the application may fail, since there reliability can be provided by TSIA.

Since the resources used by the routine `program evordel`, or its equivalent in another application, are often quite robust, there effectively may be very little disadvantage to having only part of an application use TSIA. For example, imagine an application execution using a local desktop computer plus a million other computers. It may well be sufficient to use TSIA for only the application code on the million computers.

#### *3.2.7 The Compatibility of Tasks and Processes*

Subsections 3.2.4, 3.2.5 and 3.2.6 may be summarized as a demonstration of the compatibility of tasks and processes. Here a process includes anything which communicates and thus also includes routines used in the subordinate style. As illustrated in Figure 22a), processes may exist within a task. As illustrated in Figure 22b), tasks may exist within a process. The compatibility of tasks and processes also is demonstrated in subsection 3.7.4, where for efficiency tasks in the delegation style may be executed as subroutines in the subordinate style of the process model.

This compatibility in no way compromises the constraint by the TS that an application execute in terms of tasks. The compatibility merely allows for processes within a task and allows for TS and tasks within a process.

### *3.3 Recursion*

#### *3.3.1 The Addtorial Function*

The $\text{addtorial}(N)$ is a function returning the sum of all integers from $1$ to $N$ inclusive. The name addtorial is introduced for the convenience of this presentation. The addtorial, in name and function, thus is a slight variation on the factorial function which returns the product of all integers from $1$ to $N$ inclusive.

Applications which compute the addtorial allow for the convenient demonstration of several properties of TSIA. As seen in this and the next few sections, each addtorial application literally computes the sum $1 + 2 + \ldots + N$. The computed sum can be shown to be correct by comparing it to the solution $\text{addtorial}(N) = n(n+1)/2$. Since the solution exists, computing the sum obviously only is useful here as a demonstration.

A simple complete Fortran application which computes the addtorial is shown in Figure 23a) and b). The routine `addtorial(n;;r)` is not defined in a style suitable for executing the routine in terms of tasks. Of course the entire routine `addtorial(n;;r)` could be executed as a single big task, but that would not make for a very interesting demonstration. This presentation thus assumes here that a finer granularity of tasks is desired. In other words, the routine `addtorial(n;;r)` is to be partitioned into many tasks. In fact, the partition results in a very fine granularity. Thus with the examples of the addtorial function and the classic application, this presentation already spans from a very fine through to a very coarse granularity, respectively.

#### *3.3.2 Recursion is a Delegation Style Equivalent for a Loop*

For the original routine `addtorial(n;;r)` of Figure 23b), there is a partitioning which yields a task for each iteration of the loop of the original routine. The corresponding Fortran code appears in Figure 24a) and with `program addtprog` of Figure 23a) is a complete Fortran application. The partitioning uses the technique of recursion [Conditional Expressions][Recursive Techniques]. This means that a child uses the same instruction as its parent.

Though not a part of standard Fortran, recursion is supported by many different Fortran compilers. In the worst case, though not explicitly pursued in this presentation, delegation allows for a recursive application definition even when using a language implementation which does not support recursion. Admittedly, this handicap does hinder unrolled delegation and other execution properties of TSIA described later.

In the code of Figure 24a), the real work is done by routine `addt1(;i,r;)`. The routine `addtorial(n;;r)` only performs the minor initialization for the use of `addt1(;i,r;)`. The digit `1` in the name `addt1` identifies that this routine is the first in a series of such routines in this presentation.

The `addtorial(n;;r)` and the `addt1(;i,r;)` routines each obey the delegation style. The execution of Fortran does not directly support delegation. Thanks to ancestor delegation, this shortcoming easily can be circumvented by introducing a ia routine, as shown in Figure 24b) and c). The Fortran routine `addt1(;i,r;)` of Figure 24a) essentially is the same as `addt2(;i,r;)` of Figure 24b). While the former calls itself directly, the latter calls itself via the ia routine `addt2_ia(;i,r;)` of Figure 24c).

As explained in subsection 3.1.10, in order to understand the application definition, ia code may be treated as if the ia language were an imperative programming language such as Fortran. Thus the application definition is accurately represented by the complete Fortran application which uses the Fortran routine `addt2_ia(;i,r;)` of Figure 24d) instead of the ia routine of Figure 24c). While making no difference to the application definition, using the ia routine or the Fortran routine for `addt2_ia(;i,r;)` obviously determines the application execution.

Since it is current computing practice the application execution using the Fortran routine `addt2_ia(;i,r;)` requires little explanation here. For example, the application can be compiled and executed by any reader with access to a Fortran compiler. Alternatively, it is little more than a straightforward change of syntax to rewrite any of the examples in C. In essence, the execution builds up a stack of recursive `addt2(addt2_ia(addt2(...)))` calls. The stack collapses once the end of recursion is reached. In calculating $\text{addtorial}(N)$, the stack thus reaches a size of order $N$. If $N$ is large, the application may exhaust available computing resources. As demonstrated below, the application execution using the ia routine `addt2_ia(;i,r;)` does not build up such a stack, even though the recursive definition is the same.

The execution using the ia routine `addt2_ia(;i,r;)` of Figure 24c) easily is illustrated by tracing a sample execution of the application. The example assumes that `program addtprog` of Figure 23a) has called `addtorial(9;;r)` of Figure 24b). This calls `addt2(;9,0;)` which calls `addt2_ia(;8,9;)`. This places `addt2(;8,9;)` into the task pool. Since apt delegation allows neither of its inout to be delegated, `addt2_ia(;8,9;)` thus must block until both inout are evaluated. From the task pool, `addt2(;8,9;)` is executed and calls `addt2_ia(;7,17;)`. Apt delegation allows it to delegate to `addt2(;7,17;)`. In other words, `addt2_ia(;7,17;)` places `addt2(;7,17;)` into the task pool and returns to `addt2(;8,9;)`, which also returns. Ancestor delegation thus has allowed `addt2(;8,9;)` to be replaced in the task pool by `addt2(;7,17;)`. This in turn is replaced by `addt2(;6,24;)` and so on until

addt2(;1,44;) is in the task pool. Since the execution of addt2(;1,44;) performs no delegation, its two inouts are assumed to have been evaluated. Thus addt2_ia(;8,9;) unblocks and returns to addt2(;9,0;), which returns to addtorial(9;;r), which returns to program addtprog, which prints out the correct answer. In the previous sentence, the in value of the inout is used to identify the particular execution of each routine. Of course, on the return of each routine the out value of each inout is not the in value.

The next subsection further explains and demonstrates that an iteration of a loop is equivalent to an iteration of recursion.

### *3.3.3 A Simple Imitation of TSIA for the Addtorial Application*

The addtorial application of Figure 23a) and Figure 24b) and c) allows for a beautiful and simple demonstration of delegation, including ancestor delegation and apt delegation. The demonstration is possible because the execution of the addtorial application is extremely simple. For example, during its entire execution, the addtorial application never has more than one task in the task pool.

In the demonstration, the ia routine addt2_ia(;i,r;) of Figure 24c) is replaced by the Fortran imitation of Figure 25a). Thus the code of Figure 23a), Figure 24b) and Figure 25a) is a complete Fortran application. The reader is strongly encouraged to compile and execute the application and to read and work through it. This is an excellent introduction to a deep understanding of delegation.

The Fortran imitation of Figure 25a) is presented here in order to explain delegation and to demonstrate its feasibility. In order to avoid possible confusion, it is stated explicitly here that in TSIA a typical application definition does not include such imitations. The ia routine of Figure 24c) is part of an application definition for TSIA; the routine's imitation in Figure 25a) is not. Thus in TSIA, the complete addtorial application is given by the code of Figure 23a) and Figure 24b) and c).

Obviously the imitation of Figure 25a) should interest someone wanting to implement a TSIA, since the imitation essentially is a crude implementation of TSIA. Even though an application developer normally has no contact to the implementation of a TSIA, the imitation also should be of interest since it demonstrates much of application execution in TSIA. For example, such knowledge can help a developer achieve an efficient application definition.

The Fortran imitation of Figure 25a) imitates the execution of the addtorial application as if the ia routine of Figure 24c) were used in TSIA. In other words, the Fortran imitation mimics the actions of the TS and the compiled ia routine.

The Fortran imitation of Figure 25a) uses suggestive variable names in order to help relate the imitation to a real execution using TSIA. The variable full imitates whether or not a task is in the task pool. Since full thus must be static across calls to addt2_ia(;i,r;), full is in the common block taskpool. The block data initpool sets the initial value of full.

Obviously the Fortran imitation of Figure 25a) does not provide the application execution with any execution elements as would be done by a TS. Nonetheless, the essence of the execution is the same. In particular, the imitation provide delegation, including ancestor delegation and apt delegation.

The imitated execution using the Fortran routine addt2_ia(;i,r;) of Figure 25a) easily is illustrated by tracing a sample execution of the application. This trace is very similar to that presented in the subsection 3.3.2 for the normal execution using TSIA with the ia routine of Figure 24c). Throughout this trace, the in value of the inout is used to identify each execution of each routine. Assume that program addtprog of Figure 23a) has called addtorial(9;;r) of Figure 24b). This calls addt2(;9,0;) which calls addt2_ia(;8,9;). The application execution now is inside the imitation. In order to initialize apt delegation, the logical variable full initially is .true.. After setting full to .false., addt2(;8,9;) is called. This calls addt2_ia(;7, 17;), which just sets full to .true. and returns to addt2(;8,9;) which returns to addt2_ia(;8,9;). Since full is now .true., full is set to .false. and addt2(;7,17;) is called. Ancestor delegation thus has allowed addt2(;8,9;) to be replaced by addt2(;7,17;). This in turn is replaced by addt2(;6,24;) and so on until addt2(;1,44;) is called by addt2_ia(;8,9;). Since the execution of addt2(;1,44;) does not call addt2_ia(), full remains .false. when it returns to addt2_ia(;8,9;). Previously blocked by apt delegation, this now returns to addt2(;9,0;), which returns to addtorial(9;;r), which returns to program addtprog, which prints out the correct answer.

The Fortran imitation of Figure 25a) thus is equivalent in both definition and execution to the ia routine of Figure 24c). This equivalence is demonstrated by the behavior of the stack [Recursive Techniques][Stack] during the execution of the application. In order to examine the behavior of the stack, Figure 25b) contains a slightly modified version of the routine addt2(;i,r;) of Figure 24b). The modification introduces the variable s, which is allocated on the stack each time addt2(;i,r;) is executed. The function loc(), included with many Fortran compilers, is used to print the address of s and thus expose the behavior of the stack during execution. The output of the application execution for addtorial(5) is given in Figure 25c). As seen from the output addresses for s, the stack does not grow between the second and last execution of addt2(;i,r;). The initial growth of the stack, from the first to the second execution of addt2(;i,r;) is due to apt delegation.

Thanks to delegation, the stack thus does not grow, even though addt2(;i,r;) is recursively defined. Since the iteration of a loop also does not increase the size of the stack, a recursion using delegation executes like a loop [IMPERATIVE].

In terms of its execution, the routine addtorial(n;;r) thus comes full circle. The routine of Figure 23b) is defined and executes iteratively, but without a partitioning into tasks. In contrast, the routine of Figure 24b) also executes iteratively, but is defined recursively and is partitioned into tasks.

In contrast to the Fortran routine of Figure 25a), the Fortran routine of Figure 24d) is equivalent to the ia routine of Figure 24c) only in definition and not in execution. In fact the execution is completely different. As above, a demonstration is given by the behavior of the stack during the execution of the application. This again uses the modified version in Figure 25b) of the routine addt2(;i,r;). The output of the application execution for addtorial(5) is given in Figure 25d). As seen from the output addresses for s, the stack grows from the first through to the last execution of addt2(;i,r;). Thus without delegation, recursion does not execute as iteration.

### *3.3.4 Divide-and-Conquer*

The singly recursive partitioning of an application execution into tasks, as discussed in the previous subsections, allows for all execution elements but one. Single recursion does not allow for parallel computing since each task delegates to a single task. For single recursion, there is at most a single task in the task pool at any moment. Without multiple independent tasks in the task pool, there is no opportunity for parallel computing.

*Divide-and-conquer* (DC) [Divide-and-Conquer] is a recursion which allows for parallel computing. DC divides a problem into several independent parts. Each part is then solved. Given the solutions to all parts, the original problem is then solved. This may or may not require an explicit combination of the solutions to all parts.

In IA, the problem corresponds to a parent task. Each part is a child task, as is any task to combine the solutions of all parts.

If the problem is too simple to be divided into parts, then there is no need for DC. Such a problem is known as a base case and is solved directly.

In DC, each part is similar to the original problem, but is simpler to solve. Each part thus is solved as a problem in its own right. The similarity between each part and the parent underlies the recursive nature of DC. Compared to the problem, each of its parts is simpler to solve since each part is closer to the base case. Thus as in other recursion, DC ends when a base case is reached.

Because it divides a problem into multiple self-similar parts, DC is an example of multiple recursion. In contrast, the single recursion of the previous subsections keeps a problem as a single part. Single and multiple recursion also are known as linear and nonlinear recursion, respectively.

While DC may be used to divide a problem into any number of parts, typically a problem is divided into two parts. This also is known as double recursion. For example, all the examples of DC in this presentation use double recursion. Despite the predominance of double recursion, in this presentation and elsewhere, DC using more than two parts should not be forgotten. For example, increasing the number of parts is a way to coarsen the partitioning of the application execution into tasks. In other words, increasing the number of parts is a form of recursion unrolling, as described in subsection 3.7.1.

The addtorial application, used in the previous subsections to explain single recursion, also allows for a simple demonstration of DC. A complete addtorial application using DC is given by the Fortran code of Figure 23a) and Figure 26a) and the ia code of Figure 26b). In the application, DC is implemented by the ia routine `addt_dc(b,t;;r)`. The base case is identified by `b==t`. The Fortran routine `addtorial(n;;r)` merely is an easy interface to `addt_dc(b,t;;r)`. The Fortran routine `add(a,b;;c)` combines the partial solutions.

The Fortran routine `addt_dc(b,t;;r)` of Figure 26c) has the same definition as the ia routine of Figure 26b). The Fortran code of Figure 23a) and Figure 26a) and c) makes up a complete Fortran application. This allows the correctness of the definition to be verified. Of course an execution using the Fortran `addt_dc(b,t;;r)` is not like an execution using IA.

The execution using the ia routine `addt_dc(b,t;;r)` of Figure 26b) may be illustrated by examining the tasks in the task pool. For example consider the task `addt_dc(1,9;;r)`. As illustrated in Figure 27, this task will replace itself in the task pool by 3 tasks. Two of the tasks, `addt_dc(1,5;;rb)` and `addt_dc(6,9;;rt)`, are smaller than and self-similar to the original task. The results of these two tasks are combined by the third task, `add(rb,rt;;r)`. As further illustrated, each `addt_dc()` task expands into further tasks. Such expansion continues until the base cases are reached. In the figure, the items are labelled such that the dependencies between tasks are clear. The former two `addt_dc()` tasks may be executed in parallel, as may the latter four tasks. Parallelism thus is achieved.

### *3.4 Some Issues Related to Delegation*

#### *3.4.1 Tail Delegation versus Deep Delegation*

*Tail delegation* delegates to a child task **without** a subsequent task. Tail delegation is the basis of the addtorial application presented in subsection 3.3.2 using code in Figure 24.

*Deep delegation* delegates to a child task **with** a subsequent task. Deep delegation is the basis of the addtorial application presented in this subsection using code in Figure 28.

Though the examples discussed here involve recursion, neither tail delegation nor deep delegation need involve recursion.

Figure 28a) presents a version of the Fortran routine `addtorial(n;;r)` which is very similar to that of Figure 24a). Either routine may be used with `program addtprog` of Figure 23a) for a complete Fortran application. While the code of Figure 24a) leads to a tail delegation, that of Figure 28a) leads to deep delegation.

In the code of Figure 28a), the work is done by the routine `addtS(i;;r)`. The routine `addtorial(n;;r)` exists only for uniformity with the other addtorial application examples. The routine `addtS(i;;r)` is in the subordinate style; there is code `r=r+i` subsequent to the recursion. As explained in subsection 3.2.3.3, any routine in the subordinate style may be converted to the delegation style. The result of the conversion is the Fortran code of Figure 28b) and the ia code of Figure 28c). The ia code is used since the execution of Fortran does not directly support delegation. Of course the ia routine `addtS_ia(i;;r)` of Figure 28c) may be replaced by the Fortran version of Figure 28d), in order to execute the whole as a Fortran application and thus to verify the application definition.

The ia routine `addtS_ia(i;;r)` delegates to `addtS(i;;r)`. The delegation is deep since the ia routine also delegates to the subsequent task `incr(8;r;)`.

The execution using the ia routine `addtS_ia(i;;r)` of Figure 28c) easily is illustrated by tracing the tasks in the task pool. This example assumes that the task pool initially contains the task `addtS(9;;r)`. As illustrated in Figure 29a), the execution of `addtS(9;;r)` causes it to replace itself in the task pool by the tasks `addtS(8;;r)` and `incr(9;r;)`. Similarly, the execution of `addtS(8;;r)` replaces it by the tasks `addtS(7;;r)` and `incr(8;r;)`. In this as in other illustrations of the task pool, the dependencies between tasks are top to bottom, as they are in ia code.

The differences between deep delegation and tail delegation easily are illustrated by comparing their respective executions as seen from the task pool. As described above, Figure 29a) illustrates an execution of deep delegation. Using the addtorial application presented in subsection 3.3.2 with code in Figure 24, the equivalent execution of tail delegation is illustrated in Figure 29b).

For the addtorial application evaluating $\text{addtorial}(N)$, tail delegation presumably is more efficient than deep delegation. Firstly, tail delegation results in an execution with a total of only $N$ tasks. In contrast, deep delegation results in $2N$ tasks. Secondly, tail delegation results in at most $1$ task in the task pool at any moment during the execution. In contrast, deep delegation ultimately results in a stack of $N$ tasks. The word stack is used here since each task depends on the task above it. Thus no task can leave the pool or task for execution until the task above it has executed.

As explained in subsection 3.3.3, a recursion using delegation executes like a loop. From the above discussion, this result should be refined as follows. A recursion using tail delegation executes like a loop. In contrast, a recursion using deep delegation does not execute like a loop since the equivalent of a stack grows with each iteration [IMPERATIVE].

In an alternative description [Recursive Techniques][Towards], tail delegation is in iterative form, while deep delegation is not.

In general, tail delegation presumably is more efficient than deep delegation. In other words, having no subsequent task should be more efficient than having one. Thus where possible, tail delegation should be used instead of deep delegation. In other words, the use of a subsequent task should be avoided. This implies doing work prior to the delegation instead of subsequent to the delegation.

Of course in many situations it is impossible to avoid deep recursion. For example, deep recursion often is intrinsic to divide-and-conquer, a technique introduced in subsection 3.3.4.

As described in subsection 3.2.3.3, any routine in the subordinate style may be converted to the delegation style. As demonstrated by the example of this subsection, a brute force conversion can lead to poor deep delegation instead of good tail delegation. For the addtorial application, the difference is a matter of whether the code `r=r+i` occurs before or after the delegation. While this difference has little effect on the Fortran execution, this difference has a huge effect on the delegation-based execution of TSIA. In essence, the Fortran execution always is a stack-based execution. Tail delegation, but not deep delegation, effectively allows a TSIA execution to avoid the need for the stack.

The comparison of deep delegation to tail delegation is essentially equivalent to the comparison in functional computing of deep recursion to tail recursion [IMPERATIVE]. The similarity includes viewing the contents of the task pool as a continuation. The continuation grows for deep recursion, but not for tail recursion. Similarly, the contents of the task pool grow for deep delegation, but not for tail delegation.

### *3.4.2 Memory Management Related to Ancestor Delegation*

As an introduction to memory management in TSIA, consider the ia routine `dg(x;;y)` of Figure 16b). When `dg(x;;y)` delegates to the routines `h(x;;z)` and `k(z;;y)`, TSIA has to manage the memory for the item `z`. Memory management in TSIA is a topic beyond the scope of this presentation. Nonetheless, this subsection introduces a property of memory management related to ancestor delegation. This particular property is addressed in order to further illustrate delegation and to corroborate its feasibility.

As an example, consider the Fortran routine `addt3(i;r;)` of Figure 30a) and its ia routine `addt3_ia(i;r;)` of Figure 30b). The routines are similar to the Fortran routine `addt2(;i,r;)` of Figure 24b) and its ia routine `addt2_ia(;i,r;)` of Figure 24c). The difference between the pairs of routines is that `i` is an in of the former pair, while `i` is an inout of the latter pair.

Just as `addt2_ia(;i,r;)` provides `addt2(;i,r;)` with ancestor delegation, `addt3_ia(i;r;)` provides `addt3(i;r;)` with ancestor delegation. For `addt2_ia(;i,r;)` and `addt2(;i,r;)`, the delegation involves no memory management, since the items `i` and `r` in the delegation by `addt2_ia(;i,r;)` are the `i` and `r` given to `addt2(;i,r;)`. In other words, in the execution of `addt2(;i,r;)`, the items `i` and `r` come from the task pool and are returned to the task pool.

As introduced in subsection 3.2.2.1, an item is identified by a unique reference. This could be its address in memory for example. In ancestor delegation, the unique reference allows a ia routine to identify whether or not an item originates from the task pool. This merely assumes that the items of a task are accessible by any ia routine which is part of that task.

For `addt3_ia(i;r;)` and `addt3(i;r;)`, the delegation of the item `r` involves no memory management, since the item `r` in the delegation by `addt3_ia(i;r;)` is the `r` given to `addt3(i;r;)`. In contrast, the delegation of the item `i` involves memory management. The item `i` in the delegation by `addt3_ia(i;r;)` is not the `i` given to `addt3(i;r;)`. Instead, the item `i` in `addt3_ia(i;r;)` is the variable `n`, which is local to the execution of `addt3(i;r;)`. The variable `n` thus does not exist beyond the old `addt3(i;r;)` task and thus `n` cannot be used as an item delegated to a new task.

For example, consider the task `addt3(8;9;)`. This calls `addt3_ia(7;17;)`, which delegates to `addt3(7;17;)` and returns to `addt3(8;9;)`. The value `7` passed by `addt3(8;9;)` to `addt3_ia(7;17;)` is local to `addt3(8;9;)` and thus cannot be used for the task `addt3(7;17;)`. This is recognized by `addt3_ia(7;17;)`, since the address of its argument valued `7` is not the address of the value `8` which is an item of the task `addt3(8;9;)`.

Delegation thus requires `addt3_ia(i;r;)` to allocate memory for item `i` of the new task `addt3(i;r;)`. Since each `addt3(i;r;)` is the only task using its item `i`, the memory for `i` may be freed once `addt3(i;r;)` completes.

As a demonstration, Figure 30c) is a Fortran imitation of the ia routine `addt3_ia(i;r;)` of Figure 30b). In essence, the imitation of Figure 30c) adds memory management to the imitation of Figure 25a). As part of its simplification, the imitation of Figure 30c) assumes that memory management is required for item `i` of the new task `addt3(i;r;)`; the imitation does not identify whether memory management is required.

The Fortran imitation of Figure 30c) thus is equivalent in both definition and execution to the ia routine of Figure 30b). The code of Figure 23a), Figure 30a) and c) thus is a complete Fortran application with an execution like that in TSIA. Without memory management, for example if the imitation of Figure 30c) is replaced by that of Figure 25a), the application does not execute correctly.

### *3.4.3 Using Apt Delegation to Override Ancestor Delegation*

Apt delegation may be used to override ancestor delegation. As described in subsection 3.2.4, ancestor delegation assumes that each ancestor routine treats each of its arguments in the delegation style.

For whatever reason, an ancestor routine may not want to treat one of its arguments in the delegation style. In other words, the ancestor wishes to use an argument in the subordinate style.

An ancestor routine may not naively use the subordinate style for one of its arguments. This contradicts the assumption of ancestor delegation that all ancestors use the delegation style.

For example, Figure 31a) shows an example of a naive and incorrect use of the subordinate style. The Fortran routine `addtW(;i,r;)` uses the items `i` and `r` subsequent to the use of `addtW_ia(;i,r;)`. That subsequent use of the items `i` and `r` is undefined since ancestor delegation allows the ia routine `addtW_ia(;i,r;)` of Figure 31b) to delegate the items `i` and `r` to another task. The undefined values of the items `i` and `r` may be demonstrated by using for `addtW_ia(;i,r;)` the equivalent of the Fortran imitation of Figure 25a). For the imitation, the names `addt2` and `addt2_ia` have to change to `addtW` and `addtW_ia`, but otherwise everything is the same. The code of Figure 23a), Figure 25a) and Figure 31a) then is a complete Fortran application, with an execution as in TSIA. An example output is shown in Figure 31c).

In the Fortran imitation of Figure 25a), the child task always begins execution after its parent has completed. This strict ordering need not be true in TSIA, but here it helps explain the output of Figure 31c). Because of the strict ordering, the use of `addtW_ia(;i,r;)` does not change the values of `i` and `r` before their subsequent use. In each invocation of `addtW(;i,r;)`, the values of `i` and `r` in the output of Figure 31c) are as if `addtW_ia(;i,r;)` had not been called. The output requires one further explanation. Apt delegation results in a final `i=0,r=45` , instead of an initial `i=8,r=9` .

An ancestor routine may use the subordinate style for an argument, by using apt delegation to override ancestor delegation. As described in subsection 3.2.5, apt delegation identifies whether an argument to a ia routine is an item of the task. If yes, then the item may be delegated, if no, the item must be evaluated. An ancestor routine thus may use the subordinate style for an argument, by passing to its child a copy of the argument instead of the argument itself. Since apt delegation recognizes that the copy is not an item of the task, ancestor delegation is overridden. In other words, apt delegation forces the copy, and thus the argument, to be evaluated and not delegated.

For example, Figure 31d) shows an example of a correct use of the subordinate style. The Fortran routine `addtR(;i,r;)` uses the items `i2` and `r2` subsequent to the use of `addtR_ia(;i2,r2;)`. That subsequent use of the items `i2` and `r2` is defined since apt delegation forces the ia routine `addtR_ia(;i2,r2;)` of Figure 31e) to evaluate the items `i2` and `r2`. The defined values of the items `i2` and `r2` may be demonstrated by using for `addtR_ia(;i,r;)` the equivalent of the Fortran imitation of Figure 24d). For the imitation, the names `addt2` and `addt2_ia` have to change to `addtR` and `addtR_ia`, but otherwise everything is the same. The Fortran imitation of Figure 24d) always evaluates its arguments. The code of Figure 23a), Figure 24d) and Figure 31d) then is a complete Fortran application, with an execution as in TSIA. An example output is shown in Figure 31f).

Because of apt delegation, `addtR_ia(;i2,r2;)` evaluates `i2` and `r2` in each invocation of `addtR(;i,r;)`. As seen in the output of Figure 31f), the evaluation is completed once `i2` reaches `0`.

The correct use of the subordinate style is completely local to an ancestor routine. Thus a compiler could recognize and flag as an error any naive and incorrect use of subordination. Alternatively, a compiler could introduce the required copy of an argument and thus convert the incorrect use of subordination to a correct use.

### *3.5 Semantics before Syntax*

The preceding sections of this chapter have introduced some of the abilities of IA. With the meaning of IA thus established, this section emphasizes that this presentation concerns the semantics of an application definition, not its syntax. With syntax even further out of the way, the remaining sections of this chapter present additional abilities of IA.

In short, semantics precedes syntax. This presentation addresses some of the semantics of an application definition, but does not even begin to address its syntax.

#### *3.5.1 The Use of Fortran in this Presentation*

For a variety of reasons, the examples of this presentation are coded in Fortran, as opposed to another programming language. As the oldest language still in widespread use and as a relatively small and simple language, Fortran serves well as a baseline language.

A reason for coding the examples in Fortran is that Fortran is relatively easily understood. Instead of Fortran, many a reader probably would favour a different language, but Fortran seems like an acceptable compromise.

As a baseline language, Fortran emphasizes that any one of a large variety and number of other languages instead could have been used to code the examples of this presentation. The use of Fortran allows each example to be translated in a straightforward fashion to another language. A reader preferring a language other than Fortran is encouraged and free to make the translation. For example, the translation to C is particularly easy.

The straightforward translation of the examples between languages emphasizes that IA and thus TSIA and the task model largely are orthogonal and indifferent to the choice of language. As a set of definition elements, a language serves to define an application. A IA is an implementation of definition elements. A IA cares little about the origin of a definition element and thus cares little about the choice of language.

In TSIA, execution elements result from the implementation of definition elements. In implementing the definition elements, IA serves TS, which provides the execution elements. Definition elements and languages thus serve to define an application, not to execute it. Vice versa, TSIA serves to execute an application, not to define it. This property of this thesis is emphasized by the use of Fortran in this presentation. Fortran and its definition elements clearly serve to define an application, not to execute it.

Throughout its history, implementations of Fortran have provided efficient execution. TSIA thus can be seen by Fortran as just another iteration of implementation.

In contrast to Fortran, some other languages support the execution of an application. For example, some languages directly support the process model by providing threads or other techniques. In this thesis, such language support for the application execution is avoided and is considered to be a detriment to the application definition. In corroboration, years of computing practice and research would seem to indicate that the process model makes for a poor application definition.

At least in the process model, several definition elements have been proposed as having a special role for supporting the execution of an application. The most prominent of these definition elements probably is object computing [Active Objects]. Fortran includes none of these definition elements and thus emphasizes that in the task model neither these nor other definition elements have a special role in execution. This thesis does not question the use of object computing or other definition elements for defining an application. As mentioned in subsection 3.25.3, it seems that a IA might be able to implement object computing as it implements other definition elements. The use of Fortran thus emphasizes that definition elements are orthogonal to execution elements or equivalently that application definition is orthogonal to application execution.

Since Fortran is a relatively small and simple language, it is relatively transparent that there are no properties of Fortran which are poorly served by IA or worse yet, which conflict with IA. This completeness is valuable since it implies that IA is suitable for any application which can be defined in Fortran. On the basis of Fortran alone, IA and thus TSIA and the task model thus are suitable for a large variety of applications.

In contrast to those of Fortran, there are definition elements for which it is not immediately obvious whether IA serves them well or at least does not conflict with them. Examples of such definition elements include object computing, higher order functions and reflection.

As a simple language, Fortran does not include topics like delegation and non-strict evaluation and thus is well suited for introducing these topics. By using Fortran, this thesis emphasizes that these topics are not restricted to particular languages or families of languages. In addition, since their introduction using Fortran is fairly straightforward, these topics thus are made fairly accessible to the reader.

### *3.5.2 The Fortran Used in this Presentation*

Throughout this presentation, Fortran refers to Fortran 77 [Fortran]. As for a translation to another language, the examples of this presentation may be translated to an earlier or later version of Fortran.

Some of the examples of this presentation do not strictly comply with the Fortran standard. As described in subsection 3.2.4, an argument to a routine is assumed to be passed by its address. As described in subsection 3.3.2, this presentation uses recursion. Both of these features exist in many different Fortran compilers. While convenient for this presentation, neither of these features is intrinsic to the arguments of this thesis.

### *3.5.3 A Mixed Definition*

An application with a mixed definition is one defined using a variety of definition tools and techniques. A mixed definition thus might include a variety of programming languages. The alternative to a mixed definition is a uniform definition and an example would be a traditional monolithic C application. Mixed definition computing thus is the definition analogue to heterogeneous computing, which uses a variety of resources for the application execution.

This presentation introduces a mixed definition because the ia language initially has few definition elements and thus must be used with another language or languages. Such a mixed definition can be convenient for the application and can have a relatively straightforward implementation [GLU].

For example, given the Fortran routine `mult` shown in Figure 32a), the ia task declaration in Figure 32b) allows `mult` to be used in ia code. Its use from ia does not require

`mult` to be modified. Thus the use from ia does not require the source code to `mult` and instead `mult` can be supplied as a compiled object, as it can for use from Fortran. A similar mixed definition task declaration can make a ia routine callable from Fortran. For simplicity elsewhere in this presentation, such mixed definition task declarations are implicitly assumed and are not explicitly stated.

A mixed definition task declaration as in Figure 32b) could be in the source code of an application. Alternatively, the declaration could be in a header file, like those of the C programming language, and thus available at compile-time. Alternatively, the declaration or its equivalent could be made available at link-time or at run-time, as might be required for dynamic computing.

A mixed definition allows the instruction of a task to be implemented by a routine or by a process. For example, for the classic application presented in subsection 3.1.4 and in Figure 11a), the instruction `produce` obviously could be implemented by a routine. Alternatively, `produce` could be implemented by a process with a task declaration as in Figure 32c). As a transformer process, `produce` reads in a sequence of `bytes` items. Each `bytes` in to `produce` yields a `bytes` out [Funnel].

As a slight variation on the above, `produce` could be implemented by a process with a task declaration as in Figure 32d). In this case `produce` inputs a single `bytes` item and outputs a single `bytes` item. Thus the processing of each `bytes` item uses an individual `produce` transformer process [DBC][Nimrod].

A mixed definition is not intrinsic to a ia language, IA or TSIA. A ia language could include a large set of definition elements and thus not require that an application definition also use other languages. While TSIA thus does not depend on mixed definition computing, the inverse may happen since TSIA would seem to offer an opportunity for convenient mixed definition.

The general motivation for a convenient mixed definition is an invariance of the application definition and its parts to their particular definition tools and techniques. For example, each part of an application definition thus can be defined using the most appropriate definition tool or technique. Similarly, software reuse is greatly simplified if there is no requirement for a particular language.

The discussion of such a convenient mixed definition, via TSIA or otherwise, is beyond the scope of this presentation and thus is not pursued further here. Examples of the history and of the present state of mixed definition can be found elsewhere [cfortran.h][ILU][LSS][Mixed Language Programming].

### *3.6 Throttling*

As introduced in subsection 2.11.5 and Figure 9, scheduling includes throttling and mapping. In essence, throttling controls the expression of the application definition in terms of tasks.

Properties of an application execution controlled by throttling include the granularity and the task order. As described in section 3.7, the granularity describes the number of tasks making up the application execution and correspondingly the amount of the execution performed per task. As introduced in section 3.8, the task order can vary the moment in the execution at which a given task enters the task pool from the application definition.

Throttling also includes fusion and fission, which each may be considered part of the task order. As described in section 3.12, items separated in the application definition may be fused in the execution, thus allowing a structured application definition to have an efficient execution. As described in section 3.15, fission eliminates structures not needed for the execution, but which allow for a structured application definition.

Throttling also includes transparent delegation. As described in section 3.14, transparent delegation recognizes that two tasks essentially may be the same even if their delegated ins differ. Thus only one of the two tasks need execute. Instead of executing, the other task can be replaced by a copy of the outs of the executed task, after substituting the delegated items.

The definition of throttling in this presentation differs from some earlier definitions. For example, in some earlier definitions throttling refers only to a part of the task order. [Lazy Task Creation][Parallel Functional].

Throttling thus allows an application definition to be expressed in one of a variety of sets of tasks. In large part, the particular set used for a given application execution is determined by the mapping. As a first example, mapping provides the cost of resources and thus determines whether a particular fusion or a fission is worthwhile. As a part of the TS, the mapping aims to satisfy the execution elements desired by the application. Thus as a second example, mapping for reliable or reactive computing requires a sufficiently fine granularity for the partitioning of the application execution into tasks. Throttling ensures that this granularity is compatible with that of an efficient execution. As a third example, mapping which overcomes communication latency by overlapping communication and computation requires a minimum number of tasks in the task pool. By managing the task order, throttling aims to maintain that number of tasks in the task pool throughout the application execution.

Neither throttling, mapping nor the whole of scheduling generally is solvable. Thus heuristics often are used to approximate the best possible scheduling. Task autonomy is very valuable to an imperfect throttling. For example, an aggressive task order can exhaust memory or other resources. Task autonomy allows the results of some completed tasks to be thrown away and the tasks to be re-executed later. The resources thus freed allow the application execution to continue with a less aggressive task order. Similarly, aborting the execution of a task can free a resource required by a real-time task. Such backtracking can be a part of throttling, allowing it to aggressively pursue performance, without paying a high price for the occasional mistake.

In general, throttling is orthogonal to other properties of IA and its implementation of definition elements. Moreover, the various properties of throttling largely can be examined independent of one another. Thus outside of its respective section, each property of throttling generally is not pursued in the other sections of this presentation. For example, outside of section 3.7 the simplicity of the presentation demands that examples of DC applications use the simplest and smallest base case. In practice, granularity often requires throttling to introduce larger base cases.

### *3.7 Granularity*

For the partitioning of an application execution into tasks, the granularity describes the number of tasks making up the application execution and correspondingly the amount of the execution performed per task. A finely grained partitioning has many tasks, each with a small fraction of the application execution. Conversely, a coarsely grained partitioning has few tasks, each with a large fraction of the application execution.

Each task involves some overhead, which is largely independent of the amount of application execution performed by the task. For example, the overhead includes managing a task in the task pool. A too finely grained partitioning thus is undesirable, since the overhead of the many tasks consumes a significant fraction of the resources intended for the application execution. Conversely, a too coarsely grained partitioning also is undesirable. With too few or too large tasks, the TS has insufficient flexibility to achieve the desired execution elements. For example, the lengthy execution of a large task might prevent another task from meeting its real-time constraints. The ideal partitioning of an application execution thus lies somewhere between too fine and too coarse.

The addtorial application, using the recursive definition of `addt2(;i,r;)` of Figure 24b), has a very fine grain. Other than the overhead of the task, each execution of `addt2(;i,r;)` performs only a few arithmetic operations. The granularity of the partitioning thus is close to the finest possible. Since this granularity often is too fine, the following subsections thus examine how the recursive partitioning of the addtorial application may be coarsened.

### *3.7.1 Recursion Unrolling*

*Recursion unrolling* essentially is the same as the well known optimization technique of loop unrolling [Compiler Transformations]. By increasing the amount of work executed during each iteration, the number of iterations is decreased. If each iteration corresponds to a task, recursion unrolling thus coarsens the partitioning of an application into tasks.

For example, Figure 33a) shows the Fortran routine `addt2(;i,r;)` of Figure 24b) after applying recursion unrolling. The unrolled version of the routine can be used everywhere in place of the original version. For example, the code of Figure 23a), Figure 25a), Figure 33a) and `addtorial(n;;r)` of Figure 24b) is a complete Fortran application with an execution like that in TSIA. The unrolling triples the amount of work performed in each iteration and task. Correspondingly, the application execution only has a third of the original number of iterations and tasks. Obviously any factor can be used for the unrolling.

Since it essentially replicates the body of the iteration, recursion unrolling generally does not make for nice source code. For example, the routine `addt2(;i,r;)` of Figure 24b) is more readable and compact than that of Figure 33a). Fortunately, it may not be necessary for an application definition to include recursion unrolling. Just as loop unrolling can be performed by a compiler, it would seem that recursion unrolling also could be performed by a compiler. In addition, with a compiler the unrolling factor easily is changed. Such flexibility helps to tune the application execution performance. For example, adaptive computing allows a routine to be compiled with an unrolling suitable for the execution situation.

If for iteration, a loop is more efficient than a recursion, then recursion unrolling also is achieved when a recursion is replaced by a loop. For example, Figure 33b) shows the Fortran routine `addt2(;i,r;)` of Figure 24b) after replacing recursion by a loop. As for the routine of Figure 33a) discussed above, also this unrolled version of the routine can be used everywhere in place of the original version. Similarly, the code of Figure 23a), Figure 25a), Figure 33b) and `addtorial(n;;r)` of Figure 24b) is a complete Fortran application with an execution like that in TSIA.

In the code of Figure 33b), the amount of unrolling is given by the variable `maxunrl`. Since `maxunrl=20`, the unrolling coarsens the granularity of the tasks by a factor

20. If `maxunrl=1`, then there is no unrolling and the `addt2(;i,r;)` of Figure 33b) essentially is the original of Figure 24b).

The best partitioning of an application execution into tasks often depends on the execution situation. Thus ideally `maxunrl` would be a variable whose value would be adjusted by the TS during the course of the application execution.

### *3.7.2 Delegation Unrolling*

In *delegation unrolling*, a delegation to a task is replaced by the execution of the task. The latter is like the synchronous call to a routine as done in an imperative execution. Of course if a task is delegated, it eventually is executed from the task pool. Delegation unrolling thus merely avoids the delegation and thus the overhead of using the task pool.

With delegation unrolling, what would have executed as two tasks, parent and child, thus executes as one task. Delegation unrolling thus increases the amount executed in a task and correspondingly decreases the number of tasks of an application execution. Delegation unrolling thus coarsens the partitioning of an application execution into tasks.

In order to benefit the application execution, delegation unrolling requires that the overhead of executing the task is less than that of a delegation to the task. This would seem to be a safe assumption, since a delegated task eventually is executed.

Delegation unrolling obviously is similar to loop or recursion unrolling. These increase the amount executed in each iteration and correspondingly decrease the number of iterations. The description of delegation unrolling in this subsection thus is similar to that of recursion unrolling in the previous subsection.

The techniques of delegation unrolling and recursion unrolling are independent. The techniques thus may be used singly or in combination in order to coarsen the partitioning of an application into tasks.

An example of delegation unrolling is shown in Figure 34a). It shows the Fortran routine `addt2(;i,r;)` of Figure 24b) after applying delegation unrolling. In the original `addt2(;i,r;)` of Figure 24b), each iteration of the recursion corresponds to a task. In the delegation unrolled `addt2(;i,r;)` of Figure 34a), four iterations of recursion correspond to each task. The unrolling quadruples the amount of work performed in each task. Correspondingly, the application execution only has a fourth of the original number of tasks. Obviously any factor can be used for the unrolling.

The unrolled version of the routine can be used everywhere in place of the original version. For example, the code of Figure 23a), Figure 25a), Figure 34a) and `addtorial(n;;r)` of Figure 24b) is a complete Fortran application with an execution like that in TSIA. As in subsection 3.3.3, a nice demonstration of the application execution is given by the behavior of the stack. The output of the application execution for addtorial(13) is given in Figure 34b). As seen from the output addresses for `s`, the stack grows for three iterations of `addt2(;i,r;)` and on every fourth iteration collapses back to the original size. Obviously each task contains four iterations of the recursion.

As an aside, the previous demonstration also is nice example of ancestor delegation. In particular, it demonstrates that ancestor delegation allows a ia routine to have an arbitrary number of ancestors within the task. In the demonstration, the ia routine `addt2_ia(;i,r;)` typically has four ancestors. In this case, each ancestor is an execution of the routine `addt2(;i,r;)`. Of course in general each ancestor can be an execution of a different instruction.

Delegation unrolling generally does not make for nice source code. For example, the routine `addt2(;i,r;)` of Figure 24b) is more readable and compact than that of Figure 34a). Fortunately, it may not be necessary for an application definition to include delegation unrolling. As demonstrated in the next subsection, TSIA can provide delegation unrolling. This also allows the unrolling factor to be easily changed.

### *3.7.3 Introducing Unrolled Delegation*

*Unrolled delegation* is the ability of a ia routine to provide delegation unrolling. By replacing a delegation to a child task with the execution of the child task, unrolled delegation allows a ia routine to increase the granularity of the parent task.

Like apt delegation and ancestor delegation, unrolled delegation is a property of the implementation of delegation within a task. Also unrolled delegation is not visible outside of a task. These properties of the implementation of delegation thus maintain the black box nature of a task.

For example, unrolled delegation can be provided by the ia routine `addt2_ia(;i,r;)` of Figure 24c). As a demonstration, Figure 35a) introduces unrolled delegation to the ia routine as imitated by the Fortran routine of Figure 25a). With unrolled delegation, a delegation to a task is replaced by the execution of the task. In the original imitation of Figure 25a), each execution of `addt2_ia(;i,r;)` delegates to the task `addt2(;i,r;)`. With unrolled delegation, as imitated in Figure 35a), only every fifth execution delegates to the task `addt2(;i,r;)`. The other four executions of `addt2_ia(;i,r;)` execute `addt2(;i,r;)`. Obviously any factor can be used for the unrolling.

The Fortran imitation of Figure 35a) thus is equivalent in both definition and execution to the ia routine of Figure 24c), including the implementation of unrolled delegation. The code of Figure 23a), Figure 35a), Figure 25b) and `addtorial(n;;r)` of Figure 24b) thus is a complete Fortran application with an execution like that in TSIA. As in subsection 3.3.3, the application execution is nicely demonstrated by the behavior of the stack. The output of the application execution for addtorial(13) is given in Figure 35c). As seen from the output addresses for `s`, the stack grows for four iterations of `addt2(;i,r;)` and on every fifth iteration collapses back to the original size. Clearly each task contains five iterations of the recursion.

Obviously unrolled delegation can provide any amount of unrolling. In the Fortran imitation of Figure 35a), the amount of unrolling is determined by the variable `maxunrl`.

The ideal partitioning of an application execution into tasks may depend on the execution situation. Unrolled delegation can accommodate such a dynamic partitioning. For example, in the Fortran imitation of Figure 35a), the amount of unrolling can be varied by having the TS adjust the value of `maxunrl` during the course of the application execution. Though `maxunrl` appears as a global value in the imitation of the ia routine, TSIA can be implemented such that the TS adjusts a unique and individual `maxunrl`, or its equivalent, for each ia routine. A different and differently varying amount of delegation unrolling thus can be enjoyed by each ia routine.

Unrolled delegation also can be implemented in a fashion different than that imitated in Figure 35a). For example, the imitation of unrolled delegation in Figure 35b) allows a TS to request that a task yield its execution and complete as soon as possible. This is an example of the seventh variation from a normal task execution as discussed in subsection 2.11.7.

In the imitation of Figure 35b), the variable `yield` is set to a random number for the purposes of the demonstration. In reality, `yield` is set by the TS, but the value of the variable `yield` effectively is random as seen from `addt2_ia(;i,r;)`. As above for the variable `maxunrl`, the TS can share a unique and individual variable `yield` with each task.

In the imitation of Figure 35b), the task continues to execute as long as the value of the variable `yield` is `0`. In this case, `addt2(;i,r;)` is executed. Once the value of yield is not `0`, then `addt2_ia(;i,r;)`delegates to the task `addt2(;i,r;)`.

As for the Fortran imitation of Figure 35a), the Fortran imitation of Figure 35b) is equivalent in both definition and execution to the ia routine of Figure 24c), including the implementation of unrolled delegation. The code of Figure 23a), Figure 35b), Figure 25b) and `addtorial(n;;r)` of Figure 24b) thus is a complete Fortran application with an execution like that in TSIA. As above, a nice demonstration of the application execution is given by the behavior of the stack. The output of the application execution for addtorial(13) is given in Figure 35d). As seen from the output addresses for `s`, the stack grows for random iterations of `addt2(;i,r;)` and on the other iterations collapses back to the original size. Obviously each task contains a random number of iterations of the recursion.

As an aside, the above demonstration using the Fortran imitation of Figure 35b) is a simple example of how an application with a determinate definition can have an indeterminate execution, as described in subsection 3.1.8. Since its result is indeterminate, printing the address of `s` in order to observe the stack behavior is assumed to not be a part of the application definition.

### *3.7.4 Unrolled Delegation*

Subsection 3.7.3 introduced unrolled delegation for a very simple ia routine. Unrolled delegation is the ability of a ia routine to replace a delegation to a task by an execution of the task. This subsection argues that unrolled delegation can be provided for any routine.

The example used in this subsection is the ia routine `addt_dc(b,t;;r)` of Figure 26b) and subsection 3.3.4. The example assumes that each ia routine is implemented by IA with the `yield` variable, as discussed in subsection 3.7.3. Each ia routine executes its child tasks, unless asked to yield the execution; in which case the task delegates to its child tasks.

The example begins with the task `addt_dc(1,9;;r)`. Since it is to execute its child tasks, the execution initially proceeds through the code of `addt_dc(b,t;;r)` of Figure 26b) as if it were an imperative execution. Thus `addt_dc(1,5;;rb)` is executed, which in turn executes `addt_dc(1,3;;rbb)`. The items are labelled such that the dependencies between tasks are clear. The execution continues and eventually `addt_dc(1,3;;rbb)` completes with the result `rbb=6`.

At this point, inside the execution of `addt_dc(1,5;;rb)`, the example assumes that the value of `yield` has changed. The task thus must yield the execution. Thus `addt_dc(1,5;;rb)` completes its execution by delegating to the tasks `addt_dc(4,5;;rbt)` and `add(6,rbt;;rb)`. The execution then returns to

addt_dc(1,9;;r), which completes its execution by delegating to the tasks addt_dc(6,9;;rt) and add(rb,rt;;r).

Figure 36 shows the complete execution of the task addt_dc(1,9;;r), as seen from the task pool. The transition from addt_dc(1,9;;r) to the new tasks truly is a single transition since it is the result of a single task. There are no intermediate stages.

As demonstrated by the example, unrolled delegation easily is implemented by IA for any ia routine. The delegation to a task thus is replaced by the execution of the task. Thanks to unrolled delegation, tasks are not unnecessarily delegated. In this achievement, the application definition remains completely determinate, even though unrolled delegation results in an indeterminate application execution. Adopting the description of subsection 3.1.8, this is because the application definition is independent of the unrolled delegation.

The value of the yield variable of course may change at any point during the execution of a task. If the value does not change, then there are two extreme possibilities for unrolled delegation. In the first, the yield variable always requires that a task yield the execution. In other words, each task must delegate to its child tasks. There thus is no delegation unrolling. For example, the execution of the task addt_dc(1,9;;r) then is like that explained in subsection 3.3.4 and illustrated in Figure 27.

In the other extreme possibility, the yield variable never requires that a task yield the execution. In other words, each task executes its child tasks. There thus is full unrolling. For example, the execution of the task addt_dc(1,9;;r) would complete with r=45. A fully unrolled execution is like an imperative execution.

Though probably not the only possible unrolled delegation which may be implemented by IA, the unrolled delegation of this subsection is a simple implementation. Unrolled delegation is very similar to lazy task creation [Lazy Task Creation]. With the caveat that it originates from the process model, not the task model, lazy task creation may suggest alternative implementations of unrolled delegation in IA. Similarly, the efficiency of lazy task creation, as argued elsewhere [Cilk-5], also would seem to hold for unrolled delegation.

### *3.7.5 Minimal Delegation*

An *evaluation task* for an item is simply a name for a task which evaluates the item. For example, given the instruction two(;;o){o=2;}, then two(;;t) is an evaluation task for the item t. Similarly, a *delegation task* for an item is simply a name for a task which delegates the item.

A delegation task thus guarantees that the evaluation of an out requires at least two tasks: the delegation task and its child. In contrast, an evaluation task guarantees that the evaluation of an out requires exactly one task: the evaluation task. Using an evaluation task instead of a delegation task thus reduces the number of tasks of an application execution. The use of the evaluation task, instead of the delegation task, thus coarsens the partitioning of the application execution into tasks. If the finer partitioning into tasks is not required, then such use of an evaluation task is an example of *minimal delegation.*

In general, given a choice of tasks, each of which leads to the evaluation of an item, minimal delegation is the choice which yields just enough partitioning. Due to the overhead of each task, minimizing the number of tasks improves the efficiency of the application execution. On the other hand, the flexibility for the execution is proportional to the number of tasks. Minimal delegation thus trades unneeded flexibility for efficiency. In other words, minimal delegation aims to achieve the best possible partitioning of the application execution into tasks.

Unrolled recursion and unrolled delegation, as discussed in the previous subsections, are examples of techniques used for minimal delegation. The amount of unrolling provides a variety of partitions into tasks.

Furthermore, unrolled recursion and unrolled delegation allow minimal delegation for a determinate application definition. Regardless of the amount of unrolling, even if changing during the execution, the application definition remains the same since it does not depend on the unrolling. Unrolling thus demonstrates that minimal delegation can be compatible with a determinate application definition.

#### *3.7.5.1 Other Techniques*

There are other techniques for minimal delegation. The example discussed here uses the subsection 3.3.4 divide-and-conquer (DC) definition of the addtorial application. The DC ia routine `addt_dc(b,t;;r)` of Figure 26, modified for minimal delegation, is shown in Figure 37b). In the example, evaluation of `r` occurs when `t-b<mindc`. Instead of the delegation task `addt_dc(b,t;;r)`, then the evaluation task `addt_ev(b,t;;r)` is used. The Fortran routine for the evaluation task is shown in Figure 37a).

The Fortran routine `addt_dc(b,t;;r)` of Figure 37c) has the same definition as the ia routine of Figure 37b). The Fortran code of Figure 23a), Figure 26a) and Figure 37c) thus makes up a complete Fortran application. This allows the correctness of the definition to be verified. Of course an execution using the Fortran `addt_dc(b,t;;r)` is not like an execution using IA.

The execution using the ia routine `addt_dc(b,t;;r)` of Figure 37b) may be illustrated by examining the tasks in the task pool. For example consider the task `addt_dc(1,35;;r)`. As illustrated in Figure 38, this task will replace itself in the task pool by three tasks: `addt_dc(1,18;;rb)`, `addt_dc(19,35;;rt)` and `add(rb,rt;;r)`. As further illustrated, each of the two `addt_dc()` delegations tasks replaces itself by an evaluation task `addt_ev()`.

In the example, the value of the variable `mindc` may be adjusted in order to best achieve minimal delegation. For example, in the code of Figure 37, `mindc` has the value `20`. Compared to the original DC of Figure 26, the DC of Figure 37 thus has only $1/20$ of the tasks. The larger the value of `mindc` is, the coarser is the application partitioning into tasks.

The smallest possible value for `mindc` is `1`. Below this value the DC algorithm does not work. If `mindc=1`, then the ia routine `addt_dc` of Figure 37b) essentially is the same as that of Figure 26b). In both cases, the base case of DC is reached when `b==t`.

For the above example of DC, minimal delegation changes the amount of execution performed in the base case. For simplicity, the other examples of DC in this chapter use the simplest and smallest base case. For minimal delegation, larger base cases of course could be introduced.

In the above example, if `mindc` is too large, then the granularity of the application execution is too coarse to meet the execution elements. For example, too few tasks hinders parallelism for an application execution. Similarly, tasks that are too large can spoil real-time constraints. With an extremely large value for `mindc`, the above addtorial applica-tion essentially is the iterative solution of Figure 23b). At this extreme, there are no tasks and thus the TS can provide few execution elements.

#### *3.7.5.2 A Definition Dependent on the Minimal Delegation*

In contrast to the unrolled delegation of subsection 3.7.4, the particular implementation of minimal delegation of the subsection 3.7.5.1 demonstrates that an application definition can depend on the minimal delegation. The application definition there is determinate only if the minimal delegation, via the value of `mindc`, is determinate. Correspondingly, the application definition is indeterminate if the value of `mindc` is indeterminate. As introduced in subsection 3.1.8, sacrificing the determinate application definition may allow for a better application execution performance.

The application definition is dependent on the minimal delegation because `mindc` forces the decision between two different algorithms: the base case iteration of Figure 37a) or the DC of Figure 37b). For example, if `mindc=10`, then `addt_dc(1,4;;r)` results in `r=(((0+1)+2)+3)+4`. In contrast, if `mindc=1`, then `r=(1+2)+(3+4)`.

An indeterminate `mindc` and the resulting indeterminate algorithm may or may not cause the application definition of subsection 3.1.8 to be indeterminate since integer arithmetic is associative and commutative, except in cases of overflow or underflow. In contrast, a similar change for floating point arithmetic, which due to finite precision is neither associative nor commutative [Compiler Transformations], would almost certainly cause an application definition to be indeterminate. Other applications, for example sorting, also may have a different DC algorithm and base case algorithm, which thus similarly could lead to an indeterminate application definition.

As described in subsection 3.1.8, an indeterminate application definition can cause much grief. Thus a technique which causes an indeterministic application definition should be a last resort for achieving minimal delegation. Thus in the above example, `mindc` should be determinate, unless the application execution performance demands otherwise.

The best partitioning of an application execution into tasks often depends on the execution situation. Thus in the above example, if the value of `mindc` is indeterminate, then the value ideally would be adjusted by the TS during the course of the application execution. Perhaps more straightforwardly, since `addt_dc(b,t;;r)` and `addt_ev(b,t;;r)` have the same arguments, the TS could choose directly between the two tasks. In any case, the application developer thus is spared adjusting the value of `mindc` in order to achieve minimal delegation.

### *3.8 The Task Order*

The *task order* refers to the entry of the tasks into the task pool from the application definition. Since a task entering the task pool is the child of a previously executed task, the task order equivalently refers to the order of execution of tasks in the task pool.

Since an application definition is expressed in terms of tasks in the task pool, throttling can change the task order. The possibility of a different task order is part of the indeterminate application execution in TSIA. Since every task order must obey the dependencies between tasks, the application definition of course remains determinate.

As illustrated in Figure 39, the task order of an application execution includes possibilities across several orthogonal axes. As described in section 3.9, depth-first and breadth-first are the two extremes of one axis of the task order. As described in section 3.11, supply-driven and demand-driven are the extremes of another axis of the task order. As described in section 3.20, speculation and conservation are the extremes of a sub-axis of the supply- versus demand-driven axis. As described in sections 3.12 and 3.15 respectively, fusion and fission each affect the task order and thus each can be considered as yet another axis of the task order. In practice, the task order generally lies somewhere between the extremes of each axis.

None of the above axes for the task order uses any information about the interior of any task. Thus as for other properties of TSIA, a task is treated as a black box. Similarly, none of the axes use any information derived from previous executions of similar tasks. Looking into the black box or at the execution history could introduce even more axes of possibilities for the task order, but none of these axes are pursued in this presentation.

The task order has no effect on the granularity of any task. Other than speculative execution, which might execute ultimately irrelevant tasks, the task order thus has no effect on the total number of tasks of the complete application execution. Though as demonstrated below, the order can have a huge effect on the number of tasks in the task pool at any one instant of the application execution.

### *3.9 Depth-First versus Breadth-First Order*

Depth-first and breadth-first are the two extremes of one axis of possibilities for the task order [ADAM]. On this axis, the criteria for the task order depends on the position of a task in the parent-child hierarchy. As demonstrated below, this criteria very effectively manages the number and responsibility of the tasks in the task pool.

A depth-first task order first executes deep into the parent-child hierarchy of tasks making up the application execution. The execution of a task thus is followed by the execution of the first child of that task. The execution of any second child thus follows the execution of the first child and all of its children; similarly for any third child and so on. A depth-first order is illustrated in Figure 40a).

A depth-first order thus executes tasks in the same order as a traditional imperative execution. Since an imperative execution obeys the dependencies between tasks, so does a depth-first execution.

The divide-and-conquer (DC) definition of subsection 3.3.4 of the addtorial application may execute depth-first as illustrated in Figure 41. The depth-first order is the same as that of the traditional imperative execution of the addtorial application using the Fortran code of Figure 23a) and Figure 26a) and c).

As illustrated in Figure 41, the collection of tasks in the task pool may be treated like a stack. This emphasizes the similarity of the depth-first execution to the traditional stack-based imperative execution. The treatment as a stack also clearly illustrates why a depth-first execution also is known as a last-in-first-out (LIFO) execution. In a depth-first execution, the last task to enter the task pool is the first to be executed.

In contrast to a depth-first order, a breadth-first order first executes broadly across the task hierarchy. The execution of a task thus is followed by the execution of a sibling task, which is a task with the same parent. After the last sibling task, a cousin task is executed. Cousin tasks have the same grandparent. The execution continues from siblings to cousins and so on. Only after all the tasks of a generation are executed does a task of the next generation execute. A breadth-first order is illustrated in Figure 40b).

Of course a breadth-first execution only occurs to the extent allowed by the dependencies between tasks. Thus the execution first of the tasks of a generation does not include any task which depends on the child of a sibling. Such a dependence on a nephew arises when a task depends on an item of a sibling, which delegates the item to a child. For example, in Figure 40b), the task `4` is assumed to not depend on the tasks `5`, `6`, `7`, `8` or `9`.

The DC definition of the addtorial application may execute breadth-first as illustrated in Figure 42. As illustrated, treating the collection of tasks in the task pool like a stack reveals why a breadth-first execution also is known as a first-in-first-out (FIFO) execution. In a breadth-first execution, the first task to enter the task pool is the first to be executed.

As explained above, the task hierarchy may be divided into generations. A depth-first order executes only a single task of a generation before executing a task of the next generation. In contrast, a breadth first order executes every task of a generation before executing a task of the next generation.

The difference between a depth-first and a breadth-first execution is dramatic if the execution of a task results in more than one new tasks. This construction of the task hierarchy ends at the base cases. With each generation, a depth-first execution thus causes a linear growth of the number of tasks in the task pool. In contrast, a breadth-first execution causes an exponential growth.

A simple example is the execution of the task `addt_dc(1,32768;;r)` of the DC addtorial application. The task hierarchy has $\log_2 32768 = 15$ generations. At any moment during a depth-first execution, there are at most on the order of $15$ tasks in the task pool. In other words, each generation adds a few tasks to the task pool. In contrast, a breadth-first execution will have at one moment on the order of $32768$ tasks in the task pool. In addition to all the `add()` tasks, these include `addt_dc(1,1;;r1)`, `addt_dc(2,2;;r2)` through to `addt_dc(32768,32768;;r32768)`.

If a single computer processor is to execute a task hierarchy, then a depth-first execution generally is optimal, since the number of tasks in the task pool is kept to a minimum throughout the execution. Since the single processor eventually executes all tasks in the task hierarchy, it does not matter what tasks are in the task pool at any moment during the execution. Using a depth-first execution is natural and familiar; as explained above, it is very much like a traditional imperative execution. This choice is corroborated by the fact that unrolled delegation is depth-first.

If an unlimited number of computer processors are to execute a task hierarchy, then a breadth-first execution generally is optimal, since the number and responsibility of the tasks in the task pool is kept to a maximum throughout the execution. As described in subsection 3.2.2, the responsibility of a task includes all of its items and corresponds to the work of the task and of all of its children and further descendants. With the number and responsibility of the tasks at a maximum, parallel computing is maximized. For simplicity, task overhead and other details are ignored here. For example, given at least 32768 processors, then the evaluation of $\text{addtorial}(32768)$ requires an elapsed time on the order of 15 cycles, where a cycle is the time to execute one task. In contrast, an elapsed time on the order of 32768 cycles is required if only a single processor is used. Of course, the total amount of execution is the same in all cases.

In practice, only a limited number of computer processors are available. In addition, task overhead and other details effectively set a minimum limit of responsibility for a task which may execute at another processor. A parallel execution thus often lies between a breadth-first and a depth-first execution. A simple example implementation [ADAM][Cilk-NOW][Lazy Task Creation] is summarized below. The LIFO end and the FIFO end of the stack also are known as the head and the tail of a double-ended queue (deque) [Cilk-NOW]. The deque also is known as the lazy task queue [Lazy Task Creation].

Each processor has its own task pool. Each processor executes depth-first since that generally optimizes the local execution. Thus if the collection of tasks in the task pool is treated like a stack, then the local processor executes at the LIFO end of the stack. At the LIFO end, the local processor removes a task for execution and adds any resulting tasks.

If a task pool runs out of tasks, then it steals a task from the task pool of another processor. The task is stolen from the FIFO end of the stack. In other words, stealing is performed breadth-first. In general, at the FIFO end of the stack are the tasks with the maximum responsibility. Thus maximized is the execution of the stolen task and all of its children and further descendants.

For example, a task may be stolen from the third state of the task pool shown in Figure 41 for the addtorial application. Stealing from the FIFO end of the stack yields more execution than stealing from the LIFO end since `addt_dc(6,9;;rt)` has more responsibility than `addt_dc(1,2;;rbbb)`.

Maximizing the execution resulting from each stolen task minimizes the number of steals for the complete application execution. Since there presumably is some overhead associated with each steal, breadth-first stealing thus generally optimizes the global execution. The efficiency of such stealing is argued in detail elsewhere [Cilk-NOW].

### *3.10 Benefiting from the Various Uses of an Item*

The use of an item by a task refers to whether the item is evaluated, delegated or ignored. These various uses of an item are introduced in section 3.2 in introducing delegation. In addition to the properties of delegation demonstrated up until this point of the presentation, the various uses of an item also allow for many other benefits to an application.

A simple candidate for such a benefit is in the ia code of Figure 43a), which has an execution illustrated in Figure 43b). The task `f(;y;)` replaces itself by the tasks `a(;;x)` and `b(x;y;)`. Because of the apparent dependence on `x`, only after `a(;;x)` completes can `b(x;y;)` begin execution. It then replaces itself by the task `c(x;y;)`, whose completion completes the responsibility of the original task `f(;y;)`.

In the ia language of this presentation, by default an item of a task is evaluated or delegated. Thus the declaration `b(x;y;)` in Figure 43a) is correct, but does not provide the most precise possible definition. An equivalent, but more precise declaration is `b(del x;del y;)` as in the ia code of Figure 43c). Other than the introduction of the keyword `del`, the ia code of Figure 43c) is identical to that of Figure 43a).

In the ia language of this presentation, the keyword `del` declares that an item is delegated and thus is neither evaluated nor ignored. More generally, the use of an item may be declared explicitly using one of the keywords described in Figure 44. This *use declaration* for an item must agree with the actual use by the task. For example, the routine `a(;;del x){x=2;}` is in error since `x` is evaluated not delegated.

In a ia language different than that of this presentation, instead of an explicit use declaration for an item, the use could be implicitly determined from the code of the routine. For the routine `a(;;x){x=2;}` for example, a ia compiler could recognize that `x` is evaluated.

A more precise application definition can increase the flexibility for the execution. The precise definition of Figure 43c) can have the same execution illustrated in Figure 43b) as for the imprecise definition of Figure 43a). In addition, the precise definition can execute as illustrated in Figure 43d).

The illustration explicitly includes the use declaration `del`. This allows the execution in the task pool to be followed without having to refer to the ia code.

In the execution illustrated in Figure 43d), the task `f(;del y;)` replaces itself by the tasks `a(;;x)` and `b(del x;del y;)`. Because it delegates `x`, the task `b(del x;del y;)` has no dependence on `a(;;x)` and thus can execute independent of `a(;;x)`. In this execution, before `a(;;x)` even begins execution, `b(del x;del y;)` executes and replaces itself by the task `c(x;y;)`. Because of the dependence on `x`, only after `a(;;x)` completes can `c(x;y;)` begin execution and its completion completes the responsibility of the original task `f(;del y;)`.

Via the increased flexibility for the execution, a more precise definition thus may allow TS to better satisfy the execution elements required by the application. In the above example, the more precise definition allows `a(;;x)`, and `b(del x;del y;)` to execute in parallel, as illustrated in Figure 43e). Similarly, in a sequential execution `a(;;x)` and `b(del x;del y;)` may execute in a task order best suited to the resources used for the execution. A larger and more realistic example involving these issues is given below by the N-queens application of subsection 3.10.3.

The following sections demonstrate that the various uses of an item allow for many benefits to an application. As in the above example of this section, some of the benefits help the application execution. Other benefits help the application definition. For example, the precision offered by the various uses allows for some application definitions which are not expressed easily otherwise. Many of the demonstrated benefits help both the application definition and execution.

### *3.10.1 Delegation is a Part of Non-Strict Evaluation*

As demonstrated by the above example of Figure 43, delegation allows a task to execute before an item is evaluated. In functional computing, executing a function before an argument is evaluated is known as non-strict evaluation [Parallel Functional]. Delegation thus is a part of non-strict evaluation. As described in subsection 3.19.1, the conditional in is the other part of non-strict evaluation.

In functional computing, the alternative to a non-strict evaluation is known as a strict evaluation. In this case, an argument is evaluated before executing the function. An evaluated item in IA thus is a variation of strict evaluation.

Figure 45 illustrates the above correspondences between functional computing and TSIA. Strict evaluation corresponds to evaluation. Non-strict evaluation corresponds to delegation and the conditional in.

In functional computing, implementations of non-strict evaluation include lazy evaluation [Unboxed values] and future order [Future Order]. The implementation of non-strict evaluation in TSIA seems to have interesting similarities with other implementations, especially with some examples [Lazy Evaluation], but a such comparison of the implementations is beyond the scope of this presentation.

### *3.10.2 Use Declarations versus Strictness Annotations*

Some functional computing programming languages provide strictness annotations [Parallel Functional]. An application definition thus can influence the task order or its equivalent.

Since each can influence the task order, a strictness annotation of functional computing is similar to a use declaration of TSIA. The degree of similarity depends on the particular example of strictness annotation.

A use declaration is very similar to a global strictness annotation [Concurrent Clean]. Each influences the execution order for all uses of a function or task, since each appears in the definition of the function or task. A similar example of a global strictness annotation would seem to be the unboxed value of non-strict functional computing [Unboxed values].

A use declaration is less similar to other types of strictness annotations. In contrast to a use declaration or a global strictness annotation, a local strictness annotation appears in the use of a function and thus influences the execution order of only that single use of the function [Concurrent Clean]. A future construct is another example of a local strictness annotation [Lazy Task Creation][Multilisp][Speculative].

A use declaration or strictness annotation generally is used to move away from the default strictness. Thus, where the default is strict, as in Scheme or as in the ia language of this presentation, a future construct or a use declaration allows a non-strict evaluation. Similarly, where the default is non-strict, as with lazy evaluation, an unboxed value allows a strict evaluation.

### *3.10.3 The N-Queens Problem*

The N-queens problem demands that N elements of an N*N board be chosen such that each row, column and diagonal of the board contains at most one chosen element. The N-queens problem derives from the game of chess, where a queen can move on the board within its row, column or diagonals.

Figure 46 shows a Fortran application which determines the number of solutions for the N-queens problem. Similar N-queens applications may be found elsewhere [Cilk-5] [Structured Programming]. Instead of recording just the number of solutions, the application of Figure 46 easily could be extended also to record each of the solutions.

Because any solution contains one element in each row, a solution simply can be stored in a one-dimensional array, such as `board[1:n]`. Then each element of the solution corresponds to an element of the array, with the index `i` and value `board[i]` corresponding to the row and column, respectively.

The N-queens application thus makes use of arrays, but the use is simple since each of its arrays can be treated as an item. Full IA support for an array is described beginning in section 3.13.

In the N-queens application, the algorithm first determines the solutions for 1 element on a 1*N board. Obviously there are N possibilities and obviously each is a solution. For each of the solutions to the 1*N problem, the algorithm then determines the solutions for 2 elements on a 2*N board. For each of the solutions to the 1*N problem, obviously there are N possibilities for the 2*N board but only some are solutions. For each of the solutions to the 2*N problem, the algorithm then determines the solutions for 3 elements on a 3*N board. This recursion continues until the algorithm has determined the solutions for N elements on a N*N board.

The N-queens application of Figure 46 thus consists principally of the two routines `nattempts` and `attempt`. The divide-and-conquer routine `nattempts` is given a solution to the `b_size*N` problem. If `b_size` is N then the attempt is recorded as a solution. If not, `nattempts` attempts the N possibilities for solutions to the `(b_size+1)*N` problem. Each of these attempts is dealt with by the routine `attempt`. If the attempt is not safe for a solution, it is abandoned. Otherwise, the attempt is passed to `nattempts` for the next recursion.

The Fortran routines `nqsols`, `nattempts` and `attempt` of Figure 46 are rewritten in ia in Figure 47. As always, the rewriting from Fortran to ia preserves the application definition, while yielding possibilities for the execution. In particular, the declaration of the delegation of `sols` in `nattempts(...;del sols;)` and `attempt(...; del_ign sols;)` provides a more precise application definition than that in Fortran. As described below, the parallelism inherent in the algorithm of the N-queens application provides for a very dramatic demonstration of the benefits of the precise application definition in ia.

The ia N-queens application, consisting of the ia code of Figure 47 and the remaining Fortran routines of Figure 46, has a simple execution. Upon execution, `nqsols(n;;sols)` replaces itself by `nattempts(n,0,0;sols;)`, which replaces itself by `n` pairs of the tasks `testsafe` and `attempt(...;del_ign sols;)`. Each pair of tasks is an attempt for a solution. Since `sols` is delegated by each of these pairs, the attempts are independent and thus can execute in parallel. Each execution of `testsafe` merely evaluates `safe`. Assuming the attempt is safe for a solution, each `attempt(...;del_ign sols;)` replaces itself by the pair of tasks `makeboard` and `nattempts(...;del sols;)`. Within the pair, `makeboard` and `nattempts(...;del sols;)` are independent and thus also can execute in parallel, but this parallelism is minor compared to that introduced by the execution of `nattempts(...;del sols;)`. Since `nattempts(...;del sols;)` delegates `sols`, the attempts continue to be independent. Each execution of `makeboard` merely makes a copy of a `b_size`*N solution. The recursion continues as each `nattempts(...;del sols;)` expands each solution for `b_size`*N into N attempts for `(b_size+1)`*N. The attempts continue to be independent.

The only property of the attempts which is not independent is when `nattempts(...;del_ign sols;)` has a solution to the N*N problem and thus replaces itself by `incr(;sols;)`. Obviously the `incr(;sols;)` tasks cannot execute in parallel.

In order to provide further freedoms to the execution, one could imagine replacing the declaration `incr(;int a;);` in the ia code of Figure 47 by `incr(;commut int a;);`. The use declaration `commut` indicates that the operation is commutative and thus the `incr(;sols;)` tasks can execute in any order. This also requires the modified declarations `attempt(...;del commut sols;);` and `nattempts(... ;del commut sols;);`. Such refinements of TSIA are beyond the scope of this presentation.

Thus other than the execution of `incr(;sols;)`, all properties of the attempted solutions are independent. Thus the ia N-queens application unleashes the exponential parallelism inherent in the algorithm of the N-queens application. For example, in an ideal world and ignoring the time for the `incr(;sols;)` tasks, the ia N-queens application can complete in time proportional to N if given $N^{M-1}$ computers, where M increases with increasing N. In contrast, the sequential execution on a single computer requires a time proportional to $N^M$.

### *3.11 Supply-Driven versus Demand-Driven Order*

Demand-driven and *supply-driven* are the two extremes of one axis of possibilities for the task order. On this axis the task order obeys to various degrees the dependencies between the tasks of the application. In previous presentations supply-driven is known as data-driven.

Lazy evaluation is an example of the demand-driven order [Lazy Evaluation] [Unboxed values]. Future order is an example of the supply-driven order [Future Order]. While dataflow once was synonymous with supply-driven [ADAM][Coarse Grain Data-flow][Dataflow Architectures], dataflow can include examples of both the demand-driven and the supply-driven order [Dataflow]. The implementation of the supply-driven and the demand-driven order in TSIA is described below.

The supply-driven order executes any task for which the items are supplied or available. Some items, such as an evaluated in or a computer processor, often are not available, at least not immediately. Other items, such as an out or a delegated in, are almost always available. Since it is the minimum possible criteria for the execution of any task, the supply-driven order effectively corresponds to imposing no additional constraints on the task order. The supply-driven order thus executes any task that can be executed. This task order thus is driven by the availability of the data of a task, or more generally, by the availability of the items of a task.

The demand-driven order executes any task which is demanded. Such a demanded task has at least one of its out demanded as an evaluated in of another demanded task. Alternatively, at least one out of the task is demanded as an evaluated out of the application. The requirement for execution that a task be demanded is of course in addition to the requirement that its items are available. Thus as for every other task order, a demand-driven order places constraints on the task order beyond those of the supply-driven order.

An in of a task is an out of a preceding task. While an evaluated in is a demand for the preceding task, a delegated in is not. Thus the ability to delegate an item introduces the distinction between the demand-driven and the supply-driven order. Without delegation, all in are evaluated and thus all task orders satisfy the demand-driven order. Thus without delegation, the demand-driven order imposes no additional constraints on the task order and thus the demand-driven order is the same as the supply-driven order.

The execution of the example application fragment of section 3.10 and Figure 43c) allows for a demonstration of a demand-driven order and of a supply-driven order.

Given the task `f(;del y;)`, Figure 43d) illustrates a demand-driven order. A demand for the out `y` is assumed. The task `f(;del y;)` replaces itself by the tasks `a(;;x)` and `b(del x;del y;)`. Since there is no task which demands `x`, the task `a(;;x)` is not yet executed. The task `b(del x;del y;)` does not demand `x`, since it delegates, not evaluates, `x`. Since there is a demand for the out `y`, the task `b(del x;del y;)` executes and replaces itself by the task `c(x;y;)`. Since there now is a demand for x, the task `a(;;x)` executes and when it completes the task `c(x;y;)` can execute.

Given the task `f(;del y;)`, Figure 43e) illustrates an execution with a supply-driven order. The task `f(;del y;)` replaces itself by the tasks `a(;;x)` and `b(del x;del y;)`. Since all the items are available, the tasks `a(;;x)` and `b(del x;del y;)` execute in parallel. The task `c(x;y;)`, resulting from the execution of the task `b(del x;del y;)`, can then execute.

The above supply-driven order assumes that there are enough computing resources available for the parallel execution of the tasks `a(;;x)` and `b(del x;del y;)`. This may not be the case. For example, the application may execute on a single computer processor. In this case the supply-driven order is not constrained by the items of the application definition and instead is constrained by the items of the resources. Vice versa, there may be more resources available than tasks available from the application definition. Then the supply-driven order is constrained by the items of the application definition and not by the items of the resources.

Even if the constraints of a given task order are satisfied, a task only can execute if its items are available. The items include those of the application definition as well as those of the resources. By definition, an item in short supply is not immediately available to all tasks requiring the item. For example, if the example application fragment of Figure 43c) is executed using one computer, the supply-driven order has no preference between the task order of Figure 43b), where `a(;;x)` executes before `b(del x;del y;)`, or that of Figure 43d), where `b(del x;del y;)` executes before `a(;;x)`. In general, since its constraints already are satisfied, a task order offers no prescription when an item is in short supply. Instead the task order must be based on additional criteria, which may include another task order.

For example, given an infinite supply of resources and assuming that time is the only cost, then the supply-driven order is the best task order since it minimizes the time for an application execution. Each task executes as soon as its items are available. Any other task order, including the demand-driven order, only can do worse, since the additional constraints may prevent a task from executing as soon as its items are available. Unfortunately in reality, resources rarely are infinite in supply and time rarely is the only cost. A task order such as the demand-driven order can help live within these limited means.

If an application is considered as a sequence of tasks, then the supply-driven order has a simple first-to-last behavior. The ins of the application allow the first task to execute, which in turn allows the second task to execute and so on until the last task executes and provides the outs of the application.

In contrast, the demand-driven order initially has a last-to-first behavior. The outs of the application demand the last task to execute, which in turn demands the second last task to execute and so on until the first task demands the ins of the application. Of course, once this chain of demand is established, then the dependencies between the tasks result in the obligatory first-to-last execution.

In short, the demand-driven order maximally obeys the dependencies between tasks. In contrast, the supply-driven order minimally obeys the dependencies between the tasks. Thus demand-driven and supply-driven are the two extremes of one axis of possibilities for the task order. On this axis, the task order obeys to various degrees the dependencies between the tasks of the application.

As introduced in section 3.20, the position of the task order on this axis has a particularly large impact for an application which involves ignored ins or outs.

### *3.12 Fusion*

Items separated in the application definition may be fused in the execution [Fusion]. Fusion thus allows a structured application definition to have an efficient execution.

Fusion introduces an axis of possibilities for the task order. On this axis, the task order depends on the locality of items shared by tasks.

Fusion is introduced here by comparing an execution with fusion to one without. The application definition fragment of Figure 48a) does not allow for fusion, while that of Figure 48b) does. The two fragments are identical, except that `del` use declarations define the latter fragment more precisely. Though their definitions are very similar, the two fragments can have very different executions.

The tasks `a2(;x,y;)` and `b2(;x,y;)` of Figure 48a) don't allow fusion. The execution of the two tasks `a2(;x,y;); b2(;x,y;)` is illustrated in Figure 48c). The implicit default `del_eva` use declaration allows `a2(;x,y;)` and `b2(;x,y;)` to evaluate `x` and `y`. Thus `b2(;x,y;)` can only begin execution once `a2(;x,y;)` and all its descendants have completed.

In contrast, the tasks `a2(;del x,del y;)` and `b2(;del x,del y;)` of Figure 48b) allow fusion. The execution of the two tasks `a2(;del x,del y;); b2(;del x,del y;)` is illustrated in Figure 48d). Because each of them delegates `x` and `y`, the execution of `b2(;del x,del y;)` is independent of that of `a2(;del x,del y;)`. In their respective executions, the task `a2(;del x,del y;)` replaces itself by the tasks `a1(;x;)` and `a1(;y;)`; the task `b2(;del x,del y;)` replaces itself by the tasks `b1(;x;)` and `b1(;y;)`. In Figure 48d), the rearrangement of tasks in the task pool illustrates the fusion of `a1(;x;)` with `b1(;x;)` and of `a1(;y;)` with `b1(;y;)`.

Of course, the precise definition of Figure 48b) also can have the execution illustrated in Figure 48c), as for the imprecise definition of Figure 48a).

As demonstrated in subsections 3.12.2 through 3.12.4, fusion can improve locality and thus can improve the efficiency of the execution.

#### *3.12.1 Delegation Increases the Possibilities for Fusion*

As demonstrated in the above example of Figure 48, delegation increases the possibilities for fusion by leaving an item free to be fused. A delegated item has the freedom to not necessarily be combined with any other item of the routine. In subsection 3.15.4, this is described as the fission of the routine. In contrast to a delegated item, an evaluated item implicitly is combined with all other evaluated items of the routine.

#### *3.12.2 Fusion for Efficient Parallel Computing*

A parallel execution of the imprecisely defined tasks `a2(;x,y;)` and `b2(;x,y;)` of Figure 48a) is inefficient since fusion is unavailable. The execution of `a2(;x,y;); b2(;x,y;)` illustrated in Figure 48c) is shown in Figure 48e) as a parallel execution, assuming the use of two computers.

In Figure 48e), the execution is inefficient since it involves the item `y` in much unnecessary communication. The communication is unnecessary since it is an artifact of the imprecise application definition.

In Figure 48e), the items `x` and `y` are assumed to be *gathered* initially on `COMPUTER 1`, as required for the execution of `a2(;x,y;)`. The items `x` and `y` then are *scattered* across `COMPUTER 1` and `2` for the execution of `a1(;x;)` and `a1(;y;)`, respectively. The items `x` and `y` then are gathered on `COMPUTER 1` for the execution of `b2(;x,y;)`. The items `x` and `y` then are scattered across `COMPUTER 1` and `2` for the execution of `b1(;x;)` and `b1(;y;)`, respectively.

In contrast to the above inefficient execution, the precise application definition of Figure 48b) allows for fusion and thus for an efficient execution. The execution of `a2(;del x,del y;)`; `b2(;del x,del y;)` illustrated in Figure 48d) is shown in Figure 48f) as a parallel execution, again assuming the use of two computers. The execution is efficient since it involves no unnecessary communication.

In Figure 48f), the items `x` and `y` are assumed to be scattered initially across `COMPUTER 1` and `2`, respectively. Since both tasks `a2(;del x,del y;)` and `b2(;del x,del y;)` delegate both items `x` and `y`, these items needs not be gathered for the execution of these tasks. Also no communication is required for the execution of the fused tasks `a1(;x;)` and `b1(;x;)`on `COMPUTER 1`. Similarly no communication is required for the execution of the fused tasks `a1(;y;)` and `b1(;y;)`on `COMPUTER 2`.

Since it avoids unnecessary communication, fusion thus allows for efficient parallel computing.

### *3.12.3 Fusion for Efficient Hierarchical Storage*

Fusion allows for efficient hierarchical storage.

The example presented here assumes only two levels of hierarchical store. As usual, the primary store is smaller and faster than the secondary store. In reality, the two stores could be any pair in the usual hierarchy stretching across registers, multiple levels of cache, physical memory, virtual memory, disk and tape storage.

The example also assumes that an item is in the primary store when it is evaluated by a executing task. Furthermore, the example assumes that only the item `x` or the item `y`, not both, can fit into the primary store. An execution evaluating the items `x` and `y` thus must swap `x` and `y` between the primary and the secondary store.

Since fusion is unavailable, an execution of the imprecisely defined tasks `a2(;x,y;)` and `b2(;x,y;)` of Figure 48a) makes inefficient use of a hierarchical store. The execution of `a2(;x,y;)`; `b2(;x,y;)` illustrated in Figure 48c) is shown in Figure 48g) using a primary and a secondary store.

In Figure 48g), the execution is inefficient since it swaps the items `x` and `y` more often than necessary. The extra swapping is unnecessary since it is an artifact of the imprecise application definition.

In Figure 48g), the items `x` and `y` are assumed to be initially in the primary and the secondary store, respectively. This is unchanged by the execution of the task `a2(;x,y;)` since it does not actually evaluate either `x` or `y`. The store also remains unchanged by the execution of the task `a1(;x;)`. The execution of the task `a1(;y;)` requires swapping the items `x` and `y`. The store then remains unchanged by the execution of the tasks `b2(;x,y;)` and `b1(;y;)`. The execution of the task `b1(;x;)` requires swapping the items `x` and `y`. Thus in total `x` and `y` are swapped twice.

In contrast to the above inefficient execution, the precise application definition of Figure 48b) allows for fusion and thus for an efficient execution. The execution of `a2(;del x,del y;);b2(;del x,del y;)` illustrated in Figure 48d) is shown in Figure 48h) again using a primary and a secondary store. The execution is efficient since it involves no unnecessary swapping.

In Figure 48h), the items `x` and `y` again are assumed to be initially in the primary and the secondary store, respectively. This is not changed by the execution of the tasks `a2(;del x,del y;)` and `b2(;del x,del y;)`, nor by the execution of the fused tasks `a1(;x;)`and `b1(;x;)`. The execution of the fused tasks `a1(;y;)` and `b1(;y;)` requires swapping the items `x` and `y`. Thus in total `x` and `y` are swapped only once.

The symmetry of items in TSIA raises another example of fusion for efficient hierarchical storage. In the execution of `a2(;del x,del y;);b2(;del x,del y;)` as in Figure 48, what if `x` and `y` both fit into the primary store, while only one of the instructions `a1` or `b1` fits into the primary store? The hierarchical store is best used by the execution illustrated in Figure 48c) or equivalently in Figure 48g). Since `a1(;x;)`and `a1(;y;)`remain fused as do `b1(;x;)`and `b1(;y;)`, this execution swaps `a1` and `b1` only once, which is the minimum possible. In contrast the execution illustrated in Figure 48h) uses the hierarchical store inefficiently since it swaps `a1` and `b1` twice.

The above example illustrates that fusion may or may not rearrange the application definition for its execution.

The above example also illustrates that not all fusions are simultaneously possible. The fusion of `a1(;x;)`with `a1(;y;)`and of `b1(;x;)`with `b1(;y;)`does not simultaneously allow the fusion of `a1(;x;)`with `b1(;x;)`and of `a1(;y;)`with `b1(;y;)`. Thus the execution efficiency sometimes has to choose the most valuable fusions and forego other fusions. In this example the choice is based on whether the locality of `a1` and of `b1` is more valuable the locality of `x` and of `y`.

The above examples are brought full circle by the following example. Again assuming that only `a1` or `b1` fits into the primary store, fusion ensures that `a1` and `b1` are swapped only once in the execution of the two tasks `ab1(;del x;);ab1(;del y;)` for `ab1(;del x;){a1(;x;);b1(;x;);}` and `ab1(;del y;){a1(;y;); b1(;y;);}`. The fusion of `a1(;x;)` with `a1(;y;)` and of `b1(;x;)` with `b1(;y;)` rearranges the application for its execution.

### *3.12.4 Fusion for Efficient Parallel Input/Output*

Input/output (I/O) is the transfer of data or other items. An example of I/O is the transfer between the permanent storage of a disk and the volatile storage of a computer. One example of parallel I/O thus involves multiple computers and/or multiple disks [Passion].

Fusion allows for efficient parallel I/O. For example, this may fuse a task with part of a data file. This is consistent with fusion as the possibility of rearranging the application definition for its execution, since a data file is a part of the application definition. The symmetry of items in TSIA sees little difference between a file of instructions and a file of data.

An example candidate for parallel I/O is the application given by the ia code of Figure 49 and Figure 48b). In the example, `xy.dat` is a file containing the two items `x` and `y`. With parallel I/O, `x` and `y` can be stored on the disks of two different computers.

With fusion, the example can execute as illustrated in Figure 48f). The execution is efficient since neither `x` nor `y` is communicated.

Without fusion, the execution would be inefficient due to much unnecessary communication. This again assumes parallel I/O has stored `x` and `y` on the disks of two different computers. The `x` and `y` then first have to be gathered onto one computer for the execution of `a2(;x,y;)`. The execution then would proceed as illustrated in Figure 48e).

### *3.12.5 Fusion for Locality*

As demonstrated in the above three subsections, fusion may rearrange an application definition. Fusion thus can improve locality and thus improve the efficiency of the application execution.

Improving locality avoids unnecessary communication. As demonstrated above, examples of such communication can be between computers or between the different storage hierarchies.

While the above demonstrations fuse only pairs of tasks, of course an arbitrary number of tasks may be fused.

As discussed in subsection 3.15.5, fusion is very valuable for a structured application definition. This section has demonstrated the value of fusion for routines. Subsection 3.13.3 demonstrates the value of fusion for arrays.

## *3.13 Arrays*

An application definition consists of items. A structure is a convenient collection of items. Section 3.17 provides a general description of structures and their role in a structured application definition.

An array is a kind of structure. This section demonstrates that a IA can support the one-dimensional arrays of an application. Extending the support to two- and higher-dimensional arrays would seem to be straightforward.

The IA support for arrays demonstrated in this presentation makes heavy use of divide-and-conquer algorithms. There probably are other ways for a IA to support arrays, but such alternatives are not pursued in this presentation. For example, an alternative technique to those of this thesis is the use of expression templates [Blitz++]. It would seem that expression templates are complementary to the techniques of this presentation.

As introduced in subsection 3.1.1, the item is the fundamental unit of application definition. The structure is not. While an entire structure may be treated as an item, support of a structure implies that its individual items are supported. Thus for example, the array `a[3]` often is supported as the three items `a[1]`, `a[2]` and `a[3]`.

In order to support a structure, IA thus must understand how items make up the structure. Such an understanding includes expressing a structure in terms of smaller sub-structures. For example, this includes recognizing that the array `a[1:6]` can consist of the two sub-arrays `a[1:3]` and `a[4:6]`.

The understanding of a structure by a IA begins in the ia language. Thus a language must provide a means to identify the items and sub-structures of a structure. Given the array `a[10]` for example, then in the ia language of this presentation `a[3:7]` identifies a sub-array with five elements. The greater the means in a language for such identification, the greater the ability for a IA to support a structure. Other than simple means for routines and for one-dimensional arrays, the possibilities for such identification are not explored in this presentation.

### 3.13.1 The Array-Based Addtorial Definition

A Fortran definition of the addtorial is shown in Figure 50a) and b). The code may be combined with `program addtprog` of Figure 23a) for a complete Fortran application. The `addtorial(n;;del r)` routine of Figure 50a) consists of an intermediate array `a[n]` and two routines which pass over the array. The divide-and-conquer (DC) routine `vseq(1,n;;del a[n])` sets the array elements `a[1]` through `a[n]` to the numbers `1` through `n`, respectively. The DC routine `vsum(n,del a[n];;del r)` sums all the elements of the array `a[n]`, returning the result in the scalar out `r`.

As explained in section 3.16, an array-based application definition is defined in terms of arrays. Because of its intermediate array, the addtorial definition of Figure 50 thus is named here the array-based addtorial definition.

Figure 50c) shows `addtorial`, `vseq` and `vsum` rewritten as ia routines. The rewriting from Fortran to ia does not change the application definition. Instead, the rewriting introduces new possibilities for the application execution. Beginning in subsection 3.13.3, some of these possibilities for array execution are presented.

### 3.13.2 Array Notation in this Presentation

The ia language of this presentation allows some conveniences for identifying elements and sub-arrays of arrays. In essence, information for a sub-array available implicitly need not be repeated explicitly. Thus for example, given the array `a[n]` with `n=1`, then in ia code `a` is `a[1]` is `a[1:1]`. Similarly, given the declaration of `vseq` in Figure 50c), then in ia code `vseq(w,k,a)` is `vseq(w,k,a[1:k])` and likewise `vseq(w,n-k,a[k+1])` is `vseq(w,n-k,a[k+1:n])`.

In contrast to its ia code, this presentation does repeat information when discussing individual tasks. Referring to `vseq(1,9,a[1:9])` instead of to `vseq(1,9,a[1])`, explicitly presents all the items of a task, including those of an array. All the items of a task thus are explicit, without having to refer to the definition of the routine.

As already evident in the ia code of Figure 11a) for the classic application and as repeated in Figure 50c), the ia language of this presentation numbers array elements `1` through `N`, as in Fortran. This simplifies the relationship between Fortran and the ia language of this presentation. Despite this choice for this presentation, the author generally considers more natural the `0` through `N-1` numbering, as in the C programming language. Similarly, for multi-dimensional arrays in this presentation, the left-most index is fastest varying index, as in Fortran. Outside of this presentation, the author generally prefers the right-most index as the fastest varying index, as in C and as in the digits of a number.

### 3.13.3 Fusion for an Efficient Array Execution

Fusion allows for the efficient execution of an application involving arrays. Fusion is described in section 3.12. As demonstrated in this subsection, fusion easily incorporates arrays since an array merely is a structure of items.

The array-based addtorial definition of Figure 50 demonstrates for an array the efficient execution resulting from fusion. The execution of the task `addtorial(9;;del r)` is illustrated in Figure 51a). The task replaces itself by the tasks `vseq(1,9;;del a[1:9])` and `vsum(9,del a[1:9];;del r)`. The execution of these DC tasks and their descendants eventually results in the base cases, like the task `set(1;;a[1])`, which each evaluate an individual array element.

In the ia language of this presentation, a use declaration for a structure declares the use of each item of the structure. Thus the instruction `vseq(1,n;;del a[n])` delegates each element of `a[n]` to a child task.

As illustrated in Figure 51a), the tasks of the array-based addtorial definition may be rearranged, fusing tasks which evaluate the same array element. The fallacy in treating a structure as indivisible long is known [Multilisp]. For example, `set(1;;a[1])` thus is fused with `set(a[1];;r1)`. Originally named `r`, the out `r1` and others have been renamed for clarity in the presentation.

The value of fusion for an efficient execution is illustrated in Figure 51b). There `addtorial(9;;r)` has a parallel execution using nine computers. Having one computer for each array element simplifies the demonstration, but otherwise is incidental. As a result of `vseq(1,9;;del a[1:9])`, the array elements `a[1]` through `a[9]` are scattered across the nine computers for the execution of the nine base case tasks `set(1;;a[1])` through `set(9;;a[9])`, respectively. Fusion allows the scattered array also to be used by the nine base case tasks `set(a[1];;r1)` through `set(a[9];;r9)`, which result from `vsum(9;;del a[1:9];;del r)`. The execution thus is efficient since it does not involve any unnecessary communication. Possible inefficiencies of other executions are discussed in section 3.12 in discussing fusion.

In addition to the equivalent of the above parallel computing example, section 3.12 also demonstrates fusion for an execution involving a hierarchical store. The hierarchical store example also can be made for the array-based addtorial definition. In such an execution, fusion allows each array element, `a[1]` through `a[9]`, to be swapped only once into the primary store. Once swapped in, both evaluations of the array element occur. For example, for the array element `a[1]` the two evaluations are the tasks `set(1;;a[1])` and `set(a[1];;r1)`. Since swapping thus is minimized, fusion again has allowed for an efficient execution. The efficient use of a hierarchical store is important for an application using large arrays [CFD].

In short, the two passes of `vseq(1,9;;del  a[1:9])` and `vsum(9;;del a[1:9];;del r)` across the array `a[9]` have been fused to a single pass. Thus fusion allows for the efficient execution of an application using arrays.

### *3.13.4 Supporting Arrays*

IA allows one or more arrays to be elements of a routine. IA support for such array routines can be divided into two parts. The first part concerns the use of an array routine. The second part concerns the definition of such an array routine and is described in the next subsection.

Since it can be used like any other routine, the use of an array routine is convenient in IA. In particular, an array routine provides encapsulation and performance. Encapsulation allows an array routine to be used without knowing the internal details of the routine. The use of an array routine does not trade convenience for performance. As explained in sections 3.12 and 3.15, an array routine has an efficient execution thanks to fusion and fission.

### 3.13.5 Using Routines for Partitioning

The definition of an array routine is convenient in IA. As introduced in the previous subsection, an array routine has one or more arrays as elements. In order to provide an efficient execution, including fusion and fission and parallelism, TSIA requires an array routine to be partitioned. Partitioning refers to the division of the responsibility of the parent task across multiple child tasks. Since partitioning is a constraint, it can only decrease the convenience of the application definition. The decrease in convenience is small, since IA includes a very powerful mechanism to partition an array routine. Thus despite partitioning, the definition of an array routine is convenient in IA.

The routine is a powerful mechanism in IA to partition an array routine. An array routine thus is partitioned into other routines.

The routine of course has been used to partition applications since the beginning of computing and presumably will continue to do so until the end of computing. The routine thus is an old and familiar mechanism with many powerful characteristics. The routine allows for arbitrarily complex partitions. The routine allows the use of a traditional debugger to examine each partition. The routine encapsulates each partition. The routine defines precisely all items belonging to each partition. This precision also applies to any structure which is an element of the routine.

The requirement by TSIA for partitioning of course applies to the entire application definition, not just array routines. Of course, also this partitioning is performed using routines. For example, in section 3.3 recursive routines partition iteration. The routine as a mechanism for partitioning is emphasized here because array routines demonstrate how delegation extends the power of routines for partitioning.

Partitioning does not evaluate an item. Instead, partitioning eventually yields the tasks that will do the evaluation. A partitioned item thus is a delegated item and thus need not be combined with any other item of the task. In contrast, an evaluated item implicitly is combined with all other evaluated items of the task. Combining the evaluated items for the execution of the task is a form of communication. Such communication is unnecessary if evaluation is not required. Efficient partitioning thus requires that the partitioned items of a routine are identified and treated efficiently. Use declarations provide such identification and TSIA treats partitioned items efficiently. For example, for the parallel execution of the array-based addtorial definition illustrated in Figure 51b), once array `a[9]` is scattered by the partitioning of `vseq(1,n;;del a[n])`, it need not be unnecessarily gathered for the partitioning of `vsum(n,del a[n];;del r)`.

### 3.13.6 Convenient and Efficient Reduction

A reduction reduces multiple items to fewer items. The reduction is achieved by applying some operation to the original items. A reduction also is known as a fold. Often the original items are the elements of a structure. Often only a single item results from a reduction. For example in the array-based addtorial definition of Figure 50, the task `vsum(9;;del a[1:9];;del r)` reduces the array `c[n]` to the number `r` by summing the elements of the array.

IA allows for the convenient definition of a reduction since it can be programmed like any other routine.

As illustrated in Figure 51b) for the parallel execution of the array-based addtorial definition, TSIA also can efficiently execute reduction. The execution is efficient since it involves a minimum of communication between the computers of the execution. The binary tree reduction resulting from the DC algorithm of `vsum(9;;del a[1:9] ;;del r)` can be very efficient. For example it can maximize parallelism. Thus in the parallel execution of Figure 51b), while `add(r1,r2;;r12)` involves `COMPUTER 1` and 2, `add(r3,r4;;r34)` involves `COMPUTER 3` and 4 and so on.

*3.13.7 Jacobi's Iterative Relaxation Solves Laplace's Equation*

In a boundary value problem, the solution $\Phi$ must satisfy some equation in the region inside a boundary and $\Phi$ must have the given fixed values on the boundary. The solution $\Phi$, including the boundary values, may be described by an array `a`. To solve Laplace's equation, $\nabla^2\Phi = 0$, Jacobi's iterative relaxation may be used. Initially the array `a` contains the boundary values and arbitrary values for the region inside the boundary. Each iteration relaxes the array `a` towards the solution by updating each element inside the boundary. For a two-dimensional region and array, the relaxation is given by $a_{i+1}(j,k)=(a_i(j-1,k)+a_i(j+1,k)+a_i(j,k-1)+a_i(j,k+1))/4$, where the subscript numbers the iteration. Obviously the relaxation propagates the boundary values throughout the region. The solution given by the array `a` has converged when an iteration does not significantly change the values for the region. Jacobi's iterative relaxation converges very slowly and thus is not practical, but its simplicity and its similarity to practical methods makes for a good demonstration.

Figure 52 shows a Fortran definition of Jacobi's iterative relaxation for a one dimensional (1D) region and array. The 1D definition is nonsensical because it has an explicit solution. Given the boundary values `a(1)` and `a(n)`, the elements inside the region have the solution `a(k)=k*(a(n)-a(1))/(n-1)`. Nonetheless, the 1D definition makes for a good demonstration since it is simpler than the very similar 2D or higher dimensional definitions.

The Jacobi definition consists principally of the two routines `jacobi` and `relax` of Figure 52b). The routine `jacobi` simply returns if convergence or the maximum number of iterations has been achieved. If not, `jacobi` recurses by replacing itself by the tasks `relax` and `jacobi`.

The DC routine `relax` performs the relaxation $a_{i+1}(k)=(a_i(k-1)+a_i(k+1))/2$ on each element. In order to naively test for convergence, the routine `relax` also determines the maximum absolute change of any element due to the relaxation. In order to partition itself into two independent tasks, `relax` makes copies `mk` and `pk` of the elements at the partition boundary. Because of these copies, `relax` takes the boundary elements as explicit arguments `m` and `p`, instead of as elements of the array `a[n]`.

The Fortran routines `jacobi` and `relax` of Figure 52b) are rewritten in ia in Figure 53. As always, the rewriting from Fortran to ia preserves the application definition, while yielding possibilities for the execution. The Jacobi definition of Figure 52a) and c) and Figure 53 introduces three array properties of TSIA, beyond those of the array-based addtorial definition of subsection 3.13.1 and Figure 50.

Firstly, the Jacobi definition demonstrates that the partitioning of an array may have some elements present in more than one partition. As described above, `relax` makes a copy of such an element. Figure 54 illustrates a parallel execution using four computers for an array `a[1:42]`. As illustrated, the boundary elements of the sub-array on each computer are copied to the neighboring computer for each iteration of the relaxation. Since only these required elements are communicated, the execution is efficient.

Secondly, the Jacobi definition demonstrates that arbitrarily many tasks may be fused and that the number need not be known at compile-time. As illustrated in Figure 54, the sub-array on each computer is used by arbitrarily many iterations of `relax`.

Thirdly, the Jacobi definition demonstrates further support by TSIA for reduction. The routine `relax` uses reduction to determine the maximum absolute change of any element due to the relaxation. This demonstrates that reduction can originate within another routine. In the array-based addtorial definition reduction is a stand-alone routine.

### *3.13.8 Quicksort*

Hoare's quicksort is a sorting algorithm which uses recursion. From the array to be sorted, an element is chosen as the pivot. The remaining elements are partitioned into two sub-arrays, those sorted before the pivot and those sorted after the pivot. The elements of the before sub-array, the pivot and the elements of the after sub-array thus are sorted with respect to each other. Sorting the original array thus is reduced to sorting the before sub-array and sorting the after sub-array. Each sub-array is sorted by applying the above quicksort algorithm. This divide-and-conquer (DC) recursion ends when a sub-array has fewer than two elements. Further information on quicksort is well described elsewhere [C].

Figure 55b) shows a Fortran definition of quicksort. The use of the sorting routine is demonstrated by the simple Fortran application of Figure 55a).

Like other examples in this presentation, the quicksort of Figure 55b) is written in the delegation style. Rewriting the Fortran code in ia thus involves little more than minor changes in syntax. As always, the ia code has the same definition as the Fortran code, but allows for execution elements. Since the rewriting from Fortran to ia is straightforward and has been demonstrated for many previous examples, the corresponding code is not given for the quicksort example. Instead, Figure 56 shows the ia declarations for the routines of the quicksort definition. The items of each task thus are clearly and explicitly defined. For many of the subsequent examples of this presentation, the rewriting from Fortran to IA is treated in a similarly abbreviated fashion.

The definition of quicksort of Figure 55b) sorts an array of integers into ascending order. More general implementations, for arbitrary types of arrays and for arbitrary sorting orders, are described in subsection 3.25.2 and elsewhere [C].

The quicksort of Figure 55b) begins with the routine `dcqsort`. The name `dcqsort` avoids conflict with the routine `qsort` of standard libraries. For simplicity in this presentation, `dcqsort` uses the first element of the array as the pivot. If a different pivot `a[?]` were somehow chosen, it simply could be swapped with the first element using the task `swap(;a[?],a[1];)`. In order to obey the delegation style, `dcqsort` uses the subsequent routine `dcqsort2`. For simplicity in the code of `dcqsort2` when `m.eq.0`, then `swap(a(1),a(1))` is assumed to perform correctly, despite violating the no alias rule of Fortran. Similarly, when `m.eq.n-1`, the call to `dcqsort(0,a(n+1))` assumes there is no checking of array bounds.

The routine `dcqsort` and its subsequent routine `dcqsort2` perform the quicksort algorithm exactly as described above.

The quicksort algorithm specifies the result of the partitioning of the original array, but does not specify an algorithm for the partitioning. Other than the arbitrary choice of the pivot, thus only the definition of the partitioning differs between a traditional quicksort definition and the delegation style definition of Figure 55b). Here the partitioning routine `dcpart` defines a DC algorithm. The divided parts of `dcpart` are combined by the routine `flop`. This in turn uses the routine `dcswap`, which also defines a DC algorithm. Since the quicksort algorithm itself also is a DC algorithm, the quicksort definition of Figure 55b) includes three DC algorithms.

The quicksort definition of Figure 55b) demonstrates that an application using TSIA can partition an array based on the values of its elements. The fact that the array is partitioned is very obvious in the ia declarations of Figure 56; the entire array is never an evaluated item of any task of the quicksort definition.

The value partitioning of an array is an array property of TSIA beyond those demonstrated by the array-based addtorial definition and by the Jacobi definition. In those applications, the partitioning of an array uses only the indices of the elements, not the values of the elements. To be more exact, it is the `dcpart` routine of the quicksort definition which value partitions an array. The `dcqsort` and the `dcswap` routines each index partition an array, like the array-based addtorial definition and like the Jacobi definition.

Having partitioned the array and its array routines, fusion allows for an efficient execution. As always, this implies that the execution involves no unnecessary communication. A simple demonstration is the sorting of an arbitrarily large file of integers stored across multiple computers. The ia code for this application is shown in Figure 57 and uses the quicksort definition of Figure 55b) rewritten in ia. An example of the efficient communication is the movement of the elements towards their sorted positions. All the movement occurs via the task `swap(a[i],a[j])`, where `a[i]` and `a[j]` are elements to be interchanged. TSIA can fuse `swap(a[i],a[j])` to a computer where either `a[i]` or `a[j]` is stored at that moment of the sort. Thus for any movement, the only communication is the required communication between the two involved computers.

### *3.13.9 Mergesort*

Mergesort is a sorting algorithm. Like quicksort, mergesort also uses divide-and-conquer (DC) recursion, but the two algorithms are very different. In mergesort, the array to be sorted is divided in two. Each half is sorted by recursively applying the mergesort algorithm. The recursion ends when a sub-array has fewer than two elements. The two halves are not sorted with respect to each other, but since each half is sorted, the two halves easily are merged in order to yield the sorted original array.

Figure 58 and Figure 59a) show a delegation style Fortran definition of mergesort. The ia declarations for the routines are shown in Figure 60. Like the quicksort definition of Figure 55b), the mergesort definition sorts an array of integers into ascending order.

By replacing the use of `dcqsort` by `msort`, the simple Fortran application of Figure 55a) can demonstrate the use of the mergesort definition. Similarly, `msort` coded in ia can be used in the application of Figure 57 to sort an arbitrarily large file of integers.

The mergesort definition of Figure 58 and Figure 59a) begins with the routine `msort` which performs the mergesort algorithm exactly as described above. The mergesort algorithm specifies the result of merging the two sorted halves of the original array, but does not specify an algorithm for the merging. Thus only the definition of the merging differs between a traditional mergesort definition and the delegation style definition of Figure 58 and Figure 59a). Here the merging routine `merge` defines a DC algorithm. Details may be found in the code and comments of Figure 58 and Figure 59a). The routine `merge` uses the routines `find` and `rcycle`. Each routine uses linear recursion. The routine `find` is very similar to binary search, presented in subsection 3.13.10.

The routine `rcycle` cycles the elements of an array and its code is shown in Figure 59a). In order to obey the delegation style, the routine `rcycle` uses the routines `rcycle2` and `rcycle3`. The iterative routine `cycle` is the origin of the routines `rcycle`, `rcycle2` and `rcycle3`. The routine `cycle` is shown in Figure 61a) in order to clarify the relationship between `rcycle` and `rcycle2` and `rcycle3`. The routines also demonstrate how an iterative routine like `cycle` may be rewritten in the delegation style. Figure 59b) illustrates an execution of `rcycle(6,2;a[1:6];)`. The essentially identical execution of `cycle(6,2;a[1:6];)` is illustrated in Figure 61b).

The mergesort definition partitions its array and array routines, both by index and by value of the array elements. Fusion and other properties of TSIA thus allow for an efficient execution of mergesort.

The mergesort definition does not introduce any significant array properties of TSIA beyond those of the array-based addtorial definition, the Jacobi definition and the quicksort definition. Instead, the mergesort definition merely demonstrates that the array properties of TSIA are suitable for a variety of array applications.

### *3.13.10 Binary Search*

Binary search is a search algorithm for sorted items. This presentation assumes that the items an array of integers in ascending order. In binary search, the argument to be found is compared to the middle element of the array. If the comparison is successful, the search ends. Otherwise, based on the result of the comparison, either the first half or the second half of the remaining array is searched. The search ends with a successful comparison or when the ultimate remaining half has no elements. Further information on binary search can be found elsewhere [C].

Figure 62b) shows a recursive delegation style Fortran definition of binary search. The ia declarations for the routines are shown in Figure 62c). The use of the search routine is demonstrated by the simple Fortran application of Figure 62a).

The binary search definition of Figure 62b) consists of the two routines `bs(m,p,del a[m:p],del v;;del i)` and `bs1(del m,del p,del a[m:p],v,k,ak;;del i)`. By consisting of two routines, not one routine, the definition evaluates in each iteration of the algorithm just the array element `a[k]`, i.e. `ak`, compared against the argument `v`. For a search of an N element array, binary search compares the argument to at most $\ln_2 N$ of the elements. The binary search definition thus evaluates only these $\ln_2 N$ elements of the N element array.

Thus like the above array-based addtorial, jacobi, quicksort and mergesort definitions, at no point does the binary search definition evaluate the entire array. Thus for example, just as the above quicksort or mergesort definitions can efficiently sort an arbitrarily large array, the binary search definition can efficiently search an arbitrarily large array. In contrast, evaluating the entire array is unnecessary and would prevent such an efficient execution.

Like the mergesort definition, the binary search definition does not introduce any significant array properties of TSIA beyond those of the array-based addtorial definition, the Jacobi definition and the quicksort definition. Instead, the binary search definition also merely demonstrates that the array properties of TSIA are suitable for a variety of array applications.

### *3.14 Transparent Delegation*

As introduced in section 2.6, due to referential transparency the same outcome results from executions of the same task or from essentially the same tasks. For example, the execution of `a(b;;c)` and `a(b;;d)` result in `c` and `d` having the same value since the only ins, `a` and `b`, are the same for the two tasks. After executing `a(b;;c)`, executing `a(b;;d)` thus is unnecessary. Instead, a TS can give `d` the value of `c`.

*Transparent delegation* is the recognition that a delegated in of a task does not affect the outcome of its execution. Since a delegated in is not evaluated, two tasks essentially may be the same even if their delegated ins differ. Delegated ins thus increase the applicability of referential transparency.

A simple example of transparent delegation is allowed by the routine `f(del x;;del y){g(x;;z);h(z;;y);}`. The execution of the two tasks `f(del r;; del s)` and `f(del t;;del u)` essentially results in the same outcome. The task `f(del r;;del s)` results in the children `g(r;;z)` and `h(z;;s)`, while `f(del t;;del u)` results in the children `g(t;;w)` and `h(w;;u)`. Since the single in of the two tasks is the same `f`, the two tasks must result in the same outcome. Thus after executing `f(del r;;del s)`, executing `f(del t;;del u)` is unnecessary. Instead, `f(del t;;del u)` can be replaced by a copy of the children of `f(del r;;del s)` with the delegated ins `r` and `s` replaced by the delegated ins `t` and `u`.

Transparent delegation thus is concerned only with the outcome of tasks. Transparent delegation need not know how the execution of a task arrives at its outcome. Thus as for other properties of TSIA, transparent delegation can treat a task as a black box.

#### *3.14.1 A Motivation for Transparent Delegation*

The array-based addtorial definition of subsection 3.13.1 and Figure 50 includes the divide-and-conquer (DC) routine `vsum(n,del a[1:n];;r)`. This subsection uses this routine to motivate transparent delegation for DC algorithms involving arrays.

For example, the task `vsum(6,del a[1:6];;r)` replaces itself in execution by the three child tasks `vsum(3,del a[1:3];;r13)` and `vsum(3,del a[4:6] ;;r46)` and `add(r13,r46;;r)`. Since in the execution of a task there is no distinction between delegated ins, the two `vsum` child tasks essentially are the same. This is even more obvious if the sub-array `a[4:6]` is renamed `b[1:3]`, for example.

Continuing with the execution, `vsum(3,del a[1:3];;r13)` replaces itself by the three tasks `vsum(1,del a[1:1];;r11)` and `vsum(2,del a[2:3];;r23)` and `add(r11,r23;;r13)`.

Since `vsum(3,del a[1:3];;r13)` essentially is the same as `vsum(3,del a[4:6];;r46)`, the latter need not execute. Instead, transparent delegation allows the latter task to be replaced by the outcome of the former task, with the delegated ins `a[1:3]` replaced by `a[4:6]`. Transparent delegation requires the items of the array to be considered individually. Without being executed, the task

vsum(3,del a[4:6];;r46) thus can be replaced by the three tasks vsum(1,del a[4:4];;r44) and vsum(2,del a[5:6];;r56) and add(r44,r56;;r46). The items r13, r46, and so on have been renamed for the convenience of this presentation.

### *3.14.2 Symmetric-Divide-and-Conquer (SDC)*

The transparent delegation demonstrated above for vsum is made more feasible and valuable if the original DC vsum of Figure 50 is rewritten using a *symmetric-divide-and-conquer* (SDC) algorithm as in Figure 63a). An SDC algorithm is symmetric since it divides the original problem into essentially equivalent parts. Since the code of Figure 63a) is given in Fortran, it may be combined with the code of Figure 50a) and b) and Figure 23a) for a complete Fortran application. If rewritten in ia, the SDC vsum of Figure 63a) obviously has the same prototype as the original DC vsum of Figure 50c).

The routine vsum of Figure 63a) makes transparent delegation very feasible. Its child tasks vsum(k,del c[1:k];;r1) and vsum(k,del c[k+1:2k+1];;r2) identify in source code that the two tasks essentially are the same. The two tasks have the same evaluated ins, vsum and k. Recognizing that the two tasks essentially are the same thus does not involve the value of any item. A compiler for the ia language thus may identify the transparent delegation at compile-time. Such identification only requires recognizing that the exact same items are evaluated by more than one task. Given such a compilation, when a vsum task is executed, its children and its transparent delegation information enter the task pool.

The routine vsum of Figure 63a) makes transparent delegation very valuable. Since it follows a DC algorithm, a vsum task results in generations of vsum tasks, with an exponentially increasing number of vsum tasks in each generation. Because of the symmetric division, all vsum tasks of a generation essentially have the same outcome. With transparent delegation, the SDC routine vsum of Figure 63a) thus requires the execution of only one vsum of each generation. The outcome of each executed vsum task also essentially is the outcome of each of the other vsum tasks of that generation.

An example of such generations is demonstrated by the execution of vsum(6,del a[1:6];;r) illustrated in Figure 63b). In the first generation there is only one vsum task and obviously it must execute. In the second generation only one of the two vsum tasks must execute. In the third generation, only one of the four vsum tasks must execute. Thus in total, of the seven vsum tasks in the illustrated execution, only three must execute. The outcome of the other four is given by transparent delegation.

In the execution illustrated in Figure 63b), transparent delegation replaces the task vsum(3,del a[4:6];;r46) by its three child tasks vsum(1,del a[4:4];;r44) and vsum(1,del a[5:5];;r55) and add3(a[6:6],r44,r55;;r46). Transparent delegation then replaces the first two children by the tasks set(a[4:4];; r44) and set(a[5:5];;r55), respectively. In an alternative execution, transparent delegation could directly replace the original task vsum(3, del a[4:6];;r46) by its three ultimate tasks set(a[4:4];; r44) and set(a[5:5];;r55) and add3(a[6:6]r44,r55;;r46). In general, transparent delegation, allows but does not require the manipulation of intermediate tasks like vsum(1,del a[4:4];;r44) and vsum(1,del a[5:5];;r55).

In general, a DC algorithm for $N$ items requires on the order of $N$, that is $\Theta(N)$, tasks to divide the work. If the algorithm is SDC, transparent delegation requires only $\Theta(\ln(N))$ of these tasks to be executed.

The value of transparent delegation is at least twofold. Firstly, the execution of $\Theta(\ln(N))$ tasks provide a TS with a compact description of the work to be performed on $N$ items. Secondly, replacing a task by its outcome presumably usually requires less effort than executing the task.

Of course, the ultimate work performed by a task and its descendants is left unchanged by transparent delegation. In the above example of `vsum`, the work is performed by `set`, `add` and `add3` tasks. Delegation is not work; evaluation is work. Delegation merely is the coordination or partitioning of work. In this sense, the task `vsum` performs no work, instead it partitions work. Transparent delegation thus merely reduces the effort of partitioning work.

The earlier or higher in the task hierarchy transparent delegation occurs, the greater the reduction in effort of partitioning work. For example, the routine `relaxtwice` of subsection 3.15.3 and Figure 65 includes two child tasks `relaxonce(n-2,a[1],a[n],del a[2:n-1];;del b[2:n-1])` and `relaxonce(n-2,a[1],a[n],del b[2:n-1];;del a[2:n-1])`. Since their evaluated ins are the same, these two child tasks essentially are the same. For one of the `relaxonce` tasks, transparent delegation thus can avoid the entire effort of partitioning work. Instead, the outcome of the effort is copied from the execution of the other `relaxonce` task. Transparent delegation thus also is important outside of DC routines.

### *3.14.3 Symmetric-Divide-and-Conquer (SDC) Algorithms*

As described in the previous subsection, transparent delegation for a DC algorithm is better if the algorithm is a symmetric-divide-and-conquer (SDC) algorithm. This subsection demonstrates that a variety of DC algorithms may be replaced by corresponding SDC algorithms.

Obviously, a DC algorithm involving arrays is an SDC algorithm if the arrays are restricted to having $N=M^K$ elements. Here M is the division performed by each generation of the algorithm. Then K is the number of generations required to divide the array. Achieving SDC algorithms through the use of such restrictions is not further pursued in this presentation.

In the previous subsection, the DC `vsum` routine of Figure 50 is replaced by the SDC `vsum` routine of Figure 63a). Three further examples of similar replacements are given below. As for `vsum`, the code for each SDC routine is given in Fortran and thus may replace the corresponding original DC Fortran routine in a complete Fortran application. Also as for `vsum`, if rewritten in ia, each SDC routine has the same prototype as the original DC routine.

The Jacobi definition of subsection 3.13.7 and Figure 52 includes the DC routine `relax(n,del m,del p;del a[1:n];del e)`. A corresponding SDC `relax` is given in Figure 64a). Its children of interest are two `relax` tasks. If `n` is even then `relax(k,del m,del pk;del a[1:k];del em)` and `relax(k,del mk,del p;del a[k+1:n];del ep)` are the same, other than their delegated items. Similarly, if `n` is odd then `relax(k,del m,del ok;del a[1:k];del em)` and

relax(k,del ok,del p;del a[k+2:n];del ep) are the same, other than their delegated items.

The second example uses the quicksort definition of subsection 3.13.8 and Figure 55, which includes the DC routine dcpart(n,del p;del a[1:n];del m). A corresponding SDC dcpart is given in Figure 64b). Its children of interest, dcpart(k,del p;del a[1:k];del m1) and dcpart(k,del p;del a[k+1:2*k];del m2), are the same, other than their delegated items.

The quicksort definition also includes the DC routine dcswap(n;del a[1:n],del b[1:n]). A corresponding SDC dcswap is given in Figure 64c). Its children of interest, dcswap(k;del a[1:k],del b[1:k]) and dcswap(k;del a[k+1:2*k],del b[k+1:2*k]), are the same, other than their delegated items.

### *3.15 Fission*

In a structured application definition, a given structure may not need to exist in its entirety in the application execution. *Fission* is the recognition of such a structure and its exploitation for an efficient application execution [Deforestation][Deforestation Short Cut][Fusion].

This presentation introduces the name fission. Unlike the current name deforestation [Deforestation], the name fission does not imply any particular kind of structure. As described in subsection 3.15.5, fission and fusion are related and thus the name fission is well matched to the name fusion.

Because it affects the task order, fission can be considered as yet another axis of possibilities for the task order. On this axis, the task order is affected by the independence of items of a structure.

#### *3.15.1 Fission for the Array-Based Addtorial Definition*

The fission of an array easily is demonstrated using the array-based addtorial definition of subsection 3.13.1 and Figure 50. As explained in subsection 3.13.3 and Figure 51, in the execution the task set(1;;a[1]) may be fused with the task set(a[1];;r1), the task set(2;;a[2]) with set(a[2];;r2) and so on. Neither the array a[1:9] nor any of its elements are an item of any other task. After the above fusion, fission recognizes that there is no relationship between the individual array elements a[1] through a[9]. Each array element is just a short-lived item between two tasks. The fact that the elements belong to an array is convenient for the definition, but after fusion is irrelevant for the execution. Since it is irrelevant during the execution, the relationship between the elements need not be obeyed. Thus the array a[1:9] need not exist in its entirety in the application execution.

For example, the above tasks could be rewritten as set(1;;x1) with set(x1;;r1), set(2;;x2) with set(x2;;r2) and so on. Thus in the parallel execution of Figure 51b), the execution obviously does not maintain any relationship between x1 and x2 and x3 and so on, since they are the usual independent items. Equivalently, there is no relationship between the original items a[1] and a[2] and a[3] and so on. For all effective purposes the array a[1:9] need not exist in the execution. In this example, fission thus achieves an efficient parallel execution, since no resources are wasted scattering an unnecessary array a[1:9] across the computers.

Similarly, the above tasks of the array-based addtorial definition could be rewritten as `set(1;;y)` with `set(y;;r1)`, `set(2;;y)` with `set(y;;r2)` and so on. Thus in a sequential execution of Figure 51a), the execution reuses the same item `y` for each pair of tasks. Fission thus has removed the `array a[1:9]` and has replaced it by the single item `y`. Thus with fission `addtorial(n;;r)` only requires space for `y`, independent of the value of `n`. In contrast, without fission `addtorial(n;;r)` requires space proportional to the value of `n` for the array `a[1:n]`. Fission thus achieves an efficient execution, wasting no resources on an unnecessary array `a[1:n]`. Thanks to fission, despite using an intermediate array `a[1:n]` in the definition, the array-based addtorial definition has an execution as efficient, in particular as space efficient, as the other addtorial definitions of this presentation, including the iterative definitions.

### *3.15.2 Fission for the Classic Application*

A simple demonstration of fission also is allowed by the classic application of subsection 3.1.4 and Figure 11. Fission recognizes that there is no relationship between the individual array elements of the array `a[1:N]` nor between those of the array `b[1:N]`. The fact that the elements belong to the array `a[1:N]` and `b[1:N]` is convenient for the application definition since it compactly describes the input file `a.dat` and the output file `b.dat`, respectively. Thanks to fission, neither array need exist in entirety in the application execution. For example, in an execution using a single computer, the execution requires only enough memory for a single element `a[k]` and a single element `b[k]` of each array. The same memory can be re-used for the task `produce(a[k];;b[k])` for each of the `N` elements.

### *3.15.3 Fission for the Array-Based Jacobi Definition*

Like the above two demonstrations, the array-based Jacobi definition of Figure 65 also allows for a simple demonstration of fission. The array-based definition of Figure 65 is a variation on the Jacobi definition of subsection 3.13.7 and Figure 52. The array-based Jacobi definition of Figure 65 uses an intermediate array `b[1:n-1]` which may be fissioned. Since such a use of an intermediate array is common in present-day mainstream programming, the array-based Jacobi definition provides an example of a strong motivation for fission.

For the array-based Jacobi definition, Figure 66 illustrates an execution of the task `relaxtwice(n;del a[1:50];e)`. As illustrated, its execution and the execution of its divide-and-conquer (DC) children results in seven tasks. After fusion, three of the tasks use the items `b[2:25]` of the intermediate array. The remaining items `b[26:49]` are used by three other tasks. The remaining task does not use the array `b`. Fission recognizes that the use of the items `b[2:25]` is completely independent of the use of the items `b[26:49]`. During the execution, there thus is no need for the entire array `b[2:49]`.

For example, in a sequential execution it is sufficient to create the sub-array `b[2:25]` for the first three tasks and to reuse the sub-array for the second three tasks. Similarly, in a parallel execution using two computers, `b[2:25]` used on the first computer for the first three tasks needs no relationship whatsoever to `b[26:49]` used on the second computer for the second three tasks.

Since all six tasks are DC and all delegate `b[]`, only an arbitrarily small sub-array of `b[]` need be created for a sequential execution. The size of such a sub-array depends on the execution situation. Similarly, for a parallel execution, `b[]` can be broken into arbitrarily many fully independent sub-arrays. Thus the execution of `relaxtwice(n;del a[1:50];e)` for `n` arbitrarily large, only requires an arbitrarily small intermediate array `b[]`.

As demonstrated above, fission allows for the efficient execution of an application using an intermediate array. This efficiency ultimately supports a structured application definition. Using an intermediate array to structure an application definition is part of the array-based definition described in section 3.16.

### *3.15.4 Delegation allows the Fission of a Routine*

Now that fission has been introduced, the result of the delegation of an item by a routine can be described as the fission of that routine.

For an example this presentation returns to the routine `b(del x;del y;)` of section 3.10 and Figure 43c) which introduced the declaration of a delegated item. Because of its delegated items, the routine `b(del x;del y;)` need not exist in its entirety in the application execution. In other words, the items `b`, `x` and `y` need not be assembled for the execution of the routine. Instead, the item `b` need only be assembled with references to `x` and `y`. A contrasting example is the routine `b(x;y;)` of Figure 43a) which requires the items `b`, `x` and `y` to be assembled for its execution.

Because the delegation of the items `x` and `y` is declared, IA can recognize that the routine `b(del x;del y;)` need not exist in its entirety for execution. As described in sections 3.10 and onwards, IA exploits such routines for an efficient application execution. This treatment by IA of the routine `b(del x;del y;)` corresponds exactly to the definition of fission given at the beginning of this section. The delegation of an item by a routine thus allows the fission of that routine.

Though not named fission there, sections 3.10 and onwards thus include many examples of the fission of a routine.

### *3.15.5 The Cooperation between Fusion and Fission*

In a structured application definition, an item may be an element of many structures. For example, in the routine `a2(;del x,del y;){a1(;x;);a1(;y;);}` of section 3.12 and Figure 48, the item `x` is an element of the routine `a2(;del x,del y;)` and of the routine `a1(;x;)`.

As described above in this section, fission recognizes and exploits a structure whose execution does not require its items to be assembled. For example, the execution of the routine `a2(;del x,del y;)` does not require the assembly of the items `a2`, `x` and `y`. Instead, the execution only requires the assembly of `a2` with references to `x` and `y`.

As described in section 3.12, fusion combines items in order to improve the locality and thus the efficiency of the execution. The execution of the routines `a2(;del x,del y;)` and `b2(;del x,del y;){b1(;x;);b1(;y;);}` is described in that section. The fission of the routines `a2(;del x,del y;)` and `b2(;del x,del y;)` allows the tasks `a1(;x;)` and `b1(;x;)` to be fused, thus improving the locality of the item `x`.

Roughly speaking, fission thus frees an item from a structure of the definition. Possibilities thus are created for the fusion of that item with other items for an efficient execution.

Similarly, fusion can create possibilities for fission. For example, subsection 3.15.1 describes an example where the fusion of the tasks `set(1;;a[1])` with `set(a[1];;r1)` and so on allows the complete fission of the array `a[1:9]` into its nine constituent elements.

In short, fission and fusion cooperate to help provide a structured application definition with an efficient execution. Fission splits structures of the definition into items. Fusion combines items for an efficient execution. Fission and fusion thus allow items conveniently structured in the definition to be re-arranged for an efficient execution. At least in spirit, the cooperation between fission and fusion would seem to have at least one precursor [Fusion].

### *3.16 An Array-Based Definition*

An array-based application definition uses arrays to structure the application definition. A simple example is the array-based addtorial definition of subsection 3.13.1 and Figure 50. The use of the intermediate array `a[1:n]` allows for the structured definition `addtorial(n;;r){vseq(1,n;;a);vsum(n,a;;r);}`. A motivation for such a definition is the modularity gained by separating the generation of the sequence `1` through `n` from its summation. A similar example, expressed in ia code as `{fill_primes(n;;p[1:n]);   print_numbers(n,p[1:n];;);}`, is well motivated elsewhere [Structured Programming].

Another example is the array-based Jacobi definition of subsection 3.15.3 and Figure 65. The intermediate array `b[1:n-1]` of the routine `relaxtwice` separates the routine `relaxonce`, performing the Jacobi relaxation, from the routine `dcabsdiff`, determining the change between relaxations. Again such modularity is a motivation for an array-based definition.

More motivation for an array-based application definition may be found elsewhere [APL]. In addition, many applications with a list-based definition can have a similar array-based definition [Series]. Thus much of the motivation for a list-based application definition [Deforestation][Deforestation Short Cut][Fusion][Series], also motivates an array-based definition. For example, such a list-based or array-based application definition may implement a multi-pass algorithm [Fusion].

Respectively, subsection 3.13.3 and section 3.15 demonstrate that fusion and fission support the efficient execution of an array-based application definition. The need for fusion and fission also is described elsewhere [APL][Array Operation Synthesis][SAC]. Similarly, fusion and fission support the efficient execution of a list-based application definition [Deforestation][Deforestation Short Cut][Fusion][Series]. For example, fusion and fission reduce a multi-pass algorithm to a single-pass execution [Fusion].

A particular variation of an array-based application definition is described below in subsection 3.16.1. A different variation is described in section 3.24.

#### *3.16.1 A Deductive Array-Based Definition*

In a *deductive* array-based definition the elements of an array are defined in terms of other elements of the array or in terms of the elements of another array. These two alternatives are described in the following two subsections, respectively.

#### *3.16.1.1 Defined in Terms of Other Elements of the Array*

If the elements of an array are defined in terms of other elements of the array, then a deductive array is a variation on recursion. An example is the ia code of the deductive array-based addtorial definition shown in Figure 67a).

In the ia language of this presentation, `add(1,a[i-1];;a[i])` of Figure 67a) declares tasks which deductively evaluate the elements of the array `a[0:n]`. Since the index `i` is local to the declaration, `add(1,a[i];;a[i+1])` would be an equivalent declaration. In order to initialize the deduction, the elements `a[0]` and `s[0]` each are declared to have the value `0`. Such initialization corresponds to the base case of recursion. As a result of the deduction, for the array `a[0:n]` each element `a[k]` has the value `k`. For the array `s[0:n]` each element `s[k]` has the value `0+1+2+..+k`.

It would seem that a TSIA can support a deductive array-based definition since the dependencies between the elements of an array seem little more complicated than the other dependencies managed by a TSIA.

A Fortran imitation of the deductive array-based addtorial definition is shown in Figure 67b). The code may be combined with `program addtprog` of Figure 23a) for a complete Fortran application. As throughout this chapter, while the ia and the Fortran definitions are similar, their executions can be very different.

An execution of the ia definition is illustrated in Figure 67c) for the task `addtorial(9;;r)`. Its execution results in tasks and the items `a[0]` and `s[0]` in the task pool. As illustrated, `a[0]` and `add(1,a[i-1];;a[i])` yield `a[1]`. Since `a[0]` no longer is required it can be discarded. Then `a[0]`, `s[0]` and `add(a[i],s[i-1] ;;s[i])` yield `s[1]` and then `s[0]` may be discarded. Figure 67c) thus illustrates how the execution advances from `s[0]` to `s[1]`. Such advances are repeated until `s[9]` is evaluated and the execution completes with `r is s[9]`.

The illustrated execution is one of many different possible executions. The illustrated execution fuses `add(1,a[i-1];;a[i])` and `add(a[i],s[i-1];;s[i])`. In this execution fission can recognize and exploit that only two elements of the array `a[0:9]` and only two elements of the array `s[0:9]` need exist at any point of the execution.

In an alternative execution the entire array `a[0:9]` first could be evaluated and only then the array `s[0:9]`. Other alternative executions lie between these extremes. The particular execution chosen by TSIA depends on the execution situation. For example, the execution of `addtorial(n;;r)` is unlikely to first evaluate `a[0:n]` if `n` is so large that `a[0:n]` would consume a large fraction of the available memory.

For any of the different possible executions fission can recognize and exploit that neither the array `a[0:n]` nor the array `s[0:n]` need have its elements in consecutive locations in memory. The relative locations of `a[i-1]` and `a[i]` are irrelevant to `add(1,a[i-1];;a[i])`. Similarly, the relative locations of `s[i-1]` and `s[i]` are irrelevant to `add(a[i],s[i-1];;s[i])`.

As another example of a deductive array-based definition, Figure 68a) shows the ia code of a deductive array-based Fibonacci definition. The ia code with `add(a[i-2] ,a[i-1];;a[i])` and `a[0]=0` and `a[1]=1` corresponds exactly to the definition of the Fibonacci numbers. Each element `a[n]` of the deductive array corresponds to the `n`th Fibonacci number. A Fortran imitation of the deductive array-based Fibonacci definition is shown in Figure 68b).

#### 3.16.1.2 *Defined in Terms of the Elements of Another Array*

A deductive array-based definition may be defined in terms of the elements of another array. An example is the ia code shown in Figure 69 for the deductive array-based classic application. The definition of Figure 69 is essentially that of subsection 3.1.4 and Figure 11, except that deduction makes implicit the iteration over all elements.

## 3.17 *A Structured Application Definition*

An application definition consists of items. A *structured application definition* has its items in structures. This section argues that a IA allows for a structured application definition.

### 3.17.1 *Structures*

A structure is a convenient collection of items. A routine, an array, an object, a tuple, a list, a stack, a queue, a record, a tree and a set are examples of kinds of structures.

As demonstrated in section 3.13, a IA can support the arrays of an application. As described below in subsection 3.17.2, IA support for routines is demonstrated throughout this chapter. Subsection 3.17.3 considers extending IA support of routines and arrays to other kinds of structures.

While the ultimate elements of a structure are items, the immediate elements of a structure may be other structures. For example, an array may be one of the elements of a routine.

The introduction to arrays at the beginning of section 3.13 also applies to other kinds of structures supported by a IA.

### 3.17.2 *Routines*

A routine is a kind of structure. The elements of a routine include the instruction and its arguments. As described in sections 3.26 through 3.29, other items also may be elements of a routine.

In many ways a routine is a structure like any other. Perhaps most importantly, like other kinds of structures, a routine consists only of its elements.

As a structure, a routine can be supported like other kinds of structures. Three examples follow. Section 3.12 describes fusion for routines while subsection 3.13.3 describes fusion for arrays. Section 3.15 describes fission for arrays while subsection 3.15.4 describes fusion for routines. Section 3.24 implements streams using arrays or the arguments of routines.

In one respect, a routine differs from other kinds of structures. Once all its elements are assembled, a routine executes. Other kinds of structures do not execute. In its execution, a routine may change the values of its out elements and it may replace itself by other structures. Whether or not a structure executes seems to be orthogonal to the properties shared by the different kinds of structures. A routine thus may be treated as a structure.

The symmetry between routines and other kinds of structures simplifies and empowers TSIA. An example of the simplification is that structures are the only collections of items within TSIA.

The power of the symmetry is demonstrated by the extension of a property of one kind of structure to another kind. Fusion, fission and streams are examples of properties span-ning different kinds of structures. Given the symmetry between the kinds of structures, what might a property of one kind mean for another kind? For example, a sparse array raises the notion of a sparse routine. Is such a sparse routine one whose instruction provides default values for items, thus allowing the corresponding arguments to be omitted?

### *3.17.3 Other Kinds of Structures*

In this thesis, IA supports routines and arrays. Extending a IA to support other kinds of structures seems feasible, but such support is not demonstrated in this thesis.

This thesis claims that a IA can support structures and a structured application definition. Since it is only demonstrated for routines and arrays, this claim includes only routines and arrays. Nevertheless, it seems that the claim can be extrapolated also to include other kinds of structures.

For example, because an object consists only of its elements, extending a IA to also support objects seems feasible. The elements of an object include routines, as well as items and other structures [Structured Programming : Objects].

### *3.17.4 A Structured Application Definition*

A structured application definition has its items in structures.

An application definition can use various kinds of structures. Motivation for an array-based definition and its support by a IA is described in section 3.16. Since routines are ubiquitous, no explicit motivation for routines is given in this presentation. IA support for routines is woven throughout this chapter. Since a IA can support a variety of structures, it seems that a IA allows each part of an application definition to use an appropriate kind of structure. A IA thus supports a structured application definition.

This presentation introduces the name structured application definition or equivalently structured definition. At least in this presentation, a structured definition only implies that the items of an application are in structures.

A structured definition thus is the part of structured programming concerned with routines [Structured Programming], objects [Structured Programming : Objects] and other structures [Structured Programming : Data]. Structured programming also includes other parts. Each of these other parts is orthogonal to the part corresponding to a structured definition.

For example, structured programming also concerns itself with constructs such as the `if then` conditional construct, the `case` choice construct and the `while` loop construct [Structured Programming]. Such constructs usually do not concern a structured definition. Such constructs usually are internal to instructions and a structured definition is not concerned with the internals of instructions nor of any other item.

Another part of structured programming concerns types [Structured Programming : Data]. As stated already in subsection 3.1.12, types are beyond the scope of this presentation. Examples of more recent developments concerning types and other supports for a convenient application definition are described elsewhere [CLOS].

An implicit hope of this thesis is that a structured application definition, as supported by a IA, is compatible with types, with the other parts of structured programming and with other supports for a convenient application definition.

### 3.18 The Conditional Item

A *conditional item* of a task is an item which may or may not be ignored by the task. Up until this point of the presentation, a task has delegated or evaluated its items. A task has not yet ignored an item. Section 3.19 introduces the possibility that a task ignores an in. Section 3.22 introduces the possibility that a task ignores an out.

To be more honest, up until this point of the presentation, some tasks have ignored some items. Each of these previous conditional ins or conditional outs is of little consequence since its task is neither the sole destination nor the ultimate origin of the item, respectively. In other words, such a previous conditional item does not have its existence affected by being ignored. In contrast, in section 3.19 the conditional in is the sole destination for the item and thus if ignored the item need not exist. Similarly, in section 3.22 the conditional out is an origin for the item and thus if ignored the item must originate from another task.

For example, unlike an in or an out, an inout is neither the ultimate destination nor origin of an item, respectively. Thus unlike a conditional in or a conditional out, a conditional inout has relatively little repercussion on the other tasks of the application. Hence, the task `attempt(...;del_ign sols;)` of the N-queens application of subsection 3.10.3 introduced the conditional inout with no fanfare.

An optional item is another example of a conditional item with relatively little repercussion on the other tasks of the application. Such an item is a conditional out of one task and a conditional in of a subsequent task. As an optional item, either both or neither of the tasks ignore the item.

### 3.19 The Conditional In

A *conditional in* of a task is an in which may or may not be ignored by the task. The benefits of a conditional in long have been recognized [Conditional Expressions]. For example, the task `xif(m,del_ign n1,del_ign n2;;del w)` of Figure 70a) has two conditional ins, `n1` and `n2`. Depending on whether the in `m` has the value `true` or the value `false`, the out `w` is the in `n1` or the in `n2`, respectively. The other in is ignored.

In the ia language of this presentation, the keyword `is`, as in the statement `w is n1;`, causes the item `w` to refer to the item `n1`. The statement changes only the item `w`; the item `n1` is not changed.

Since either the in `n1` or the in `n2` is ignored, it is not necessary to evaluate both before executing the task `xif(m,del_ign n1,del_ign n2;;del w)`. Thus declaring the `del_ign` use of the items `n1` and `n2` allows for an efficient execution. In contrast, relying on the default declaration `del_eva` would require both `n1` and `n2` to be evaluated, which could cause a very inefficient execution.

A need for the conditional in is demonstrated by the routine `g(;;del q)` of Figure 70b), which has an execution illustrated in Figure 70c). The illustrated execution assumes the out `x` of `a(;;x)` to be `false`. The routine `g(;;del q)` cannot delegate to the routine `a(;;x)`. Instead, the routine `a(;;x)` executes in the subordinate style. The task `g(;;del q)` requires the value of `x` in order to replace itself by either the task `b(;;q)` or the task `c(;;q)`.

As described in subsection 3.2.3.3, the subordination of `a(;;x)` can be changed to delegation by rewriting `g(;;del q)` such that it uses a subsequent task. In general, such brute force rewriting can be unattractive. For example, the rewritten task may have a inconveniently large number of items used first by the parent and then by the subsequent task.

In this example, as an alternative to the use of a subsequent task, the use of a conditional in also allows the delegation of `a(;;x)`. The rewritten routine `g(;;del q)` is in Figure 70d), with an execution illustrated in Figure 70e). As for the execution illustrated in Figure 70c) for the original routine `g(;;del q)`, the illustrated execution assumes the out `x` of `a(;;x)` to be `false`. In the execution, the routine `g(;;del q)` replaces itself by the tasks `a(;;x)`, `b(;;y)`, `c(;;z)` and `xif(x,del_ign y,del_ign z;;del q)`. Once `a(;;x)` has executed, then `xif(x,del_ign y,del_ign z;;del q)` can execute, with the result that `q is z`. At this point the task `b(;;y)` is irrelevant since its out `y` no longer is an in of any subsequent task. As illustrated in Figure 70e), the task `b(;;y)` thus is crossed out of the task pool at this point. Only the task `c(;;q)` thus remains in the task pool.

As demonstrated above, a conditional in can introduce a task eventually irrelevant to the execution. As also demonstrated, recognizing an irrelevant task is straightforward in TSIA. By definition, none of the out of an irrelevant task is an in of any subsequent task.

The ins of such an irrelevant task are ignored and thus can result in further irrelevant tasks and further ignored ins and so on. In short, ignoring an in may result in an arbitrarily large dependency chain of ignored ins and irrelevant tasks.

Another example of a conditional in is `n2` of the task `or_sc(n1,del_ign n2;;w)` of Figure 71a). The task is a slight variation of the task `xif(x,del_ign y,del_ign z;;del q)` of Figure 70a). The task `or_sc(n1,del_ign n2;;w)` defines the logical OR operation, with short circuit evaluation. As seen in the ia code of Figure 71a), if the in `n1` is true, then the out `w` is true, regardless of the value of the other in `n2`. The in `n2` thus need not be evaluated, hence the name short circuit evaluation.

Analogous to the task `or_sc(n1,del_ign n2;;w)`, the task `and_sc(n1, del_ign n2;;w)` of Figure 71b) defines the logical AND operation, with short circuit evaluation.

The task `h(del i;;del r)` of Figure 71c) is a small example which uses the task `and_sc(n1, del_ign n2;;w)`. The execution illustrated in Figure 71d) assumes that `x` of `a(i;;x)` evaluates to `false`. At this point in the execution, the task `b(i;;y)` is irrelevant, since its out `y` no longer is an in of any subsequent task.

### *3.19.1 The Conditional In is a Part of Non-Strict Evaluation*

As demonstrated by the examples of this section, a conditional in allows a task to execute before the in is evaluated. As introduced in subsection 3.10.1, in functional computing, executing a function before an argument is evaluated is known as non-strict evaluation. The conditional in thus is a part of non-strict evaluation. As described in subsection 3.10.1 and illustrated in Figure 45, delegation is the other part of non-strict evaluation.

TSIA distinguishes between delegation and the conditional in, thus allowing for delegation without necessarily allowing for the conditional in. Thus TSIA allows for fusion, fission, supply- versus demand-driven order and other benefits of delegation, without necessarily allowing for the conditional in.

### 3.19.2 Binary Search Using the Conditional In

The routines `bs` and `bs1` of Figure 62b) for the binary search definition of subsection 3.13.10 are rewritten in Figure 72a) as `bs` and `bsci` using the conditional ins `im` and `ip`. The execution of `bsci` may ignore `im` or `ip` or both.

The code of Figure 72a) and the code of Figure 62a) define a complete Fortran application demonstrating binary search.

As already noted long ago [Conditional Expressions], the use in Fortran of a conditional in is inefficient. Like any other in, a conditional in is an out of a preceding routine or task. In Fortran, the preceding routine always evaluates the out, regardless of whether or not it is ignored in the subsequent routine as a conditional in. This inefficiency in Fortran is demonstrated by the child tasks of `bs` of Figure 72a). Among the child tasks, `bs(m,k-1,a(m),v,im)` always evaluates `im`, regardless of whether or not `im` is subsequently ignored by `bsci(v,k,a(k),im,ip,i)`. Similarly, `bs(k+1,p,a(k+1),v,ip)` always evaluates `ip`.

The Fortran routines `bs` and `bsci` of Figure 72a) are rewritten in ia in Figure 72b). Though the definitions essentially are identical, the ia code can have an efficient execution, not the above inefficient execution of the Fortran code. Because of the use declarations `del_ign im` and `del_ign ip` of `bsci`, IA need not evaluate `im` nor `ip` before executing `bsci`. Better yet, IA is aware that evaluating `im` or `ip` before executing `bsci` may well be wasted effort. Of course, as for the binary search definition of subsection 3.13.10 and Figure 62b), the ia code of Figure 72b) need not evaluate more than the $\ln_2 N$ elements required by the binary search algorithm for an N element array.

The execution of the binary search example and of other applications involving the conditional in is further described in section 3.19.

### 3.19.3 The N-Queens Solution Using the Conditional In

Subsection 3.10.3 introduced the N-queens problem. There Figure 46 shows a Fortran application which determines the number of solutions for the N-queens problem. In this subsection, the applications determine a solution to the N-queens problem. Figure 73 shows such a Fortran application. It uses the Fortran routines of Figure 46b). The Fortran routines `iterate`, `nattempts` and `nqans` are rewritten in ia in Figure 74.

The above ia application, while correct, does not allow much flexibility for the execution. It does not expose the parallelism inherent in the algorithm of the solution.

Figure 75 also shows a Fortran application which determines a solution to the N-queens problem. It uses the Fortran routines of Figure 46b) and Figure 73a). The Fortran routines `first` and `nattempts` are rewritten in ia in Figure 76.

The Fortran definition of Figure 75 determines the exact same solution as the Fortran definition of Figure 73b). These are the same solutions as determined by the ia definition of Figure 74 or the ia definition of Figure 76.

In contrast to the ia application of Figure 74, the ia application of Figure 76 exposes the parallelism inherent in the algorithm of the solution. This flexibility for the execution is achieved by using the conditional in `nansr[n,r]` of the task `first`. The task `nattempts` thus replaces itself by N pairs of the tasks `testsafe` and `attempt`. Each pair stores its solution in the array `nans[n,n]`. The pairs thus can, but need not, execute in parallel. The task `first` chooses from `nans[n,n]` the solution from the pairs of tasks earliest in the order of pairs. The conditional in `nans[n,n]` allows `first` to do so without requiring the entire array `nans[n,n]` to be evaluated.

### *3.19.4 Two Numerical Algorithms Using the Conditional In*

As argued elsewhere [Why], non-strict evaluation can help achieve a structured application definition. A possibility is the use of streams, as described in section 3.24. In order to introduce some of the issues of that section, including its use of conditional items, this subsection introduces two of its example applications. The applications are described in greater detail elsewhere [Why].

#### *3.19.4.1 The Newton-Raphson Algorithm*

The Newton-Raphson algorithm computes the square root of a number `n`. Starting from an initial approximation $a_0$, the algorithm iteratively computes a better approximation using $a_{i+1}=(a_i+n/a_i)/2$. The subscript numbers the iteration. The algorithm terminates once a given tolerance is achieved. For example, a Fortran definition of the Newton-Raphson algorithm is shown in Figure 77a).

Figure 77b) shows a ia definition of the Newton-Raphson algorithm. In an execution of the routine `cisqrt(n,a0,eps;;a)`, the child `cisqrt(n,a1,eps;;a2)` may or may not execute since `a2` is a conditional in of the child `ciwithin(a0,a1, del_ign a2,eps;;a)`.

The ia routines `next`, `ciwithin` and `cisqrt` of Figure 77b) are rewritten in Fortran in Figure 77c). The Fortran code includes `program newton` and thus is a complete Fortran application. The definitions of `next` and of `ciwithin` essentially are identical in ia and in Fortran. In contrast, the ia and Fortran definitions of `cisqrt` differ. A straight translation of the ia `cisqrt` would yield a Fortran routine which would never exit its recursion. Hence the Fortran definition of `cisqrt` of Figure 77c) imitates the demand-driven execution required for the ia `cisqrt` of Figure 77b).

Given the same ins `n`,`a0`,`eps` the out `a` is identical for the Fortran routine `fsqrt`, the ia routine `cisqrt` and the Fortran routine `cisqrt` of Figure 77a), b) and c), respectively. While the three codes have very different executions, they have the same definition.

#### *3.19.4.2 A Numerical Differentiation Algorithm*

The second numerical algorithm is a numerical differentiation algorithm. It has a definition very similar to that of the Newton-Raphson algorithm. Thus the above comments on the definition and the execution of the Newton-Raphson algorithm also apply to the numerical differentiation algorithm.

The numerical differentiation algorithm differentiates a function at a given point. The initial approximation $a_0$ is the slope of the straight line between the value $f(x)$ of the function at the given point $x$ and the value $f(x+h_0)$ of the function at a nearby point $x+h_0$. In order to avoid rounding errors, the initial $h_0$ is reasonably large. The algorithm iteratively computes a better approximation $a_i$ by halving $h_i$ each iteration. Thus the algorithm is given by $a_i=(f(x+h_i)-f(x))/h_i$, where $h_{i+1}=h_i/2$. The subscript numbers the iteration. The algorithm terminates once a given tolerance is achieved. For example, a Fortran definition of the numerical differentiation algorithm is shown in Figure 78a).

Figure 78b) shows a ia definition of the numerical differentiation algorithm. In an execution of the routine `cidiff(f,x,h0,a0,eps;;a)`, the child `cidiff(f,x,h1,a1,eps;;a2)` may or may not execute since `a2` is a conditional in of the child `ciwithin(a0,a1,del_ign a2,eps;;a)`.

The ia routines `halve`, `easydiff`, `ciwithin` and `cidiff` of Figure 78b) are rewritten in Fortran in Figure 78c). The Fortran code includes `program testdiff` and `subroutine cfu` and thus is a complete Fortran application. The definitions of `halve`, of `easydiff` and of `ciwithin` essentially are identical in ia and in Fortran. In contrast, the ia and Fortran definitions of `cidiff` differ. A straight translation of the ia `cidiff` would yield a Fortran routine which would never exit its recursion. Hence the Fortran definition of `cidiff` of Figure 78c) imitates the demand-driven execution required for the ia `cidiff` of Figure 78b).

Given the same ins `f`,`x`,`h0`,`a0`,`eps` the out `a` is identical for the Fortran routine `fdiff`, the ia routine `cidiff` and the Fortran routine `cidiff` of Figure 78a), b) and c), respectively. While the three codes have very different executions, they have the same definition.

### *3.20 Speculative versus Conservative Order*

For the task order as introduced in section 3.11, speculation and conservation are the extremes of a sub-axis of the supply- versus demand-driven axis. By maximally obeying the dependencies between the tasks of the application, as in a demand-driven order, a *conservative order* executes only tasks relevant for the application execution. As for any task order, speculation also executes the tasks relevant for the application execution. In addition, by minimally obeying the dependencies between the tasks of the application, as in a supply-driven order, a *speculative order* may execute tasks eventually irrelevant for the application execution.

As described in section 3.19, a conditional in introduces the possibility of an eventually irrelevant task. Thus the ability to ignore an item introduces the possibility for speculation. If no item is ignored, then all items ultimately are evaluated and thus all tasks are relevant and thus all task orders are conservative.

As described in section 3.11, the demand-driven order executes only demanded tasks. Such a demanded task has at least one of its out demanded as an evaluated in of another demanded task. In other words, a demanded task is a relevant task. Since it thus executes only relevant tasks, the demand-driven order is a conservative order. As an example of the demand-driven order, lazy evaluation is a conservative order [Concurrent Clean][Lazy Evaluation][Unboxed values].

Also as described in section 3.11, the supply-driven order executes any task for which the items are available. Since it does not prevent the execution of eventually irrelevant tasks, the supply-driven order is a speculative order. As an example of the supply-driven order, future order is a speculative order [Future Order][Multilisp][Speculative].

The examples of the conditional in of section 3.19 allow for demonstrations of speculation and conservation.

Figure 70e) illustrates a conservative order for the execution of the task `g(;;del q)` of Figure 70d). The execution assumes the out `x` of `a(;;x)` to be `false`. The execution is conservative since it includes no tasks irrelevant for the execution of `g(;;del q)`. In particular, `b(;;y)` is the only task eventually irrelevant and it is not executed.

In contrast, Figure 79a) illustrates a speculative order for the execution of the same task `g(;;del q)` of Figure 70d). Again, the execution assumes the out `x` of `a(;;x)` to be `false`. The execution is speculative since it includes tasks irrelevant for the execution of `g(;;del q)`. In particular, `b(;;y)` is executed even though it is eventually irrelevant.

The execution of the task `h(i;;r)` of Figure 71c) allows for a very similar demonstration. Figure 71d) illustrates a conservative order. Figure 79b) illustrates a speculative order. Both executions assume the out `x` of `a(i;;x)` to be `false`.

Similar demonstrations can be made using the tasks `bs` and `bsci` of the binary search definition using the conditional in of Figure 72 and subsection 3.19.2. Due to its use of recursion, the binary search definition repeatedly makes tasks available for a speculative order, with each task eventually either relevant or irrelevant. The executions illustrated here assumes `i is im` and then `i is ipm`. Figure 80a) illustrates a conservative order; no eventually irrelevant tasks are executed. Figure 80b) illustrates a speculative order; the eventually irrelevant task `bs(...;;ip)` is executed as is the eventually irrelevant task `bs(...;;imm)`. Similar demonstrations can be made using the tasks of the N-queens solution using the conditional in of Figure 76 and subsection 3.19.3.

In Figure 80b) the speculation is one level deep; a parent task is speculatively executed but its child tasks are not. Figure 80c) illustrates a speculative order with speculation two levels deep; a parent task and its child tasks are speculatively executed. In general, executing a task `bs` increases the depth of speculation by one, while executing a task `bsci` leaves the depth of speculation unchanged.

By the nature of the binary search algorithm, only a half of the speculative tasks one level deep eventually are relevant. Similarly, only a quarter of the speculative tasks two levels deep eventually are relevant. In general, for the speculative tasks L levels deep, only $2^{-L}$ eventually are relevant.

All other things being equal, it generally is best if the depth of speculation is similar across the tasks at any given point of the application execution. For example, for the above binary search definition, each `bs` child has a speculative depth of one, while each of the four `bs` grandchildren has a speculative depth of two. Once the first `bs` child speculatively executes, the next task to speculatively execute should be the other `bs` child since it maintains a speculative depth of one for the application execution. In contrast, if one of the two `bs` grandchildren were speculatively executed, then there would be a speculative depth of one and two across the application execution. The speculative execution of the other `bs` child should precede that of the either `bs` grandchild, since the former has half a chance of eventually being relevant, while the latter each have just a quarter of a chance.

A major motivation for the speculative and conservative order is support for the conditional in, which is a very powerful definition element. In addition, achieving an efficient execution provides several motivations for speculation. The most obvious of these motivations is the introduction of additional parallelism. A general example is the use of the speculative order for machine instruction level parallelism [Compiler Transformations]. A specific example is offered by the execution of `h(i;;r)`, illustrated in Figure 71d), where speculation allows the tasks `a(i;;x)` and `b(i;;y)` to execute in parallel. Similarly, for `xif(m,del_ign n1,del_ign n2;;del w)` of section 3.19 and Figure 70a), `m` and `n1` and `n2` may be evaluated in parallel [Speculative]. An execution situation also can raise other motivations for speculation. In another execution of

h(i;;r), if i is huge, y is small, b is fast, and_sc is slow, time is plentiful and space is not, then it may be efficient to speculatively execute b(i;;y) in order to minimize the time for space used for item i. A final example using the execution of h(i;;r) assumes the instruction and_sc must first be transferred from a remote computer. While waiting for and_sc to arrive, it may be efficient to overlap communication and computation by speculatively execution b(i;;y).

Similar examples motivating the speculative and conservative order can be provided by the binary search definition already discussed above. For example, for the child tasks of bs(...;;i), if the item a[k] is not yet available, then the child bsci cannot execute, but the other two children bs(...;;im) and bs(...;;ip) can. If the other elements of the array are available, the latter two tasks and their descendants even may complete their responsibility and evaluate im and ip before a[k] becomes available. Obviously, once a[k] becomes available, then bsci(v,k,a[k],im,ip;;i) can execute and evaluate i, thus immediately completing the responsibility of the original task bs(...;;i).

Similarly the speculative order can be motivated by the massive parallelism of the N-queens solution using the conditional in of Figure 76 and subsection 3.19.3.

The above motivations for the speculative order of course are balanced against the availability of computers, memory, bandwidth and other resources. When resources are scarce, it generally is best to conservatively execute tasks known to be relevant rather than to speculatively execute tasks not yet known to be relevant.

### *3.21 Multi-Origin Out*

A *multi-origin out* is an out originating from any one of several different tasks. In this section and in the next section, the out of the tasks are identical. The possibility that the out are not identical is introduced in section 3.23. The multi-origin out is introduced at this point of the presentation because the conditional out, introduced in the next section, often is used for a multi-origin out.

In Figure 81a), the out f of the task double(e;;f) is an example of a multi-origin out. The out f may originate from either the child task doubleadd or the child task doublemult. In the ia language of this presentation, the appearance of doubleadd before doublemult in the ia code of double in no way indicates that one is somehow preferred over the other. In the execution of double illustrated in Figure 81b), the child doublemult is assumed to evaluate f before the child doubleadd. In other words, in this execution the multi-origin out f originates from the task doublemult. At this point, any other possible origins of the out f are irrelevant to the execution. Thus the task doubleadd is crossed out of the task pool.

In the trivial example of Figure 81, only a single item, the instruction doubleadd versus the instruction doublemult, differs between the possible origins of the multi-origin out f. In general, any number of items could differ between the possible origins.

As described in section 2.6, due to the symmetry of items of a task, the computer executing the task is for TSIA like any other item of the task. Thus an example of a multi-origin out is the execution by TSIA of otherwise identical tasks on two different computers in the hope that one will complete significantly faster than the other.

A motivation for using multi-origin out are different tasks for the same out, where the tasks require large and very different amounts of resources and the amounts are unknown before executing the tasks. In the worst case, the amount of resources used by such a task is known, even approximately, only after executing the task. In contrast, if the amounts of resources are known prior to execution then the application definition can use the appropriate task and thus does not need the multi-origin out. In short, when the best task among several tasks to provide an out is unknown, a multi-origin out allows the application definition to defer the choice of task to TSIA.

While the `double` example of Figure 81 illustrates the workings of a multi-origin out, the example provides no motivation for its use. The tasks `doubleadd` and `doublemult` require only small, similar and known amounts of resources. Examples of multi-origin out using tasks with large, different and unknown amounts of resources are presented in sections 3.22 and 3.23.

As described in section 3.22, another motivation for using multi-origin out is provided by the conditional out. A task may ignore a conditional out and thus rely on another task to provide the out. In the worst case, as for the resources used by a task, whether a task ignores a conditional out is known only after executing the task. In contrast, if a task is known to ignore an out then the application definition does not use that task and thus does not need the multi-origin out. Again in short, when the appropriate task among several tasks to provide an out is unknown, a multi-origin out allows the application definition to defer the choice of task to TSIA.

The task order for a multi-origin out raises issues beyond the scope of this presentation. In the remainder of this section, this presentation thus only offers some comments on the task order.

The task order for a multi-origin out is related to the speculative order and the conservative order, but also goes beyond these orders.

As defined in section 3.20, a conservative order executes only tasks relevant for the application execution. A conservative order thus executes only one of the tasks of a multi-origin out. The problem is to identify which of the tasks to execute. Generally the desirable task to execute is the one requiring the least amount of resources. Unfortunately, the amount of resources is not known, or is only poorly known, before executing any of the tasks. The problems of a conservative order are increased if the multi-origin out involves a conditional out. In this case, if the task chosen for execution ignores the out, then the nominally conservative order thus has executed an irrelevant task.

As defined in section 3.20, a speculative order may execute tasks eventually irrelevant for the application execution. A speculative order thus executes, more accurately begins the execution of, any number of the tasks of a multi-origin out. If time is the only scarce resource and all other resources are plentiful, then by executing all tasks in parallel the speculative order yields the out in the shortest possible time. Once the fastest task completes, the remaining tasks are aborted. Unfortunately time is rarely the only scarce resource.

A speculative order does not necessarily imply parallel computing. For example, the execution of the tasks of a multi-origin out may be multiplexed on a single computer [Future Order].

The difficulties of the task order easily are illustrated using a multi-origin out with two tasks. First assume that each of the two tasks has an execution requiring 10 units of resources. An extremely conservative order thus evaluates the out using 10 units of resources, since only one of the two tasks is executed. In contrast, an extremely specula-tive order evaluates the out using 20 units of resources, since both of the tasks are executed.

Now assume instead that the execution of one of the tasks requires 2 units of resources while the other requires 18 units. Assuming that one can do no better than to randomly pick one of the two tasks to execute, the extremely conservative order requires on average 10 units of resources. In contrast, the extremely speculative order evaluates the out using 4 units of resources, since the slower task is aborted once the faster task completes.

Thus in the first situation the extremely conservative order is better, while in the second situation the extremely speculative order is better. In these and other situations, some point between the extremes may be best.

In this subsection, TSIA is provided with no information about the tasks of a multi-origin out. The execution situation need not be quite so extreme. As described in subsection 3.23.4, the application definition may indicate to TSIA the most promising tasks of a multi-origin out.

### *3.22 The Conditional Out*

A *conditional out* of a task is an out which may or may not be ignored by the task. Like any other out, a conditional out is an in of a subsequent task. Thus if a conditional out is ignored by its task, some other task or mechanism must supply the out. Because it is a powerful use of the conditional out, the mechanism considered in this section is the multi-origin out described in the previous section.

A definition of multiplication can illustrate the use of a conditional out as part of a multi-origin out [Future Order]. The task `mult(del_ign a, del_ign b;; del c)` of Figure 82a) is such a definition. Any of the three child tasks of `mult` can yield the out `c`. If `a==0` then the first child `mult1(a;;c)` yields that `c is 0`. In this case, the in `b` of `mult` is ignored. Vice versa, if `b==0` then the second child `mult1(b;;c)` yields that `c is 0`. In this case, the in `a` of `mult` is ignored. In all other cases, the third child `mult2(a,b;;c)` yields `c=a*b`. Obviously the third child also is valid if either `a` or `b` is `0`.

Thanks to the conditional and multi-origin out `c`, in the above definition of multiplication, if either of the arguments `a` or `b` is `0`, then the other argument can be ignored and thus need not be evaluated. Thus if one of the arguments is `0`, this situation can be of value to the application definition since the multiplication can succeed even if the other argument is unknown or undefined. Similarly, this situation can be of value to the application execution since the multiplication can complete as soon as either of the arguments is known to be `0`. At this point, the evaluation of the other argument can be avoided entirely or aborted, thus saving resources. As described in section 3.20, ignoring an argument or in also may allow other ins and tasks to be ignored, thus increasing the benefits described here.

As another example, a conditional and multi-origin out can be useful when a fast algorithm occasionally solves a problem which otherwise requires a slow algorithm [Future Order]. The task `solve` of Figure 82b) codes such a situation.

A definition of the logical OR operation also can illustrate the use of a conditional and multi-origin out [Speculative]. The task `or_co(del_ign a, del_ign b;; del c)` of Figure 82c) is such a definition. The task `or_co` is very similar to the task `mult` of Figure 82a) and they thus share the same benefits.

The definition of the logical OR operation by the task `or_co(del_ign a, del_ign b;; del c)` of Figure 82c) may be compared with that by the task `or_sc(n1,del_ign n2;;w)` of Figure 71a). The short circuit evaluation provided by `or_co` is symmetric across its arguments `a` and `b`. If either `a` or `b` is true, then `c` is true, and the other argument can be ignored. In contrast, the short circuit evaluation provided by `or_sc` is asymmetric across its arguments `n1` and `n2`. If `n1` is true, `n2` is ignored. In contrast, `n1` never can be ignored, even if `n2` is true.

The task `or_d` of Figure 82d) defines the logical OR operation in a fashion very similar to that of the task `or_co` of Figure 82c). The third child of each task is the only difference between the definitions. For `or_co`, its third child `or2(a,b;;c)` has no additional constraints. In contrast, for `or_d`, its third child `set(false;; default c)` only may execute and thus evaluate `c` if all the other tasks have ignored the multi-origin out `c`. In the ia language of this presentation, the keyword `default` is used to define this constraint for a task of the multi-origin out.

Analogous to the tasks `or_co` and `or_d`, the tasks `and_co` and `and_d` of Figure 82e) and f) each define the logical AND operation.

### *3.23 An Indeterminate Definition due to a Multi-Origin Out*

If the outs of its tasks are not identical, then a multi-origin out is indeterminate and so is the application definition. An indeterminate application definition, including its general motivation and its support by TSIA, is described in subsection 3.1.8.

#### *3.23.1 An Indeterminate Search of an Unsorted Array*

Searching an unsorted array of integers allows for a simple demonstration of an indeterminate application definition due to a multi-origin out. Similar examples are discussed elsewhere [Speculative]. Figure 83a) shows `us(n,a,v,i)`, a Fortran definition of such a sort. The Fortran definition approximates the desired ia code. After replacing `call bs(1,n,a,j,i)` by `call us(n,a,j,i)`, `program bstest` of Figure 62a) may be combined with the code of Figure 83a) for a complete Fortran application which demonstrates and tests the search code.

If the line `if(i.ne.-1)return` is removed from the Fortran code of Figure 83a), then its routine `us(n,a,v,i)` remains correct, albeit less efficient. Instead of returning the first match, the modified Fortran code returns the last match in the array.

The Fortran routines `match` and `us` of Figure 83a) are rewritten in ia in Figure 83b). Though the definitions are similar, the executions are very, very different. In the execution of the ia code, the task `us(n,a[n],j;;indet i)` replaces itself by `n` tasks `match(a[k],j,k;;i)`, where `k` ranges from `1` through `n`, in addition to the task `set(-1;;default i)`. Since it can be evaluated differently by any of the `match` tasks, the keyword `indet` declares that the multi-origin out `i` may be indeterminate.

There are special cases where `i` or similar search results using a multi-origin out are not indeterminate. In the above example, if all elements in the array are unique, then `i` is determinate, since there is only one possible answer for any search.

While a determinate application definition generally is preferred, the above indeterminate definition of search offers advantages otherwise unavailable. Some advantages concern the application definition. For example, the indeterminate definition offers a perfect short-circuit behavior. If any of the elements match, then none of the other elements need be evaluated. The other elements even may be undefined or ill-defined. Other advantages concern the application execution. For example, the definition exposes all the parallelism inherent in the search.

### *3.23.2 N-Queens with an Indeterminate Solution*

As demonstrated in subsections 3.10.3 and 3.19.3, the N-queens problem can be solved by what essentially is a simple search. Thus just like the indeterminate search of an unsorted array of integers of subsection 3.23.1, the solution to the N-queens problem also allows for a simple demonstration of an indeterminate application definition due to a multi-origin out. Figure 84a) shows `nattempts(n,b_size,board,ans)`, a Fortran definition of such a solution. The complete Fortran application uses the routines of Figure 46b) and Figure 73a). The Fortran definition of Figure 84a) determines the exact same solution as each of the two solutions of subsection 3.19.3 in Figure 73b) and in Figure 75.

If the line `if(ans(1).ne.-1)return` is removed from the Fortran code of Figure 84a), then its routine `nattempts(n,b_size,board,ans)` remains correct, albeit less efficient. Instead of returning the first solution, the modified Fortran code returns the last solution for the N-queens problem.

The Fortran routines `nattempts` and `nqans` of Figure 84a) are rewritten in ia in Figure 84b). Though the definitions are similar, the executions are very, very different. In the execution of the ia code, the task `nattempts(n,b_size,board[b_size] ;;indet ans[n])` replaces itself by `n` pairs of the tasks `testsafe` and `attempt`. The multi-origin conditional out `ans[n]` is indeterminate since it can be evaluated differently by any of the pairs of tasks.

The ia code of Figure 84b) for N-queens with an indeterminate solution is very similar to the ia code of Figure 76 for N-queens with a determinate solution using the conditional in. In essence, the indeterminate solution simply eliminates the array `nans[n,n]` and the task `first` which make determinate the determinate solution. Since the indeterminate solution thus does not have this overhead of the determinate solution, the indeterminate solution should have a more performant execution.

### *3.23.3 An Indeterminate Definition of Binary Search*

A determinate definition of binary search using the conditional in, as described in subsection 3.19.2 with ia code in Figure 72b), has a very performant execution, as described in section 3.20. Furthermore, the efficient application definition evaluates only the $\ln_2 N$ elements required by the binary search algorithm for an N element array. As described in subsection 3.1.8, a determinate application definition is in many ways preferred to an indeterminate application definition. The excellent execution and definition of this determinate definition thus generally must be exceeded by an indeterminate definition of binary search, if the indeterminate definition is to be of interest.

The ia code for an indeterminate definition of binary search is shown in Figure 85b). It is a slight modification of the ia code of Figure 72b) for the determinate definition. The determinate definition of the task `bs(...;;i)` of Figure 72b) requires the search result `i` to be the out of its child task `bsci(...;;i)`. In contrast, the indeterminate definition of the task `bs(...;;indet i)` of Figure 85b) allows the search result `i` to be a multi-origin out of any of its three child tasks `bs(...;;i is im)`, `bs(...;;i is ip)` or `bsci(...im,ip;;i)`.

The expression `i is im` as an out declares that the item `i` is the out if and only if the out is evaluated or delegated by that task or its descendants.

If the task `bs(...;;i is im)` evaluates or delegates its out, as for any other multi-origin out, then `i` is no longer required from the other two child tasks. Since no other out is required from these two tasks, both may be crossed out of the task pool.

If the task `bsco(v,k,ak,im,ip;;i)` evaluates or delegates `i`, as for any other multi-origin out, then `i` is no longer required from either of the other two child tasks. If the task `bsco` resulted in `i is k`, then no out is required from the other two tasks and both may be crossed out of the task pool. In contrast, if `bsco` resulted in `i is im`, then `im` is required from the task `bs(...;;i is im)` which now may be treated as `bs(...;;im)`. Since no out is required from the third child task `bs(...;; i is ip)`, it may be crossed out of the task pool.

Except for the declarations `del_ign i` versus `del i`, the task `bsco(...;; del_ign i)` of Figure 85b) is identical to the task `bsci(...;;del i)` of Figure 72b). The declaration `del_ign i` allows `bsco` to result in `i is im`, even if `im` is ignored. The out `i` then also is ignored.

Because the determinate ia code of Figure 72b) is very similar to the indeterminate ia code of Figure 85b), the Fortran code of Figure 72a) roughly approximates either definition. In addition, the Fortran code of Figure 85a) also roughly approximates the indeterminate definition of Figure 85b). Combined with that of Figure 62a), the Fortran code of Figure 85a) or of Figure 72a) thus each define a complete Fortran application approximating the indeterminate binary search definition of Figure 85b). In addition, either the `if(im.ne.-1)` block or the `if(ip.ne.-1)` block may be removed from the Fortran code of Figure 85a) to obtain two more approximations of the indeterminate binary search definition. Perhaps the indeterminate nature is best approximated by randomly removing the `if(im.ne.-1)` block and the `if(ip.ne.-1)` block. This can be achieved by guarding each block by `if(mod(irand(),2).eq.0)`.

The indeterminate binary search definition meets the requirements set in the opening paragraph of this subsection. The indeterminate definition of the ia code of Figure 85b) has a definition and an execution which exceed those of the determinate definition of the ia code of Figure 72b).

In searching an N element array, the determinate definition evaluates a fixed sequence of up to $\ln_2 N$ elements. In contrast, the indeterminate definition offers a perfect short-circuit behavior. This is the same achievement as that of the indeterminate search of an unsorted array of integers of subsection 3.23.1. If any of the elements match, then none of the other elements need be evaluated. The indeterminate definition of binary search thus has a definition which exceeds that of the determinate definition.

For the determinate definition, an execution of the task `bsci` only can complete the search once all the ancestor `bsci` tasks, up to $\ln_2 N$ in number, have executed. In contrast, for the indeterminate definition, an execution of the task `bsco` can complete the search regardless of any other task.

For the determinate definition or the indeterminate definition, if the matching element is not found by an execution of the task `bsci` or `bsco`, then the task eliminates from the search either the elements before or after its array element `a[k]`. This eliminates tasks and items from the task pool. For the determinate definition, the elimination is restricted to its local array elements, as illustrated in Figure 86a). In Figure 86, the array `a[1:15]` is searched. The task bsci(v,4,a[4],...) or bsco(v,4,a[4],...) is assumed to be the first bsci or bsco task executed. In addition, v<a[4] is assumed, and thus any successful search result must be among the elements a[1:3]. For the determinate definition, the task bsci(v,4,a[4],...) eliminates just the elements a[4:7]. In contrast, for the indeterminate definition, the elimination applies is not restricted to its local array elements. Instead, the task bsco(v,4,a[4],...) effectively eliminates the elements a[4:15], as illustrated in Figure 86b).

For the above two reasons, the indeterminate definition of binary search thus has an execution which exceeds that of the determinate definition.

As introduced in subsection 3.23.1, if all the elements of the searched array are unique, then even an indeterminate definition of a search provides a determinate result. For binary search, unique array elements thus offer both the advantages of a determinate application definition and the above advantages of the indeterminate definition.

### *3.23.4 An Indeterminate Solution of the 8-Puzzle*

The 8-puzzle consists of eight square tiles numbered 1 through 8 occupying eight of the nine positions of a 3*3 board. An orthogonally adjacent tile may be slid into the unoccupied position. Such moves can rearrange the positions of the tiles on the board. The object of the 8-puzzle is to find the moves from a given original arrangement to a given desired arrangement of tiles.

For convenience, the unoccupied position is known as the blank tile or as the tile numbered 0.

The 8-puzzle and the techniques for its solution, including all the techniques of this presentation, are described in greater detail elsewhere [8-Puzzle].

The 8-puzzle is used in this presentation to demonstrate that an application definition may indicate to TSIA the most promising tasks of a multi-origin out. A similar application for the 8-puzzle may be found elsewhere [Speculative].

Figure 87 and Figure 88 show a Fortran application to solve the 8-puzzle. The output of the application is shown in Figure 89a).

Its brute force algorithm for obtaining a solution is encoded in the routine move of Figure 88b). The algorithm simply tries many series of moves until a series is found which solves the 8-puzzle. Since any solvable 8-puzzle can be solved in maxdepth=31 moves or less, any longer series is abandoned as a solution.

The algorithm makes use of a very powerful heuristic, heuristic=distance+depth. The depth is the number of moves applied so far to the original board. The resulting board requires at least distance additional moves to reach the desired board. Thus if heuristic.gt.maxdepth, then the series of moves is abandoned as a solution.

The distance is the sum of the so-called manhattan distance of all the eight tiles. The manhattan distance is the number of moves required to move a tile to its desired position, while pretending that the other seven tiles do not exist and thus are not in the way. Since the manhattan distance for the board obviously never overestimates the number of additional moves required to reach the desired board, the heuristic never abandons a valid series of moves.

The Fortran routine move of Figure 88b) is rewritten in ia in Figure 89b). Unless the series of moves is a solution or is abandoned, a move task replaces itself by one, two or three further move tasks. The application execution thus results in many move tasks in the task pool. Any of the move tasks ultimately may yield the multi-origin out final_moves. The declaration move(...,int heuristic;;...promise(-heuristic) final_moves) indicates to TSIA the most promising tasks for the multi-origin out final_moves.

For the routine move, the item heuristic is an in like any other, though in this case heuristic is not used within the routine. For consistency, the Fortran routine move of Figure 88b) also has the argument heuristic, though it is completely irrelevant to the Fortran routine move.

The declaration promise(-heuristic) indicates that the promise of a solution increases with decreasing values of heuristic. In other words, since heuristic has positive values for the 8-puzzle, a task with a smaller values of heuristic is more likely to rapidly yield a solution. TSIA thus can use promise(-heuristic) in choosing which move task to execute next.

If the order given by heuristic is strictly obeyed, the solution found is guaranteed to have the smallest possible number of moves. This is known as the A* algorithm. Although the resulting number of moves is determinate, the solution itself may be indeterminate since there may be several possible solutions with the smallest possible number of moves.

In choosing which task to execute next, there is a range of possibilities for how TSIA can use promise(-heuristic). If TSIA strictly obeys promise(-heuristic), then TSIA provides the A* algorithm. Maintaining this strict ordering for the move tasks of course consumes resources. For example, this cost probably increases if the execution uses multiple computers. It thus may be more efficient for TSIA to abandon the strict ordering and instead merely use promise(-heuristic) as a guideline in choosing which move task to execute next. Of course in this case, the A* algorithm is lost. Thus the result is not guaranteed to have the shortest possible solution, but this may well be acceptable to the application definition. A third possibility, between the above two, is for the result to strictly obey promise(-heuristic) as in the A* algorithm, but leaving TSIA free to speculatively execute move tasks before they are known to be required by the A* algorithm.

This presentation makes no claims for the effectiveness of the A* algorithm for solving the 8-puzzle. Other algorithms, often derivatives of A*, may well be more effective. Similarly, other than pointing out that TSIA could obey promise(-heuristic) to varying degrees, this presentation makes no claim for the effectiveness of any of these degrees. Instead, the 8-puzzle and some of its issues are used in this presentation to demonstrate how TSIA might be able to support such a heuristic search and other search methods. Introductions to search methods may be found elsewhere [Search Methods]. Artificial intelligence and operations research are two areas which rely heavily on such search methods.

### *3.24 Streams*

A stream is any kind of structure which includes a potentially infinite number of elements. Such an infinitely large structure is feasible because most of its elements are conditional items of tasks and ultimately are ignored. The other elements of the stream ultimately are evaluated, but often need never exist simultaneously, thus further increasing the feasibility.

From the beginning [Streams], streams generally have been lists or have been very similar to lists [Monads][SICP][Why]. A stream list can imitate a stream tree [Why] and presumably also other structures.

In this thesis a stream can be any kind of structure. Thus for example, a stream array is an array, just like a sparse array is an array. This presentation demonstrates stream arrays and stream arguments.

Streams are compared to other definition elements elsewhere [Iters][Series]. Such comparisons also would seem to suggest applications for streams, including applications not mentioned in this presentation. For example, streams would seem to be an alternative to co-routines [Streams].

### *3.24.1 Introducing Stream Arrays*

Subsection 3.19.4.1 introduced the Newton-Raphson algorithm to compute the square root of a number `n`. That subsection described an application definition in Figure 77b) which uses a conditional in. This section defines the same algorithm, but using a stream array. The stream-based definition is very similar to one given elsewhere [Why].

Figure 90a) shows a stream-based Newton-Raphson definition consisting of the routines `next`, `swithin` and `ssqrt`. The routine `next` is that of Figure 77b). The routines `swithin` and `ssqrt` are similar to the routines `ciwithin` and `cisqrt` of Figure 77b), respectively.

In execution, the routine `ssqrt` of Figure 90a) replaces itself by the stream array `x` and the tasks `next` and `swithin`. The following examines each of these in turn.

Within the routine `ssqrt`, the declaration `x[0:]=a0` for the stream array serves three purposes in the ia language of this presentation. Firstly, the element `x[0]` may be used within the routine `ssqrt`. Secondly, `x[0]` is given the value `a0`. Thirdly, the trailing semi-colon declares the array to be a stream array with a potentially infinite number of elements `x[1]`, `x[2]` and so on beyond `x[0]`.

Within the routine `ssqrt`, the task `next(n,x[i];;x[i+1])` deductively evaluates the elements of the array `x[0:]`. This follows the explanation given in subsection 3.16.1 for a deductive array-based definition. The value of each element `x[k]` thus is the result of the `k`th iteration of the Newton-Raphson algorithm.

Within the routine `ssqrt`, the task `swithin(x[0],eps;;a)` has the stream array `x[0:]` as an in.

The ia routines `next`, `swithin` and `ssqrt` of Figure 90a) are rewritten in Fortran in Figure 90b). The Fortran code includes `program snewton` and thus is a complete Fortran application. The definitions of `next` and of `swithin` essentially are identical in ia and in Fortran. In contrast, the ia and Fortran definitions of `ssqrt` differ. The Fortran definition of `ssqrt` imitates the deductive evaluation of the stream array `x[0:]`.

Given the same ins `n`,`a0`,`eps` the out `a` is identical for the ia routine `ssqrt` and the Fortran routine `ssqrt` of Figure 90a) and b), respectively. Furthermore, the out `a` is identical with that of the Fortran routine `fsqrt`, the ia routine `cisqrt` and the Fortran routine `cisqrt` of Figure 77a), b) and c), respectively. While the five codes have very different executions, they have the same definition.

### 3.24.2 An Execution Involving Streams

For `ssqrt(n,a0,eps;;a)` of Figure 90a), an execution is illustrated in Figure 91. The execution assumes that the in `n,a0,eps` are such that the Newton-Raphson algorithm converges on the third iteration with the result that `a is x[3]`.

In the task pool of Figure 91, the elements shown for the stream array `x[0:]` are those that have been evaluated. In the execution, the fission of `x[0:]` is implicit. At no point does `x[0:]` exist in its infinite entirety. Furthermore, the routine `swithin` only demands that pairs of elements of `x[0:]` have consecutive memory locations. By occasionally copying or moving an element, IA allows the elements of `x[0:]` to not require consecutive memory locations. The routine `next` makes no demands on the locations of the elements of `x[0:]`.

The particular execution illustrated in Figure 91 interleaves the execution of the tasks `next` and `swithin`. In such an execution the corresponding `next` and `swithin` are fused for each element of the array `x[0:]`.

Of course, executions other than that illustrated in Figure 91 also are possible. For example, instead of fusing the corresponding `next` and `swithin`, multiple executions of `next` could be followed by multiple executions of `swithin`. In other words, multiple elements of the array `x[0:]` could be evaluated before knowing whether the latter elements will be ignored. As noted elsewhere [Speculative], this is an example of the speculative task order described in section 3.20. In contrast, the particular execution illustrated in Figure 91 is a conservative order.

### 3.24.3 A Stream-Based Numerical Differentiation Definition

Subsection 3.19.4.2 introduced a numerical differentiation algorithm to differentiate a function at a given point. That subsection described an application definition in Figure 78b) which uses a conditional in. This section defines the same algorithm, but using a stream array. The stream-based definition is very similar to that given elsewhere [Why].

Figure 92a) shows a stream-based numerical differentiation definition consisting of the routines `halve`, `easydiff`, `swithin` and `sdiff`. The routines `halve` and `easydiff` are those of Figure 78b). The routines `swithin` and `sdiff` are similar to the routines `ciwithin` and `cidiff` of Figure 78b), respectively. Actually, `swithin` is that of Figure 90a) of the above stream-based Newton-Raphson definition.

The stream-based numerical differentiation definition introduces no issues beyond those already introduced by the above stream-based Newton-Raphson definition. The same holds true for their executions. Thus neither the stream-based numerical differentiation definition nor its execution receives further explanation here.

The ia routines `halve`, `easydiff`, `ciwithin` and `cidiff` of Figure 92a) are rewritten in Fortran in Figure 92b). The Fortran code includes `program testsdiff` and `subroutine cfu` and thus is a complete Fortran application. The definitions of `halve`, of `easydiff` and of `swithin` essentially are identical in ia and in Fortran. In contrast, the ia and Fortran definitions of `sdiff` differ. The Fortran definition of `sdiff` imitates the deductive evaluation of the stream arrays `h[0:]` and `y[0:]`.

Given the same ins `f`,`x`,`h0`,`eps` the out `a` is identical for the ia routine `sdiff` and the Fortran routine `sdiff` of Figure 92a) and b), respectively. Furthermore, the out `a` is iden-tical with that of the Fortran routine fdiff, the ia routine cidiff and the Fortran routine cidiff of Figure 78a), b) and c), respectively. While the five codes have very different executions, they have the same definition.

### *3.24.4 The Benefits of Streams for an Application Definition*

The benefits of streams for an application definition are elegantly argued elsewhere [Why] . The above stream-based numerical differentiation definition begins one of the examples from that argument. This section presents the example in IA.

In order to have the stream-based numerical differentiation definition closely match the original definition [Why], the routine sdiff of Figure 92a) is rewritten as sdiff1 and differ in Figure 92c). In sdiff1, the stream array y[0:] is an out of differ and a conditional in of swithin. The execution of sdiff1 and its child task differ results in the exact same tasks and stream arrays as the execution of sdiff of Figure 92a).

For sdiff1 of Figure 92c), the stream array y[0:] is an out of the routine easy-diff and is a sequence of first order approximations which converges quite slowly. An improved sequence is the second order approximation z[i]=(y[i+1]*2**n[i]-y[i])/(2**n[i]-1), where n[i]=nint(log((y[i]-y[i+2])/(y[i+1]-y[i+2])-1)/log(2)). The function nint rounds to the nearest integer. The second order algorithm is defined by sdiff2 in Figure 93a).

The ia routines of Figure 93a) are rewritten in Fortran in Figure 93b). Adding the routines of Figure 92b) results in a complete Fortran application. In Figure 93, except for the routine sdiff2, the routines essentially are the same in ia and in Fortran. The Fortran definition of the routine sdiff2 imitates the deductive evaluation of the stream arrays h[0:], y[0:] and z[0:].

Given the same ins f,x,h0,a0,eps the out a is identical for the ia routine sdiff2 and the Fortran routine sdiff2 of Figure 93a) and b), respectively. While the two codes have very different executions, they have the same definition.

The improvement applied above to the first order sequence, thus yielding the second order sequence, also can be applied to the second order sequence, thus yielding a third order sequence. Similarly, the third order sequence can be improved to yield a fourth order sequence and so on. For example, the fourth order algorithm is defined by the routine sdiff4 in Figure 94a).

The routines sdiff1 of Figure 92c), sdiff2 of Figure 93a) and sdiff4 of Figure 94a) each define a fixed order algorithm. An alternative is the increasing order algorithm defined by the routine sdiffsuper of Figure 94b), which is described in the remainder of this subsection.

In sdiff2 of Figure 93a), the stream arrays y[0:] and z[0:] are the first and second order sequences, respectively. Similarly, in sdiff4 of Figure 94a), the stream arrays y[0:], v[0:], w[0:] and z[0:] are the first, second, third and fourth order sequences, respectively. Instead of using an individual one-dimensional stream array for the sequence of each order, a two-dimensional stream array can store the sequences of all orders. In the routine sdiffsuper of Figure 94b), s[0:,0:] is such two-dimensional stream array. The first order is s[0:,0] and is an out of the task differ(f,x,h0;;s[0:,0]). For the first order sequence, the left index of s[0:,0] is streamed, while the right index is fixed at 0.

The second order is `s[0:,1]`. As an improvement of the first order, the second order could be defined as `improve(s[i:i+2,0];;s[i,1])`. For example, this is the equivalent of `improve(y[i];;v[i])`, or more explicitly `improve(y[i:i+2] ;;v[i])`, in routine `sdiff4` of Figure 94a).

In the ia language of this presentation, if a one-dimensional piece of a two-dimensional array is an argument of a task, then the elements of the piece must be defined explicitly. Thus the above improvement of the first order to yield the second order could not be defined as `improve(s[i,0];;s[i,1])` in the ia language of this presentation.

The above improvement of the first order to yield the second order can be generalized to define all orders as `improve(s[i:i+2,k];;s[i,k+1])`, as in routine `sdiffsuper` of Figure 94b).

The second term of each order is defined by the sequence `s[1,0:]`. In the routine `sdiffsuper`, this sequence is an in of the task `within(s[1,0:],eps;;a)`. The second term of each order is used in this presentation in order to match the algorithm of the original definition [Why]. The first term received by `swithin` thus is `s[1,0]` and is first order. The second term is `s[1,1]` and is second order. In general, each successive term received by `swithin` is one order higher.

The routine `sdiffsuper` of Figure 94b) thus defines an increasing order algorithm.

The routines `sdiff1` of Figure 92c), `sdiff2` of Figure 93a), `sdiff4` of Figure 94a) and `sdiffsuper` of Figure 94b) nicely illustrate the benefits of streams for the convenient definition of an algorithm involving sequences [Why]. Similar examples also are part of similar arguments elsewhere for the benefits of streams [SICP].

### *3.24.5 The Sieve of Eratosthenes*

The sieve of Eratosthenes is an algorithm to generate the sequence of prime numbers. Like the examples of subsection 3.24.4, the sieve of Eratosthenes illustrates the benefits of streams for the convenient definition of an algorithm involving sequences [Iters][SICP].

By removing the multiples of the prime numbers from the sequence of integers `2,3,4,5,...`, the sieve of Eratosthenes yields the sequence of prime numbers `2,3,5,7,11,...`. A detailed algorithm follows.

In the initial sequence `2,3,4,5,...` the first number is `2` and is the first number in the sequence of primes. From the remaining sequence `3,4,5,6,...`, the multiples of `2` are removed yielding `3,5,7,9,...`. For this sequence the first number is `3` and again is prime and thus is appended to the sequence of primes. From the remaining sequence `5,7,9,11,...`, the multiples of `3` are removed yielding `5,7,11,13...`. For this sequence the first number is `5` and again is prime and thus is appended to the sequence of primes. From the remaining sequence `7,11,13,17,...`, the multiples of `5` are removed yielding `7,11,13,17,...`. Repeated infinitely, the algorithm yields the infinite sequence of prime numbers `2,3,5,7,11,...`. The above algorithm is defined in the ia code of Figure 95a).

The ia code of Figure 95a) is imitated in Fortran in Figure 95b). The Fortran code includes `program eras` and thus is a complete Fortran application.

### 3.24.6 *Introducing Stream Arguments*

Subsection 3.24.5 introduced the sieve of Eratosthenes to compute the sequence of prime numbers. That subsection described an application definition in Figure 95 which uses stream arrays. This section defines the same algorithm, but using stream arguments.

Figure 96 shows a stream argument-based sieve of Eratosthenes definition. The routines `rem_mults`, `sieve` and `primes` are similar to those of Figure 95a).

For the routine `primes(;;del p:)`, the argument `p:` declares that `p` is the first element of a stream argument. Similarly, for the routine `next_integer(x;; y: del z:)`, the argument `y:del z:` declares that `y` is the first and `z` is the second element of a stream argument. In general, a routine declares those elements of the stream that are used in the routine.

In the ia language of this presentation, if the type of an element of a stream argument is not declared, then the type is assumed to be that of the previous element. For example, `int y:z` corresponds to `int y:int z`.

In the ia language of this presentation, the use of each element of the stream is declared, as for any other item of the routine. As usual, if no use is declared, the default `del_eva` is assumed. For example, `y:del z` corresponds to `del_eva y:del z`.

In execution, `primes(;;p:)` replaces itself by the tasks `next_integer(1;; a:)` and `sieve(a:;;p:)`. Given the declarations of these routines, the tasks execute as `next_integer(1;; a:b:)` and `sieve(a:b:;;p:q:)`. The IA introduces and manages the stream elements `b` and `q`. In essence, such IA support for stream elements is the IA support of stream arguments.

For the simple algorithm of the sieve of Eratosthenes, the stream argument-based definition is very similar to the stream array-based definition of subsection 3.24.5 and Figure 95. The stream argument-based definition thus receives no further explanation here.

The general similarity between a stream array and a stream argument is demonstrated by the two routines of Figure 97 which convert between a stream array and a stream argument.

### 3.24.7 *Stream Arguments for Numerical Differentiation*

The previous subsection used the sieve of Eratosthenes to introduce stream arguments as an alternative to stream arrays. In order to further demonstrate the generality of stream arguments, the numerical differentiation definitions of subsection 3.24.4 are redefined in this subsection using stream arguments instead of stream arrays.

The original `sdiff1` and its associated routines using stream arrays are defined in Figure 92c). Using stream arguments, the routines are redefined in Figure 98a).

Similarly, the original `sdiff2` and its associated routines using stream arrays are defined in Figure 93a). Using stream arguments, the routines are redefined in Figure 98b). As demonstrated by the element `d` passed by `improve` to its children `order` and `elimerror`, in the ia language of this presentation, the elements of a stream argument can be the elements of an array. Alternatively, the routines could be defined as `order(d1:d2:d3;;o)` and `elimerror(n,d1:d2;;e)`.

Similarly, the original `sdiff4` using stream arrays is defined in Figure 94a). Using stream arguments, the routine is redefined in Figure 98c).

Similarly, the original `sdiffsuper` using stream arrays is defined in Figure 94b). Using stream arguments, the routine is redefined in Figure 99.

### *3.24.8 Streams for Input/Output*

Streams may be used to provide an application with input/output (I/O) or other interaction [Monads][Streams]. For such interaction, streams are an alternative to the interaction items described in section 3.28.

Figure 100 defines a complete ia application. It uses a stream to copy character-by-character its input to its output. Similar stream-based applications are described elsewhere [Monads]. Two demonstrations of the benefits of streams for I/O follow.

Firstly, a stream easily may be modified by passing it through a transformer. For example, `main(){gs(;;c);tolower(;c;);ps(c;;);}` is an application which changes any uppercase letter in the input to lowercase in the output. As in the programming language C [C], if the inout `c` is an uppercase character, `tolower(;c;)` changes it to lowercase, otherwise `c` is not changed. Such transformers on streams are discussed further in the next section.

Secondly, a speculative task order for `gs(;;c);ps(c;;);` allows many `gs` tasks to execute before the corresponding `ps` tasks are executed. Such buffering of I/O can improve the performance of the execution. In addition, filling a buffer can occur in parallel with the emptying of an earlier buffer.

### *3.24.9 Streams for Signal Processing*

A stream is a natural representation for a signal in digital form [Signal Processing]. Each element of the stream corresponds to a sample of the signal at a discrete point. Signal processing is a part of speech analysis, image analysis, seismic exploration and other applications.

For simplicity, this presentation is restricted to uniformly-sampled signals. Because the points of the signal are equally spaced, the position of an element in the stream identifies its point. For example, the index of an element in a stream array identifies its point. In contrast, a non-uniformly sampled signal requires the elements of the stream to identify the points of the signal. The use of streams for non-uniformly sampled signal is examined elsewhere [Asynchronous Streams].

The benefits of streams for signal processing are elegantly argued elsewhere [Signal Processing]. The following subsections present in IA three of the examples from that argument. Other uses for steams as signals are presented elsewhere [SICP].

#### *3.24.9.1 Averaging Samples of a Signal*

Figure 101a) defines a complete ia application for signal processing. Each element of the resulting stream is the average value of two adjacent elements of the original stream.

For simplicity, the signal processing examples of this presentation assume that each sample is a real number. Of course, each sample could be any type of item or kind of structure.

In the ia code of Figure 101a), the routine `gsr` inputs a stream array of `real` items in a fashion similar to the routine `gs` of Figure 100, which inputs a stream of `char` items. The same similarity exists between `psr` of Figure 101a) and `ps` of Figure 100 for the output of a stream.

In the ia language of this presentation, `gsr(;;real r[0: del_ign 1:])` of Figure 101a) defines that `r[0:]` is a stream array with the element `r[0]` either delegated or evaluated in `gsr`, via the default `del_eva`, and the element `r[1]` either delegated or ignored.

The simple ia application of Figure 101a) processes the signal with the single transformer `avg(a[i],a[i+1];;b[i])`. In general, a ia application can chain together many transformers. For example, replacing the above single transformer by the two transformers `avg(a[i],a[i+1];;x[i]);avg(x[i],x[i+1];;b[i])` results in a different application.

Because it is deductively evaluated, the transformer `avg(a,b;;c)` in Figure 101a) is a traditional routine, unaware that it is used for a stream. For example, such a traditional routine could be a Fortran routine.

Figure 101b) shows a ia application very similar to that of Figure 101a). While the transformer `avg` of Figure 101a) creates a second stream, the transformer `avginout` of Figure 101b) modifies the original stream. Also the transformer `avginout(a;b;)` is a traditional routine which for example could be a Fortran routine.

Figure 101c) shows a different ia application very similar to that of Figure 101a). While the ia code of Figure 101a) uses stream arrays, that of Figure 101c) uses stream arguments.

#### *3.24.9.2 A Rate Changer*

Figure 102a) defines a ia application which converts a signal to a different sampling rate. The transformer `FourForThree` produces four samples for every group of three samples in the original stream. The sampling rate thus is increased by a factor 4/3.

Figure 102b) shows a ia application very similar to that of Figure 102a). While the transformer `FourForThree` of Figure 102a) is aware that it is used for a stream, that of Figure 102b) is unaware and instead is traditional routine. As already mentioned in the previous subsection, such a traditional routine could be a Fortran routine.

In the ia language of this presentation, the task `gsr(;;a[1:3,0:])` of Figure 102b) defines a stream of length-3 vectors `a[1:3,0]`, `a[1:3,1]` and so on. This presentation assumes that the routine `gsr(;;r[0: del_ign 1:])` first fills `a[1,0]`, then `a[2,0]`, then `a[3,0]`, then `a[1,1]` and so on. This assumption is unnecessary if instead a routine like `gsrn(n;;r[1:n,0: del_ign 1:])` is defined and used. Similar comments apply to the routine `psr` of Figure 102b).

#### *3.24.9.3 A Two Dimension Filter*

Figure 103 defines an application which applies a three-by-three array `filter` to each sample of a stream of vectors. Since it is deductively executed in the ia code of Figure 103a), the transformer `TwoDimFilter` is a traditional routine which may be a Fortran routine as shown in Figure 103b).

### *3.25 Defining the Elements of a Routine*

Up until this point of the presentation, a single method is used to define the elements of a routine. So far, the elements of a routine are defined by the parent of the routine. For example, in the ia code `f(a;;b){g(a;;x);h(x;;b);}` the elements `g`, `a` and `x` of the routine `g(a;;x)` are defined by the parent `f(a;;b)`. This method for defining the elements of a routine is sufficient for many applications, including all those demonstrated so far.

The remainder of this chapter demonstrates other methods to define the elements of a routine.

### *3.25.1 A Variation on Currying*

An alternative method to define the elements of a routine is a variation on currying [Curry : C][Curry : Teaching]. For example, given the routine `add(a,b;;c){c=a+b;}`, then the statement `incr(i;;j) is add(1,i;;j);` defines `add` and `1` to be elements of the routine `incr(i;;j)`. The items `i` and `j` remain to be defined elsewhere. In this context, the keyword `is` may be read as "curries".

In functional computing, currying is the conversion of a function with N arguments into a nest of N functions of one argument each. For example, if `((a(x))(y))(z)=b(x,y,z)` for all `x,y,z` then the function `a` is a curried version of the function `b`. The function `a` is called a curried function. A curried function thus expects exactly one argument and returns either the desired result or another curried function.

Currying in TSIA is similar, but not identical, to currying in functional computing. In TSIA, any routine may be a curried routine. In contrast to functional computing, a routine in TSIA needs no conversion for currying. For example, given the *curry* `incr(i;;j) is add(1,i;;j);`, then `add` is the *curried routine* and `incr` is the *currying routine*.

In general, a curried routine can be used like any other routine. Thus any combination of its items may be curried. An example is the curry `x(c,e;;g) is a(b,c,d,e,f;;g);` where `a,b,d,f` are previously defined. A curry even may leave the instruction free. For example, the curry `args(f;;x) is f(2,3;;x);` subsequently may be used as `args(mult;;y)` or `args(add;;z)`, where `mult(a,b;;c)` and `add(a,b;;c)` are routines.

Like a curried routine, also a currying routine generally can be used like any other routine. Thus a currying routine of course may be curried. For example, the above currying routine `incr` may be curried in the definition `two(;;t) is incr(1;;t);`. Other examples for a currying routine are illustrated in Figure 104. Each of the four illustrations in Figure 104a) through d) uses the same currying routine `incr` in a different fashion to define effectively the same routine `incrincr`.

As demonstrated in Figure 104a) and b), the currying routine `incr` can be defined inside or outside of the routine `incrincr`, respectively. A currying routine thus can be nonlocal or local, respectively. In Figure 104a) the currying routine `incr` is global since it is defined at the topmost level of the application definition, just like the definition of most routines in this presentation.

In Figure 104c), `incr` is an in of the routine `twice(incr,k;;l)`. While `incr` is local in Figure 104c), it alternatively could be nonlocal. In Figure 104d), `incr` is an out of the routine `getincr(;;incr)`.

The feasibility of currying in TSIA easily is demonstrated. In the illustrations of Figure 104a) and b), the curry `incr(i;;j)is add(1,i;;j);` can be a textual substitution which occurs at compile-time. For example, this is similar to the macro substitution performed by the C preprocessor [C]. Thus in the illustration of Figure 104a), the original incrincr(k;;l){incr(k;;m);incr(m;;l);} can be compiled as incrincr(k;;l){add(1,k;;m);add(1,m;;l);}.

For the illustrations of Figure 104c), the currying routine incr involves a similar substitution, but one which cannot occur at compile-time. Instead, the substitution occurs at run-time in the task pool. An execution is illustrated in Figure 105a). Similarly, Figure 105b) illustrates an execution of the code of Figure 104d).

As illustrated in Figure 105a) and b), currying in TSIA places definitions like incr(i;;j)is add(1,i;;j); into the task pool. This is feasible since coordinating the elements of such a curry, here add and 1, is no more difficult for the TS than coordinating the items of a task. Of course this coordination includes managing any dependencies introduced by the curry.

Like the dependencies on any other item or structure in the task pool, the dependencies on a currying routine are explicit. Thus, in the execution of Figure 105a), the currying routine incr, along with its elements add and 1, is removed by the TS from the task pool as soon as nothing in the task pool depends on incr. Similarly for the currying routine g in Figure 105b).

The remaining paragraphs introduce some additional issues concerning currying.

The above introduction of currying only involves the ins of routines. Symmetry raises the possibility of currying the outs of routines. This possibility is introduced below in subsection 3.25.3.

It would seem that a ia language could use the same syntax for currying as for the definition of a usual routine. For example, the above curry incr(i;;j) is add(1,i;;j); then would be defined as incr(i;;j){add(1,i;;j);}. A ia compiler then would recognize each definition as a macro, a curry or a usual routine. This presentation uses the is notation in order to make currying explicit. The unification of macros, curries and usual routines is not pursued in this presentation.

Since the TS coordinates the elements of a currying routine, an element can be any item or structure in the task pool. In the above example of the currying routine incr, the elements add and 1 simply are constants. In contrast, in the example routine incrn(n;;i(a;;b)){i(a;;b) is add(n,a;;b);}, the element n of the currying routine i is an item of the parent incrn. Similarly, an element of a curry could be an out of a child task. For example, in the routine incr2n(n;;i(a;;b)) {add(n,n;;nn);i(a;;b)is add(nn,a;;b);}, the item nn of the currying routine i is an out of the child add(n,n;;nn).

Just as child tasks may depend on items evaluated by the parent, a curry may have similar dependencies. For example, f(n;;i(a;;b)){i(a;;b) is (n<0?g:h)(n,a;;b);} curries g(n,a;;b) or h(n,a;;b) depending on the value of n. This example makes use of the ?: conditional operator as in the C programming language [C].

A curry may have multiple curried routines. A simple example is the curry add3(a,b,c;;d) is {add(a,b;;x); add(x,c;;d);}. Whether a curry involves one or more routines makes little difference to the TS coordinating the elements of the currying routine.

The order of the ins, inouts and outs of the curried routine of course need not match those of the currying routine. For example, the curry new(a,b;c,d;e,f) is old(b,a;d,c;f,e) just changes the order of the elements.

### 3.25.2 Currying for Configuring and Combining Routines

Some of the benefits offered by currying in defining an application are described elsewhere [Curry : C]. For example, currying enables configuring and combining routines and thus allows for more concise and reusable software. This motivation is demonstrated in this subsection by examples similar to some from the description mentioned above.

The examples of this subsection are based on the use of a general sorting routine. The routine is similar to the `qsort` routine provided by the C programming language [C]. Some implementations of such a sort routine are described in subsections 3.13.8 and 3.13.9. Here the routine `intsort(n,cmp(k,e;;c);a[n])` sorts into ascending order an array of `n` elements `a[1]`,...,`a[n]`. For the comparison function `cmp`, the ins `k` and `e` will be individual elements of the array `a[n]`. The out `c` must be negative if the in `k` is less than the in `e`, zero if equal, and positive if greater. The meaning of less, equal and greater for the array elements is defined by `cmp`. For example, if the array is to be sorted into ascending order by value, then the comparison routine is `subtract(x,y;;z){z=x-y;}`, as illustrated in Figure 106.

Currying allows a comparison function to be conveniently configured. An example is illustrated in Figure 106 using the comparison routine `divcmp`. The curry `cdivcmp(x,y;;z) is divcmp(10,x,y;;z);` configures `divcmp` with `10` for the divisor `d`. Obviously such a curry could configure `divcmp` with any other value or item.

In Figure 106, the configuration allowed by currying may be contrast with the routine `f100divcmp` which provides no possibility for configuration. With such an unconfigurable routine, a different value for the divisor `d` of `divcmp` requires using another routine.

An alternative to currying for configuration uses a global item. This alternative is illustrated in Figure 106 by the routine `gdivcmp` and its global item `globalm`. Such configuration by global items is less convenient than that by currying since a global item needlessly exposes the configuration information to the application. In contrast, currying keeps such information local and thus provides encapsulation. IA support for global items is introduced in section 3.26.

The configuration allowed by currying extends through to combining routines in order to create another routine. This is illustrated in Figure 106 by the curry `mdivcmp(x,y;;z) is{modulus(x,10;;xm);modulus(y,10;;ym);subtract(xm,ym;;z);}`. Currying thus allows a a small variety of elementary routines to be combined into a large variety of comparison functions.

### 3.25.3 An Application Defined Type (ADT)

A programming language typically defines several built-in types such as `int`, `real` and array of `real`. An *application defined type* (ADT) is a type defined by the application, with a use as convenient or nearly so as the use of a built-in type. In previous presentations an ADT is known as an abstract data type or as a user defined type. The benefits of an ADT are described elsewhere [C++][CLU].

An example of an ADT is `ran(;;real)`, the type of the out of the routine `rangen(int;;ran(;;real))` in Figure 107a). The routine `rangen` is used to define an item of that ADT. In the code fragment of Figure 107a), the task `rangen(66;;c)` defines an item named `c`. Such an ADT item is similar to an item of a built-in type. For example, given the built-in type `int` and the routine `add(int a,int b;;int c)`, then the task `add(5,7;;d)` defines an `int` item `d`.

The ADT `ran(;;real)` of Figure 107a) is a pseudo-random number generator. The ADT is based on the routine `random(;int seed;real num)`, which can be regarded as an interface to a sequence of pseudo-random numbers, with `seed` as an index to that sequence. Inside `random`, the in value of `seed` thus determines the value of the out `num`. The value of `seed` then is updated to correspond to the next number in the sequence.

The definition of an ADT includes all operations on an item of that type. The type `ran(;;real)` of Figure 107a) is unusual since it has only one operation. The task `c(;;r1)` of Figure 107a) is an example of an operation on the item `c` of the type `ran(;;real)`.

In `rangen(int;;ran(;;real))`, the ADT `ran(;;real)` is defined by the curry `ran(;;r) is random(;s;r)`. This demonstration introduces how currying supports an ADT. Since IA supports currying, IA thus supports an ADT.

The execution of the code fragment in Figure 107a) is illustrated in Figure 107b). The execution of the task `rangen(66;;r)` replaces it in the task pool by the item `s is 66` and the curry `c(;;r) is random(;s;r)`. For the tasks `c(;;r1)` and `c(;;r2)`, the curry then substitutes `random(;s;r1)` and `random(;s;r2)`, respectively.

If the item `c(;;r)` is not used beyond the code fragment of Figure 107a), it then may be removed from the task pool as illustrated in Figure 107b). Similarly, the item `s` may be removed from the task pool once `random(;s;r1)` and `random(;s;r2)` have executed. Thus an ADT item may be removed from the task pool once that item no longer is an in of any subsequent task. In this sense at least, an ADT thus is a type like any other in TSIA.

In the curry `c(;;r) is random(;s;r)`, the item `s` is an inout. This curry thus introduces the possibility of currying an out or inout of a routine.

The execution illustrated in Figure 107b) demonstrates that some dependencies between tasks in the task pool only become apparent after the substitutions due to currying. In particular, the task `c(;;r1)` must execute before the task `c(;;r2)` because of their mutual dependence on the item `s` hidden in their ADT `ran(;;real)`. This dependence only becomes apparent only after currying replaces the tasks by `random(;s;r1)` and `random(;s;r2)`, respectively.

The items of a type are independent. For example, the tasks `add(5,7;;d)` and `add(11,13;;e)` are independent and yield the independent `int` items `d` and `e`. Similarly, the tasks `rangen(77;;a)` and `rangen(88;;b)` in Figure 107c) are independent and yield the independent items `a` and `b` of the type `ran(;;real)`. Because the items `a` and `b` are independent, the tasks `a(;;a1)` and `b(;;b1)` of the code fragment in Figure 107c) are independent.

The execution in the task pool of the code fragment in Figure 107c) is illustrated in Figure 107d). The execution of the task `rangen(77;;a)` replaces it in the task pool by the item `sa is 77` and the curry `a(;;r) is random(;sa;r)`. Similarly, `rangen(88;;b)` is replaced by the item `sb is 88` and the curry `b(;;r) is random(;sb;r)`. The tasks `rangen(77;;a)` and `rangen(88;;b)` may execute in any order, including in parallel, since the tasks are independent. In particular, the items `a` and `b` are independent. In the continuing execution, for the task `a(;;a1)` the curry `a(;;r)` the curry substitutes `random(;sa;a1)`. Similarly, the curry `b(;;r)` replaces the task `b(;;b1)` with the task `random(;sb;b1)`. Since the tasks `random(;sa;a1)` and `random(;sb;b1)` are independent they may execute in any order.

In order to illustrate their independence, the items `sa` and `sb` have been relabelled from the original item `s` hidden in the `rangen` routine of Figure 107a). The item `s` is an example of an *item element*. Each item has its own independent item elements which maintain the state of the item between operations on that item. For the type `ran(;;real)`, the only item element is `s`, which maintains its value across calls to `random(;s;r)`. In contrast to an item element, a *type element* is shared across all items of a particular type. Type elements are introduced in subsection 3.26.5.

The ADT `ran(;;real)` of Figure 107a) is expanded to `record Ran{ran(;;real),seed(;;int)}` in Figure 107e), thus making available the value of the seed. A `Ran` item is defined using the routine `Rangen(int;;Ran)`. A `record` is similar to a `struct` of the C or C++ programming languages [C][C++]. A record is a convenient structure for a type. Alternatively, the ADT could be defined using `Rangen(int;;ran(;;real),seed(;;int))`, but this might not be convenient in some situations.

In `Rangen(int i;;Ran g)`, the order in which the two operations `g.ran(;;r) is random(;s;r);` and `g.seed(;;e) is set(s;;e);` are defined is irrelevant to their mutual dependence on the item element `s`. Instead, the dependencies on `s` are given by the uses of the operations in the application. These dependencies are illustrated by the code fragment in Figure 107e).

An item of an ADT can be an ADT. In other words, a IA does not distinguish between items and ADTs. The ADT `ran(;;real)` is the type of the out of the routine `rangen(int;;ran(;;real))` in Figure 107a). This example is extended in Figure 108. There the ADT `rangen(int;;ran(;;real))` is the type of the out of the routine `genrangen(gen(;int;real);;rangen(int;;ran(;;real)))`. As in Figure 107a), the ADT `an(;;real)` of Figure 108 is a pseudo-random number generator. The ADT `rangen(int;;ran(;;real))` of Figure 108 is a generator of pseudo-random number generators.

The basic support for an ADT consists of facilities for defining a set of operations for a type and for restricting the access to an item of the type to that set of operations [C++] [CLU]. This subsection has demonstrated that IA can provide such support. The next subsection introduces nested routines and thus extends IA support for an ADT.

Support for an ADT is one part of the support for object computing [C++]. Among other things, this part maintains the state of an item between operations on that item. The ability for IA to maintain such state for an ADT, suggests that IA also might be able to do so for object computing. While demonstrated here for an ADT defined using a routine, the ability for IA to maintain state would seem to be more general. For example, it seems that IA might be able to support object computing using classes as in C++.

### *3.25.4 Nested Routines*

Currying allows IA to support nested routines. Example of nested routines are given in Figure 109a). There the routines `c.get(;;int)` and `c.inc(;;)` are nested within the routine `count(int;;Count c)` with `record Count{get(;;int),inc(;;)}`.

The routine `c.get(;;g){g=x;}` uses the item `x` which is not local to the routine. Similarly, `x` also is a nonlocal item within the routine `c.inc(;;){x=x+1;}`.

In order to support nested routines with nonlocal items, IA can adopt and extend a solution described elsewhere [CPCPS]. A nonlocal item of a routine can be made a local item by adding that item to the arguments of the routine. The extra argument then can be satisfied by currying. A routine without nonlocal items is said to be closed. Unnesting a closed routine at most requires renaming the routine in order to avoid clashing with the names of other routines. In short, currying can close any nested routine and such a routine can be trivially unnested.

The routine `count` of Figure 109a) thus can be rewritten as in Figure 109b). For example, the original nested routine `c.inc(;;){x=x+1;}` is rewritten as the curry `c.inc(;;) is cinc(;x;)`. The curried routine `cinc` is a closed routine and is not nested. Such rewriting could be an early stage of a compiler. The routine `count` thus is reduced to the form supported in the previous subsection for an application defined type (ADT). The IA support for an ADT thus now extends through to the use of nested routines. The routine `count` and its ADT `Count` thus implement a simple counter whose use is illustrated in Figure 109c).

Another example of an ADT using nested routines is `Stack` of Figure 109d). Like in the real world, items pushed onto a stack are popped off in LIFO order. The implementation of a stack is described elsewhere [C]. The `Stack` ADT allows an application to use many independent stacks. Each `Stack` item defined by `stack(int max;;Stack s)` holds up to `max` integers. Alternatively, IA support for arbitrarily large arrays allows for a `Stack` which holds arbitrarily many integers. Such a `Stack` is implemented in Figure 109e).

The above examples demonstrate that a IA can support nested routines. The examples are ADTs, but nested routines also have other uses. One such use is demonstrated in the next subsection.

### *3.25.5 Unnamed Routines*

Usually a routine is named, but occasionally it may be convenient for a routine to remain unnamed. For example, the routine `add(x,y;;z){z=x+y;}` is named `add`, while the routine `(x,y;;z){z=x+y;}` is unnamed. In another example the curry `incr(a;;b)is add(1,a;;b)` has a currying routine named `incr`. In contrast, the curry `(a;;b)is add(1,a;;b)` is unnamed. IA support for named currying routines and for named nested routines is described in subsections 3.25.1 and 3.25.4, respectively.

The code of Figure 110 demonstrates the use of unnamed routines for currying routines and for nested routines. The code of Figure 110 derives from that of Figure 106, which is described in subsection 3.25.2.

Figure 110 first demonstrates the use of a named currying routine. The curry `cdivcmp(x,y;;z) is divcmp(10,x,y;;z)` permits the task `intsort(n, cdivcmp;a;)`. Instead of introducing the name `cdivcmp`, the unnamed curry can be an argument of the task `intsort(n,(x,y;;z) is divcmp(10,x,y;;z);a;)`. Such a use of an unnamed curry trivially is implemented by rewriting it using a named curry. For example, it can be rewritten using the named curry `cdivcmp` described above. Such rewriting could be an early phase of a compiler.

Figure 110 then demonstrates the use of a named nested routine. The nested routine `n10divcmp(x,y;;z){z=x/10-y/10;}` permits the task `intsort(n, n10divcmp;a;)`. Instead of introducing the name `n10divcmp`, the unnamed nested routine can be an argument of the task `intsort(n,(int x,int y;;int z){z=x/10-y/10;}; a;)`. Such a use of an unnamed routine trivially is implemented by rewriting it using a named nested routine. For example, it can be rewritten using the nested routine `n10divcmp` described above. Such rewriting could be an early phase of a compiler.

From the above descriptions, the use and implementation of unnamed routines is very similar for curries and for nested routines.

The argument types for an unnamed routine can be taken from the prototype of the called routine. In Figure 110 for example, the prototype `intsort(int,cmp(int, int;;int);int [n];)` permits the task `intsort(n,(x,y;;z){z=x/10-y/10;}; a;)`, with a fairly compact definition of an unnamed nested routine. Even more compact definitions for an unnamed nested routine may be possible. Similarly, compared to that shown above, more compact definitions for an unnamed curry may be possible.

Up until this point of the presentation, all curried routines have been unnested routines. Also nested routines may be curried. Figure 110 concludes with a demonstration of an unnamed curry where the curried routine `ldivcmp` is nested.

### *3.25.6 The while Loop Construct as a Routine*

As mentioned elsewhere [Discipline : Preface], a routine can imitate the `while` loop construct. Unnamed routines allow a close imitation. An example imitation is the routine `whiledo(c(;;boolean l),b(;;);;)` of Figure 111a). If the out `l` of the conditional routine `c(;;boolean l)` is true, then the body routine `b(;;)` is executed and `whiledo(c,b;;)` again executes recursively.

A use of the `while` loop construct is demonstrated in Figure 111b) by the routine `addcon(int n;;int r)` which calculates the addtorial. The same calculation is performed in Figure 111c) by the routine `addrou(int n;;int r)`, but using the routine `whiledo`. The use of the `while` loop construct is closely imitated by using two unnamed routines as the arguments to the task `whiledo((;;l){l=i<=n;}, (;;){r+=i;i++;};;)`.

As described in subsection 3.25.5, an unnamed routine may be implemented by rewriting it as a named routine. As demonstrated in Figure 111d), an example rewrite of the routine `addrou` has the named routines `cond(;;boolean l){l=i<=n;}` and `body(;;){r+=i;i++;}` as arguments to the task `whiledo(cond,body;;)`.

In Figure 111d), the nested routine `cond(;;boolean l){l=i<=n;}` uses the nonlocal items `i` and `n`. Similarly, the nested routine `body(;;){r+=i;i++;}` uses the nonlocal items `i` and `r`. As described in subsection 3.25.4, currying allows nonlocal items to be made local and the routines to be unnested. A corresponding rewrite of the routine `addrou` is demonstrated in Figure 111e). The previously nested routines thus are given by the curries `cond(;;l)is ccond(i,n;;l)` and `body(;;)is cbody(;r, i;)`, respectively.

The rewritten addtorial definition of Figure 111e) is imitated by the Fortran definition in Figure 111f). The code may be combined with `program addtprog` of Figure 23a) for a complete Fortran application. As throughout this chapter, while the ia and the Fortran definitions are similar, their executions can be very different.

Similar to that demonstrated above for the `while` loop construct, the use of unnamed routines obviously also allows routines to closely imitate the `if then` conditional construct, the `case` choice construct and other constructs.

As explained in subsection 3.2.3.3, any routine in the subordinate style may be converted to the delegation style. Such conversion might benefit from the ability for routines to imitate the `while` loop construct and other constructs.

Subsection 3.3.3 demonstrates that recursion can execute like a loop. This subsection demonstrates that recursion can be defined like a loop. A loop thus can be regarded as a special case of recursion. In other words, iteration using a loop can be largely equivalent to iteration using recursion.

## *3.26 A Determinate Definition Using Nonlocal Items*

### *3.26.1 Introducing Global Items*

A global item can be a nonlocal item of any task of an application. In the code of Figure 112a) for example, the global item `gs` is an item of all `gran` tasks and all `gseed` tasks. The code is a variation on the `Ran` ADT demonstrated in subsection 3.25.3 and Figure 107e). There the ADT allows an application to use arbitrarily many `Ran` items, each an independent pseudo-random number generator. Here `gran` and `gseed` correspond to a single generator for the entire application.

The code of Figure 112a) uses `int gs=31` to define the global item `gs`. Given the routine `set(int x;;int y){y=x;}`, an alternative might use the task `set(31;;gs)` to define `gs`. Elsewhere throughout this presentation most items are defined using tasks. The possibility of using tasks to define global items is not pursued in this presentation. More generally, the definition and declaration of items, such as the possibility of exclusively using tasks to define items, is not pursued in this presentation.

### *3.26.2 A Determinate Definition Using Global Items*

The introduction of global items to the definition of an application raises several issues for the execution of such an application. In particular, global items introduce dependencies between tasks which must be obeyed by the execution if the application definition is to remain determinate. In order to provide more freedom for the application execution, an application may forego this determinate definition, but this possibility is not pursued in this presentation.

Some of the issues introduced by global items are illustrated in Figure 112b) by the execution in the task pool of the code fragment in Figure 112a). Throughout the execution, the global item `gs` is in the task pool. The illustrated execution of the tasks `gseed(;;x);gran(;;y);gseed(;;z);gran(;;w)` assumes that `gseed(;;z)` executes first. Even though the resulting task `set(gs;;z)` is the only task in the pool depending on the global item `gs`, the task may not yet execute since the preceding tasks `gseed(;;x)` or `gran(;;y)` may delegate to a child or other descendant which depends on `gs`. The remaining execution thus first must execute `gseed(;;x)` and `gran(;;y)`. Since they share no dependencies, the two tasks may execute in any order. The task `gseed(;;x)` delegates to `set(gs;;x)`. The task `gran(;;y)` delegates to

`random(;gs;y)`. Since there are no tasks in the pool which precede it, `set(gs;;x)` may execute, leaving item `x` in the task pool for some subsequent task not included in the code fragment. Now that there are no tasks in the pool preceding `random(;gs;y)`, it too may execute, leaving item `y` in the task pool. Now that `gseed(;;x)` and `gran(;;y)` and all their descendants have executed, there are no tasks in the pool preceding `set(gs;;z)`, which thus now may execute, leaving item `z` in the task pool. The task `gran(;;w)` is subsequent to the other three tasks of the code fragment and this has allowed it to be ignored in the execution described up until this point.

As illustrated by the above execution, the novelty introduced by global items is that any task may delegate to a child or descendant which ultimately evaluates a global item. In other words, its responsibility evaluates a global item. While the resulting dependencies can be obeyed by a traditional depth-first sequential execution, the above execution and its discussion below demonstrate that many other execution orders also obey the same determinate application definition.

The above execution illustrates that no dependencies are introduced to a task which delegates a global item. A delegated global item is like any other delegated item. In the execution of Figure 112b), the task `gseed(;;z)` thus may execute before the task `gran(;;y)`, even though both delegate the global item `gs` and the latter's child `random(;gs;y)` must execute before the former's child `set(gs;;z)`. In other words, task autonomy is preserved for a task which delegates a global item.

The dependencies of a task thus only are due to its evaluated items. In subsection 3.26.3, also a global item may be evaluated by a task.

The above execution also illustrates that a task which evaluates a global item only can execute after there are no preceding tasks whose responsibility evaluates the global item. By obeying these dependencies the application definition remains determinate. As described above, task autonomy is preserved for a task which delegates a global item. Thus the execution is constrained to the exact degree required to obey the dependencies due to global items.

In order to make the dependencies due to global items more explicit, an application definition might indicate whether or not a task delegates to a child or descendant which ultimately evaluates a global item, but this possibility is not pursued in this presentation. Such an indication also might provide a solution to the following problem. As described in sections 3.19 and 3.21, irrelevant tasks can result from the use of conditional ins and multi-origin outs, respectively. Such a task no longer is irrelevant if its responsibility evaluates a global item [Speculative]. If ia language allows the use of conditional ins and multi-origin outs to be combined with the use of global items, then an indication of the use of global items is required for a truly irrelevant task to be recognized.

In the above example, `gs` is an item of the tasks `set(gs;;x)`, `random(;gs;y)` and `set(gs;;z)`. The tasks themselves are unaware whether an item is a global item or not. Instead, the issues introduced by the global items are dealt with by the IA.

The dependencies between tasks are similar whether or not an item is a global item. Since the tasks `set(gs;;x)`, `random(;gs;y)` and `set(gs;;z)` each have `gs` as an evaluated item, the tasks must execute in this order. In contrast, the hypothetical tasks `dset(del gs;;x)`, `drandom(;del gs;y)` and `dset(del gs;;z)` each have `gs` as a delegated item and thus may execute in any order since the tasks share no depen-dencies. This is another example of how task autonomy is preserved for a task which delegates a global item.

In short, a IA allows a determinate application definition to use global items. As described above, the IA support allows a task to use a global item much like any other item. Within the routine `gran(;;r){random(;gs;r);}` for example, `gs` is delegated like any other item. Similarly, `gs` is an item like any other within `random`.

### *3.26.3 Tasks Evaluating Global Items*

The previous subsection demonstrated that a global item may be delegated by a task. This subsection demonstrates that a global item may be evaluated by a task.

In the code of Figure 113a) for example, the global item `gx` is evaluated by all `gget` tasks and all `ginc` tasks. The code is a variation on the `Count` ADT demonstrated in subsection 3.25.4 and Figure 109a). There the ADT allows an application to use arbitrarily many `Count` items, each an independent counter. Here `gget` and `ginc` correspond to a single counter for the entire application.

A global item evaluated by a task is a nonlocal item of that task. As described in subsection 3.25.4, a nonlocal item of a routine can be made a local item by adding that item to the arguments of the routine [CPCPS]. The extra argument then can be satisfied by currying. This section thus largely repeats subsection 3.25.4.

The routines `gget` and `ginc` of Figure 113a) thus can be rewritten as in Figure 113b). For example, the original routine `ginc(;;){gx=gx+1;}`, which evaluates the global item `gx`, is rewritten as the curry `ginc(;;) is cinc(;gx;)`. The curried routine `cinc` is a closed routine and receives the global item `gx` as an argument. Such rewriting could be an early stage of a compiler. The routines `gget` and `ginc` thus are reduced to the form supported in the previous subsection for delegated global items. The IA support for global items thus now allows a global item to be evaluated by a task. The routines `gget` and `ginc` thus implement a simple counter whose use is illustrated in Figure 113c).

Instead of being rewritten as the curry `ginc(;;) is cinc(;gx;)`, the original routine `ginc` instead could be rewritten to the routine `ginc(;;){cinc(;gx;);}`. Also this form is supported in the previous subsection for delegated global items. The choice between the two forms depends on the situation.

### *3.26.4 An Instruction is a Global Item*

An instruction is a global item. Up until this point of the presentation, the instructions are constants. A constant global item introduces no dependencies between tasks. Thus the instructions presented so far safely have ignored the issues of global items described in this section.

Now that this section has demonstrated IA support for non-constant global items, this subsection can demonstrate a non-constant instruction. For example, in Figure 114a) the global item `triple(int;;int)` is an out of the routine `byadd(;;){triple(a;;b){b=a+a+a;}}`. Similarly it is an out of the routine `bymult(;;){triple(a;;b){b=3*a;}}`. The code fragment of Figure 114b) illustrates the use of the routines `triple(int;;int)`, `byadd(;;)` and `bymult(;;)`.

As mentioned in the comments of the code fragment, in this demonstration the instruction `triple(int;;int)` initially has no value. It thus may not be used before it is given a value by `byadd(;;)` or `bymult(;;)`. Alternatively, in a different demonstration, the instruction may have an initial value, for example as given by a usual definition like `triple(int q;;int r){r=3*q;}`.

### *3.26.5 Static Items*

A static item is an item of every task using a particular instruction. For the execution of an application, a static item is very similar to a global item. A global item can be an item of any task of an application. Similarly, a task with a static item can be a child of any task of an application. Thus like a global item, a static item introduces dependencies between tasks which must be obeyed in the execution in order to preserve a determinate application definition.

In the code of Figure 115a) for example, the static item `sts` is an item of all `sran` tasks. The code is a variation on the `gran` routine demonstrated in subsection 3.26.2 and Figure 112a). Like `gran`, `sran` is a pseudo-random number generator.

A static item can be treated like a global item whose scope in the application definition is restricted to a single particular instruction. For example, the routine `sran` using the static item `sts` in Figure 115a) is rewritten in Figure 115b) to use the global item `sran_sts`. The name `sran_sts` is chosen here to imply that the scope of the global item is restricted to the routine `sran`.

Since a static item can be treated like a global item, a static item raises similar issues for the execution of an application. This can be illustrated by the execution of the code fragment `f(;x;);sran(;;a);sran(;;b)` in Figure 115c). In addition to the routine `sran` of Figure 115a), the code fragment uses the hypothetical routine `f`. Since none of the original three tasks share any dependencies, they may execute in any order. The execution of the task `sran(;;a)` results in the task `random(;sts;a)`. Since it depends on the static item `sts`, the task `random(;sts;a)` only can execute once all preceding tasks, such as `f(;x;)` and all its descendants, have completed execution. This is necessary since a preceding task could use the routine `sran` and thus use its static item `sts`. The execution described here for a static item thus is very similar to the execution described in subsection 3.26.2 for a global item.

As described in subsection 3.25.3, each item of an application defined type (ADT) has its own independent item elements which maintain the state of the item between operations on that item. In contrast to an item element, a *type element* is shared across all items of a particular ADT. A type element thus is a static item [C++].

An example of a type element is the static item `count` of the routine `Rangen` in Figure 116. The ADT `record Ran{ran(;;real),seed(;;int),id(;;int)}` and the routine `Rangen(int;;Ran)` of Figure 116 slightly extend those of subsection 3.25.3 and Figure 107e). The extension records in the type element `count` the number of `Ran` items used by the application. The present value of `count` is used to uniquely initialize the item element `myid` of each `Ran` item. The operation `g.id(;;k){k=myid;}` thus returns in `k` the unique identifier of a `Ran` item. This is demonstrated in the code fragment in Figure 116. There the tasks `f.id(;;x)` and `h.id(;;y)` return in `x` and `y` the unique identifier of the `Ran` items `f` and `h`, respectively.

*3.26.6 A Determinate Definition Using Nonlocal Items*

The previous subsections 3.26.2 through 3.26.5 describe a determinate application definition using global items. This description is generalized in this subsection to a determinate application definition using nonlocal items.

As introduced in subsection 3.26.2, global items introduce dependencies between tasks which must be obeyed by the execution if the application definition is to remain determinate. Not only a global item, but more generally a nonlocal item can introduce such dependencies between tasks.

The generalization from global items to nonlocal items is illustrated in Figure 117a) by the routine `f` which encapsulates the code of Figure 112a). The code defines and uses an interface to a pseudo-random number generator. The global item `gs` of Figure 112a) corresponds to the nonlocal, but not global, item `ns` of Figure 117a).

For the global item `gs`, Figure 112a) includes the code fragment `gseed(;;x);gran(;;y);gseed(;;z);gran(;;w)`. For the nonlocal item `ns`, the code fragment `m(;k;);f(31;;x,y,z,w)` of Figure 117b) results in the code fragment `m(;k;);gseed(;;x); gran(;;y);gseed(;;z);gran(;;w)`. The two code fragments are identical, except that the latter is preceded by the task `m(;k;)`. The execution of the code fragments in the task pool is illustrated in Figure 112b) and Figure 117c), respectively. The executions are identical, except for the task `m(;k;)` preceding the tasks of Figure 117c). The task `m(;k;)` is assumed to execute last. The two executions of the other tasks are assumed to be the same in order to emphasize that the issues of global items are more generally those of nonlocal items.

The task `m(;k;)` shares no dependencies with the other tasks of the nonlocal example. The task `m(;k;)` thus is free to execute first, last or at any other time with respect to the execution of the other tasks. In particular, the nonlocal item `ns` is local to the original task `f(31;;x,y,z,w)`. The nonlocal item `ns` thus is outside the scope of the task `m(;k;)`. Since it thus can have no dependence of `ns`, the task `m(;k;)` need not execute before any task depending on `ns`. The possibility for a task to be independent of a nonlocal item goes beyond the possibilities for a global item, since any task can depend on a global item.

As described in subsection 3.26.2, a task depending on global item like `gs` only can execute after the execution of all preceding tasks. More generally, a task depending on a nonlocal item only can execute after the execution of all preceding tasks which are in the scope of the item. This rule for nonlocal items trivially reduces to that for global items, since all tasks are in the scope of a global item.

The generalization from global to nonlocal items thus merely takes the scope of the item into account. All aspects of execution discussed in subsection 3.26.2 for a global item thus also are valid for a nonlocal item. Similarly, other issues concerning global items, for example those discussed in subsections 3.26.3 through 3.26.5, can be generalized to nonlocal items.

*3.27 Some Issues Concerning Nonlocal Items*

*3.27.1 Declaring Nonlocal Items*

In ALGOL 60 and in many other programming languages since, a nested block structure restricts the scope of a variable. For example, the code fragment `{int x; {int`

`y=2;} x=y;}` is incorrect in the C programming language, since the statement `x=y;` tries to use the variable `y` outside of its scope [C].

> **"The ALGOL 60 scope rules protect the local variables of an inner block from outside interference; in the other direction, however, they provide no protection whatsoever." [Discipline : Scope]**

For example, the C code fragment `{int a=1; {int b=++a;}}` is correct and allows the inner block to modify the nonlocal variable `a`. In order to protect nonlocal items, a IA can adopt and extend the solution proposed elsewhere [Discipline : Scope].

Nonlocal items may be protected by requiring that a block declare those that are used. For example the routine `vseq` of subsection 3.13.1 and Figure 50c) includes the block repeated in Figure 118a). With the declaration `(int w,int n;;del int a[n])` in Figure 118b), the same block protects all other nonlocal items.

The type and use of each item in the declaration of nonlocal items may be determined from the contents of the block. For example, an early phase of a compiler could generate such a declaration. Thus the above declaration could be given as `(w,n;;del a[n])` or as `(w,n;;a)` or omitted altogether. This presentation takes no sides in whether or not an application definition is well served by including such declarations in whatever form. Arguments in favor of such declarations may be found elsewhere [Discipline : Scope]. The arguments for or against such declarations must also consider issues not addressed in this presentation. For example, since an instruction may be a nonlocal item, should it be included in the declaration?

A declaration of nonlocal items conveniently summarizes the use of nonlocal items by a block. For example, the declaration trivially allows the block to be replaced by a call to a routine which contains the contents of the block. This is illustrated in Figure 118c) for the above block of the `vseq` routine. A declaration of nonlocal items thus provides encapsulation much like that provided by a routine. A block with a declaration might be regarded as an unnamed routine.

The declaration of nonlocal items is a kind of identity function, since it neither loses information nor introduces extraneous information. For example, if the block `(int w, int n;;del int a[n]){...}` is placed into a block, then the declaration remains unchanged as `(int w,int n;;del int a[n]){(int w,int n;;del int a[n]){...}...}}`. Similarly, if a single task is placed into a block, the declaration is given by the items of the task. For example, the task `vseq(w,k;;a)` results in the block `(w,k;;a){vseq(w,k;;a)}`.

The use of blocks can extend through to the topmost level of an application definition. For example, the definition of the pseudo-random number generator of Figure 112a) is placed into a block in Figure 119. The nonlocal declaration `(random(;int; real),set(int;;int);;gran(;;real),gseed(;;int))` makes clear that the block uses the routines `random` and `set` and defines the routines `gran` and `gseed`. The item `gs` within the block is protected from outside interference. Many programming languages provide a module system to support global definitions [Module][STATE]. This example might be a starting point for a module system in a ia language, but this is beyond the scope of this presentation.

### 3.27.2 Declaring the Nonlocal Items of a Routine

Just as the nonlocal items of a block can be declared, so can those of a routine. For example, the routine `c.get(;;g){g=x;}` of subsection 3.25.4 and Figure 109a) could be defined as `c.get(;;g)(x;;){g=x;}`. Similarly, the routine `gget(;;g){g=gx;}` of subsection 3.26.3 and Figure 113a) could be defined as `gget(;;g)(gx;;){g=gx;}`. As demonstrated by the items `gx` and `x` of these examples, the declared nonlocal items of a routine may or may not be global items.

The declaration of the nonlocal items of a routine seems to bear some similarity to the use of monads in functional computing [Monads].

### 3.27.3 IA Support for Nonlocal Items

As for blocks, this presentation takes no sides in whether or not an application definition is well served by declaring for routines the use of nonlocal items. Instead this presentation uses the declaration of nonlocal items in order to introduce some possibilities for the IA support of nonlocal items.

As described in subsections 3.25.4 and 3.26.3, a nonlocal item of a routine can be made a local item by adding that item to the arguments of the routine [CPCPS]. The extra argument can be satisfied by currying or by delegation. The declaration of nonlocal items suggests generalizations of this technique for the IA support of nonlocal items.

For example, the routines `gget(;;int g)` and `ginc(;;)` of subsection 3.26.3 and Figure 113a) are rewritten in Figure 120a) with declarations for the nonlocal items as `gget(;;int g)(gx;;)` and `ginc(;;)(;gx;)`, respectively. The routines `gget` and `ginc` of Figure 113a) or Figure 120a) implement a simple counter whose use is illustrated in Figure 113c) by the code fragment `gget(;;d0);ginc(;;);gget(;;d1)`.

The declaration of nonlocal items suggests that perhaps the prototypes of Figure 113c) be given as `gget(;;int g)(gx;;)` and `ginc(;;)(;gx;)`. In accordance with the statements made above, this presentation takes no sides in whether or not the application definition is well served by having the prototypes declare the nonlocal items. For example, some might argue that such a declaration needlessly exposes information internal to the routine.

As illustrated in Figure 120b), the declaration of nonlocal items allows the code fragment of Figure 113c) to appear as in the task pool as the tasks `gget(;;d0)(gx;;)`, `ginc(;;)(;x;)` and `gget(;;d1)(gx;;)`. The declarations make explicit the dependencies between tasks due to nonlocal items. When determining the task execution order, a IA thus can obey the dependencies and preserve the determinate application definition.

The declaration of a nonlocal item of a task leaves open whether or not IA manages the item. Such management includes for example the communication and checkpoints of the item. For example, the interaction items of sections 3.28 are examples of items that might not be managed by a IA. In general, any item might not be managed by a IA. For example, the global item `gx` of the above example might not be managed by a IA.

If a IA does manage a nonlocal item, then the declaration leaves open how the item is provided to the task. In this presentation, currying or delegation provides a task with a nonlocal item as a regular argument of a routine. For example, in subsection 3.26.3 the routine `ginc(;;)(;gx;){gx=gx+1;}` is rewritten as the curry `ginc(;;)is`

`cinc(;gx;)` or as the routine `ginc(;;){cinc(;gx;)}`. Other mechanisms to provide a task with a nonlocal item are not pursued in this presentation.

In short, the declaration of nonlocal items helps identify three parts to the IA support for nonlocal items. The first part is the identification of dependencies between tasks due to nonlocal items. The second part is whether or not a IA manages a nonlocal item. If a IA does manage a nonlocal item, the third part is how a nonlocal item is provided to a task.

### *3.28 Interaction*

In addition to the use of the environment via TSIA, an application may interact with the environment directly. This section demonstrates that a IA allows for such *interaction* between an application and the environment.

In particular this section demonstrates that the behavior of an application in interaction can be determinate. Thus if the behavior of the environment is determined, then the entire interaction is determined and reproducible. For example, a word processing program is expected to be determinate. Given the same initial conditions and input, the application behavior should be reproducible.

An example interaction is illustrated in Figure 121a). The interaction is indeterminate. Given the prototype `putchar(int;;int)`, there is no dependency that requires that `putchar('a';;e1)` execute before `putchar('b';;e2)`. The output of the two tasks thus is either `"ab"` or `"ba"`. The dependencies between tasks thus only are complete after the interactions of the tasks have been taken into account.

The indeterminate behavior of the above example assumes that the two tasks are used in the delegation style. One possibility for a determinate interaction instead uses the subordinate style. For example, the code fragment `putchar('a';;e1); if(e1!=EOF)putchar('b';;e2);` has the determinate output `"ab"`. Similarly, the use of the subordinate style guarantees the determinate interaction of the `getchar-putchar` example of subsection 3.24.8 and Figure 100.

Given the delegation style, a determinate interaction can be achieved by declaring the interaction. An interaction item is a nonlocal item and can be declared as described in section 3.27. The indeterminate example interaction of Figure 121a) thus becomes the determinate example of Figure 121b). Given the prototype `putchar(int;;int)(;stdout;)`, the mutual dependence on `stdout` requires that `putchar('a';;e1)` execute before `putchar('b';;e2)`, resulting in the determinate output `"ab"`.

The interaction item `stdout` of the prototype `putchar(int;;int)(;stdout;)` does not only note dependencies between `putchar` tasks. Thus in Figure 121c), the prototype `puts(string;;int)(;stdout;)` assures that the tasks `putchar('a';;e1);puts("bc";;e2);putchar('d';;e3)` yield the determinate output `"abcd"`.

For the above prototype `putchar(int;;int)(;stdout;)`, the interaction item `stdout` is the same for all `putchar` tasks. In contrast, for the routine `fputc(int, FILE f;;int)(;f;)` the interaction item `f` only is the same for the `fputc` tasks with the same in `f`. By allowing an interaction item to be a usual in, inout or out of a task, a ia language thus precisely can define the determinate interactions between an application and environment. For example in the code fragment of Figure 121d), `fputc('a',g;;e1)` must execute before `fputs("bc",g;;e2)`, but is independent of `fputs("12",`

h;;e3). In other words, tasks operating on FILE g share dependencies, but not with tasks operating on FILE h.

In contrast to the above precise definition of the interaction dependencies, there may be situations requiring a coarser definition. In the extreme, there could be a single interaction item, here named env, and every routine interacting with the environment is prototyped as f(..;..;..)(;env;). Even seemingly unrelated interactions with the environment thus are given a determinate execution order in the application.

The dependencies between tasks introduced by interaction items are very similar to those introduced by global items in section 3.26.

For an application definition involving interaction, an execution which preserves the determinate definition thus is very similar to the execution described in subsection 3.26.2 for global items. The issues described in that subsection for global items thus largely are repeated here for interaction items. For interaction, the issues are illustrated by the code in Figure 122a) and its execution in the task pool in Figure 122b). The illustrated execution of the tasks pa(;x;);b(;y;);pc(;z;); assumes that pc(;z;) executes first. Even though the resulting task c(;z;)(;e;) is the only task in the pool depending on the interaction item e, the task may not yet execute since the preceding tasks pa(;x;) or b(;y;) may delegate to a child or other descendant which depends on e. The remaining execution thus first must execute pa(;x;) and b(;y;). Since they share no dependencies, the two tasks may execute in any order. The task b(;y;) modifies its inout y and disappears from the task pool. The task pa(;x;) delegates to a(;x;)(;e;). Since there are no tasks which precede it in the pool, a(;x;)(;e;) may execute. Once a(;x;)(;e;) and all of its descendants have completed execution, there are no tasks previous to c(;z;)(;e;) in the pool and it thus may execute.

As illustrated by the above execution, any task may delegate to a child or descendant which ultimately interacts with the environment. The resulting dependencies can be obeyed by a traditional depth-first sequential execution. The above execution and its discussion below demonstrate that, as for global items, many other execution orders also obey the same determinate application definition.

The above execution illustrates that a child with interaction items does not affect the dependencies of its parent. The dependencies of the parent include only the items of the parent and not those of any descendant. Thus in the execution of Figure 122b), the parent task pc(;z;) may execute before the parent task pa(;x;), even though the latter's child a(;x;)(;e;) must execute before the former's child c(;z;)(;e;). In other words, task autonomy and its benefits are preserved for such parent tasks.

The above execution also illustrates that a task with an interaction item only can execute after there are no preceding tasks whose responsibility involves the interaction item. For example, given the prototype fputc(int,FILE f;;int)(;f;), the task fputc('d', k;;e)(;k;) may execute once no previous task in the task pool involves the item k. By obeying these dependencies the application definition remains determinate. As described above, task autonomy is preserved for a task which delegates to a child or descendant which ultimately involves the interaction item. Thus the execution is constrained to the exact degree required to obey the dependencies due to interaction.

In order to make the dependencies due to interaction more explicit, an application definition might indicate whether or not a task delegates to a child or descendant which ultimately interacts with the environment, but this possibility is not pursued in this presentation. As for global items, such an indication of interaction also might allow for the recognition of irrelevant tasks due to the use of conditional ins and multi-origin outs [Speculative].

In Figure 121d), the prototype `fputs(string,FILE f;;int)(;f;)` declares that `fputs` involves the interaction item `f`. Alternatively, `fputs` could be implemented using `fputc(int,FILE f;;int)(;f;)`. The alternate `fputs` thus no longer involves the interaction item `f` and thus has the prototype `fputs(string,FILE f;;int)`. The code fragment of Figure 121d) involving `fputs`, valid for the original `fputs`, also is valid for the alternative `fputs`. For the code `fputc('a',g;;e1); fputs("bc",g;;e2);` the original `fputs` required `fputc` to execute before `fputs` since both depend on the interaction item `f`. In contrast, for the alternative `fputs` the execution order of `fputc` and `fputs` is irrelevant since without the interaction item `f` for `fputs` the two tasks share no dependent items. Of course `fputc` must execute before the `fputc` children of the alternative `fputs`. This example demonstrates again how the interaction items of child or descendant tasks don't affect the task autonomy of the parent. Again the execution is constrained to the exact degree required to obey the dependencies due to interaction.

In short, a IA allows a determinate application definition to interact with the environment.

### *3.29 Down versus Updown Nonlocal Items*

The default nonlocal item in the ia language of this presentation is not the only possible nonlocal item. For example, the default nonlocal item is an *updown* item. This aspect of a nonlocal item is described in this section. In particular, an updown item is contrast against a *down* item.

In the ia language of this presentation, a down item is identified by the keyword `down`. For an updown item, the corresponding keyword `updown` rarely appears, since it is the default for a nonlocal item.

The codes of Figure 123a) and b) introduce the differences between an updown item and a down item. The codes of Figure 123a) and b) are identical, except for the use of the updown item `u` and the down item `d`, respectively.

When used as local items, updown items and down items have the same behavior. Thus in Figure 123a) and b), the tasks `set(1;;u);set(u;;a[1])` and the tasks `set(1;;d);set(d;;a[1])` both result in `a[1]=1`. Such use as a local item of course can involve child tasks and further descendants. For example, the results `a[1:4] =1,1,2,2` and `a[1:4]=1,1,2,1` for Figure 123a) and b), respectively, remain unchanged if the original routine `set(x;;y){y=x;}` is replaced by the routine `set(x;;y){f(z;;w){w=z;}f(x;;y);}`.

In contrast to their same behavior when used as local items, updown items and down items have different behaviors when used as nonlocal items.

As illustrated in Figure 123a), for an updown item `u`, the tasks `{..;set(2;;u); set(u;;a[3]);}set(u;;a[4])` result in `a[4]=2`. In general, the change of an updown nonlocal item propagates to all subsequent tasks. For the code of Figure 123a), Figure 123c) illustrates that the updown nonlocal item `u` propagates up and down the hierarchy of tasks.

As illustrated in Figure 123b), for a down item `d`, the tasks `set(1;;d); {..;set(2;;d);set(d;;a[3]);}set(d;;a[4])` result in `a[4]=1`. In general, the change of a down nonlocal item propagates only to the subsequent sibling tasks and their descendants. In other words, the change of a down nonlocal item propagates only within its responsibility. For the code of Figure 123b), Figure 123d) illustrates that the down nonlocal item `d` only propagates down the hierarchy of tasks.

The codes of Figure 123a) and b) use the `set` routine and are rewritten to use `=` assignment in Figure 123e) and f), respectively. The rewriting does not change the behavior of the updown item `u` and the down item `d`, respectively.

In the codes of Figure 123e) and f) the items `u` and `d` are nonlocal within a block. In Figure 123g) and h), respectively, the codes are rewritten such that the items `u` and `d` are nonlocal within the routine `nl(;;)`. The rewriting does not change the behavior of the updown item `u` and the down item `d`.

The above description of updown and down items also may be expressed in other terms. For example, all uses of an updown item, both local and nonlocal, refer to a single item. Thus throughout the routine `f` of Figure 123a), the item `u` may be considered to be a single item.

Similarly, a local use of a down item also refers to a single item. In contrast, a nonlocal use refers to a copy of the item. Thus in the code of Figure 123b), the nonlocal use of the down item `d` within the block refers to a copy of the item `d` in the remaining code of the routine `f`.

A down item obeys an aspect of the behavior of some variables in the Lisp family of languages [FUNARG][IMPERATIVE]. Similar examples of this behavior also are found elsewhere [Dynamic Binding]. For example, in the UNIX operating system, a child process receives a copy of its environment variables from its parent. Variables and environment variables of a UNIX shell thus are similar to local items and down nonlocal items, respectively.

A down nonlocal item has practical uses not offered by an updown item. Perhaps the simplest use is a limited change of a nonlocal item. A simple example, taken from elsewhere [IMPERATIVE], assumes a library of numerical algorithms which uses the nonlocal item `eps` to specify the desired tolerance of the result of each routine. For simplicity, only two routines, `g(;;eg)` and `h(;;eh)`, from such a library are considered here. Achieving the tolerance `eps` for the routine `h(;;eh)` is assumed to require the use of the routine `g(;;eg)` to the tolerance `eps*eps`.

A crude imitation of the routines `g(;;eg)` and `h(;;eh)` is given in Figure 124a). There the tolerance `eps` is an updown item. Thus the routine `h(;;eh)` does `{real oldeps=eps; eps=eps*eps; g(;;x); eps=oldeps;}` in order to use the routine `g(;;eg)` to the desired tolerance `eps*eps` and to restore `eps` to its original value. The original value for `eps` is required by the subsequent code of the routine `h(;;eh)` as well as by any subsequent user of the library of routines.

The same crude imitation of the two routines is given in Figure 124b), except that the tolerance `eps` is a down item. Thus the routine `h(;;eh)` does `{eps=eps*eps; g(;;x);}` in order to use the routine `g(;;eg)` to the desired tolerance `eps*eps`. Since it is a down item, the nonlocal use `eps=eps*eps` does not modify the original `eps`, which thus does not need to be restored. As promised above, a down item thus allows for a limited change of a nonlocal item.

### 3.29.1 A Determinate Application Definition

As described in section 3.26, nonlocal items introduce dependencies between tasks which must be obeyed by the execution if the application definition is to remain determinate. As described in subsection 3.26.6, a task depending on an updown nonlocal item thus only can execute after the execution of all preceding tasks which are in the scope of the item.

Applied to a down nonlocal item, the above rule for an updown item is more than sufficient to ensure a determinate application definition. Unlike an updown item, a down item cannot be changed by nonlocal use. A task depending on a down item thus only can execute after the execution of all preceding tasks which are in the scope of the item and which make local use of the item. In other words, a task depending on a down item is independent of all tasks which don't make local use of the item.

The differences in dependencies between updown and down items can be illustrated using the code of Figure 123a) and b). The codes are identical, except for the use of the updown item `u` and the down item `d`, respectively. As demonstrated below, the use of a down item results in less dependencies between tasks than the equivalent use of an updown item.

For the updown item `u` of Figure 123a), the execution of the routine `f(;;a[1:4])` yields in the pool the tasks `set(1;;u);set(u;;a[1]);(;u;a[3:4]){};set(u;;a[4])`. For the block `{}`, the nonlocal items are declared as `(;u;a[3:4])`, but the individual tasks are not given here. The dependencies on the updown item `u` allow the task `set(u;;a[4])` only to execute after the execution of the block `(;down u;a[3:4]){}` and all its descendants.

For the down item `d` of Figure 123b), the execution of the routine `f(;;a[1:4])` yields in the pool the tasks `set(1;;d);set(d;;a[1]);(;down d;a[3:4]){};set(d;;a[4])`. For the down item `d`, the task `set(d;;a[4])` is independent of the block `(;down d;a[3:4]){}` and all its descendants. The task and the block thus are free to execute in any order. The task and block are independent because the block `(;down d;a[3:4]){}` does not make local use of the down item `d`.

A similar illustration of the dependencies compares the execution of the code fragment in Figure 124a) using the updown item `eps` to that in Figure 124b) using the down item `eps`. In order to satisfy the dependency due to the updown item `eps`, the code fragment `h(;;ih)(;eps;);g(;;ig)(eps;;)` requires that the two tasks execute in order. In contrast, the two tasks of the code fragment `h(;;ih)(;eps;);g(;;ig)(eps;;)` may execute in any order since the down item `eps` introduces no dependencies between the two tasks.

In short, both down and updown nonlocal items allow for a determinate application definition. Compared to an updown item, a down item results in less dependencies between tasks. A down item thus allows for a greater variety of task execution orders.

### 3.29.2 The Nonlocal Items of Routines Passed as Arguments

As mentioned above, a nonlocal use of a down item may be considered to refer to a copy of the item. Assume a routine which makes nonlocal use of a down item. Furthermore, assume that the routine is passed as an argument to another routine. As described below, this situation raises the question of when a down item is copied [FUNARG][IMPERATIVE]. Inside the routine passed as an argument, the value of the nonlocal down item can depend on when the item is copied. This issue for a down item does not exist for an updown item, since the nonlocal use of an updown item does not involve a copy of the item.

The above issue is illustrated by the codes of Figure 125a) and b). The codes are identical, except for the use of the updown item `u` and the down item `d`, respectively. The codes are a simplified version of an example taken from elsewhere [FUNARG]. Since its nonlocal item `u` is a familiar updown item, the execution of the code fragment in Figure 125a) is straightforward and is not explicitly explained here.

The execution of the code fragment in Figure 125b) is just as straightforward as that of Figure 125a), but the use of the down item `d` requires the following explanation. The ia language of this presentation assumes that the nonlocal down items of a routine are copied at use-time. Here, *use-time* refers to the definition of a use of the routine. Thus in Figure 125b), the task `g(f;;z)` may be considered as `g(f(;;int)(down int d=1;;);;z)`, since this defines a use of the routine `f`. For the task `g(f;;z)`, this use of the routine `f` thus is given a copy of the down item `d` with the value `1`. The statement `d=0` within the routine `g(h(;;int);;int)` thus is irrelevant to `f` or to any other routine given as the argument `h(;;int)`. The task `g(f;;z)` thus yields the result `z=1`.

The different results `z=0` and `z=1` of Figure 125a) and b), respectively, thus demonstrate that the value of a down nonlocal item depends on when it is copied.

As explained above, the use-time refers to the definition of a use of the routine. The use-time does not refer to the *definition-time* of the routine. In Figure 125b), the definition `f(;;int y){y=d;}` is an example of such a definition-time and is not when the down item `d` is copied. Such action would be similar to the macro substitution performed by the C preprocessor [C]. In some situations the definition-time and the use-time of a routine or block coincide. The block `{}` of Figure 123b) is an example of such a coincidence. Another example is the task `g((;;int y){y=z;};;z)` which results from replacing the task `f` by an unnamed routine in the task `g(f;;z)` of Figure 125b).

As explained above, the nonlocal down items of a routine are copied at use-time. This rule also satisfies the usual use of a routine. For example, the task `nl(;;)` of Figure 123h) thus can be considered as `nl(;;)(;down int d=1;)`. To be pedantic the task `g(f;;z)` of Figure 125b) could be considered as `g(f;;z)(;;down int d=1)`, but this additional information is of little interest since the routine `g` ignores the initial value of the item `d`.

It is experience elsewhere [FUNARG][IMPERATIVE] that leads the ia language of this presentation to assume that the down nonlocal items of a routine are copied at use-time. Instead of copying at use-time, an alternative possibility would be to copy at execution-time. Here, *execution-time* refers to the execution of the routine using the nonlocal items. If copying at execution-time is assumed, then in Figure 125b) the task `h(;;x)` may be considered as `h(;;x)(down d=0;;)` since this is the execution of a routine. The execution of the task `g(f;;z)` thus yields the task `f(;;z)(down d=0;;)`, which yields the result `z=0`. The result of copying at execution-time thus is not the same as copying at use-time.

As mentioned above, all uses of an updown item, both local and nonlocal, refer to a single item. The value of an updown item thus always is the value at execution-time. The behavior of an updown item thus is closer to that of a down item copied at execution-time than to one copied at use-time. This explains why the same `z=0` results from the code of Figure 125a) using the updown item `u` as from the code of Figure 125b) using the down item `d` copied at execution time.

For the usual use of a routine, the use-time and the execution-time coincide. For example, the task `nl(;;)` of Figure 123h) is both the use-time and the execution-time of the routine. Due to the coincidence, the usual use of a routine does not raise the question of when a down item is copied. Instead, the question only arises when the use-time differs from the execution-time, such as when a routine is passed as an argument to another routine.

Instead of assuming that a down item is copied at use-time, a ia language could allow an application definition to copy a down item at definition-time, use-time, execution-time or at any time in between. This possibility is not pursued in this presentation.

The codes of Figure 125a) and b) demonstrate a routine that is an in to another routine. The codes are extended in Figure 125c) and d), respectively, to demonstrate a routine that is an out of one routine and an in to another routine.

The codes of Figure 125c) and d) are identical, except for the use of the updown item `u` and the down item `d`, respectively. The codes are a simplified version of an example taken from elsewhere [FUNARG]. Since its nonlocal item `u` is a familiar updown item, the execution of the code fragment in Figure 125c) is straightforward and is not explicitly explained here.

Due to the assumption that a down nonlocal item is copied at use-time, the statement `m is f` of Figure 125d) corresponds to `m(;;int)is f(;;int)(down d=0;;)`. The task `k(;;p)` thus results in `p=f(;;int)(down d=0;;)`. The subsequent statement `d=1` thus is irrelevant to the item `p`. The task `p(;;w)` thus corresponds to `f(;;w)(down d=0;;)` and yields the result `w=0`.

The different results `w=1` and `w=0` of Figure 125c) and d), respectively, thus demonstrate that the value of a down nonlocal item depends on when it is copied.

### *3.29.3 Down Items for Configuring Routines*

As demonstrated in subsection 3.25.2, currying can encapsulate the information configuring a routine. Currying thus enables configuring routines and thus allows for more concise and reusable software. As demonstrated below, a down item copied at use-time similarly enables configuring routines.

An alternative to currying for configuration uses a nonlocal item. This alternative is illustrated in Figure 106 by the routine `gdivcmp` and its nonlocal item `globalm`. The illustration in Figure 106 is repeated in Figure 126 by the task `intsort(n,nldivcmp;a;)` which uses the routine `nldivcmp` and its nonlocal item `nl`.

Configuration by updown nonlocal items is less convenient than that by currying since an updown item needlessly exposes the configuration information to the application. In the code of Figure 126, the updown item `nl` is exposed to the application. If for example, `nl` were changed within `intsort`, then that change would affect the routine `nldivcmp` passed to `intsort`.

In contrast, configuration by down nonlocal items is as convenient as that by currying. Both techniques can encapsulate the configuration information. Such encapsulation can be achieved in the code of Figure 126 by replacing the updown item `int nl` by the down item `down int nl`. For example, due to the resulting encapsulation, any change to `nl` within `intsort` would not affect the routine `nldivcmp` passed to `intsort`.

In addition to the example provided by the task `intsort(n,nldivcmp;a;)`, the code of Figure 126 also includes the example provided the task `intsort(n,comp;b;)`. The comparison routine `comp` is the out of the task `getcomp(100;;comp)`. If `nl` is an updown item, then the statement `nl=h` within `getcomp` affects the entire subsequent application definition. In contrast, if `nl` is a down item, then the statement `nl=h` has no effect outside `getcomp`, except of course for the out `comp`. This again demonstrates how an updown item lacks the encapsulation provided by a down item.

If `nl` is an updown item, then the tasks `intsort(n,nldivcmp;a;)` and `intsort(n,comp;b;)` of Figure 126 share a dependency on `nl` and thus must execute in order. In contrast, if `nl` is an down item, then the tasks share no dependencies and thus may execute in any order. The lack of dependencies between the two tasks is yet another example of how a down item encapsulates configuration information.

### *3.30 Summary*

Chapter 2 demonstrates that an application which executes in terms of tasks can have a definition free of execution details and can be provided by a task system (TS) with a variety of execution elements. This chapter thus is based on an application which executes in terms of tasks. As summarized below, this chapter demonstrates that an item architecture (IA) allows a structured application definition to execute in terms of tasks. In combination, the TS and the IA of TSIA thus support the execution and definition requirements of a large variety of applications.

An application execution in terms of tasks does not require an application definition in terms of tasks. A task consists of items. A IA thus allows an application to be defined in terms of items. An item of a task is either an in, inout or out. The out of one task can be the in of another.

The items of a task often include an instruction which encodes how the ins of the task are used to produce the outs. A IA places no restrictions on the instruction. An instruction thus may correspond to a single machine instruction or to a multi-million line program in an arbitrary programming language.

During its execution, a task is independent of all other tasks. Due to this task autonomy, a parent task cannot communicate with its child tasks. Instead, a parent can replace itself by its children. The responsibility for the items of the parent thus are delegated to the children. Delegation thus makes task autonomy compatible with an application definition structured in terms of the familiar hierarchy of routines.

In order to help achieve a performant application execution, throttling controls the expression of an application definition in terms of tasks. Throttling attempts to achieve a set of tasks that can be mapped well onto the resources available for the execution. Throttling has a variety of techniques at its disposal. For example, tasks may be combined in order to coarsen the granularity of the execution. Another example varies the task execution order between depth-first and breadth-first in order to adjust the number and responsibility of the tasks awaiting execution.

By declaring that an item is delegated to a child task, a IA allows for a non-strict evaluation. In other words, such an item need not be evaluated before executing the task. For such items, throttling may choose between a supply-driven and a demand-driven task exe-cution order. Furthermore, such items allow throttling to fuse for execution items separated in a structured application definition. By thus improving locality, fusion can improve application execution performance. Similarly, fission may remove some of the structures of the application definition which hinder a performant execution.

In addition to routines, this chapter demonstrates how arrays may structure an application definition. The support for routines and arrays extends through to stream arrays and arguments.

A conditional item of a task is an item which may or may not be ignored by the task. For example, a conditional in allows the `if then else` conditional construct to be provided by a task. Since it ultimately may not need to be evaluated, a conditional item allows throttling to choose between a speculative and a conservative task execution order.

A multi-origin out is an out originating from any one of several different tasks. When the best task among several tasks to provide an out is unknown, a multi-origin out allows the application definition to defer the choice of task to TSIA. If the outs of its tasks are not identical, then a multi-origin out is indeterminate and so is the application definition. Among the tasks of a multi-origin out often are tasks for which the out is a conditional out.

The demonstrated IA support for a structured application definition also includes currying, application defined types, nested routines, unnamed routines and nonlocal items. In addition, a IA allows a determinate application definition to interact with the environment. This interaction is in addition to that via TSIA.

Though the topics addressed go beyond those of the above summary, many topics concerning an application definition are not addressed by this chapter. For example, this chapter does not address defining the requirements for a reactive application. Nonetheless, the topics addressed by this chapter are sufficient to demonstrate that a IA allows a structured application definition to execute in terms of tasks.

## 4 Conclusion

Chapter 2 demonstrates that an application which executes in terms of tasks can be provided by a task system (TS) with execution elements. Chapter 3 demonstrates that an item architecture (IA) allows a structured application definition to execute in terms of tasks. A task system and item architecture (TSIA) thus can provide a structured application definition with execution elements.

This presentation announces the discovery of TSIA and results of its initial exploration. TSIA awaits implementation, use and further exploration.

**Acknowledgements**

Family, friends, colleagues and others as well as institutions and other organizations have provided me with the opportunity to discover and explore TSIA. There are many that I would like to thank by name. In order to protect the innocent, I will do so after there has been some public reaction to TSIA.

**Figures**

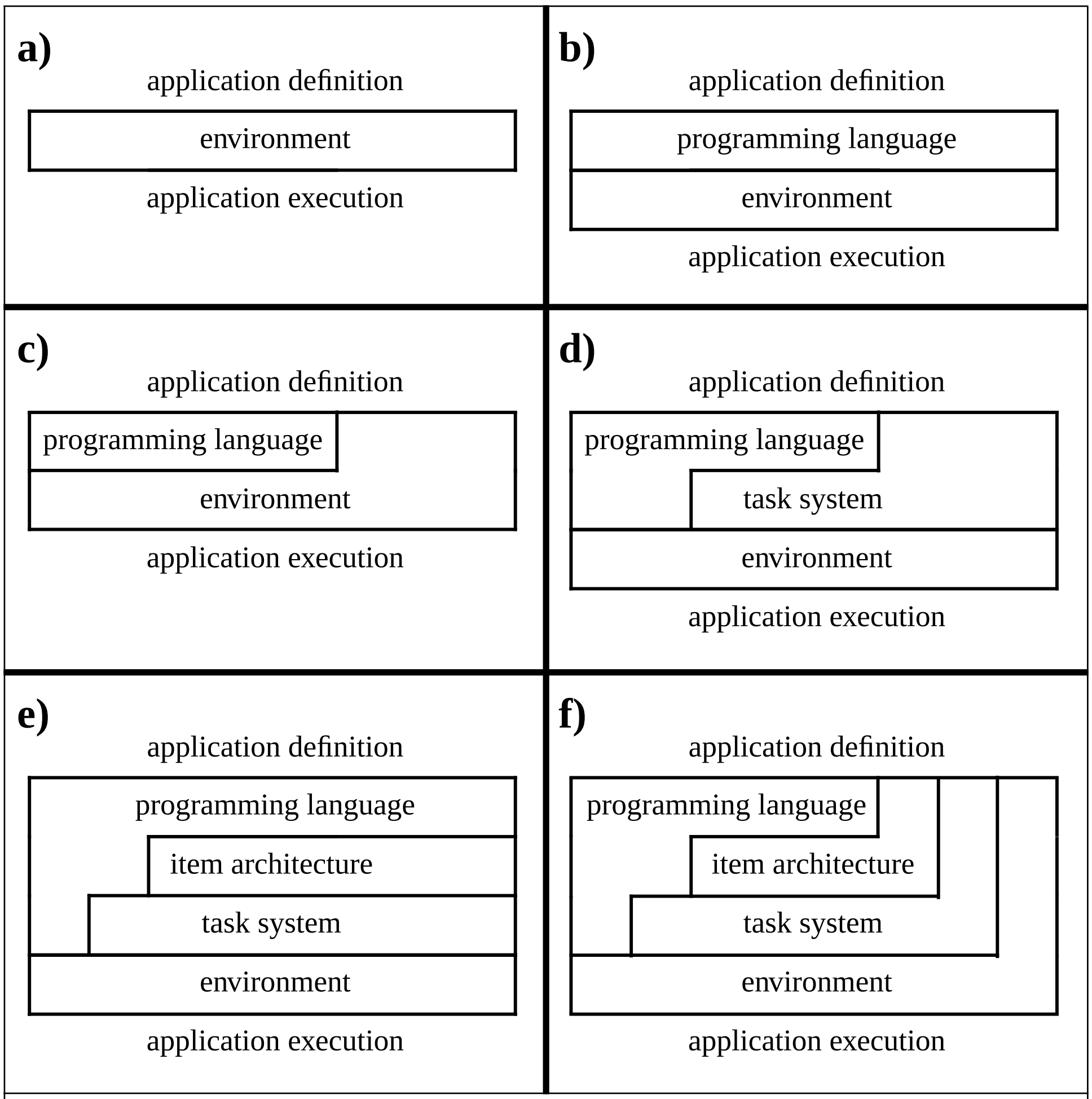

Figure 1 The relationship of the application definition, environment and application execution:
a) before the introduction of programming languages.
b) after the introduction of programming languages.
c) if the application is defined in terms of the programming language and the environment.
d) after the introduction of a task system (TS).
e) after the introduction of a task system (TS) and an item architecture (IA).
f) as in e), but without simplification.

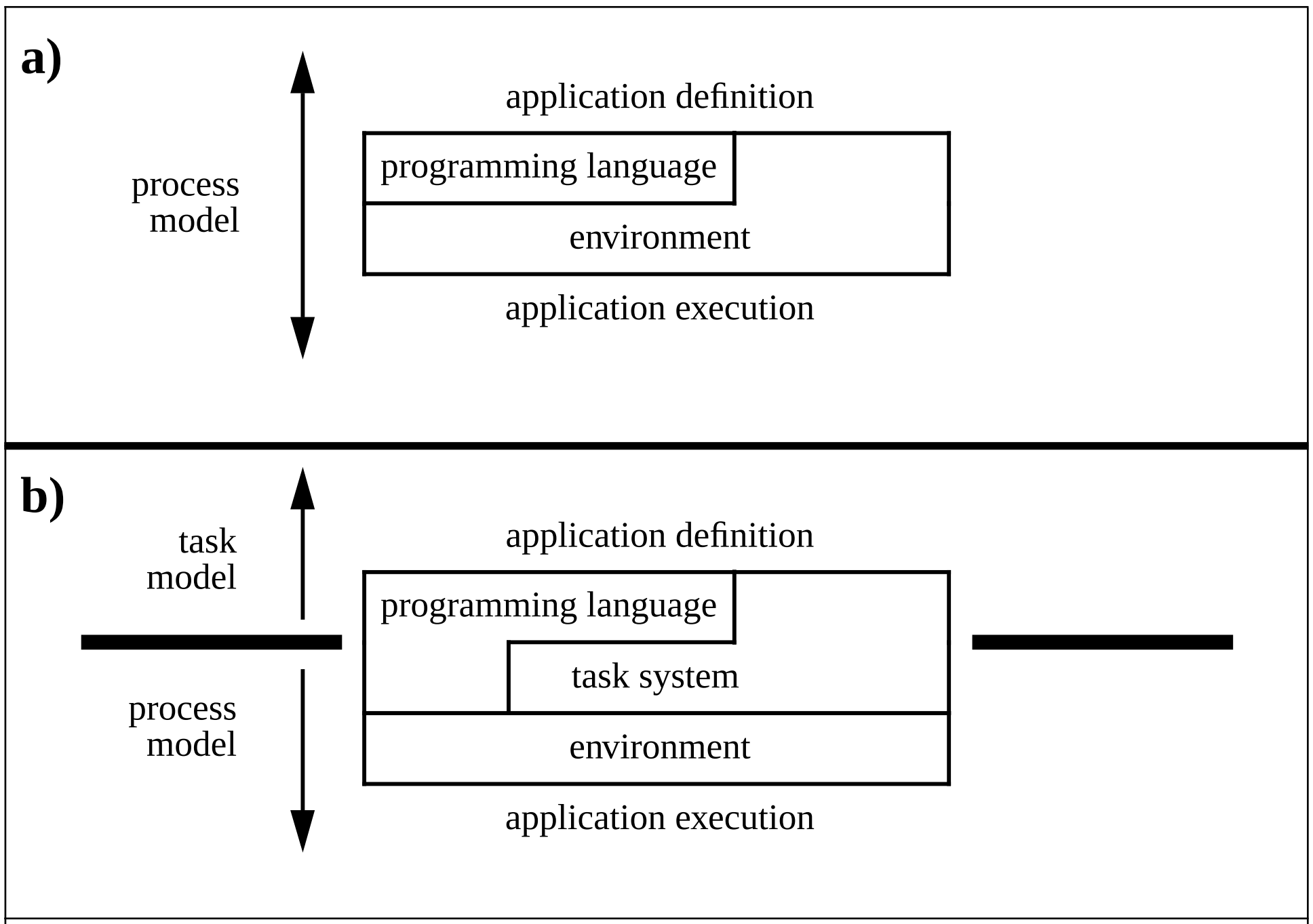

Figure 2 a) The process model pervades throughout the application execution and definition.
b) If used by the environment, the process model can be restricted to the application execution, with the task model used for the application definition.

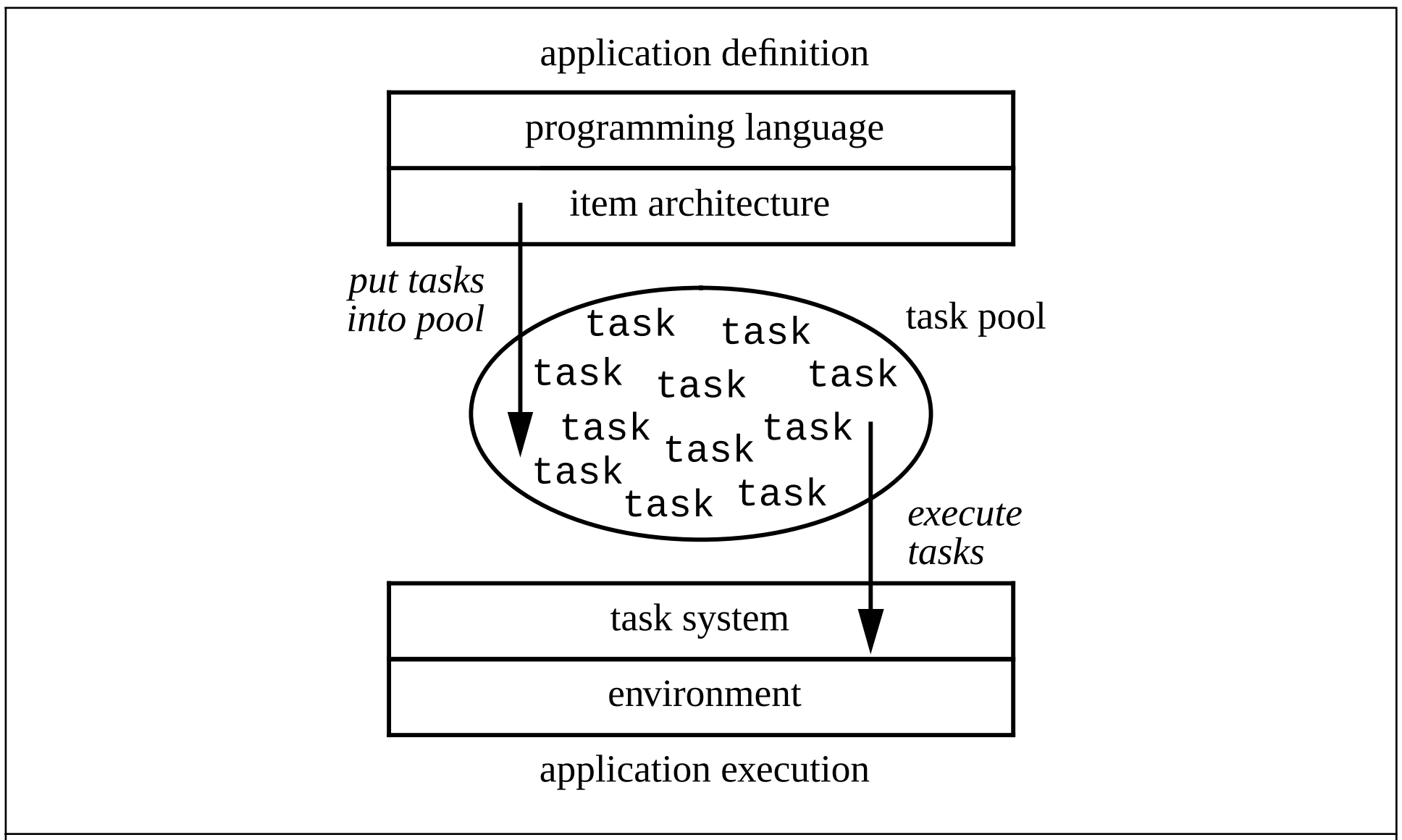

Figure 3 The tasks of an application execute via the task pool.

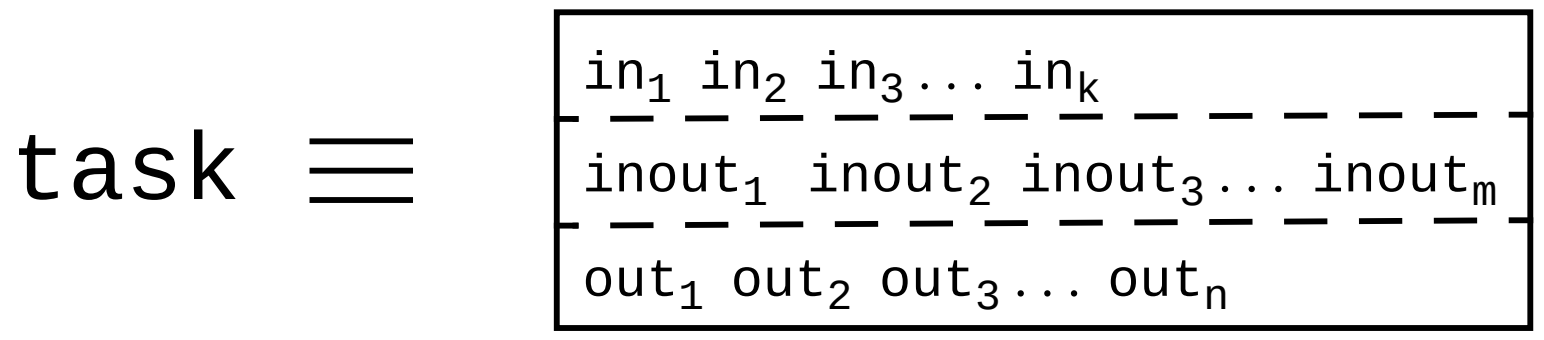

Figure 4 A task is defined by its items. An item is an in, an inout or an out.

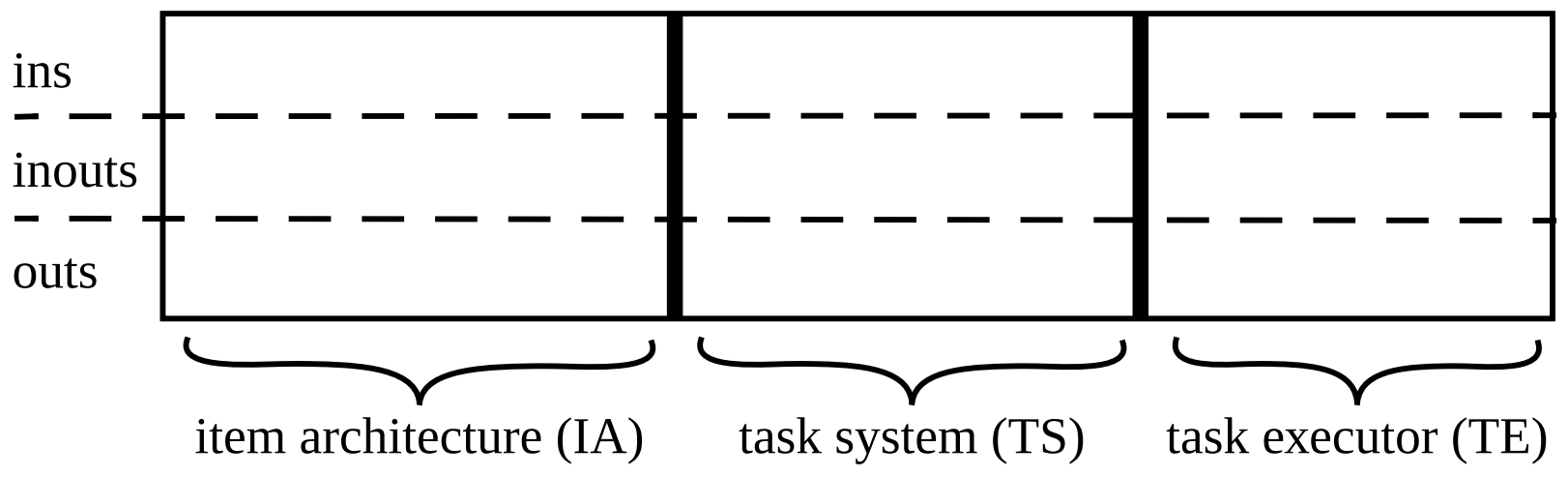

Figure 5 As in Figure 4, but identifying that each item is defined by the IA, the TS or the TE.

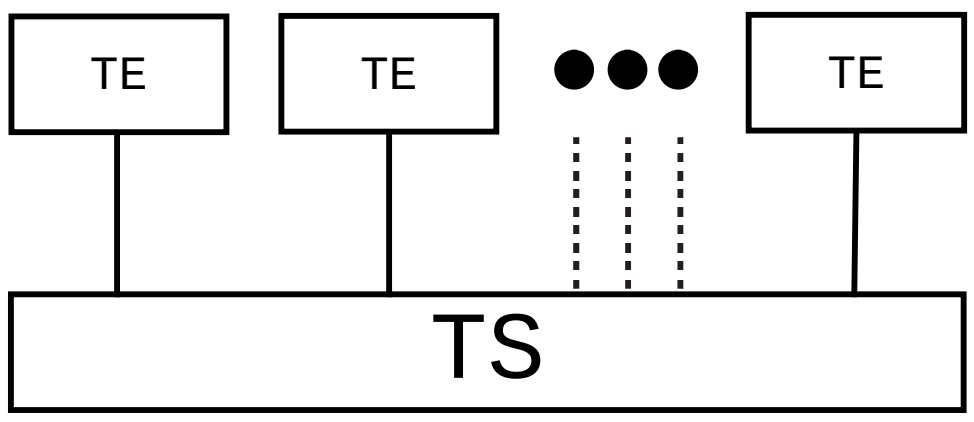

Figure 6 Each task is passed by the TS to a copy of the TE for execution.

**a)**

```
PROGRAM APP
FOREVER
  READ(A)
  PRODUCE(A,B)
  WRITE(B)
END
```

**b)**

$A_1, B_1$
$A_2, B_2$
$A_3, B_3$
.
.
$A_N, B_N$

Figure 7 a) Pseudocode program for the classic application.
b) If a classic application consists of N ins, then the application is defined by N tasks.

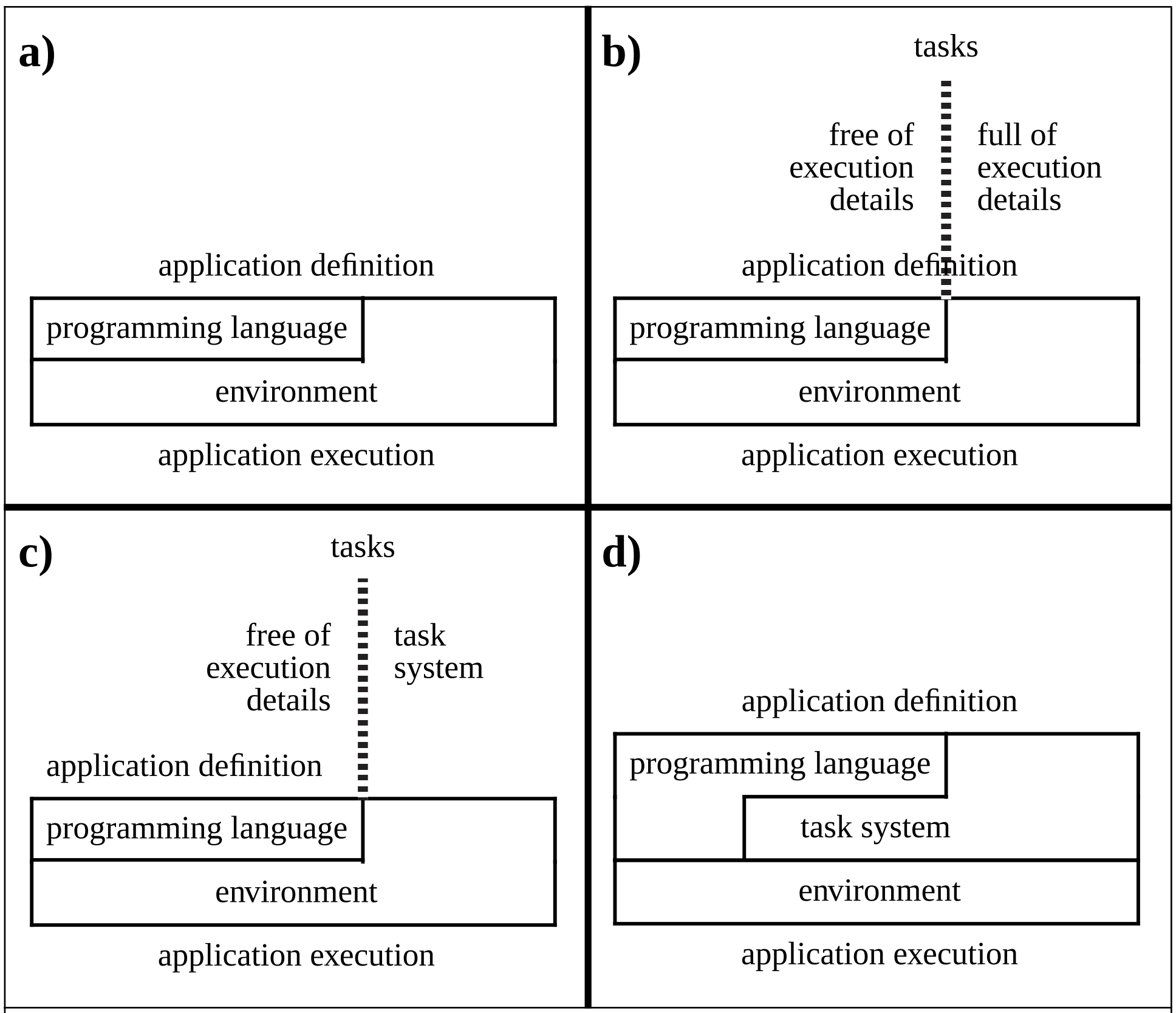

Figure 8 The relationship of the application definition, environment and application execution:
a) if the application is defined in terms of the programming language and the environment.
b) then if using tasks, the definition is divided into two parts,
one free and one full of execution details.
c) then if the part full of execution details is replaced by a task system (TS).
d) as in c), but re-arranged to be identical to Figure 1d).

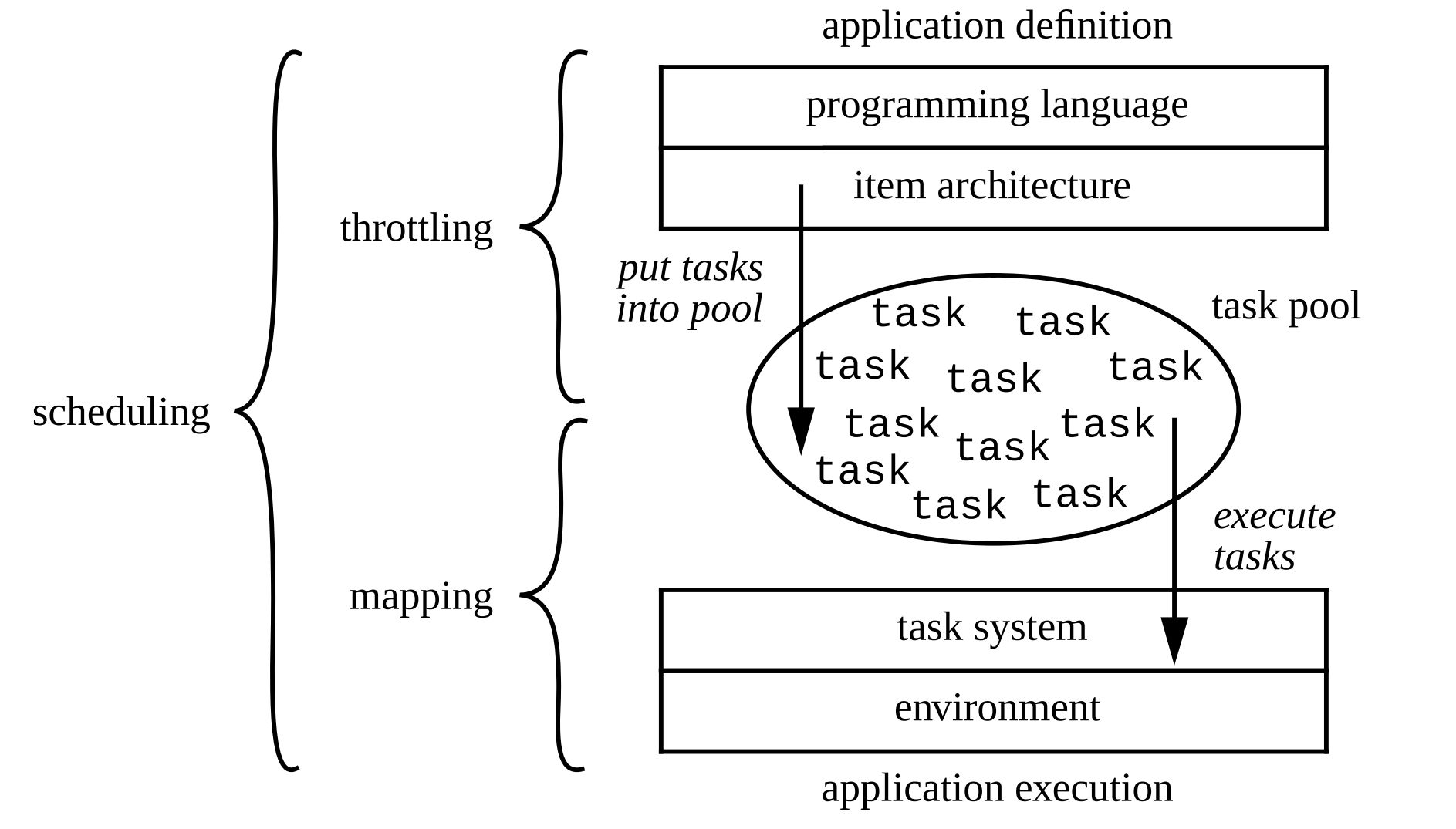

Figure 9  As in Figure 3, but also illustrating that scheduling consists of throttling and mapping.

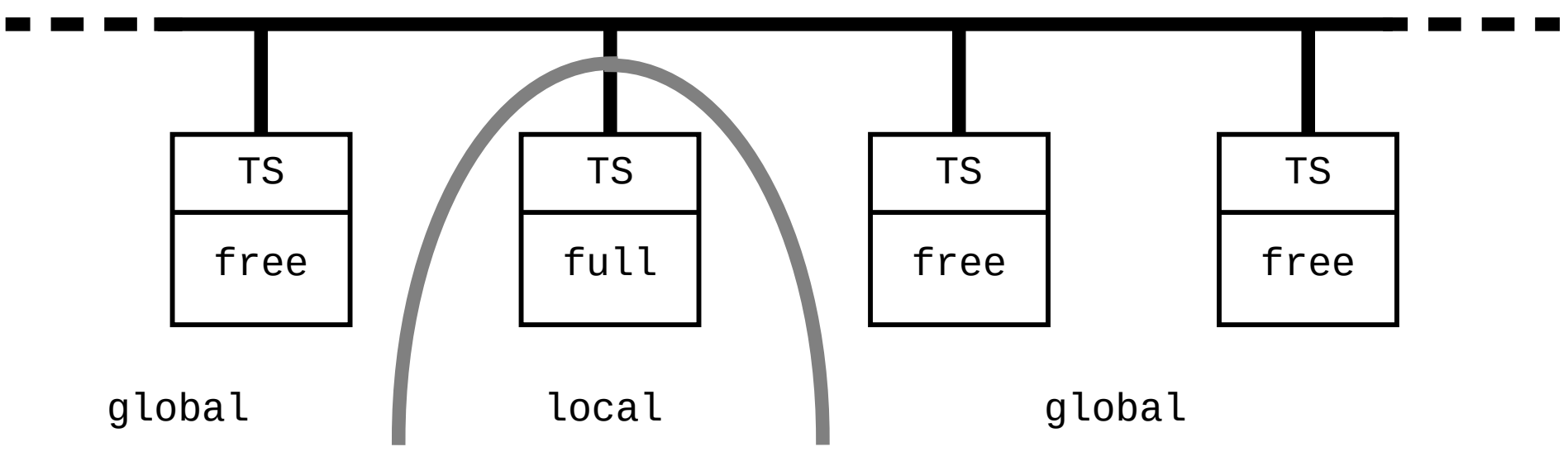

Figure 10 The part of an application definition full of execution details only can use local resources. The other part of the application definition is free of execution details and can use local and global resources. For example, the resources may be networks of computers.

**a)**

```
// Declare the instruction produce.
produce(bytes;;bytes);

// Ia routine for a classic application.
main() {
  infile  "a.dat" bytes a[N];
  outfile "b.dat" bytes b[N];
  for (int k=1; k<=N; k++) {
    produce(a[k];;b[k]);
  }
}
```

**b)**

```
//         - begin an in-line comment.
bold       - highlight keyword for reader.
bytes      - a type of item with
               arbitrarily many bytes.
int        - an integer.
{}         - enclose a block of ia code.
;          - terminates a statement.
main       - first routine of application.
[]         - a dimension of an array.
1:N        - the numbers 1 to N inclusive.
for        - loop as in C language.
t(i;b;o)   - a task with:
             t = name of instruction.
             i,b,c = in,inout,out items.
```

Figure 11 a) Ia routine for a classic application. b) Some of the syntax for ia code.

```
/* Example fragment of C code that could correspond to ia code. */
for (k=1; k<=N; k++) {
  /* submit() is pseudocode for actions submitting a task to the task pool. */
  submit(produce,a[k],b[k]);
}
```

Figure 12 Fragment of C code resulting from a translation of the ia program in Figure 11a).

```
C Simple Fortran example of evaluation.
C Declared in ia as f(a;;b).
      subroutine f(a,b)
      implicit none
      integer a,b
      b = a
      end
```

Figure 13 A simple Fortran example of evaluation.

**a)**

```
b(int x;; int y);

// A simple ia routine using delegation.
a(int x;; int y) { b(x;;y); }
```

**b)**

```
d(int x;; int y, int q);
e(int x, int y, int q;; int z);

// Another ia routine using delegation.
c(int x;; int y, int z)
{ d(x;;y,q); e(x,y,q;;z); }
```

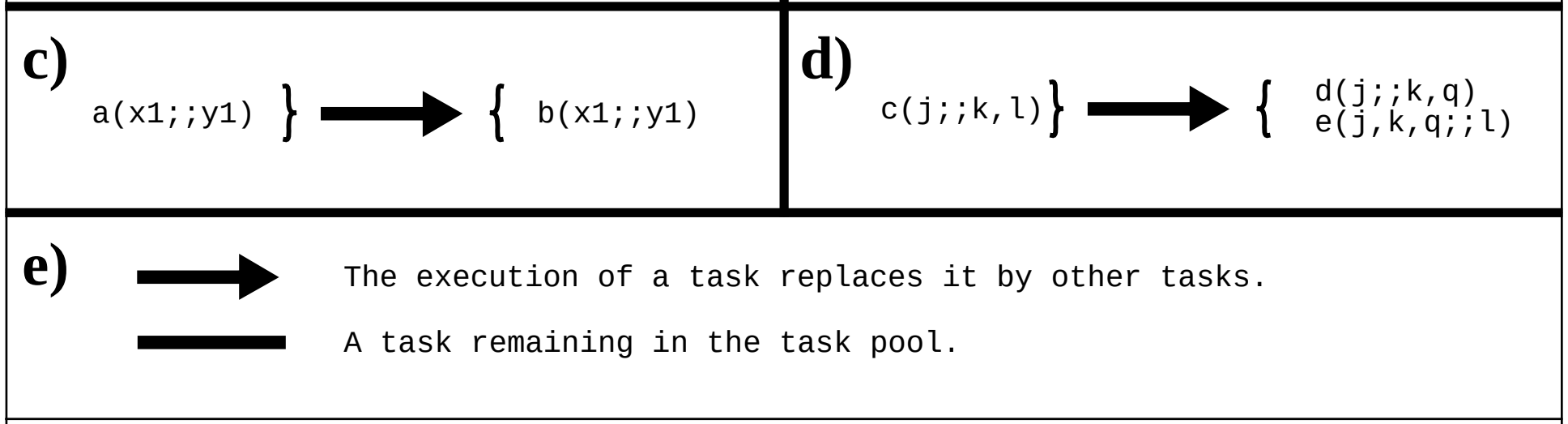

Figure 14 a) and b) Simple examples of delegation.
c) The execution of `a(x1;;y1)` from the task pool, leaves `b(x1;;y1)` in the pool.
d) The execution of `c(j;;k,l)` from the task pool,
leaves `d(j;;q,k)` and `e(j,q,k;;l)` in the pool.
e) The arrow and the line are used to describe the execution as seen from the task pool.

**a)**

```
/* Cilk-NOW code. */
thread b( cont int y, int x);

thread a( cont int y, int x)
{ spawn b(y,x); }
```

**b)**

```
// Mentat code.
mentat class eg
{ public: int a(int); int b(int); }

int eg::a(int x) { eg e; rtf(e.b(x)); }
```

Figure 15 a) and b) The ia example of Figure 14a), written in Cilk-NOW and Mentat, respectively.

**a)**

```
h(int x;; int z);

// Ia example of subordination.
sg(int x;; int y) { h(x;;z); y=2*z; }
```

**b)**

```
k(int z;; int y) { y=2*z; }
h(int x;; int z);

// Ia example of delegation.
dg(int x;; int y) { h(x;;z); k(z;;y); }
```

Figure 16 a) The child `h(x;;z)` is used by the parent `sg(x;;y)` in the subordinate style.
b) The child `h(x;;z)` is used by the parent `dg(x;;y)` in the delegation style.

**a)**

```
// Ia parent in the subordinate style.
p(..;..;..)
{
  // prior code.
  child(..;..;..);
  // subsequent code.
}
```

**b)**

```
// Ia parent in the delegation style.
pp(..;..;..)
{
  // prior code.
  child(..;..;..);
  ps(..;..;..);
}

ps(..;..;..)
{
  // subsequent code.
}
```

Figure 17 a) The child `child()` is used by the parent `p()` in the subordinate style.
b) The child `child()` is used by the parent `pp()` in the delegation style.

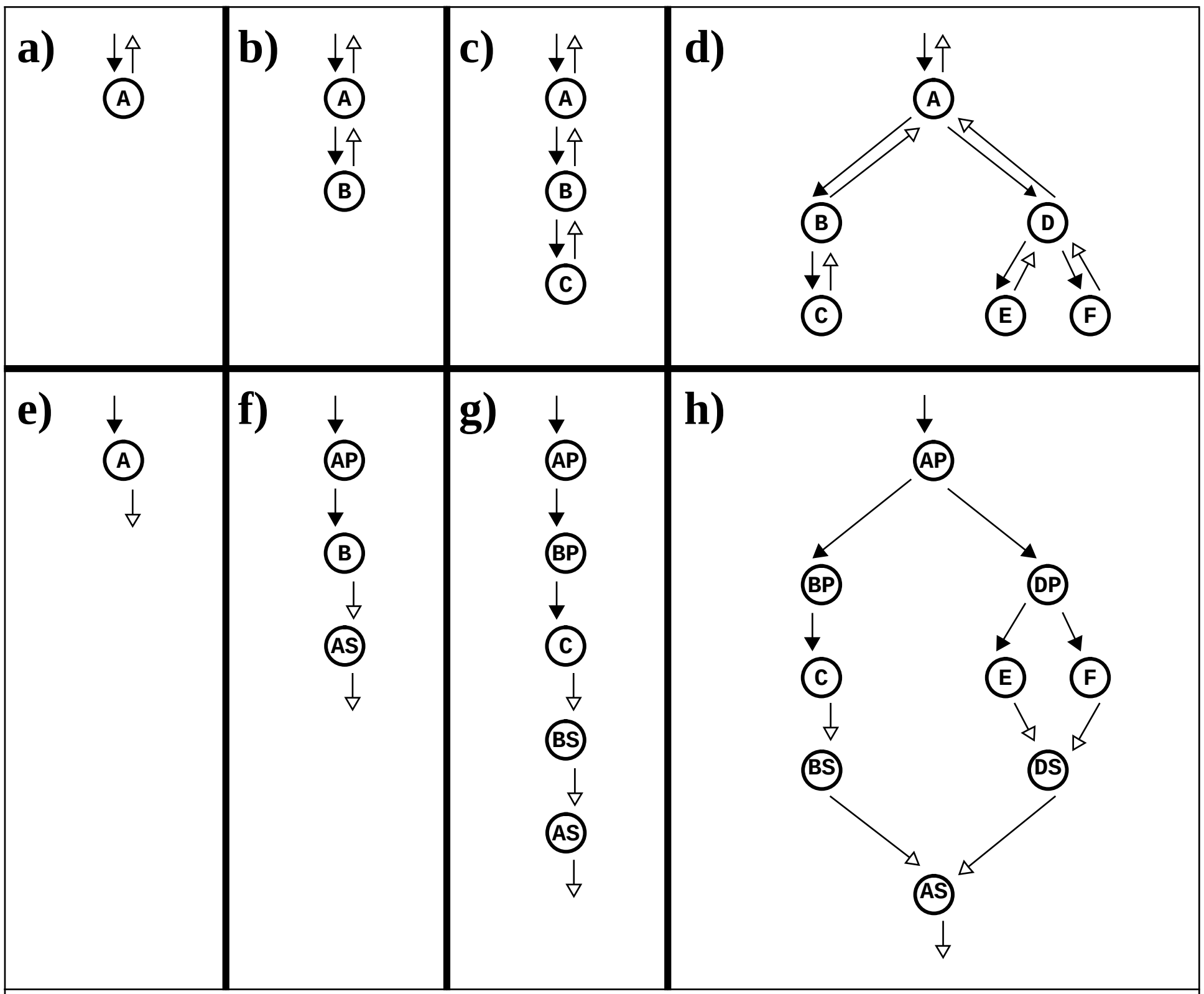

Figure 18 A solid arrow calls a routine or task. An open arrow returns from a routine or task.
a) to d) Examples of hierarchies of routines in the subordinate style.
e) to h) The corresponding hierarchies in the delegation style.
The return of a task is the call of another task.

**a)**

```
C Delegation style.Fortran.
C In ia is a(int x;; int y);
      subroutine a(x,y)
      implicit none
      integer x,y
      call b(x,y)
      end
```

**b)**

```
c(int x;; int y);

// A ia routine using delegation.
b(int x;; int y) { c(x;;y); }
```

**c)**

```
a(x1;;y1) }  ➡  { c(x1;;y1)
```

**d)**

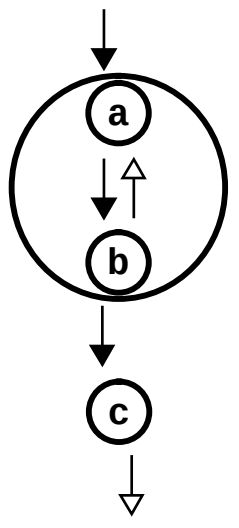

Figure 19 a) and b) Ancestor delegation using a Fortran and a ia routine, respectively.
c) The execution of `a(x1;;y1)` from the task pool, leaves `c(x1;;y1)` in the pool.
d) An illustration of the hierarchical definition.

**a)**

```
C Subordinate style Fortran.
      program evordel
      implicit none
      integer y
      call a(3,y)
      print*,y
      end

      subroutine c(x,y)
      implicit none
      integer x,y
      y = x
      end
```

**b)**

```
c(int x;; int y);

// A ia routines using delegation.
b(int x;; int y) { c(x;;y); }

a(int x;; int y) { b(x;;y); }
```

**c)**

```
b(3;;y)   }  ➡  { c(3;;y)
```

**d)**

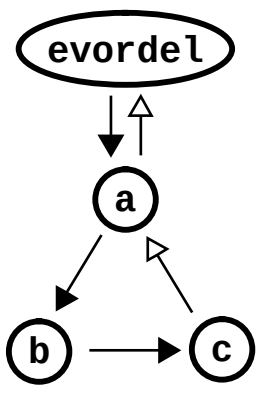

Figure 20 a) and b) Fortran and ia routines for an example of apt delegation.
c) The execution of `b(3;;y)` from the task pool, leaves `c(3;;y)` in the pool.
d) An illustration of the hierarchical definition.

**a)**

```
C An entire app.n in the delegation style.
      program begin_evordel
      implicit none
      call ia_evordel(3)
      end

      subroutine end_evordel(y)
      implicit none
      integer y
      print *,y
      end
```

**b)**

```
a(int n;;int y);
end_evordel(int y;;);

// Ia routine for begin_ and end_evordel.
ia_evordel(int n)
{
  a(n;;y);
  end_evordel(y;;);
}
```

**c)**

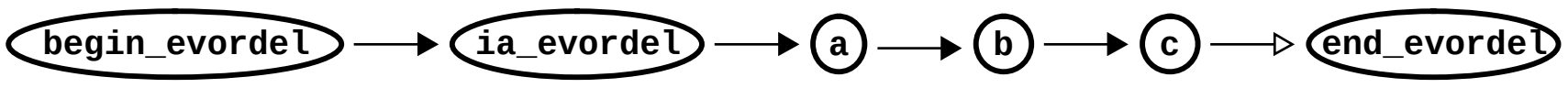

Figure 21 a) and b) A rewrite of `program evordel` of Figure 20a) in delegation style.
c) An illustration of the hierarchical definition.
The trivial return of `ia_evordel` to `begin_evordel` and of `b` to `a` is ignored.

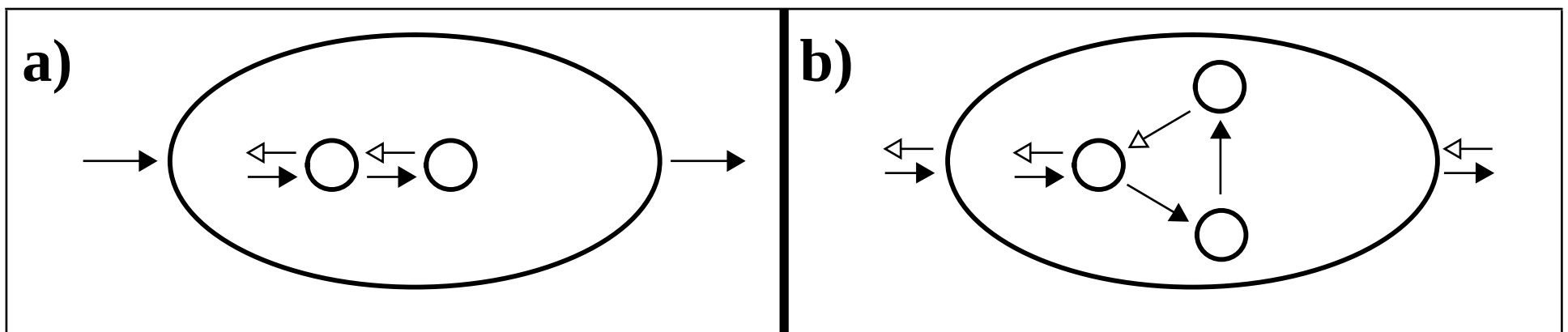

Figure 22 a) Processes are allowed within a task.
b) Tasks are allowed within a process.

**a)**

```
      program addtprog
      implicit none
      integer n,r,c
      print*,'Enter integer value for n:'
      read*,n
      call addtorial(n,r)
      c = (n+1)*n/2
      if (r .eq. c) then
        print*,'CORRECT:'
        print*,'addtorial(',n,')=',r
      else
        print*,'ERROR  :'
        print*,'addtorial(',n,')=',c,', not',r
      endif
      end
```

**b)**

```
C Ia prototypes as
C          addtorial(int n;;int r).

      subroutine addtorial(n,r)
      implicit none
      integer n,r,k
      r = 0
      do k = 1,n
        r = r+k
      enddo
      end
```

Figure 23 a) and b) A Fortran program for the addtorial function.

**a)**

```
      subroutine
     +      addtorial(n,r)
      implicit none
      integer n,r,i
      i = n
      r = 0
      call addt1(i,r)
      end

      subroutine addt1(i,r)
      implicit none
      integer i,r
      r = r + i
      i = i - 1
      if (i .gt. 0) then
        call addt1(i,r)
      endif
      end
```

**b)**

```
      subroutine
     +      addtorial(n,r)
      implicit none
      integer n,r,i
      i = n
      r = 0
      call addt2(i,r)
      end

      subroutine addt2(i,r)
      implicit none
      integer i,r
      r = r + i
      i = i - 1
      if (i .gt. 0) then
        call addt2_ia(i,r)
      endif
      end
```

**c)**

```
addt2(;int i,int r;);

addt2_ia(;int i,int r;)
{
addt2(;i,r;);
}
```

**d)**

```
      subroutine
     +       addt2_ia(i,r)
      implicit none
      integer i,r
      call addt2(i,r)
      end
```

Figure 24 a) The addtorial function in Fortran in a delegation style.
b) and c) As in a, but using a ia routine in order to use TSIA.
d) A Fortran routine equivalent in definition, but not in execution, to the ia routine of c).

**a)**

```
      block data initpool
      common /taskpool/ full
      logical full
      data full /.true./
      end

      subroutine addt2_ia(i,r)
      implicit none
      integer i,r
      common /taskpool/ full
      logical full
      if ( full ) then
 100    full = .false.
        call addt2(i,r)
        if (full) goto 100
      endif
      full = .true.
      end
```

**b)**

```
      subroutine addt2(i,r)
      implicit none
      integer i,r,s
      print *,'address of s',loc(s)
      r = r + i
      i = i - 1
      if (i .gt. 0) call addt2_ia(i,r)
      end
```

**c)**

```
Enter integer value for n:
address of s  2147429820
address of s  2147429748
address of s  2147429748
address of s  2147429748
address of s  2147429748
CORRECT:
addtorial( 5)= 15
```

**d)**

```
Enter integer value for n:
address of s  2147429820
address of s  2147429748
address of s  2147429676
address of s  2147429604
address of s  2147429532
CORRECT:
addtorial( 5)= 15
```

Figure 25 a) A functioning Fortran imitation of ia routine `addt2_ia(;i,r;)` of Figure 24d).
b) A version of the Fortran routine of Figure 24b) which shows the stack behavior.
c) and d) The output of the Fortran addtorial application when using the Fortran version of the ia routine in Figure 25a) and in Figure 24d), respectively.

**a)**

```
      subroutine
     +      addtorial(n,r)
      implicit none
      integer n,r
      r = 0
      call addt_dc(1,n,r)
      end

     subroutine add(a,b,c)
      implicit none
      integer a,b,c
      c = a + b
      end
```

**b)**

```
add(int a,int b;;int c);

addt_dc( int b, int t
      ;; int r)
{
if (b == t) r = b;
else {
  int m = (b+t)/2;
  addt_dc(b  ,m;;rb);
  addt_dc(m+1,t;;rt);
  add(rb,rt;;r);
}
}
```

**c)**

```
      subroutine addt_dc(b,t,r)
      implicit none
      integer b,t,r,m,rb,rt
      if (b .eq. t) then
        r = b
      else
        m = (b+t)/2
        call addt_dc(b  ,m,rb)
        call addt_dc(m+1,t,rt)
        call add(rb,rt,r)
      endif
      end
```

Figure 26 a) Two Fortran routines - one using and one supporting `addt_dc(b,t;;r)` of b).
b) A ia routine using divide-and-conquer to compute the addtorial function.
c) A Fortran routine equivalent in definition, but not in execution, to the ia routine of b).

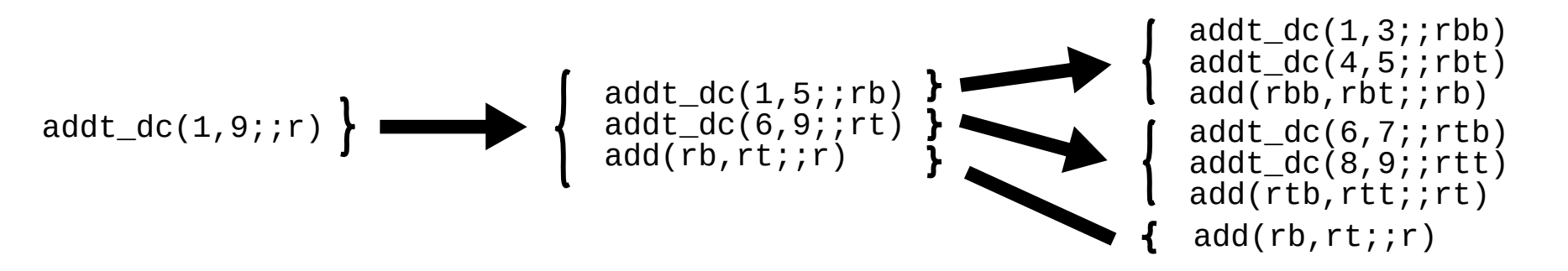

Figure 27 An evolution of tasks in the task pool for an addtorial using divide-and-conquer.

**a)**

```
      subroutine
     +      addtorial(n,r)
      implicit none
      integer n,r
      call addtS(n,r)
      end

      subroutine addtS(i,r)
      implicit none
      integer i,r
      if (i .lt. 2) then
        r = i
      else
        call addtS(i-1,r)
        r = r + i
      endif
      end
```

**b)**

```
      subroutine addtS(i,r)
      implicit none
      integer i,r
      if (i .lt. 2) then
        r = i
      else
        call addtS_ia(i,r)
      endif
      end


      subroutine incr(i,r)
      implicit none
      integer i,r
      r = r + i
      end
```

**c)**

```
addtS(int i;;int r);
incr(int i;int r;);

addtS_ia(int i;;int r)
{
addtS(i-1;;r);
incr(i;r;);
}
```

**d)**

```
      subroutine
     +        addtS_ia(i,r)
      implicit none
      integer i,r
      call addtS(i-1,r)
      call incr(i,r)
      end
```

Figure 28 a) The addtorial function in Fortran in a subordinate style.
b) and c) As in a, but converted to the deep delegation style and using a ia routine.
d) A Fortran routine equivalent in definition, but not in execution, to the ia routine of c).

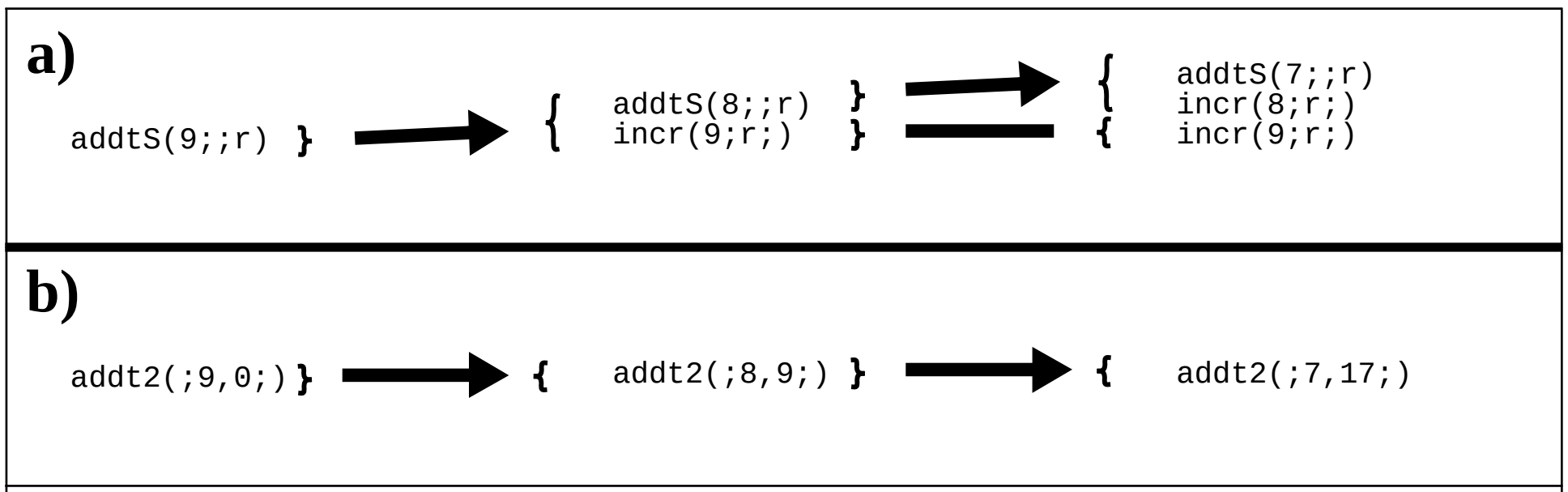

Figure 29 a) An evolution of tasks in the task pool for an addtorial with deep delegation.
b) As in a), but with tail delegation.

**a)**

```
      subroutine addtorial(n,r)
      implicit none
      integer n,r
      r = 0
      call addt3(n,r)
      end

      subroutine addt3(i,r)
      implicit none
      integer i,r,n
      r = r + i
      n = i - 1
      if (i .gt. 0) then
        call addt3_ia(n,r)
      endif
      end
```

**b)**

```
addt3( int i; int r;);

addt3_ia(int i; int r;)
{
addt3(i;r;);
}
```

**c)**

```
      block data initpool
      common /taskpool/ full,heap
      logical full
      integer heap
      data full /.true./
      end

      subroutine addt3_ia(i,r)
      implicit none
      integer i,r,n
      common /taskpool/ full,heap
      logical full
      integer heap
      if ( full ) then
        n = i
 100    full = .false.
        call addt3(n,r)
        if (full) then
          n = heap
          goto 100
        endif
      endif
      heap = i
      full = .true.
      end
```

Figure 30 a) and b) Respectively, the code of Figure 24b) and c), but with `i` as an in, not as an inout. c) The Fortran imitation of b). Thus it is Figure 25a) modified to manage an in.

**a)**

```
subroutine
+      addtorial(n,r)
implicit none
integer n,r,i
i = n
r = 0
call addtW(i,r)
end

subroutine addtW(i,r)
implicit none
integer i,r
r = r + i
i = i - 1
if (i .gt. 0) then
  call addtW_ia(i,r)
endif
print*,'i=',i,',r=',r
end
```

**b)**

```
addtW(;int i,int r;);

addtW_ia(;int i,int r;)
{ addtW(;i,r;); }
```

**c)**

```
Enter integer value for n:
i = 7, r = 17
i = 6, r = 24
i = 5, r = 30
i = 4, r = 35
i = 3, r = 39
i = 2, r = 42
i = 1, r = 44
i = 0, r = 45
i = 0, r = 45
CORRECT:
addtorial( 9)= 45
```

**f )**

```
Enter integer value for n:
i = 0, r = 45
i = 0, r = 45
i = 0, r = 45
i = 0, r = 45
i = 0, r = 45
i = 0, r = 45
i = 0, r = 45
i = 0, r = 45
i = 0, r = 45
CORRECT:
addtorial( 9)= 45
```

**d)**

```
subroutine
+      addtorial(n,r)
implicit none
integer n,r,i
i = n
r = 0
call addtR(i,r)
end

subroutine addtR(i,r)
implicit none
integer i,r,i2,r2
r2 = r + i
i2 = i - 1
if (i2 .gt. 0) then
 call addtR_ia(i2,r2)
endif
r = r2
i = i2
print*,'i=',i,',r=',r
end
```

**e)**

```
addtR(;int i,int r;);

addtR_ia(;int i,int r;)
{ addtR(;i,r;); }
```

Figure 31 a) The naive subordinate style has undefined behavior due to ancestor delegation.
b) The ia routine for the Fortran routine of a).
c) An example of the undefined output when using a) and b).
d) Subordination with defined behavior. Apt delegation overrides ancestor delegation.
e) The ia routine for the Fortran routine of d).
f) An example of the defined output when using d) and e).

**a)**

```
      subroutine mult(a,b,c,n)
      implicit none
      integer n, i
      real a(n),b(n),c(n)
      do i = 1,n
        c(i) = a(i)*b(i)
      enddo
      end
```

**b)**

```
mult(n, a, b;; c)
  is Fortran mult( real a(n), real b(n),
                   real c(n), integer n);
```

**c)**

```
produce(bytes;;bytes) is transformer
  produce < bytes[] > bytes[];
```

**d)**

```
produce(bytes;;bytes) is transformer
  produce < bytes > bytes;
```

Figure 32 a) and b) A simple Fortran routine and its task declaration in ia.
c) A task declaration for a transformer process for a stream of `bytes` items.
As in a UNIX shell, `<` and `>` identify input and output streams, respectively.
d) As in c), but for a single `bytes` item.

**a)**

```
      subroutine addt2(i,r)
      implicit none
      integer i,r
      goto (1,2,3) i
C Default is here.
      r = r + i
      i = i - 1
      r = r + i
      i = i - 1
      r = r + i
      i = i - 1
      call addt2_ia(i,r)
      return
 3    r = r + i
      i = i - 1
 2    r = r + i
      i = i - 1
 1    r = r + i
      i = i - 1
      end
```

**b)**

```
      subroutine addt2(i,r)
      implicit none
      integer i,r,j,m,maxunrl
      maxunrl = 20
      m = min(i,maxunrl)
      do j = 1,m
        r = r + i
        i = i - 1
      enddo
      if (i .gt. 0) call addt2_ia(i,r)
      end
```

Figure 33 a) The Fortran routine of Figure 24b) after applying recursion unrolling.
b) As in a), but replacing recursion by looping.

**a)**

```
      subroutine addt2(i,r)
      implicit none
      integer i,r,s
      print *,'address of s',
     +        loc(s)
      r = r + i
      i = i - 1
      if (mod(i,4).ne.0) then
        call addt2(i,r)
      else if (i .gt. 0) then
        call addt2_ia(i,r)
      endif
      end
```

**b)**

```
Enter integer value for n:
address of s  2147429820
address of s  2147429740
address of s  2147429692
address of s  2147429644
address of s  2147429596
address of s  2147429740
address of s  2147429692
address of s  2147429644
address of s  2147429596
address of s  2147429740
address of s  2147429692
address of s  2147429644
address of s  2147429596
CORRECT:
addtorial( 13)= 91
```

Figure 34 a) The Fortran routine of Figure 24b) modified to demonstrate delegation unrolling.
b) The output of the Fortran addtorial application when using a)
with the Fortran imitation in Figure 25a) of the ia routine `addt2_ia(;i,r;)`.

**a)**

```
      block data initpool
      common/taskpool/full,unroll,maxunrl
      logical full
      integer unroll,maxunrl
      data full /.true./
      data unroll,maxunrl /5,5/
      end


      subroutine addt2_ia(i,r)
      implicit none
      integer i,r
      common/taskpool/full,unroll,maxunrl
      logical full
      integer unroll,maxunrl

      unroll = unroll + 1
      if (unroll .lt. maxunrl) then
        call addt2(i,r)
        return
      endif
      unroll = 0

      if ( full ) then
 100    full = .false.
        call addt2(i,r)
        if (full) goto 100
      endif
      full = .true.
      end
```

**b)**

```
      block data initpool
      common /taskpool/ full,yield
      logical full
      integer yield
      data full /.true./
      end


      subroutine addt2_ia(i,r)
      implicit none
      integer i,r
      common /taskpool/ full,yield
      logical full
      integer yield
      integer irand

      yield = mod(irand(),2)
      if (yield .eq. 0) then
        call addt2(i,r)
        return
      endif

      if ( full ) then
 100    full = .false.
        call addt2(i,r)
        if (full) goto 100
      endif
      full = .true.
      end
```

**c)**

```
Enter integer value for n:
address of s  2147429820
address of s  2147429748
address of s  2147429676
address of s  2147429604
address of s  2147429532
address of s  2147429460
address of s  2147429748
address of s  2147429676
address of s  2147429604
address of s  2147429532
address of s  2147429460
address of s  2147429748
address of s  2147429676
CORRECT:
addtorial( 13)= 91
```

**d)**

```
Enter integer value for n:
address of s  2147429820
address of s  2147429740
address of s  2147429660
address of s  2147429580
address of s  2147429580
address of s  2147429580
address of s  2147429580
address of s  2147429500
address of s  2147429580
address of s  2147429500
address of s  2147429420
address of s  2147429580
address of s  2147429580
CORRECT:
addtorial( 13)= 91
```

Figure 35 a) The Fortran imitation of Figure 25a), modified to provide unrolled delegation.
b) As in a), but modified to continue evaluation only with the permission of the TS.
c) and d) Output of the Fortran addtorial application when using a) and b), respectively.

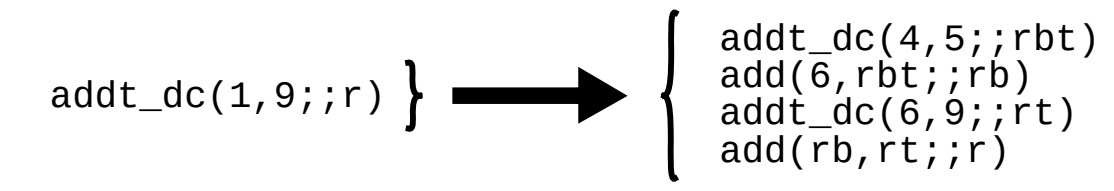

Figure 36  As in Figure 27, but with unrolled delegation at work.

**a)**

```
      subroutine
     +      addt_ev(b,t,r)
      implicit none
      integer b,t,r,i
      r = 0
      do i = b,t
        r = r + i
      enddo
      end
```

**b)**

```
add(int a,int b;;int c);
addt_ev( int b, int t
      ;; int r);

addt_dc( int b, int t
      ;; int r)
{
int mindc=20;
if (t-b < mindc) {
  addt_ev(b  ,t;;r );
} else {
  int m = (b+t)/2;
  addt_dc(b  ,m;;rb);
  addt_dc(m+1,t;;rt);
  add(rb,rt;;r);
}
}
```

**c)**

```
      subroutine addt_dc(b,t,r)
      implicit none
      integer b,t,r,m,rb,rt
      integer mindc
      mindc = 20
      if(t-b.lt.mindc) then
        call addt_ev(b  ,t,r )
      else
        m = (b+t)/2
        call addt_dc(b  ,m,rb)
        call addt_dc(m+1,t,rt)
        call add(rb,rt,r)
      endif
      end
```

Figure 37  a) An evaluation task alternative to the delegation task `addt_dc(b,t;;r)` of b).
b) A ia routine using divide-and-conquer to compute the addtorial function.
It also demonstrates minimal delegation.
c) A Fortran routine equivalent in definition, but not in execution, to the ia routine of b).

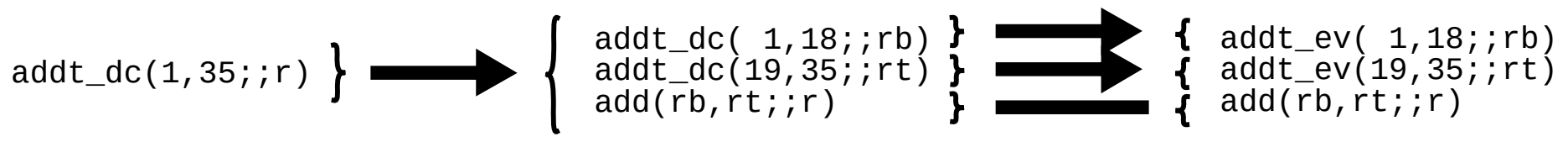

Figure 38  As in Figure 27, but with minimal delegation at work.

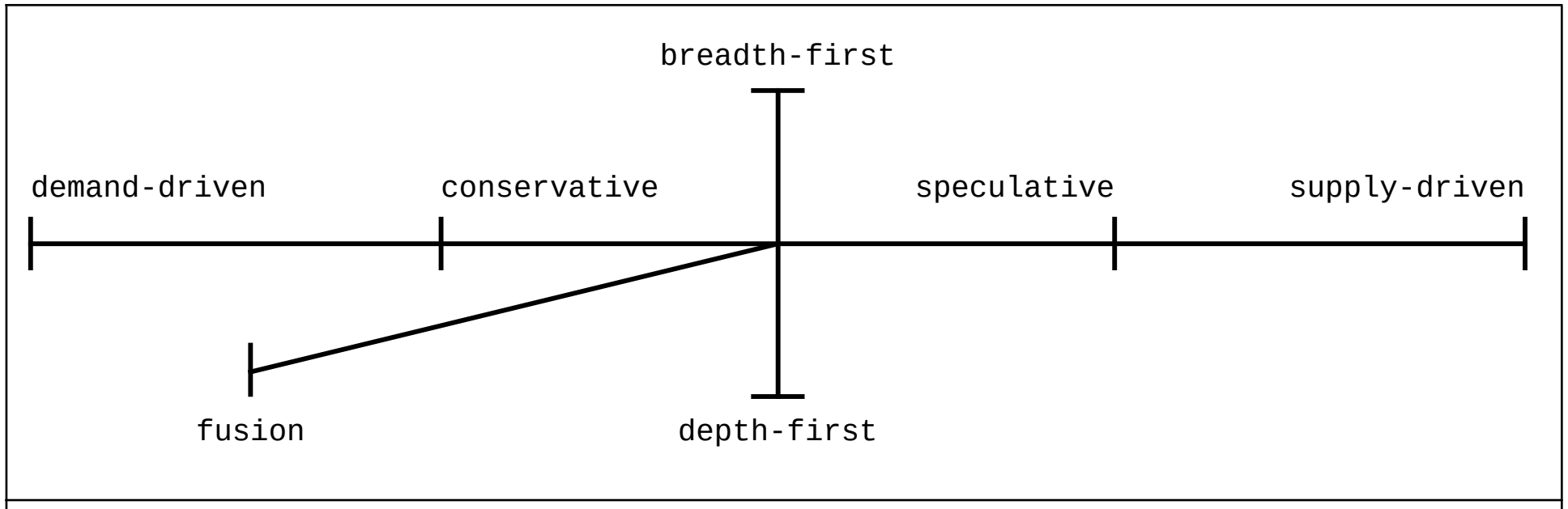

Figure 39 The task order of an application execution includes possibilities across several orthogonal axes.

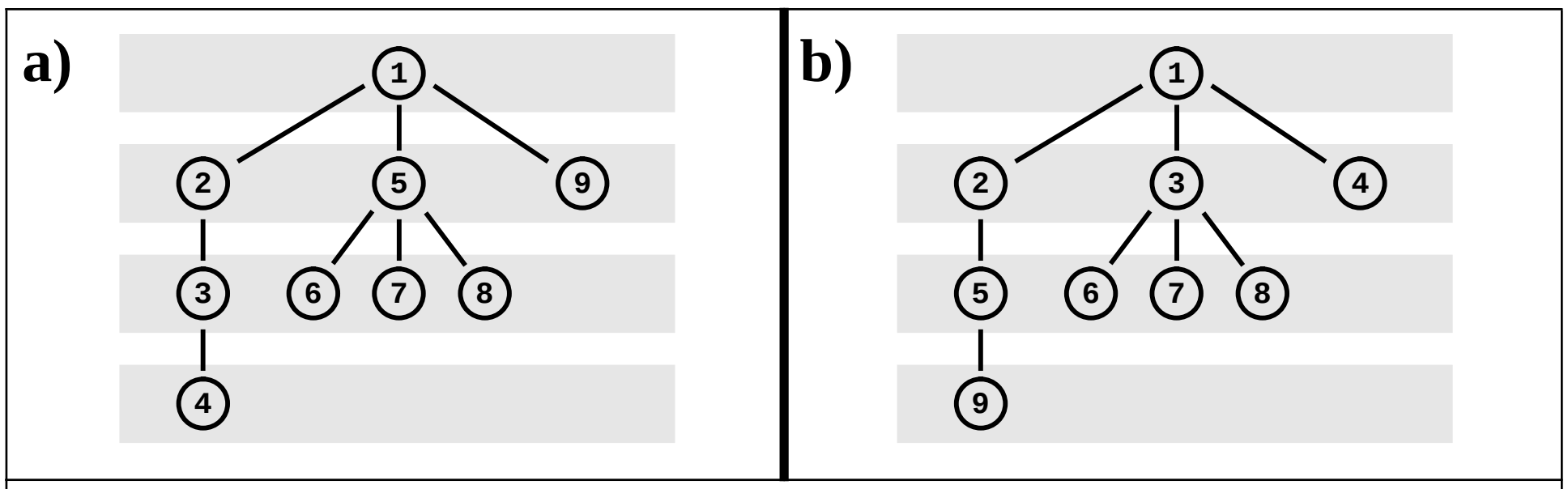

Figure 40 An illustration of task order. Each circle is a task and an arc identifies its parent.
Each gray band identifies a generation of tasks.
The number identifies the order of the task execution.
a) A depth-first execution. b) A breadth-first execution.

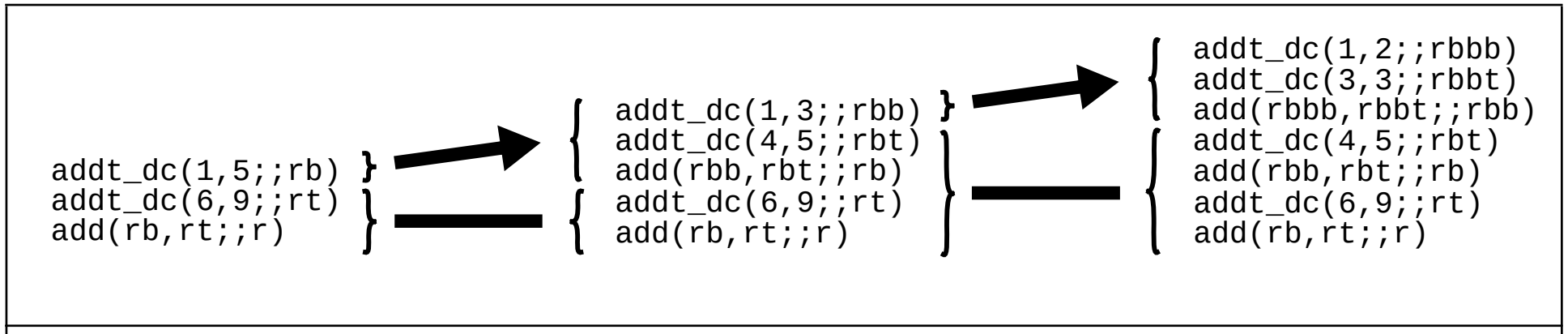

Figure 41 As in Figure 27. There the execution is breadth-first; here it is depth-first and LIFO.

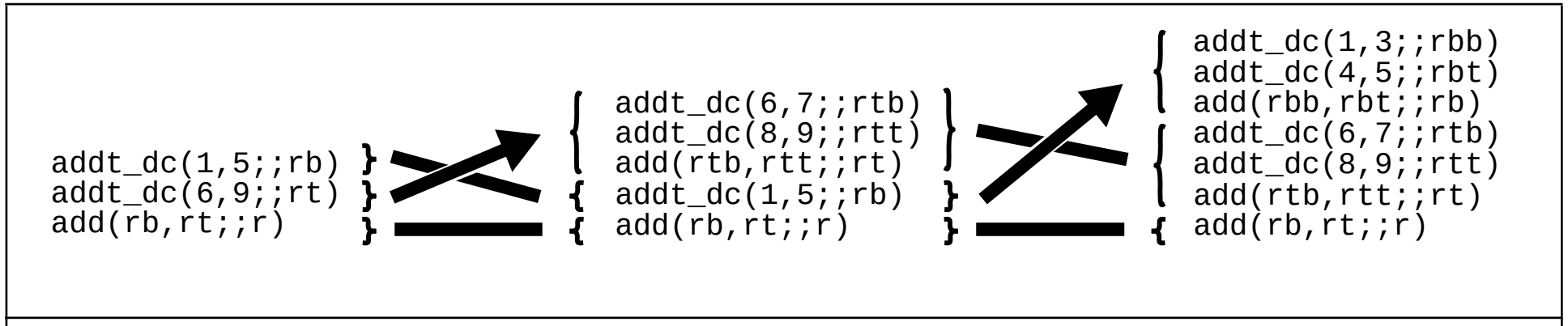

Figure 42 As in Figure 27, but better illustrating that a breadth-first execution is a FIFO execution.

**a)**

```
a(;;int x);
c(int x;int y;);

b(int x;int y;)
{ c(x;y;); }

f(;int y;)
{ a(;;x); b(x;y;); }
```

**b)**

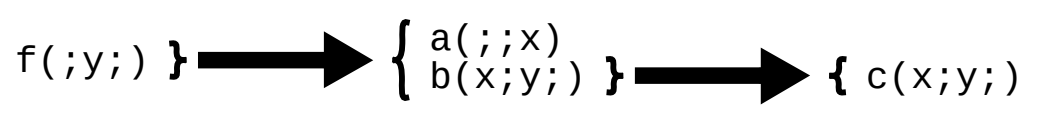

**c)**

```
a(;;int x);
c(int x;int y;);

b(del int x;del int y;)
{ c(x;y;); }

f(;del int y;)
{ a(;;x); b(x;y;); }
```

**d)**

**e)**

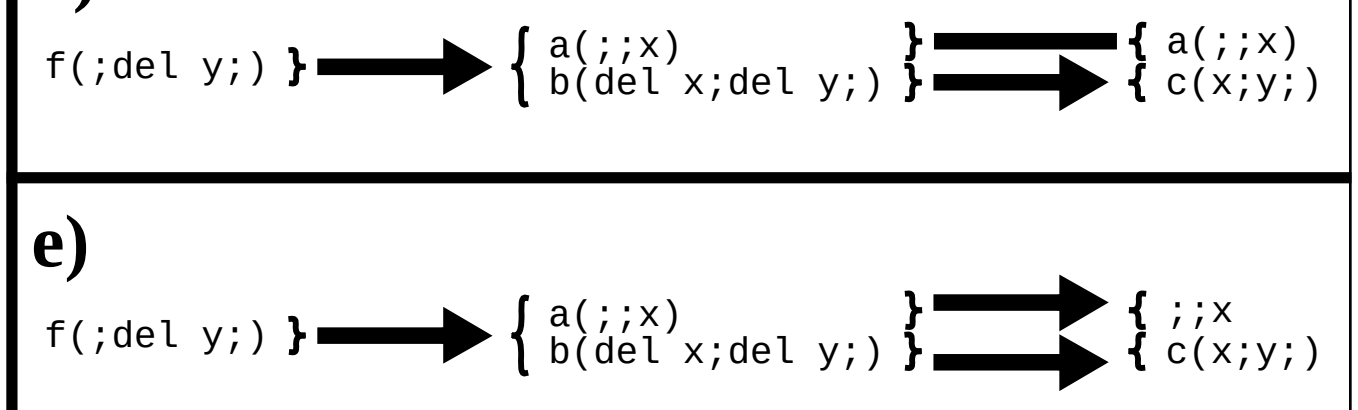

Figure 43 a) Ia code for the routines `b(x;y;)` and `f(;y;)`.
b) The execution of the code in a), as seen in the task pool.
c) The same definition as in a), but made more precise by the keyword `del`.
d) and e) Possible executions of the code in c), as seen in the task pool.
Also possible for the code in c) is the execution illustrated in b).

```
del          - the item is delegated.
eva          - the item is evaluated.
ign          - the item is ignored.
del_eva      - the item is delegated or evaluated. The default!
del_ign      - the item is delegated or ignored.
eva_ign      - the item is evaluated or ignored.
del_eva_ign  - the item is delegated or evaluated or ignored.
```

Figure 44 In the ia language of this presentation, a keyword declares the use of an item in a task.

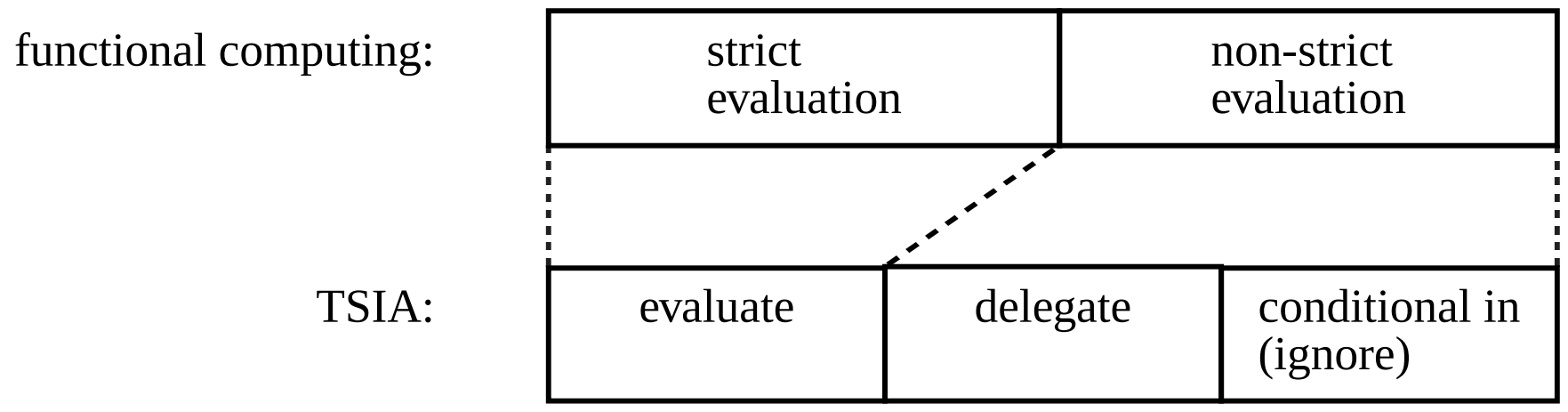

Figure 45 The correspondence of strict and non-strict evaluation in functional computing to evaluate, delegate and ignore in TSIA.

**a)**

```
      program testnqsols
      implicit none
      integer n,sols
      do n = 1,10
        call nqsols(n,sols)
        print *,n,' queens has',sols,
     +          ' solutions.'
      enddo
      end

      subroutine nqsols(n,sols)
      implicit none
      integer n,sols
      sols = 0
      call nattempts(n,0,0,sols)
      end

      subroutine nattempts(n,b_size,
     +                     board,sols)
      implicit none
      integer n,b_size,board(b_size),
     +       sols, i
      logical safe
      if (n.eq.b_size) then
        call incr(sols)
      else
        do i = 1,n
          call testsafe( b_size,board,i,
     +                   safe)
          call attempt(n,b_size,board,i,
     +                   safe,sols)
        enddo
      endif
      end

      subroutine incr(a)
      implicit none
      integer a
      a = a + 1
      end
```

**b)**

```
      subroutine testsafe(size,board,
     +                    new,safe)
      implicit none
      integer size,board(size),new,k
      logical safe
      safe = .false.
      do k = 1,size
        if (    board(k)
     +      .eq.new           ) return
        if (    iabs(board(k)-new)
     +      .eq.iabs(size+1-k)) return
      enddo
      safe = .true.
      return
      end

      subroutine attempt(n,b_size,board,
     +                   new,safe,sols)
      implicit none
      integer n,b_size,board(b_size),
     +       new,sols
     integer next_board(10)!Assume n<=10.
      logical safe
      if (safe) then
        call makeboard(b_size,board,new,
     +             b_size+1,next_board)
        call nattempts(n,b_size+1,
     +                 next_board,sols)
      endif
      end

      subroutine makeboard(b_size,board,
     +           new,n_size,next_board)
      implicit none
      integer b_size,board(b_size),new,
     +       n_size,next_board(n_size),i
      do i = 1,b_size
        next_board(i)       = board(i)
      enddo
        next_board(b_size+1) = new
      end
```

Figure 46  A Fortran application for the number of solutions to the N-queens problem.

```
makeboard(int b_size,int board[b_size],
          int new, int n_size;;
          int next_board[n_size]);

incr(; int a;);

testsafe(int size, int board[size],
         int new;; boolean safe);

nattempts(int n,int b_size,
          del int board[b_size];
          del int sols;);


attempt(int n,int b_size,
        int board[b_size], del int new,
        boolean safe; del_ign sols;)
{
if (safe)
  makeboard(b_size,board,new,
            b_size+1;;next_board);
  nattempts(n,b_size+1,next_board;sols;);
}
}
```

```
nattempts(int n,int b_size,
          del int board[b_size];
          del int sols;)
{
if (n==b_size) {
  incr(;sols;);
} else {
  for (int i=1; i<=n; i++) {
    testsafe( b_size,board,i;;safe);
    attempt(n,b_size,board,i,safe;sols;);
  }
}
}

nqsols(int n;; int sols)
{
  sols = 0;
  nattempts(n,0,0;sols;);
}
```

Figure 47 The Fortran definition of N-queens of Figure 46 rewritten in ia.

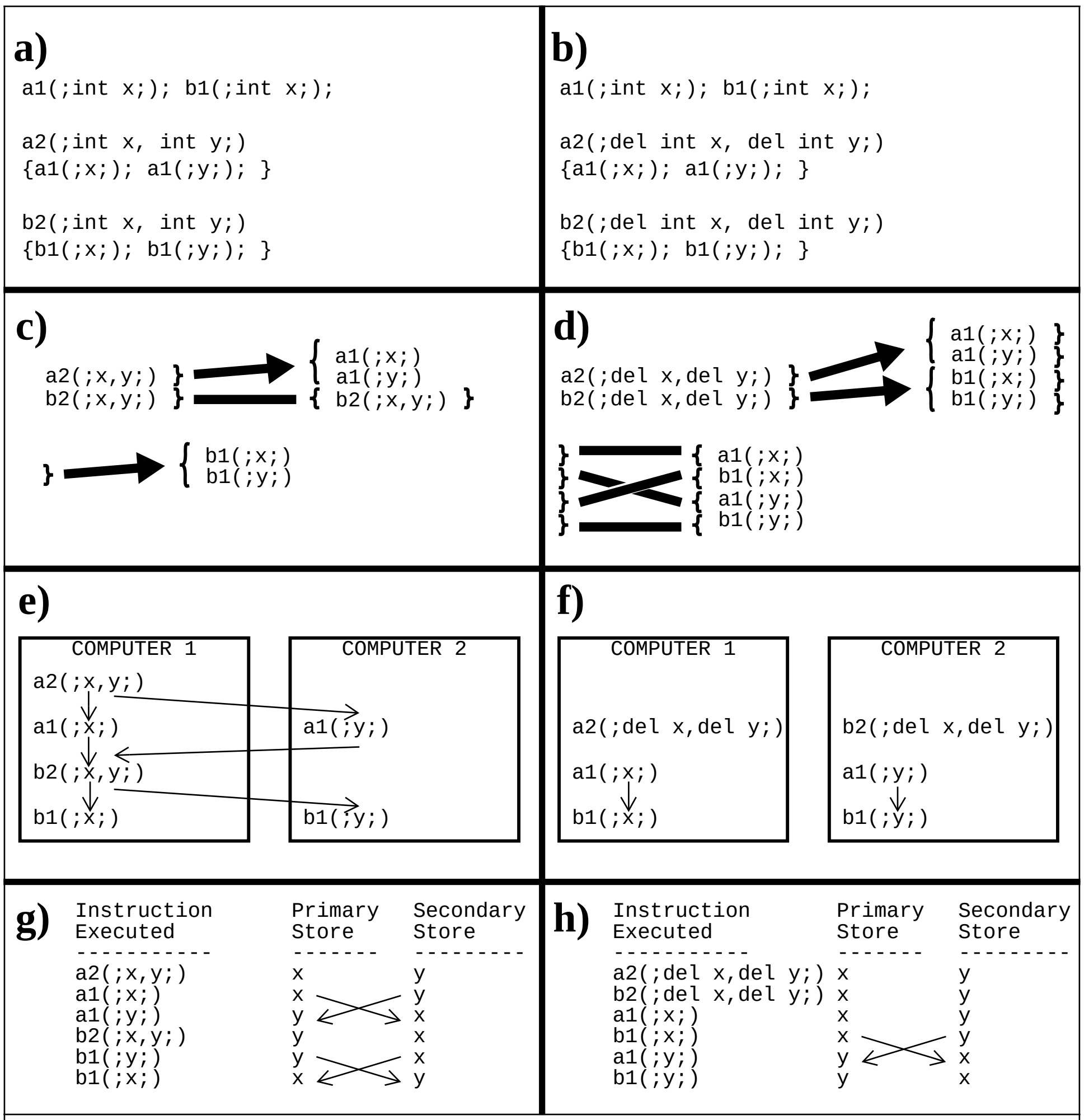

Figure 48 a) Ia code for the routines `a2(;x,y;)` and `b2(;x,y;)`.
b) The same definition as in a), but made more precise by the keyword `del`.
c) and d) As seen in the task pool, the execution of the code in a) and b), respectively.
e) and f) Parallel executions of the code in a) and b), respectively.
g) and h) Executions using a hierarchical store for the code in a) and b), respectively.

```
a2(; del x, del y;); b2(; del x, del y;);

main () { inoutfile "xy.dat" int x,y; a2(;x,y;); b2(;x,y;); }
```

Figure 49 A candidate application for parallel I/O. It uses the code of Figure 48b).

**a)**

```
      subroutine addtorial(n,r)
      implicit none
      integer n,r,a(1000) ! Assume n<1000
      call vseq(1,n,a)
      call vsum(n,a,r)
      end

! vseq does a(1 to n) = w to w+n-1
      subroutine vseq(w,n,a)
      integer w,n, a(n), k
      if (n .eq. 1) then
        call set(w,a)
      else
        k = n/2
        call vseq(w  ,k  ,a     )
        call vseq(w+k,n-k,a(k+1))
      endif
      end

! vsum does r = c(1) + c(2) + ... + c(n)
      subroutine vsum(n,c,r)
      implicit none
      integer n,c(n),r, k,r1,r2
      if (n .eq. 1) then
        call set(c,r)
      else
        k = n/2
        call vsum(k  ,c     ,r1)
        call vsum(n-k,c(k+1),r2)
        call add(r1,r2,r)
      endif
      end
```

**b)**

```
      subroutine set(a,b)
      integer a,b
      b = a
      end

      subroutine add(a,b,c)
      integer a,b,c
      c = a + b
      end
```

**c)**

```
set(int a;; int b);
add(int a, int b;; int c);

vseq(int w, int n;; del int a[n]) {
if (n == 1)
  set(w;;a);
else {
  int k = n/2;
  vseq(w  ,k  ;;a     );
  vseq(w+k,n-k;;a[k+1]);
}
}

vsum(int n, del int c[n];; del int r) {
if (n == 1)
  set(c;;r);
else {
  int k = n/2;
  vsum(k  ,c     ;;r1);
  vsum(n-k,c[k+1];;r2);
  add(r1,r2;;r);
}
}

addtorial(del int n;; del int r) {
vseq(1,n;;a);
vsum(n,a;;r);
}
```

Figure 50 a) and b) Fortran code using an intermediate array to calculate the addtorial.
c) Ia routines replacing the Fortran routines in a).

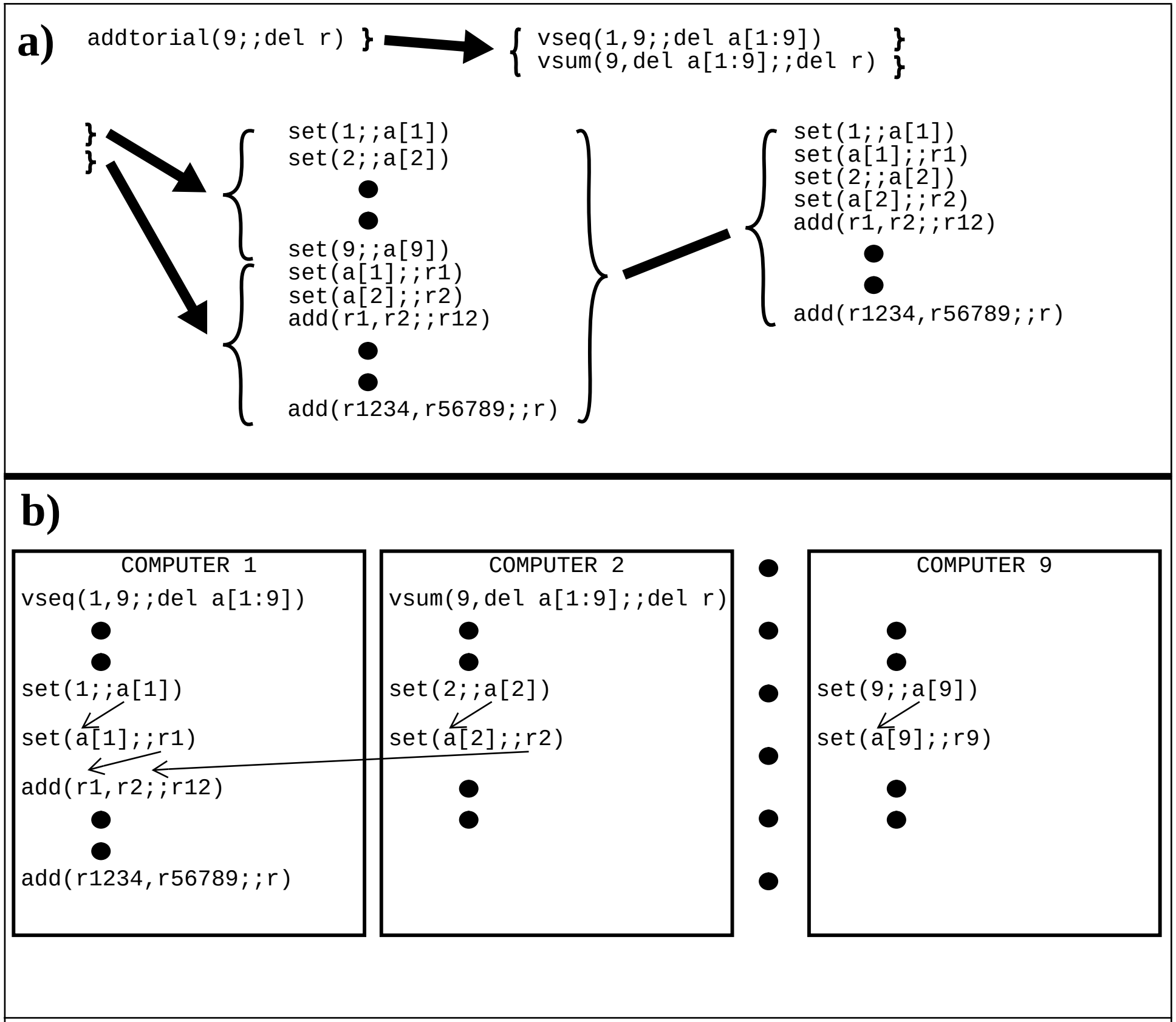

Figure 51 a) The evolution in the task pool for the array-based addtorial definition of Figure 50.
b) The code and execution of a) using nine computers.

**a)**

```
      program laplace
      implicit none
      integer n,imax,i
      parameter (n=8)
      real a(n),emax,e,new
      imax = 1000   ! Example
      emax = 0.0001 ! convergence cond.s.
      e = 2*emax ! Init. for jacobi().
      a(1) = 1.     ! Example
      a(n) = n      ! boundary cond.s.
      do i = 2,n-1  !
        a(i) = 0.   ! Define array.
      enddo         !

      call jacobi(n,emax,e,imax,a)

      print *,e,' was highest change.'
      print *,imax,' iterations remained.'
C Check jacobi() by checking convergence.
      do i = 2,n-1
        new = (a(i-1)+a(i+1))/2
        if (abs(new-a(i)) .gt. emax) then
          print *,'a(',i,') no converge.'
        endif
      enddo
      end
```

**b)**

```
C a(1) and a(n) are fixed.
C Relax a(2:n-1) for imax iterations
C or until convergence:
C        abs(new-old)<emax for each a(i).
C Relax means a(i) = (a(i-1)+a(i+1))/2.
C Initially requires e>emax.

      subroutine jacobi(n,emax,e,imax,a)
      implicit none
      integer n,imax
      real a(n),e,emax
C Convergence or max iterations?
      if (e .lt. emax) return
      imax = imax - 1
      if (imax .lt. 0) return
C Otherwise another relaxation iteration.
      call relax(n-2,a(1),a(n),a(2),e)
      call jacobi(n,emax,e,imax,a)
      end

      subroutine relax(n,m,p,a,e)
      implicit none
      integer n,k
      real a(n),m,p,e,mk,pk,olda,em,ep
      if (n .eq. 1) then
        call set(a,olda)
        call avg(m,p,a)
        call absdiff(olda,a,e)
      else
        k = n/2
        call set(a(k  ),mk)
        call set(a(k+1),pk)
        call relax(k  ,m ,pk,a     ,em)
        call relax(n-k,mk,p ,a(k+1),ep)
        call maxi(em,ep,e)
      endif
      end
```

**c)**

```
      subroutine set(a,b)
      real a,b
      b = a
      end

      subroutine maxi(a,b,c)
      real a,b,c
      if (a .gt. b) then
        c = a
      else
        c = b
      endif
      end

      subroutine avg(a,b,c)
      real a,b,c
      c = (a + b)/2.
      end

      subroutine absdiff(a,b,c)
      real a,b,c
      c = abs(b-a)
      end
```

Figure 52  Fortran application for Jacobi iteration in 1D.

```
set(    real a        ;; real b);
maxi(   real a, real b;; real c);
avg(    real a, real b;; real c);
absdiff(real a, real b;; real c);

relax(int n, del real m, del real p;
      del real a[n]; del real e) {
if (n == 1) {
  set(a;;olda);
  avg(m,p;;a);
  absdiff(olda,a;;e);
{
  int k = n/2;
  set(a(k  );;mk);
  set(a(k+1);;pk);
  relax(k  ,m ,pk;a     ;em);
  relax(n-k,mk,p ;a(k+1);ep);
  maxi(em,ep;;e);
}
}
```

```
jacobi(int n, real emax;
       real e, int imax, del real a[n];) {
// Convergence or max iterations?
if (e < emax) return;
imax = imax - 1;
if (imax < 0) return;
//Otherwise another relaxation iteration.
relax(n,a[1],a[n];a[2:n-1];e);
jacobi(n,emax;e,imax,a;);
}
```

Figure 53  For Jacobi iteration in 1D, ia routines replacing the Fortran routines of Figure 52b).

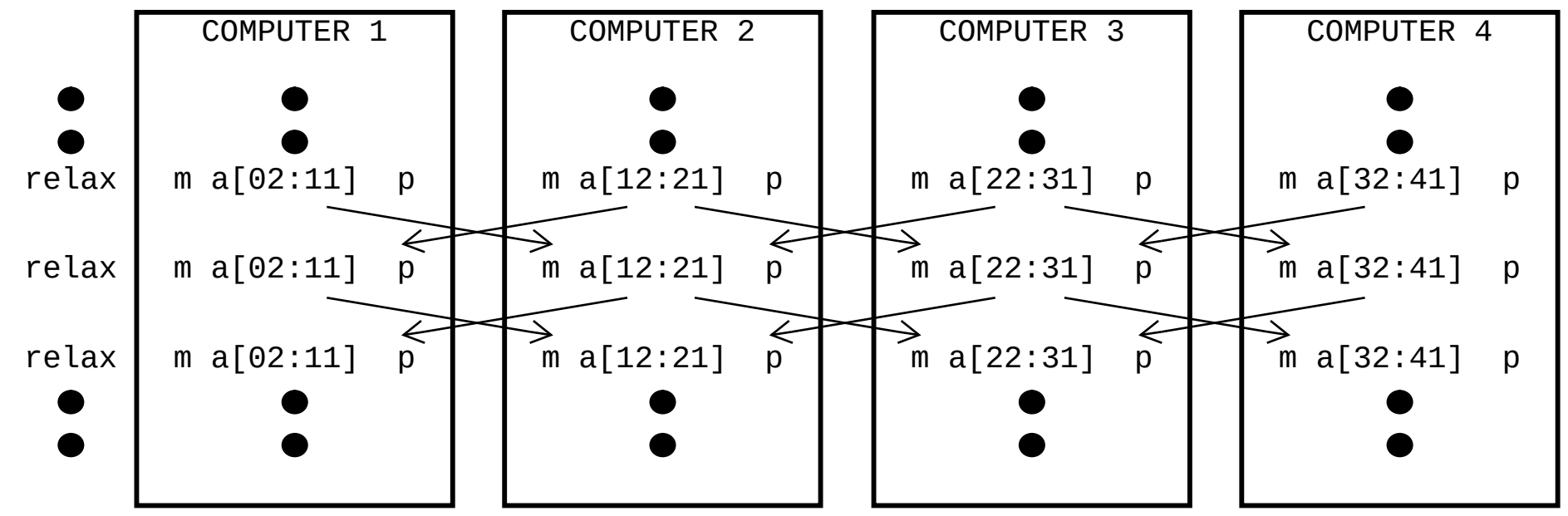

Figure 54  For the 1D Jacobi iteration code of Figure 52a) and b) and Figure 53,
a parallel execution using four computers.

**a)**

```
      program sortprog
      implicit none
      integer n,i,irand
      parameter ( n=1000 )
      integer a(n)
      do i = 1,n
       a(i) = irand()   ! Random array.
      enddo

      call dcqsort(n,a) ! Or msort(n,a).

C Check that a(1:n) is now ascending.
      do i = 2,n
       if ( a(i-1) .gt. a(i) ) then
        print *,'ERROR: a(',i-1,')=',
     +      a(i-1),' .gt. a(',i,')=',a(i)
       endif
      enddo
      end
```

**b)**

```
C quicksort a(1:n) into ascending order.
      subroutine dcqsort(n,a)
      implicit none
      integer m,n,a(n)
      if (n .lt. 2) return
      call dcpart(n-1,a(1),a(2),m)
      call dcqsort2(n,m,a)
      end

      subroutine dcqsort2(n,m,a)
      implicit none
      integer n,m,a(n)
C Have a(1),a(2:m+1),a(m+2:n)
C for pivot,before el.s,after el.s.
C swap() provides a(1:m),a(1),a(m+2:n)
C as req.d for before,pivot,after.
      call swap(a(1),a(m+1))
      call dcqsort(m    ,a     )
      call dcqsort(n-m-1,a(m+2))
      end


      subroutine swap(a,b)
      implicit none
      integer a,b,c
      c = a
      a = b
      b = c
      end

C On return, a(1:m)<p and a(m+1:n)>=p.
      subroutine dcpart(n,p,a,m)
      implicit none
      integer n,a(n),m,p,k,m1,m2
      if (n .eq. 1) then
        call part1(p,a,m)
      else
        k = n/2
        call dcpart(k  ,p,a     ,m1)
        call dcpart(n-k,p,a(k+1),m2)
C Now a(1  :m1  )<p a(m1+1:k)>=p
C     a(k+1:k+m2)<p a(m1+1:k)>=p.
C Since elements are otherwise unordered,
C only i=min(k-m1,m2) elements have
C incorrect position,
C Fix by swapping a(m1+1    :m1+1+i-1)
C            with a(k+m2-i+1:k+m2    ).
        call flop(n,k,m1,m2,a)
C Thus a(1:m1+m2)<p and a(m1+m2+1:n)>p.
        call add(m1,m2,m)
      endif
      end

      subroutine part1(p,a,m)
      integer p,a,m
      if (a .lt. p) then
        m = 1
      else
        m = 0
      endif
      end

      subroutine flop(n,k,m1,m2,a)
      integer n,k,m1,m2,a(n),i
      i = min(k-m1,m2)
      if (i.gt.0) then
       call dcswap(i,a(m1+1),a(k+m2-i+1))
      endif
      end

C Swap values of a(1:n) with b(1:n).
      subroutine dcswap(n,a,b)
      integer n,a(n),b(n),k
      if (n .eq. 1) then
        call  swap(a,b)
      else
        k = n/2
        call dcswap(k  ,a     ,b     )
        call dcswap(n-k,a(k+1),b(k+1))
      endif
      end

      subroutine add(a,b,c)
      integer a,b,c
      c = a + b
      end
```

Figure 55 a) A simple Fortran application using a sorting routine.
b) A Fortran definition of quicksort.

```
dcqsort(int n; del int a[n];);
dcqsort2(int n, int m; del int a[n];);
swap(;int a, int b;);
dcpart(int n, del int p; del int a[n]; del int m);
part1(int p, int a;; int m);
flop(int n, int k, int m1, int m2; del a[n];);
dcswap(int n; del int a[n], del int b[n];);
add(int a, int b;; int c);
```

Figure 56  Ia declarations for the routines of quicksort of Figure 55b).

```
dcqsort(int n; del int a[n];);

main() {
inoutfile "a.dat" int a[n];
dcqsort(n;a;);    // Or msort(n;a;);
}
```

Figure 57  An application using `dcqsort` of Figure 55b) to sort an arbitrarily large file of integers.

```
      subroutine msort(n,a)
      implicit none
      integer n,a(n),k
      if (n .lt. 2) return
      k = n/2
      call msort(k  ,a     )
      call msort(n-k,a(k+1))
      call merge(n,k,a)
      end


C Entry: a(1:m), a(m+1:n) each are sorted.
C Exit : a(1:n) is sorted.
      subroutine merge(n,m,a)
      implicit none
      integer n,m,a(n),x,y
      if (m .eq. 0 .or. m .eq. n) return
      x = m/2
      if ( x .eq. 0 ) x = 1
C a(x) is an elem. near middle of a(1:m).
      call find(a(x),m+1,n-m,a(m+1),y)
C a(y) is the largest element in a(m+1:n)
C      with a(y)<a(x).
      call merge2(n,m,x,y,a)
      end

C Entry: a(1:m), a(m+1:n) each are sorted.
C a(1:x) a(x+1:m) a(m+1:y) a(y+1:n)
C has a(m+1:y)<a(x) and a(y+1:n)>=a(x).
C Exit : a(1:n) is sorted.
      subroutine merge2(n,m,x,y,a)
      integer n,m,a(n),x,y
      if (x.eq.1 .and. y .eq. n) then
        ! Cycle a(m+1:y) to front of a().
        call rcycle(n,n-m,a)
      else
        call rcycle(y-x,y-m,a(x+1))
C After switch of a(x+1:m) and a(m+1:y),
C a(1:x+y-m)<a(x+y-m+1:n) for all elem.s,
C so can deal with each part indep.tly:
C 1. a(1:x+y-m) has a(1:x)
C    and a(x+1:x+y-m) each sorted.
        call merge(  (x+y-m),  x,a)
C 2. a(x+y-m+1:n) has a(x+y-m+1:y)
C    and a(y+1:n) each sorted.
        call merge(n-(x+y-m),m-x,
     +             a(x+y-m+1))
      endif
      end
```

```
C For array a(offset:offset+n-1),
C want y for a(y)<v<=a(y+1).
C Thus offset-1 <= y <= n+offset-1.
      subroutine find(n,offset,v,a,y)
      implicit none
      integer n,offset,v,a(n),y,k
      k = (n+1)/2
      call rfind(n,offset,v,a,k,a(k),y)
      end

      subroutine rfind(n,offset,v,a,k,ak,
                 y)
      implicit none
      integer n,offset,v,a(n),k,ak,y
      if (n .eq. 1) then
        if (ak .lt. v) then
          y = offset
        else
          y = offset -1
        endif
      else if (ak .lt. v) then
        ! y must be in a(k+1:n).
        call find(n-k,offset+k,v,a(k+1),y)
      else
        ! y must be in a(1:k).
        call find(k  ,offset  ,v,a     ,y)
      endif
      end


      subroutine set(a,b)
      implicit none
      integer a,b
      b =  a
      end
```

Figure 58 With the code of Figure 59a), a Fortran definition of mergesort.

**a)**

```
C For each a(1:n) cycle a(i) to a(mod(i-1+cc,n))+1
      subroutine rcycle(n,cc,a)
      implicit none
      integer n,a(n),cc,c
      if (n .lt. 2) return
      c = mod(mod(cc,n)+n,n)  ! Ensure 0<c<n.
      if (c .eq. 0) return
      call rcycle2(n,c,0,0,a)
      end

      subroutine rcycle2(n,c,moved,start,a)
      implicit none
      integer n,c,moved,start,a(n),tmp
      tmp = a(start+1)
      call rcycle3(n,c,moved,start+1,tmp,start+1,
     +             mod(start+1-c+n-1,n)+1,a)
      end

      subroutine rcycle3(n,c,imoved,start,
     +                   tmp,inew,iold,a)
      implicit none
      integer n,c,imoved,start,tmp,inew,iold,a(n)
      integer moved,new,old
      moved = imoved  ! Allow these to be in
      new   = inew    ! not inout, in Fortran.
      old   = iold

      call set(a(old),a(new))
      moved = moved + 1            ! 1:
      new   = old                  ! new index.
      old   = mod(new-c+n-1,n)+1   ! old index.
      if (old .ne. start) then
        call rcycle3(n,c,moved,start,tmp,new,old,a)
      else
        call set(tmp,a(new))
        moved = moved + 1          ! 2:
        if (moved .lt. n) then
          call rcycle2(n,c,moved,start,a)
        else if (moved .ne. n) then
          print *,'cycle: IMPOSSIBLE: moved .gt. n'
          call exit()
        endif
      endif
      end
```

**b)**

```
An example illustrating the
execution of rcycle(n,cc,a).

Let a(1:6) = abcdef
and cycle it by two positions,
i.e. call rcycle(6,2,a).


Then after either 1: or 2: of
moved = moved + 1,
of rcycle3() have the following
values for the items:

           mov sta
   a        ed  rt tmp new old
   ------  --- --- --- --- ---
1: ebcdef  1   1   a   1   5
1: ebcdcf  2   1   a   5   3
2: ebadcf  3   1   a   3   1
1: efadcf  4   2   b   2   6
1: efadcd  5   2   b   6   4
2: efabcd  6   2   b   4   2
```

Figure 59 a) With the code of Figure 58, a Fortran definition of mergesort.
b) An illustration of the routines `rcycle`, `rcycle2` and `rcycle3` of a).

```
msort(int n; del int a[n];);
find(int n, int offset, del int v, del int a[n];; del int y);
rfind(int n, int offset, int v, del int a[n], int ak;; int y);
merge(int n, int m; del int a[n];);
merge2(int n, int m, int x, int y; del int a[n];);
rcycle(int n, int cc; del int a[n];);
rcycle2(int n, int cc, int moved, int start; del int a[n];);
rcycle3(int n, int cc, int moved, int start, int i; del int a[n];);
swap(;int a, int b;);
```

Figure 60 Ia declarations for the routines of msort of Figure 58 and Figure 59.

**a)**

```
C For each a(1:n) cycle a(i) to a(mod(i-1+cc,n))+1
      subroutine cycle(n,cc,a)
      implicit none
      integer n,cc,a(n), c,moved,start,tmp,new,old
      if (n .lt. 2) return
      c = mod(mod(cc,n)+n,n)  ! Ensure 0<c<n.
      if (c .eq. 0) return

      moved = 0
      start = 0
 2    continue
        start = start + 1   ! Do gcd(n,c) starts.
        call set(a(start),tmp)
        new   = start
        old   = mod(new-c+n-1,n)+1
 3      continue
          call set(a(old),a(new))
          moved  = moved + 1      ! 1:
          new    = old                ! new index.
          old    = mod(new-c+n-1,n)+1 ! old index.
        if (old .ne. start) goto 3
        call set(tmp,a(new))
        moved  = moved + 1        ! 2:
      if (moved .lt. n) goto 2
      if (moved .ne. n) then
        print *,'cycle: IMPOSSIBLE: moved .gt. n'
        call exit()
      endif
      end
```

**b)**

```
An example illustrating the
execution of cycle(n,cc,a).

Let a(1:6) = abcdef
and cycle it by two positions,
i.e. call cycle(6,2,a).

Then after either 1: or 2: of
moved = moved + 1,
of cycle() have the following
values for the items:

           mov sta
   a        ed  rt tmp new old
   ------  --- --- --- --- ---
1: ebcdef  1   1   a   1   5
1: ebcdcf  2   1   a   5   3
2: ebadcf  3   1   a   3   1
1: efadcf  4   2   b   2   6
1: efadcd  5   2   b   6   4
2: efabcd  6   2   b   4   2
```

Figure 61 a) The iterative routine `cycle` is the origin of the recursive routines `rcycle`, `rcycle2` and `rcycle3` of Figure 59.
b) An illustration of the routine `cycle`. The illustration essentially is the same as that of Figure 59b) for the routines `rcycle`, `rcycle2` and `rcycle3` of Figure 59.a).

**a)**

```
      program bstest
      implicit none
      integer n
      parameter (n=100)
      integer a(n),i,j
      do j = 1,n  ! set-up array.
        a(j) = 2*j
      enddo
      do j = 1,2*n  ! test bs using array.
        call bs(1,n,a,j,i)
        if (mod(j,2).eq.0) then
          if (2*i.ne.j )
     +      print *,'ERROR: j i = ',j,i
        else
          if (  i.ne.-1)
     +      print *,'ERROR: j i = ',j,i
        endif
      enddo
      end
```

**b)**

```
      subroutine bs(m,p,a,v,i)
      implicit none
      integer m,p,a(m:p),v,i,k
      if (m.gt.p) then
        i = -1       ! no elements.
      else
        k = (m+p)/2  ! middle index.
        call bs1(m,p,a,v,k,a(k),i)
      endif
      end


      subroutine bs1(m,p,a,v,k,ak,i)
      implicit none
      integer m,p,a(m:p),v,k,ak,i
      if      (v.lt.ak) then
        call bs(m  ,k-1,a(m  ),v,i)
      else if (v.gt.ak) then
        call bs(k+1,p  ,a(k+1),v,i)
      else
        i = k
      endif
      end
```

**c)**

```
bs( int m, int p, del int a[m:p], del int v;; del int i);
bs1(del int m, del int p, del int a[m:p], int v, int k, int ak;; int i);
```

Figure 62 a) A Fortran application to test a search routine for a sorted array.
b) A delegation style Fortran definition of binary search.
c) Ia declarations for the routines of b).

a)

```
subroutine add(a,b,c)
implicit none
integer a,b,c
c = a + b
end

subroutine add3(a,b,c,d)
implicit none
integer a,b,c,d
d = a + b + C
end
```

```
subroutine vsum(n,c,r)
implicit none
integer n,c(n),r, k,r1,r2
if (n .eq. 1) then
  call set(c,r)
else
  k = n/2
  call vsum(k,c     ,r1)
  call vsum(k,c(k+1),r2)
  if (mod(n,2) .eq. 0) then
    call add(r1,r2,r)
  else
    call add3(c(n),r1,r2,r)
  endif
endif
end
```

b)

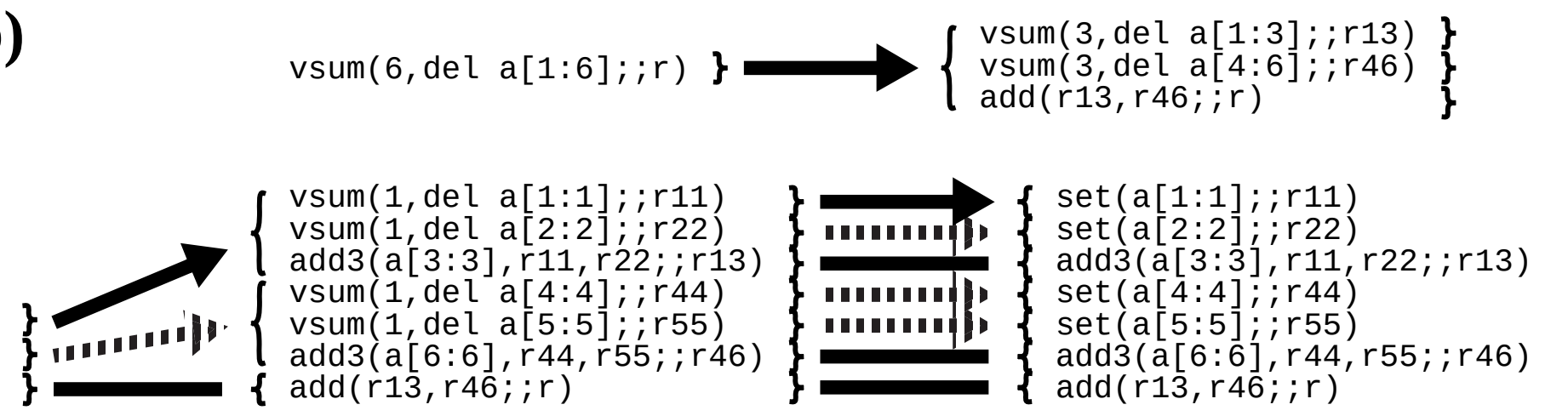

Figure 63 a) vsum as in Figure 50, but with a symmetric-divide-and-conquer (SDC) algorithm.
b) The evolution in the task pool for such a routine.
A dashed arrow indicates task replaced by its outcome using transparent delegation.

**a)**

```
      subroutine maxi(a,b,c)
      implicit none
      real a,b,c
      if (a .gt. b) then
        c = a
      else
        c = b
      endif
      end

      subroutine maxi3(a,b,c,d)
      implicit none
      real a,b,c,d,e
      call maxi(a,b,e)
      call maxi(e,c,d)
      end

      subroutine relax(n,m,p,a,e)
      implicit none
      integer n,k
      real a(n),m,p,e,mk,pk,olda,em,ep
      real ok,eo
      if (n .eq. 1) then
        call set(a,olda)
        call avg(m,p,a)
        call absdiff(olda,a,e)
      else
        k = n/2
        if (mod(n,2) .eq. 0) then
          call set(a(k  ),mk)
          call set(a(k+1),pk)
          call relax(k,m ,pk,a     ,em)
          call relax(k,mk,p ,a(k+1),ep)
          call maxi(em,ep,e)
        else
          call set(a(k  ),mk)
          call set(a(k+2),pk)
          call set(a(k+1),ok)
          call relax(k,m ,ok,a     ,em)
          call relax(k,ok,p ,a(k+2),ep)
          call relax(1,mk,pk,a(k+1),eo)
          call maxi3(em,eo,ep,e)
        endif
      endif
      end
```

**b)**

```
C On return, a(1:m)<p and a(m+1:n)>p.
      subroutine dcpart(n,p,a,m)
      implicit none
      integer n,a(n),m,p,k,m1,m2
      if (n .eq. 1) then
        call part1(p,a,m)
      else
        k = n/2
        call dcpart(k  ,p,a     ,m1)
        call dcpart(k  ,p,a(k+1),m2)
        call flop(n,k,m1,m2,a)
        call add(m1,m2,m)

C If n odd and a(n) < p,
C then swap a(n) with first element > p.
        if (mod(n,2) .eq. 1) then
          call fixn(p,a(m+1),a(n),m)
        endif
      endif
      end

      subroutine fixn(p,amp1,an,m)
      implicit none
      integer p,amp1,an,m
      if (an .lt. p) then
        call swap(amp1,an)
        m = m + 1
      endif
      end
```

**c)**

```
      subroutine dcswap(n,a,b)
      implicit none
      integer n,a(n),b(n),k
      if (n .eq. 1) then
        call   swap(a,b)
      else
        k = n/2
        call dcswap(k,a     ,b     )
        call dcswap(k,a(k+1),b(k+1))
        if (mod(n,2) .eq. 1) then
          call swap(a(n),b(n))
        endif
      endif
      end
```

Figure 64 a) `relax` as in Figure 52, but with a symmetric-divide-and-conquer (SDC) algorithm. b) and c) `dcpart` and `dcswap` as in Figure 55, but with SDC algorithms, respectively.

**a)**

```
      subroutine jacobi(n,emax,e,imax,a)
      implicit none
      integer n,imax
      real a(n),e,emax
C Convergence or max iterations?
      if (e .lt. emax) return
      imax = imax - 2         ! Different.
      if (imax .lt. 0) return
C Otherwise two more relaxations.
      call relaxtwice(n,a,e)  ! Different.
      call jacobi(n,emax,e,imax,a)
      end


      subroutine relaxtwice(n,a,e)
      implicit none
      integer n
      real a(n),e,b(1000) ! Assume n<1001.
      call relaxonce(n-2,a(1),a(n),a(2),
                     b(2))
      call relaxonce(n-2,a(1),a(n),b(2),
                     a(2))
      call dcabsdiff(n-2,a(2),b(2),e)
      end
```

```
      subroutine relaxonce(n,m,p,a,b)
      implicit none
      integer n,k
      real a(n),b(n),m,p,mk,pk
      if (n .eq. 1) then
        call avg(m,p,b)
      else
        k = n/2
        call set(a(k  ),mk)
        call set(a(k+1),pk)
        call relaxonce(k  ,m ,pk,a     ,
                       b     )
        call relaxonce(n-k,mk,p ,a(k+1),
                       b(k+1))
      endif
      end


      subroutine dcabsdiff(n,a,b,e)
      implicit none
      integer n,k
      real a(n),b(n),e,e1,e2
      if (n .eq. 1) then
        call absdiff(a,b,e)
      else
        k = n/2
        call dcabsdiff(k  ,a     ,b     ,
                       e1)
        call dcabsdiff(n-k,a(k+1),b(k+1),
                       e2)
        call maxi(e1,e2,e)
      endif
      end
```

**b)**

```
jacobi(int n, real emax; real e, int imax, del real a[n];);
relaxtwice(int n; del real a[n]; del real e);
relaxonce(int n, del real m, del real p, del real a[n];; del real b[n]);
dcabsdiff(int n, del real a[n], del real b[n];; del real e);

set(    real a        ;; real b);
maxi(   real a, real b;; real c);
avg(    real a, real b;; real c);
absdiff(real a, real b;; real c);
```

Figure 65 a) An array-based Fortran definition of Jacobi iteration in 1D.
It uses the routines of Figure 52c).
b) Ia declarations for the routines of a) and of Figure 52c).

```
relaxtwice(n;del a[1:50];e)

relaxonce(48,a[1],a[50],del a[2:49];;del b[2:49])
relaxonce(48,a[1],a[50],del b[2:49];;del a[2:49])
dcabsdiff(48,del a[2:49],del b[2:49];;e)

relaxonce(24,a[1],pka  ,del a[ 2:25];;del b[ 2:25])
relaxonce(24,mka ,a[50],del a[26:49];;del b[26:49])
relaxonce(24,a[1],pkb  ,del b[ 2:25];;del a[ 2:25])
relaxonce(24,mkb ,a[50],del b[26:49];;del a[26:49])
dcabsdiff(24,del a[ 2:25],del b[ 2:25];;e1)
dcabsdiff(24,del a[26:49],del b[26:49];;e2)
maxi(e1,e2;;e)

relaxonce(24,a[1],pka  ,del a[ 2:25];;del b[ 2:25])
relaxonce(24,a[1],pkb  ,del b[ 2:25];;del a[ 2:25])
dcabsdiff(24,del a[ 2:25],del b[ 2:25];;e1)
relaxonce(24,mka ,a[50],del a[26:49];;del b[26:49])
relaxonce(24,mkb ,a[50],del b[26:49];;del a[26:49])
dcabsdiff(24,del a[26:49],del b[26:49];;e2)
maxi(e1,e2;;e)
```

Figure 66  a) The evolution in the task pool for the Jacobi definition of Figure 65a).

**a)**

```
add(int a, int b;; int c);

addtorial(del int n;; del int r) {
  int a[0:n] = 0;
  int s[0:n] = 0;
  add(1,a[i-1];;a[i]);
  add(a[i],s[i-1];;s[i]);
  r is s[n];
}
```

**b)**

```
      subroutine add(a,b,c)
      implicit none
      integer a,b,c
      c = a + b
      end

      subroutine addtorial(n,r)
      implicit none
C Assume n<1001 .
      integer n,r,a(0:1000),s(0:1000), i
      a(0) = 0
      s(0) = 0
      do i=1,n
        call add(1,a(i-1),a(i))
        call add(a(i),s(i-1),s(i))
      enddo
      r = s(n)
      end
```

**c)**

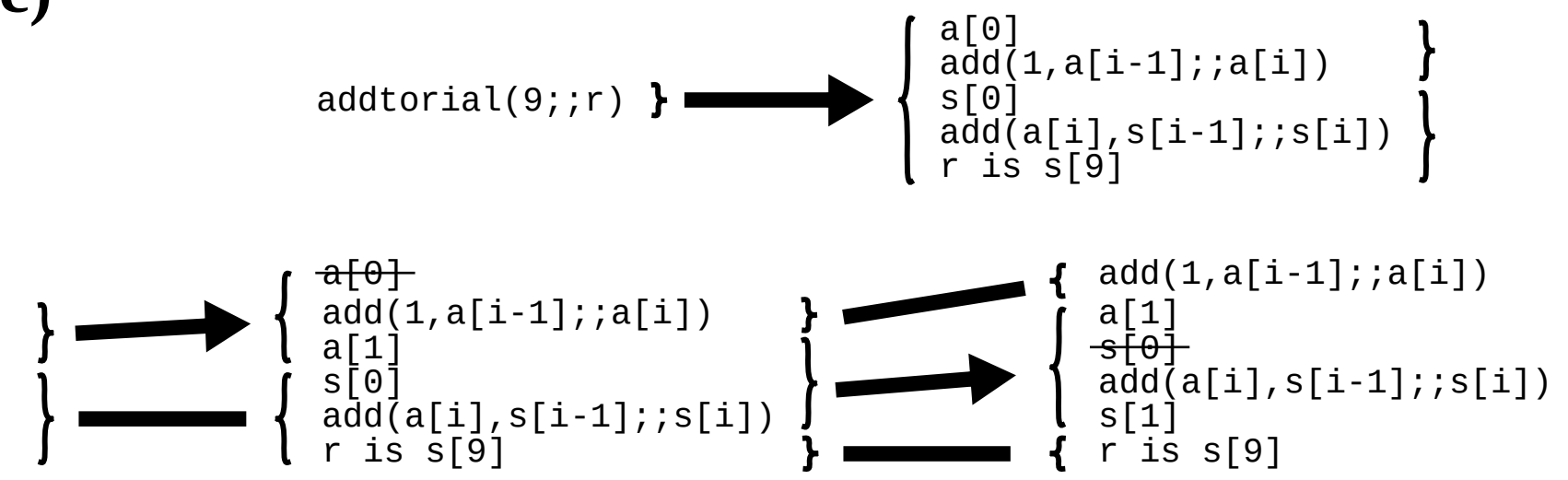

Figure 67 a) A deductive array-based addtorial definition coded in ia.
b) A Fortran imitation of the ia code in a).
c) The evolution in the task pool for the ia code of a).

**a)**

```
add(int a, int b;; int c);

fib(del int n;; del int r) {
  int a[0:n] = {0,1};
  add(a[i-2],a[i-1];;a[i]);
  r is a[n];
}
```

**b)**

```
      subroutine fib(n,r)
      implicit none
C Assume n<1001 .
      integer n,r,a(0:1000), i
      a(0) = 0
      a(1) = 1
      do i=2,n
        call add(a(i-2),a(i-1),a(i))
      enddo
      r = a(n)
      end
```

Figure 68 a) A deductive array-based fibonacci definition coded in ia.
b) A Fortran imitation of the ia code in a). The routine `add(a,b,c)` is in Figure 67b).

```
produce(bytes;;bytes);

main() {
  infile  "a.dat" bytes a[N];
  outfile "b.dat" bytes b[N];
  produce(a[k];;b[k]);
  }
}
```

Figure 69  A deductive array-based classic application. The definition is largely that of Figure 11.

**a)**

```
xif( boolean m, del_ign int n1, del_ign int n2;; del int w)
{ if (m) w is n1; else w is n2; }
```

**b)**

```
a(;;boolean x);
b(;;int x); c(;;int x);

g(;;del int q) {
  a(;;x);
  if(x) b(;;q);
  else  c(;;q);
}
```

**c)**

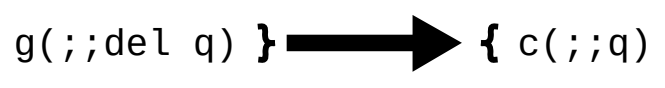

**d)**

```
a(;;boolean x);
b(;;int x); c(;;int x);

g(;;del int q) {
  a(;;x);
  b(;;y);
  c(;;z);
  xif(x,y,z;;q);
}
```

**e)**

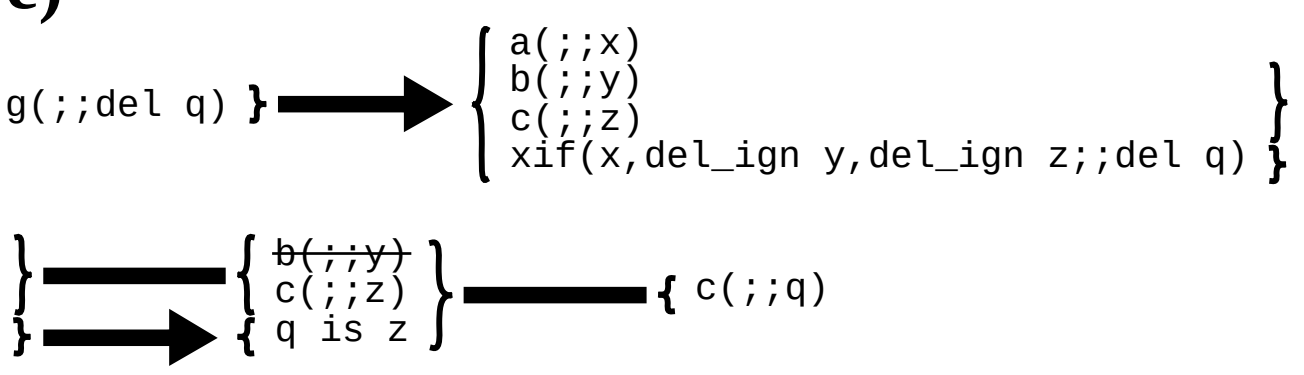

Figure 70  a) A task as an example of a conditional expression.
b) and c) The definition and execution of a simple task.
d) The same task as in b), but using the conditional expression of a).
e) An execution of the task in d), as seen from the task pool.
The task `b(;;y)` eventually is irrelevant and then is crossed out of the task pool.

**a)**

```
or_sc(   boolean n1, del_ign boolean n2
      ;; del boolean w)
{ if ( n1) w is n1; else w is n2; }
```

**b)**

```
and_sc(   boolean n1, del_ign boolean n2
       ;; del boolean w)
{ if (!n1) w is n1; else w is n2; }
```

**c)**

```
a(int;;boolean); b(int;;boolean);
and_sc(boolean n1, del_ign boolean n2;; del boolean w);

h(del int i;; del boolean r) { a(i;;x); b(i;;y); and_sc(x,y;;r); } }
```

**d)**

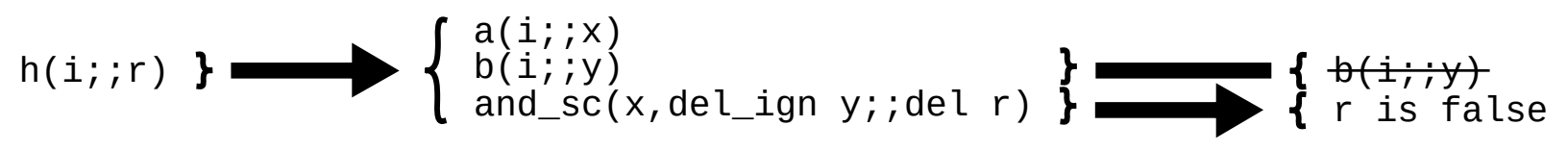

Figure 71 a) and b) Tasks providing the logical operations OR and AND, respectively.
c) A small example using the task `and_sc(n1,del_ign n2;;del w)` of b).
d) An illustration of the task in c), assuming that `x` of `a(i;;x)` evaluates to `false`.

**a)**

```
subroutine bsci(v,k,ak,im,ip,i)
implicit none
integer v,k,ak,im,ip,i

if      (v.lt.ak) then
  i = im
else if (v.gt.ak) then
  i = ip
else
  i = k
endif
end


subroutine bs(m,p,a,v,i)
implicit none
integer m,p,a(m:p),v,i ,k,im,ip
if (m.gt.p) then
  i = -1       ! no elements.
else
  k = (m+p)/2  ! middle index.
  call bs(m  ,k-1,a(m)  ,v,im)
  call bs(k+1,p  ,a(k+1),v,ip)
  call bsci(v,k,a(k),im,ip,i)
endif
end
```

**b)**

```
bsci(int v, int k, int ak,
     del_ign int im, del_ign int ip;;
     del int i)
{
  if        (v < ak) {
    i is im;
  } else if (v > ak) {
    i is ip;
  } else {
    i is k;
  }
}


bs(int m, int p, del int a[m:p],
   del int v;; int i)
{
  if (m > p) {
    i = -1;            ! no elements.
  else {
    int k = (m+p)/2;   ! middle index.
    bs(m  ,k-1,a[m]  ,v;;im);
    bs(k+1,p  ,a[k+1],v;;ip);
    bsci(v,k,a[k],im,ip;;i);
  }
}
```

Figure 72 a) The routines `bs` and `bs1` of the binary search definition of Figure 62b) rewritten as `bs` and `bsci` using conditional ins `im` and `ip`.
b) The Fortran code of a) rewritten in ia.

**a)**

```
      program testnqans
      implicit none
      integer n,ans(10)
      do n = 1,10
        call nqans(n,ans)
        if (ans(1).eq.-1) then
          print *,n,' queens has no',
     +              ' solution.'
        else
          print *,n,' queens has at',
     +              ' least solution:'
          call display(n,ans)
        endif
      enddo
      end

      subroutine nqans(n,ans)
      implicit none
      integer n,ans(n)
      ans(1) = -1
      call nattempts(n,0,0,ans)
      end

      subroutine display(n,board)
      implicit none
      integer board(n),n
      character*(10) work !Assume n<=10.
      integer i
      work=''
      do i=1,n
        work(i:i) = '0'
      enddo
C Array board has 1 entry for each row.
      do i=1,n
        work(board(i):board(i)) = '1'
        print *,work
        work(board(i):board(i)) = '0'
      enddo
      end

      subroutine vcopy(n,a,b)
      implicit none
      integer n,a(n),b(n),i
      do i = 1,n
        b(i) = a(i)
      enddo
      end
```

**b)**

```
      subroutine nattempts(n,b_size,
     +                     board,ans)
      implicit none
      integer n,b_size,board(b_size),
     +        ans(n), i
      if (n.eq.b_size) then
        call vcopy(n,board,ans)
      else
        call iterate(n,b_size,board,
     +               1,ans)
      endif
      end

      subroutine iterate(n,b_size,board,
     +                   i,ans)
      implicit none
      integer n,b_size,board(b_size),
     +        i,ans(n)
      logical safe
      if (ans(1).ne.-1) return
      call testsafe( b_size,board,i,safe)
      call attempt(n,b_size,board,i,safe,
     +             ans)
      if (i.lt.n) then
        call iterate(n,b_size,board,
     +               i+1,ans)
      endif
      end
```

Figure 73 A Fortran application for a solution to the N-queens problem.
It requires the Fortran routines of Figure 46b).

```
testsafe(int size, int board[size],
         int new;; boolean safe);

vcopy(int n, int a[n];; int b[n]);

attempt(int n,int b_size,
        int board[b_size], del int new,
        boolean safe;
        del_ign int ans[n];);


iterate(del int n,del int b_size,
        del int board[b_size], int i;
        int ans[n];)
{
  if (ans[1] != -1) return;
  testsafe( b_size,board,i;;safe);
  attempt(n,b_size,board,i,safe;ans;);
  if (i < n) {
    iterate(n,b_size,board,i+1;ans;);
  }
}
```

```
nattempts(int n,int b_size,
          del int board[b_size];
          del int ans[n];)
{
  if (n==b_size) {
    vcopy(n,board;;ans);
  } else {
    iterate(n,b_size,board,1;ans;);
  }
}


nqans(int n;; int ans[n])
{
  ans[1] = -1;
  nattempts(n,0,0;ans;);
}
```

Figure 74 The Fortran definition of N-queens of Figure 73 rewritten in ia.

```
      subroutine first(n,r,nansf,nansr,
     +                 ans)
      implicit none
      integer n,r,ans(n), nansf(n),
     +        nansr(10,r)  !Assume n<=10.
      if (nansf(1).ne.-1) then
        call vcopy(n,nansf,ans)
      else if (r.gt.0) then
        call first(n,r-1,nansr(1,1),
     +             nansr(1,2),ans)
      endif
      end
```

```
      subroutine nattempts(n,b_size,
     +                     board,ans)
      implicit none
      integer n,b_size,board(b_size),
     + ans(n),i,nans(10,10)!Assume n<=10.
      logical safe
      if (n.eq.b_size) then
        call vcopy(n,board,ans)
      else
        do i = 1,n
          call testsafe( b_size,board,i,
     +                   safe)
          nans(1,i) = -1
          call attempt(n,b_size,board,i,
     +                   safe,nans(1,i))
        enddo
        call first(n,n-1,nans(1,1),
     +             nans(1,2),ans)
      endif
      end
```

Figure 75 A Fortran application for a solution to the N-queens problem.
It requires the Fortran routines of Figure 73a) and Figure 46b).

```
testsafe(int size, int board[size],
         int new;; boolean safe);

vcopy(int n, int a[n];; int b[n]);

attempt(int n, int b_size,
        int board[b_size], del int new,
        boolean safe;
        del_ign int ans[n];);


first(del int n,int r,int nansf[n],
      del_ign int nansr[n,r];
      del_ign int ans[n];)
{
  if (nansf[1] != -1) {
    vcopy(n,nansf;;ans);
  } else if (r > 0) {
    first(n,r-1,nansr[1,1],nansr[1,2];
          ans;);
  }
}
```

```
nattempts(int n,int b_size,
          del int board[b_size];
          del_ign int ans[n];)
{
  if (n==b_size) {
    vcopy(n,board;;ans);
  } else {
    for (int i=1; i<=n; i++) {
      testsafe( b_size,board,i;;safe);
      nans[1,i] = -1;
      attempt(n,b_size,board,i,safe;
              nans[1,i];);
    }
    first(n,n-1,nans[1,1],nans[1,2];
          ans;);
  }
}


nqans(int n;; int ans[n])
{
  ans[1] = -1;
  nattempts(n,0,0;ans;);
}
```

Figure 76  The Fortran definition of N-queens of Figure 75 rewritten in ia.

**a)**

```
      subroutine fsqrt(n,a0,eps,a)
      implicit none
      real n,a0,eps,a, a1
      a  = a0     ! a0 is not an out.
 100    a1 = a
        a  = (a+n/a)/2.
      if (abs(a-a1).gt.eps) goto 100
      end
```

**b)**

```
next(real n, real a0;; real a)
{
  a = (a0+n/a0)/2.;
}

ciwithin(real a0, real a1,del_ign real a2,
         real eps;; del real a)
{
  if (abs(a0-a1) < eps) {
    a is a1;
  } else {
    a is a2;
  }
}

cisqrt(real n, real a0, real eps;;
       del real a)
{
  next(n,a0;;a1);
  cisqrt(n,a1,eps;;a2);
  ciwithin(a0,a1,a2,eps;;a);
}
```

**c)**

```
      program newton
      implicit none
      real n,a0,eps,a
      n   = 100.
      a0  = 1.
      eps = .1
      call cisqrt(n,a0,eps,a)
      print *,n,a0,eps,a
      end

      subroutine next(n,a0,a)
      implicit none
      real n,a0,a
      a = (a0+n/a0)/2.
      end

      subroutine ciwithin(a0,a1,a2,eps,a)
      implicit none
      real a0,a1,a2,eps,a
      if (abs(a0-a1).lt.eps) then
        a = a1
      else
        a = a2
      endif
      end

      subroutine cisqrt(n,a0,eps,a)
      implicit none
      real n,a0,eps,a, a1,a2
      a2 = -12345.           ! Imitate  !
      call next(n,a0,a1)
      call ciwithin(a0,a1,a2,eps,a)
C Imitate demand-driven ex.n of ia.     !
      if (a.eq.-12345) then             !
        call cisqrt(n,a1,eps,a)
      endif                             !
      end
```

Figure 77 a) A Fortran definition to compute a square root using the Newton-Raphson algorithm.
b) The same algorithm defined in ia using a conditional in.
c) The ia definition of b), imitated in Fortran.

**a)**

```
      subroutine fdiff(f,x,h0,a0,eps,a)
      implicit none
      real x,h0,a0,eps,a, a1,h1,h
      external f
      a = a0    ! a0 is not an out.
      h = h0    ! h0 is not an out.
 100    a1 = a
        h1 = h
        call halve(h1,h)
        call easydiff(f,x,h,a)
      if (abs(a-a1).gt.eps) goto 100
      end
```

**b)**

```
halve(real h;; real h2)
{
  h2 = h/2.;
}

easydiff(f(real a;;real b),
         real x, real h;; real ed)
{
  f(x;;fx);
  f(x+h;;fxh);
  ed = (fxh-fx)/h;
}

ciwithin(real a0, real a1,del_ign real a2,
         real eps;; del real a);

cidiff(f(real a;;real b), real x, real h0,
       real a0, real eps;; real a)
{
  halve(h0;;h1);
  easydiff(f,x,h1;;a1);
  cidiff(f,x,h1,a1,eps;;a2);
  ciwithin(a0,a1,a2,eps;;a);
}
```

**c)**

```
      program testdiff
      implicit none
      external cfu
      real x,h0,a0,eps,a
      x  = 1.                     !Init.
      h0 = 1.                     !
      call easydiff(cfu,x,h0,a0) !
      eps = .0001                 !
      call cidiff(cfu,x,h0,a0,eps,a)
      print *,x,h0,a0,eps,a
      end

      subroutine cfu(x,y)
      implicit none
      real x,y
      y = exp(x)
      end

      subroutine halve(h,h2)
      implicit none
      real h,h2
      h2 = h/2.
      end

      subroutine easydiff(f,x,h,ed)
      implicit none
      real x,h,ed, fx,fxh
      external f
      call f(x,fx)
      call f(x+h,fxh)
      ed = (fxh-fx)/h
      end

      subroutine cidiff(f,x,h0,a0,eps,a)
      implicit none
      real x,h0,a0,eps,a, a1,a2,h1
      external f
      a2 = -12345.          ! Imitate  !
      call halve(h0,h1)
      call easydiff(f,x,h1,a1)
      call ciwithin(a0,a1,a2,eps,a)
C Imitate demand-driven ex.n of ia.    !
C Fails if valid ever is -12345.       !
      if (a.eq.-12345) then            !
        call cidiff(f,x,h1,a1,eps,a)
      endif                            !
      end
```

Figure 78 a) A Fortran definition to differentiate a function at a given point using a numerical differentiation algorithm. The definition uses the Fortran routines `halve` and `easydiff` of c).
b) The same algorithm defined in ia using a conditional in. It uses routine `ciwithin` of Figure 77b).
c) The ia definition of b), imitated in Fortran. It uses routine `ciwithin` of Figure 77c).

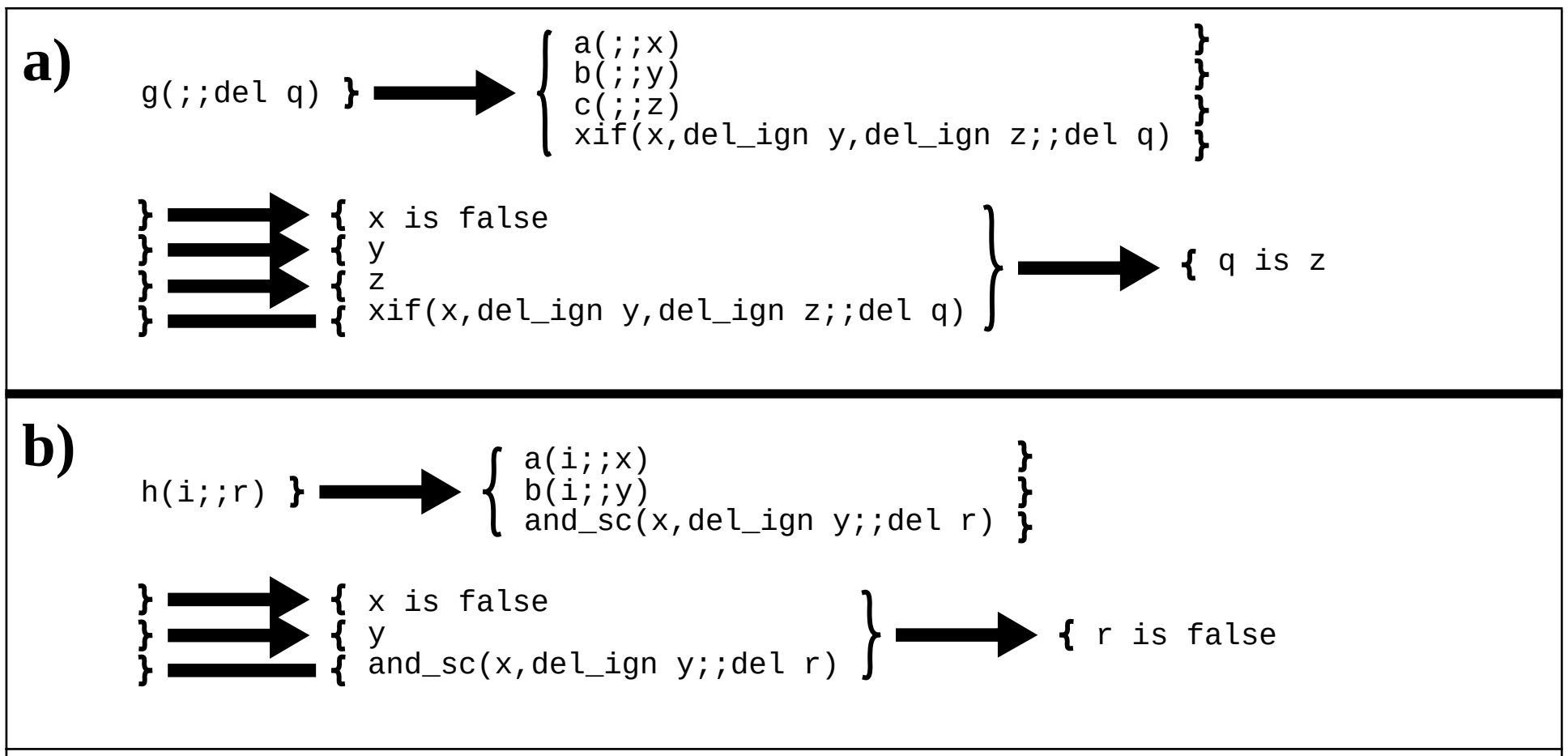

Figure 79 a) An execution with a speculative order for `g(;;del q)` of Figure 70d).
b) An execution with a speculative order for `h(i;;r)` of Figure 71c).

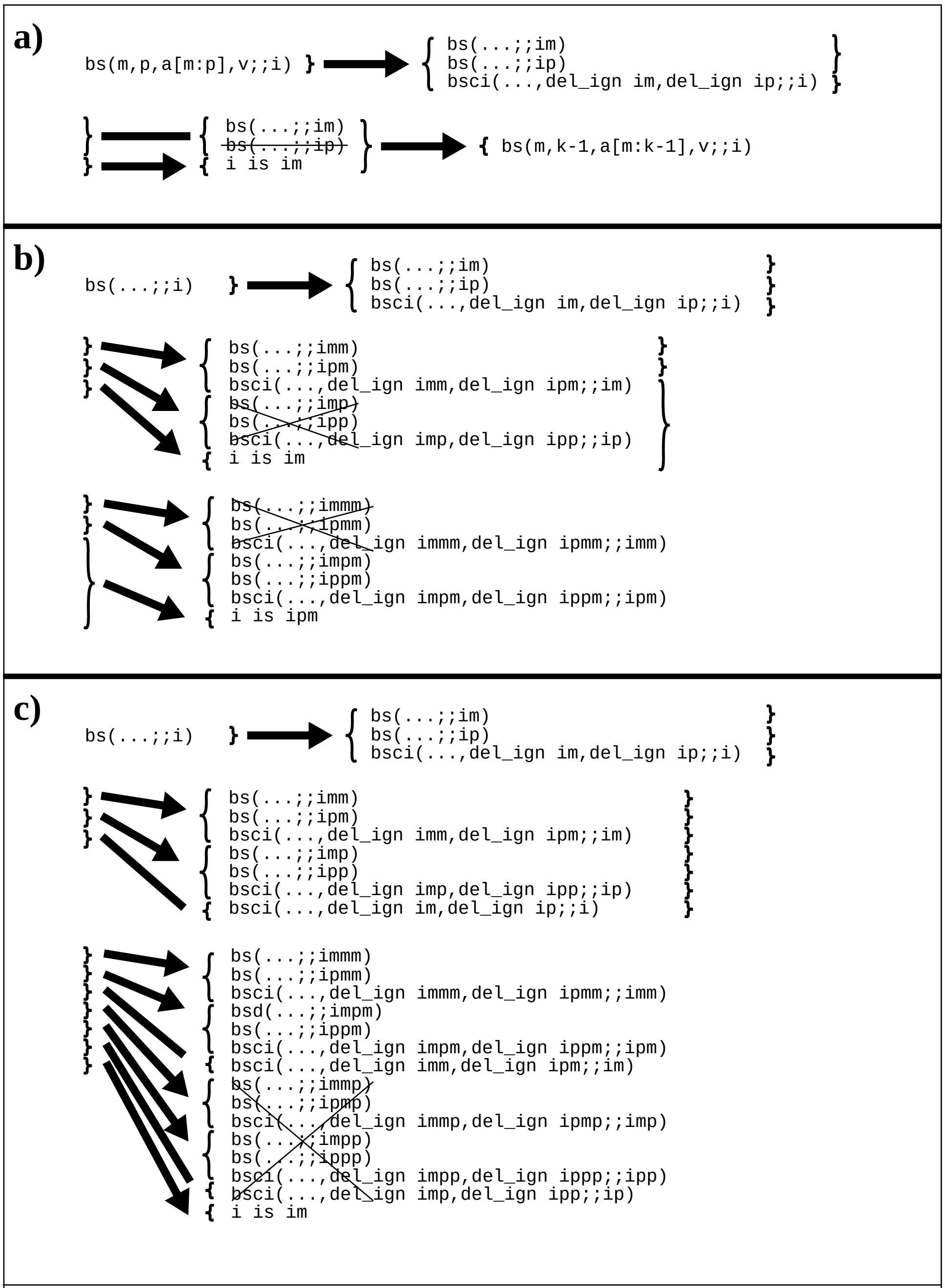

Figure 80 For the execution of `bs` and `bsci` of the binary search definition of Figure 72:
a) a conservative order, b) a speculative order, c) a different speculative order.

**a)**

```
doubleadd( int a;; int b) { b = 2*a;}

doublemult(int c;; int d) { d = c+c;}
```

```
double(int e;; int f)
{
  doubleadd( e;;f);
  doublemult(e;;f);
}
```

**b)**

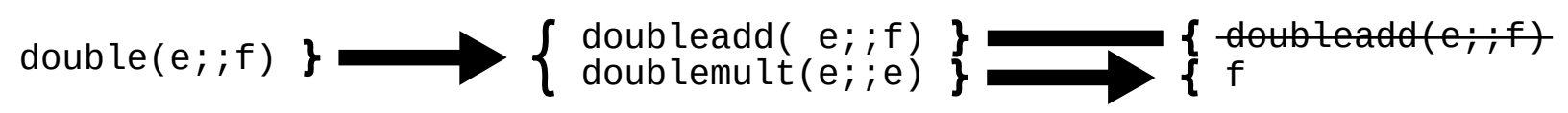

Figure 81 a) The out `f` of the task `double(e;;f)` is an example of a multi-origin out.
b) An execution of the task `double`, assuming that its child `doublemult` evaluates the multi-origin out `f` before its other child `doubleadd`.

**a)**

```
mult1( int x;; del_ign int z)
{ if (x == 0) z is 0; }

mult2( int x, int y;; int z)
{ z= x*y; }

mult( del_ign int a, del_ign int b;;
      del int c)
{
  mult1(a  ;;c);
  mult1(b  ;;c);
  mult2(a,b;;c);
}
```

**b)**

```
fast( int x;; del_ign int y);

slow( int x;; int y);

solve( int a;; del int c)
{
  fast(a;;c);
  slow(a;;c);
}
```

**c)**

```
or1( boolean x;; del_ign boolean z)
{ if (x) z is true; }

or2( boolean x, boolean y;; boolean z)
{ z = x || y; }

or_co( del_ign boolean a,
       del_ign boolean b;;
       del     boolean c)
{
  or1(a  ;;c);
  or1(b  ;;c);
  or2(a,b;;c);
}
```

**d)**

```
or1( boolean x;; del_ign boolean z)
{ if (x) z is true; }

set( boolean x;; boolean z)
{ z = x; }

or_d( del_ign boolean a,
      del_ign boolean b;;
      del     boolean c)
{
  or1(a  ;;c);
  or1(b  ;;c);
  set(false;; default c);
}
```

**e)**

```
and1( boolean x;; del_ign boolean z)
{ if (!x) z is false; }

and2( boolean x, boolean y;; boolean z)
{ z = x && y; }

and_co( del_ign boolean a,
        del_ign boolean b;;
        del     boolean c)
{
  and1(a  ;;c);
  and1(b  ;;c);
  and2(a,b;;c);
}
```

**f)**

```
and1( boolean x;; del_ign boolean z)
{ if (!x) z is false; }

set( boolean x;; boolean z)
{ z = x; }

and_d( del_ign boolean a,
       del_ign boolean b;;
       del     boolean c)
{
  and1(a  ;;c);
  and1(b  ;;c);
  set(true;; default c);
}
```

Figure 82 The use of a conditional out as part of a multi-origin out for implementing:
a) multiplication.
b) a fast occasionally successful algorithm and a slow reliable algorithm.
c) and d) the logical OR operation.
e) and f) the logical AND operation.

**a)**

```
      subroutine match(a,b,c,d)
      implicit none
      integer a,b,c,d
      if (a.eq.b) d = c
      end


      subroutine us(n,a,v,i)
      implicit none
      integer n,a(n),v,i,k
      i = -1
      do k=1,n
        call match(a(k),v,k,i)
        if (i.ne.-1) return !Fortran only
      enddo
      end
```

**b)**

```
match(int a,int b,del_ign int c;;
      del_ign int d)
{
  if (a==b) d is c;
}

set(int a;; int b) {b = a;}

us(int n,del int a[n],del int v;;
   indet del int i)
{
  for (k=1; k<=n; k++)) {
    match(a[k],v,k;;i);
  }
  set(-1;;default i);
}
```

Figure 83 a) A Fortran definition of a search for an unsorted array of integers.
b) The Fortran code of a) rewritten in ia, using a multiple-origin conditional out.

**a)**

```
      subroutine nattempts(n,b_size,
     +                     board,ans)
      implicit none
      integer n,b_size,board(b_size),
     +        ans(n), i
      logical safe
      if (n.eq.b_size) then
        call vcopy(n,board,ans)
      else
        do i = 1,n
          call testsafe( b_size,board,i,
     +                   safe)
          call attempt(n,b_size,board,i,
     +                   safe,ans)
C Following line is only in Fortran.
          if (ans(1).ne.-1) return
        enddo
      endif
      end
```

**b)**

```
testsafe(int size, int board[size],
         int new;; boolean safe);

vcopy(int n, int a[n];; int b[n]);

attempt(int n, int b_size,
        int board[b_size], del int new,
        boolean safe;;
        del_ign int ans[n]);

nattempts(int n, int b_size,
          del int board[b_size];;
          indet del_ign int ans[n])
{
if (n==b_size)
  vcopy(n,board;;ans);
else
  for (i = 1; i<=n; i++) {
    testsafe( b_size,board,i;;safe);
    attempt(n,b_size,board,i,safe;;ans);
  }
}


nqans(int n;; int ans[n])
{
  nattempts(n,0,0;;ans);
  vcopy(1,-1;;default ans[1]);
}
```

Figure 84 a) A Fortran definition for an indeterminate solution to the N-queens problem.
b) The Fortran code of a) rewritten in ia, using a multiple-origin conditional out.

**a)**

```
subroutine bsco(v,k,ak,im,ip,i)
implicit none
integer v,k,ak,im,ip,i

if      (v.lt.ak) then
  i = im
else if (v.gt.ak) then
  i = ip
else
  i = k
endif
end


subroutine bs(m,p,a,v,i)
implicit none
integer m,p,a(m:p),v,i ,k,im,ip
if (m.gt.p) then  ! No elements.
  i = -1             !Fortran only.
else
  k = (m+p)/2     ! Middle index.
  call bs(m  ,k-1,a(m)  ,v,im)
  if (im.ne.-1) then !Fortran only.
   i = im            !Fortran only.
   return            !Fortran only.
  endif              !Fortran only.
  call bs(k+1,p  ,a(k+1),v,ip)
  if (ip.ne.-1) then !Fortran only.
   i = ip            !Fortran only.
   return            !Fortran only.
  endif              !Fortran only.
  call bsco(v,k,a(k),im,ip,i)
endif
end
```

**b)**

```
bsco(int v, int k, int ak,
     del_ign int im, del_ign int ip;;
     del_ign int i)
{
  if        (v < ak) {
    i is im;
  } else if (v > ak) {
    i is ip;
  } else {
    i is k;
  }
}


bs(int m, int p, del int a[m:p],
   del int v;; indet del_ign int i)
{
  if (m > p) {
    ! i is ignored if no elements.
  else {
    int k = (m+p)/2;   ! middle index.
    bs(m  ,k-1,a[m]  ,v;; i is im);
    bs(k+1,p  ,a[k+1],v;; i is ip);
    bsco(v,k,a[k],im,ip;; i);
  }
}
```

Figure 85 a) A Fortran definition for an indeterminate binary search.
b) The Fortran code of a) rewritten in ia, using a multiple-origin conditional out.

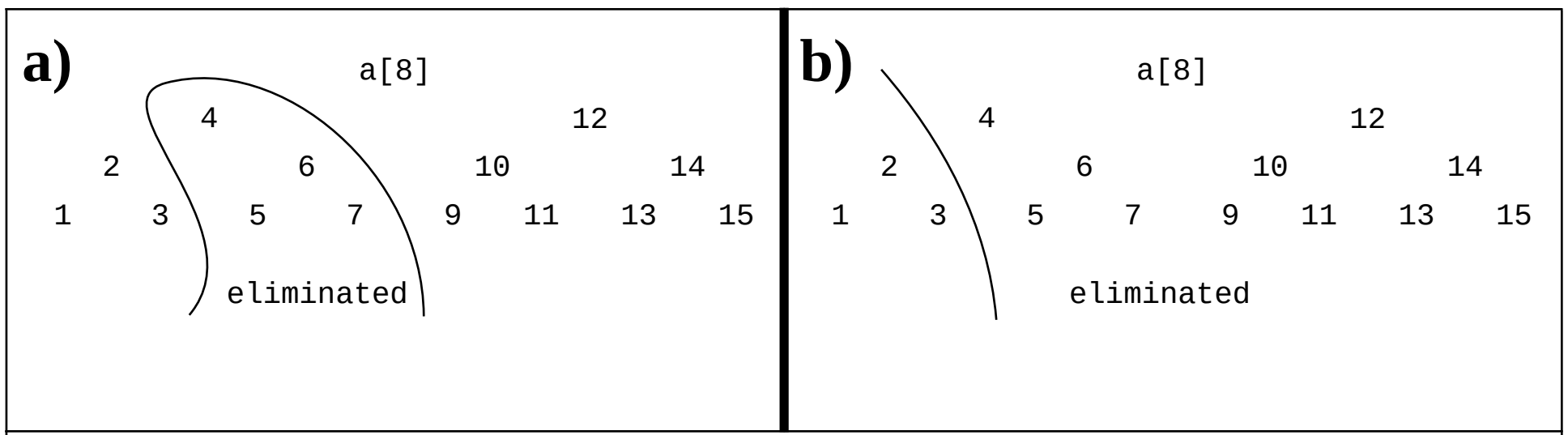

Figure 86 Elements of array `a[1:15]` eliminated when `v<a[4]` for binary search:
a) the determinate definition eliminates only local elements.
b) the indeterminate definition eliminates elements throughout the array.

```
      program eight
      implicit none
      integer maxdepth
      parameter (maxdepth=31)
      integer desired(3,3),board(3,3)
      integer moves(maxdepth),depth
      integer i,j,solved
      character*5 dir(-2:2)
      data dir /'UP','RIGHT','','LEFT',
     +          'DOWN'/

Change either for a different 8-puzzle.
      data desired /0,1,2,3,4,5,6,7,8/
      data board   /8,7,6,0,4,1,2,5,3/

      print *,'Desired board is:'
      call display(desired)
      print *,'Original board is:'
      call display(board)

C Solve the 8-puzzle.
      depth = -1
      call move(desired,maxdepth,0,0,
     +          board,0,0,depth,moves)

C VITAL TO UNDERSTANDING THIS CODE!!!!!!
C Each element of moves(depth) is a
C movement of the blank tile, i.e. 0.
C See dir(-2:2) above for encoding.

      if (depth.eq.-1) then
        print *,maxdepth,' moves cannot'
     +           ,' reach desired board.'
      else

C Display the solution.
        do i = 1,depth
          print *,i,'. Move blank (0) ',
     +            dir(moves(i))
          call make_move(moves(i),board)
          call display(board)
        enddo

Check the solution.
        solved = 1
        do i = 1,3
          do j = 1,3
            if (board(i,j).ne.
     +          desired(i,j)) solved = 0
          enddo
        enddo
        if (solved .eq. 1 ) then
          print *,'Code is good. :-) '
        else
          print *,'Code has bugs. :-( '
        endif
      endif
      end

      subroutine display(board)
      implicit none
      integer board(3,3), r,c,ichar
      character*(3) work
      character char
      do r=1,3
        do c=1,3
          work(c:c) =
     +        char(board(c,r)+ichar('0'))
        enddo
        print *,work
      enddo
      end


C Move blank tile by swap with neighbor.
      subroutine make_move(move,board)
      implicit none
      integer move, board(3,3), c,r
      call blank(board,c,r)
      if      (move .eq. +1) then
        board(c,r) = board(c-1,r  )
                     board(c-1,r  ) = 0
      else if (move .eq. -1) then
        board(c,r) = board(c+1,r  )
                     board(c+1,r  ) = 0
      else if (move .eq. +2) then
        board(c,r) = board(c  ,r+1)
                     board(c  ,r+1) = 0
      else
        board(c,r) = board(c  ,r-1)
                     board(c  ,r-1) = 0
      endif
      end


C Find col and row of blank tile.
      subroutine blank(board,col,row)
      implicit none
      integer board(3,3), col,row, r,c
      do r = 1,3
        do c = 1,3
          if (board(c,r) .eq. 0) then
            col = c
            row = r
            goto 10
          endif
        enddo
      enddo
 10   end
```

Figure 87 Combined with the code of Figure 88, a Fortran application to solve the 8-puzzle.

**a)**

```
Calculate min. distance between 2 boards.
      subroutine manhattan(a,b,distance)
      implicit none
      integer a(3,3), b(3,3), distance,
     +         r,c,k, ar(0:8),ac(0:8),
     +                br(0:8),bc(0:8)
      distance = 0
C Find row and column of each tile
C in each board.
      do r = 1,3
        do c = 1,3
          ar(a(c,r)) = r
          ac(a(c,r)) = c
          br(b(c,r)) = r
          bc(b(c,r)) = c
        enddo
      enddo
C Sum manhattan distance of all tiles.
C NOT 0,8. Blank is NOT included!
      do k = 1,8
        distance = distance
     +           + abs(ar(k)-br(k))
     +           + abs(ac(k)-bc(k))
      enddo
      end


      subroutine vcopy(n,a,b)
      implicit none
      integer n, a(n), b(n), i
      do i = 1,n
        b(i) = a(i)
      enddo
      end
```

**b)**

```
      subroutine move(desired,maxdepth,
     + adepth,amoves,aboard,next_move,
     + aheuristic,
     + final_depth,final_moves)
      implicit none
      integer desired(3,3),maxdepth,
     + amoves(maxdepth),adepth,
     + aboard(3,3),next_move,aheuristic,
     + final_moves(maxdepth),final_depth
C Should be moves(maxdepth),
C but Fortran is not so kind.
      integer moves(100),depth,board(3,3)
     +,heuristic,r,c,a,distance
C The arguments are in, so modify a copy.
      heuristic = aheuristic
      depth     = adepth
      call vcopy(depth,amoves,moves)
      call vcopy(9,aboard,board)

      if (next_move .ne. 0) then
        call make_move(next_move,board)
        depth = depth + 1
        moves(depth) = next_move
      endif

      call manhattan(desired,board,
     +               distance)
      if (distance .eq. 0) then
        call vcopy(depth,moves,
     +             final_moves)
        final_depth = depth
        return
      endif

      heuristic = distance + depth
      if (heuristic .gt. maxdepth) return

      call blank(board,c,r)

C Try a move in each of the 4 directions.
      do a=-2,+2
C Not a direction.
        if (a.eq.0              ) goto 10
C Reversing previous move is useless.
        if (a.eq.-moves(depth)  ) goto 10
C Don't fall off the board!
        if (a.eq.+1 .and. c.eq.1) goto 10
        if (a.eq.-1 .and. c.eq.3) goto 10
        if (a.eq.+2 .and. r.eq.3) goto 10
        if (a.eq.-2 .and. r.eq.1) goto 10
        call move(desired,maxdepth,depth,
     +           moves,board,a,heuristic,
     +           final_depth,final_moves)
C Following line only in Fortran, not ia.
        if (final_depth .ne. -1) return
 10     continue
      enddo
      end
```

Figure 88  Combined with the code of Figure 87, a Fortran application to solve the 8-puzzle.

**a)**

```
Desired board is:
012
345
678
Original board is:
876
041
253
           1. Move blank (0) UP
076
841
253
           2. Move blank (0) RIGHT
706
841
253
           3. Move blank (0) RIGHT
760
841
253
           4. Move blank (0) DOWN
761
840
253
           5. Move blank (0) LEFT
761
804
253

[Steps 6. through 26. not shown here]

125
034
678
          27. Move blank (0) RIGHT
125
304
678
          28. Move blank (0) RIGHT
125
340
678
          29. Move blank (0) UP
120
345
678
          30. Move blank (0) LEFT
102
345
678
          31. Move blank (0) LEFT
012
345
678
Code is good. :-)
```

**b)**

```
make_move(int move; int board;);
vcopy(int n, int a[n];; int b[n]);
manhattan(int a[3,3], int b[3,3];;
          int distance);
blank(int board[3,3];; int col, int row);


move(del int desired[3,3], int maxdepth,
     int depth, int moves[maxdepth],
     del int board[3,3], int next_move,
     int heuristic;;
     indet del_ign int final_depth,
     promise(-heuristic) indet del_ign
       int final_moves[final_depth,])
{
  // Assume that ia copies the ins
  // heuristic, depth, moves and board
  // before they are modified.

  if (next_move != 0) {
    make_move(next_move;board;);
    depth = depth + 1;
    moves[depth] = next_move;
  }

  manhattan(desired,board;;distance);
  if (distance == 0) {
    vcopy(depth,moves;;final_moves);
    final_depth is depth;
    return
  }

  heuristic = distance + depth;
  if (heuristic > maxdepth) return;

  blank(board;;c,r);

  C Try to move in all 4 directions.
  for (int a = -2; a <= +2; a++) {
    // Not a direction.
    if (a.eq.0              ) continue;
    // Reversing previous move is useless
    if (a.eq.-moves[depth]  ) continue;
    // Don't fall off the board!
    if (a.eq.+1 .and. c.eq.1) continue;
    if (a.eq.-1 .and. c.eq.3) continue;
    if (a.eq.+2 .and. r.eq.3) continue;
    if (a.eq.-2 .and. r.eq.1) continue;
    move(desired,maxdepth,depth,moves,
         board,a,heuristic;;
         final_depth,final_moves);
  }
}
```

Figure 89 a) The output of the Figure 87 and Figure 88 Fortran application to solve the 8-puzzle. b) The Fortran routine `move` of Figure 88b) rewritten in ia.

**a)**

```
next(real n, real a0;; real a)
{
  a = (a0+n/a0)/2.;
}

swithin(real x[0:1:], real eps ;;
        del real a)
{
  if (abs(x[0]-x[1]) < eps) {
    a is x[1];
  } else {
    swithin(x[1],eps;;a);
  }
}

ssqrt(real n, real a0, real eps;;
      del real a)
{
  int x[0:] = a0;
  next(n,x[i];;x[i+1]);
  swithin(x[0],eps;;a);
}
```

**b)**

```
      program snewton
      implicit none
      real n,a0,eps,a
      n   = 100.
      a0  = 1.
      eps = .1
      call ssqrt(n,a0,eps,a)
      print *,n,a0,eps,a
      end

      subroutine next(n,a0,a)
      implicit none
      real n,a0,a
      a = (a0+n/a0)/2.
      end

      subroutine swithin(x,eps,a)
      implicit none
      real x(0:1),eps,a
C Ensure that imitation is OK.        !
      if (x(1) .eq. -12345.) then      !
        print *,'Imitation fails.'     !
        call exit()                    !
      endif                            !
      if (abs(x(0)-x(1)).lt.eps) then
        a = x(1)
      else
        call swithin(x(1),eps,a)
      endif
      end

      subroutine ssqrt(n,a0,eps,a)
      implicit none
      real n,a0,eps,a, x(0:99)
      integer i                        !
C Imitate demand-driven ex.n of ia.    !
C Assume 99 iterations is sufficient.  !
      x(0) = a0
      do i = 0,97                      !
        call next(n,x(i),x(i+1))
      enddo                            !
      x(99) = -12345.    ! For safety. !
      call swithin(x(0),eps,a)
      end
```

Figure 90 a) Ia code using a stream and the Newton-Raphson algorithm to compute a square root.
b) The ia definition of a), imitated in Fortran.

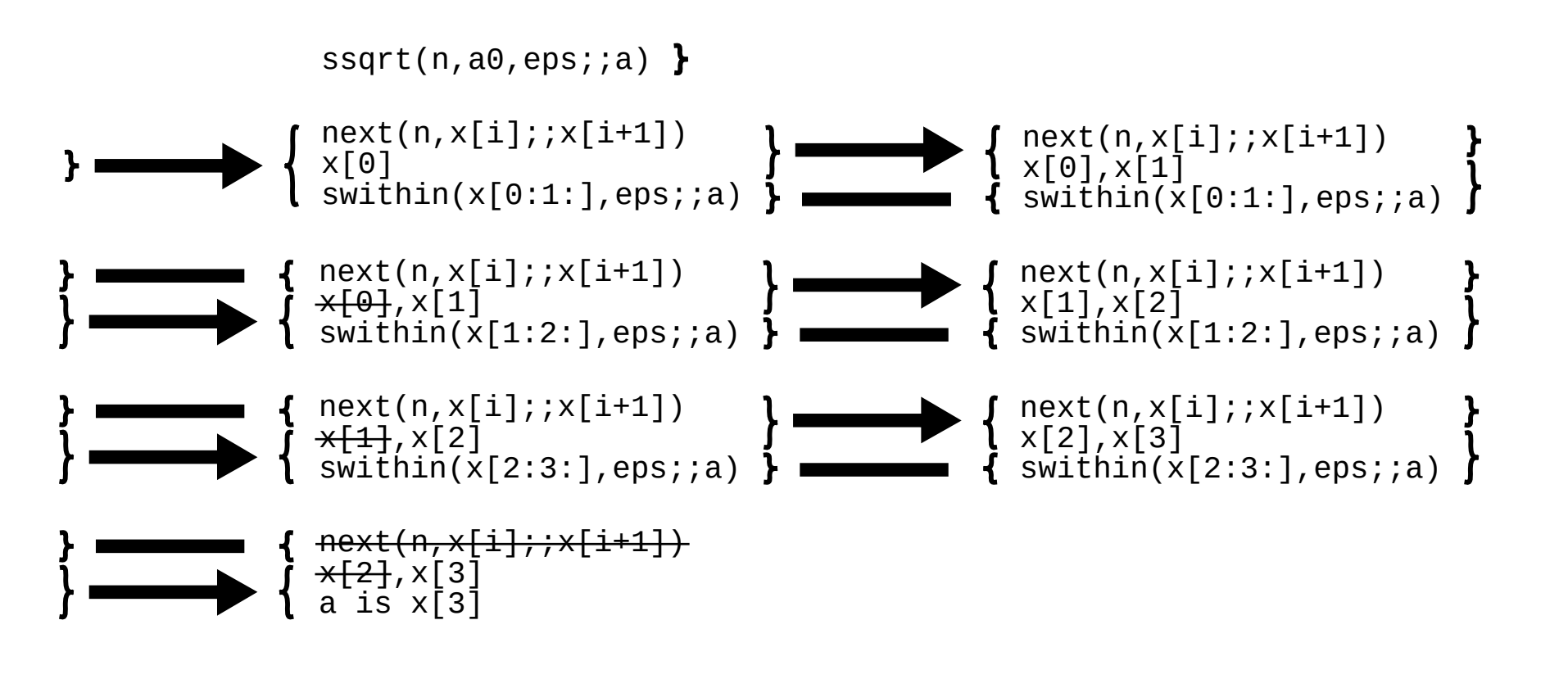

Figure 91 The evolution in the task pool for the ia code of Figure 90b).

**a)**

```
halve(real h;; real h2);

easydiff(f(real a;;real b),
         real x, real h;; real ed);

swithin(real x[0:1:], real eps ;;
        del real a);

sdiff(f(real a;;real b), real x, real h0,
      real eps;; real a)
{
  int h[0:] = h0;
  halve(h[i-1];;h[i]);
  easydiff(f,x,h[i];;y[i]);
  swithin(y[0],eps;;a);
}
```

**b)**

```
      program testsdiff
      implicit none
      external cfu
      real x,h0,eps,a
      x  = 1.                     !Init.
      h0 = 1.                     !
      eps = .0001                 !
      call sdiff(cfu,x,h0,eps,a)
      print *,x,h0,eps,a
      end

      subroutine cfu(x,y)
      implicit none
      real x,y
      y = exp(x)
      end

      subroutine halve(h,h2)
      implicit none
      real h,h2
      h2 = h/2.
      end

      subroutine easydiff(f,x,h,ed)
      implicit none
      real x,h,ed, fx,fxh
      external f
      call f(x,fx)
      call f(x+h,fxh)
      ed = (fxh-fx)/h
      end

      subroutine sdiff(f,x,h0,eps,a)
      implicit none
      real x,h0,eps,a, y(0:99),h(0:99)
      external f
      integer i                         !
C Imitate demand-driven ex.n of ia.    !
C Assume 99 iterations is sufficient.  !
      h(0) = h0
      call easydiff(f,x,h(0),y(0))      !
      do i = 1,98                       !
        call halve(h(i-1),h(i))
        call easydiff(f,x,h(i),y(i))
      enddo                             !
      y(99) = -12345.    ! For safety.  !
      call swithin(y(0),eps,a)
      end
```

**c)**

```
halve(real h;; real h2);

easydiff(f(real a;;real b),
         real x, real h;; real ed);

swithin(real x[0:1:], real eps ;;
        del real a);

differ(f(real a;;real b), real x, real h0
       ;; real y[0:])
{
  int h[0:] = h0;
  halve(h[i-1];;h[i]);
  easydiff(f,x,h[i];;y[i]);
}

sdiff1(f(real a;;real b), real x, real h0,
       real eps;; real a)
{
  differ(f,x,h0;;y[0:]);
  swithin(y[0],eps;;a);
}
```

Figure 92 a) Ia code using a stream and a numerical differentiation algorithm to differentiate a function at a given point. The routines `halve` and `easydiff` are as in Figure 78b). The routine `swithin` is as in Figure 90a).
b) The ia definition of a), imitated in Fortran. It uses routine `swithin` of Figure 90b).
c) As in a), but the routine `sdiff` is split into `sdiff1` and `differ`.

**a)**

```
halve(real h;; real h2);

easydiff(f(real a;;real b),
         real x, real h;; real ed);

swithin(real x[0:1:], real eps ;;
        del real a);

differ(f(real a;;real b), real x, real h0
       ;; real y[0:]);

order(real d[1:3];; int o)
{
  o = nint( log( (d[1]-d[3])
                /(d[2]-d[3])-1)
           /log(2.))
}

elimerror(int n, real d[1:2];; real e)
{
  e = (d[2]*2**n-d[1])/(2**n-1);
}

improve(real d[1:3];; real i)
{
  order(d;;o);
  elimerror(o,d;;i);
}

sdiff2(f(real a;;real b), real x, real h0,
      real eps;; real a)
{
  differ(f,x,h0;;y[0:]);
  improve(y[i];;z[i]);
  swithin(z[0],eps;;a);
}
```

**b)**

```
      subroutine order(d,o)
      implicit none
      real d(1:3)
      integer o
      o = nint( log( (d(1)-d(3))
     +              /(d(2)-d(3))-1)
     +         /log(2.))
      end

      subroutine elimerror(n,d,e)
      implicit none
      real d(1:2),e
      integer n
      e = (d(2)*2**n-d(1))/(2**n-1)
      end

      subroutine improve(d,i)
      implicit none
      real d(1:3),i, o
      call order(d,o)
      call elimerror(o,d,i)
      end

      subroutine sdiff2(f,x,h0,eps,a)
      implicit none
      real x,h0,eps,a, y(0:99),h(0:99)
      real             z(0:99)
      external f
      integer i                         !
C Imitate demand-driven ex.n of ia.    !
C Assume 98 iterations is sufficient.  !
      h(0) = h0
      call easydiff(f,x,h(0),y(0))      !
      do i = 1,99                       !
        call halve(h(i-1),h(i))
        call easydiff(f,x,h(i),y(i))
      enddo                             !
C This example and cpu has rounding    !
C trouble after 10 iterations.         !
      do i = 0,10                       !
        call improve(y(i),z(i))
      enddo                             !
      z(11) = -12345.    ! For safety.  !
      call swithin(z(0),eps,a)
      end
```

Figure 93 a) The ia routine `sdiff2` defines a second order algorithm which improves on the first order algorithm of `sdiff1` of Figure 92c). The routines `halve`, `easydiff`, `swithin` and `differ` are as in Figure 92c).
b) The ia definition of a), imitated in Fortran. It uses routines `testsdiff`, `cfu`, `halve`, `easydiff` and `swithin` of Figure 92b).

**a)**

```
// Prototypes as in Figure 93a).

sdiff4(f(real a;;real b), real x, real h0,
      real eps;; real a)
{
  differ(f,x,h0;;y[0:]);
  improve(y[i];;v[i]);
  improve(v[i];;w[i]);
  improve(w[i];;z[i]);
  swithin(z[0],eps;;a);
}
```

**b)**

```
// Prototypes as in Figure 93a).

sdiffsuper(f(real a;;real b), real x,
           real h0, real eps;; real a)
{
  real s[0:1:,0:];
  differ(f,x,h0;;s[0:,0]);
  improve(s[i:i+2,k];;s[i,k+1]);
  swithin(s[1,0:],eps;;a);
}
```

Figure 94 a) The ia routine `sdiff4` defines a fourth order algorithm which improves on the second order algorithm of `sdiff2` of Figure 93a).
b) The ia routine `sdiffsuper` defines an increasing order algorithm.

**a)**

```
rem_mults(int m, del int a[1:2:]
          ;; del int c[1:2:])
{
  if (mod(a[1],m) == 0) {
// a[1] is multiple of m, so throw away.
    rem_mults(m,a[2];;c[1]);
  } else {
// a[1] is not a multiple of m, so keep.
    c[1] is a[1];
    rem_mults(m,a[2];;c[2]);
  }
}

sieve(del int a[1:2:];; del int p[1:2:])
{
  p[1] is a[1];
  rem_mults(a[1],a[2];;c[1]);
  sieve(c[1];;p[2]);
}

add(int a, int b;; int c)
{
  c = a + b;
}

primes(;; del int p[1:])
{
  int a[1:] = 2;
  add(1,a[i];;a[i+1]);
  sieve(a[1];;p[1]);
}
```

**b)**

```
      program eras
      integer p(1000), i
      call primes(p(1))
      do i = 1,1000
        print *,i,p(i)
        if (p(i).eq.0) goto 10
      enddo
 10   end

      subroutine rem_mults(m,a,c)
      integer m,a(*),c(*)                 !
      if (a(1).eq. 0) then   ! Imitate  !
        c(1) = 0                          !
        return                            !
      endif                               !
      if (mod(a(1),m).eq.0) then
C a(1) is multiple of m, so throw away.
        call rem_mults(m,a(2),c(1))
      else
C a(1) is not a multiple of m, so keep.
        c(1) = a(1)
        call rem_mults(m,a(2),c(2))
      endif
      end

      subroutine sieve(a,p)
      integer a(1000),p(1000), c(1000)  !
      if (a(1).eq. 0) then   ! Imitate  !
        p(1) = 0                          !
        return                            !
      endif                               !
      p(1) = a(1)
      call rem_mults(a(1),a(2),c(1))
      call sieve(c(1),p(2))
      end

      subroutine add(a,b,c)
      integer a,b,c
      c = a + b
      end

      subroutine primes(p)
      integer a(1000),p(1000), i        !
      a(1) = 2
      do i = 1,998           ! Imitate  !
        call add(1,a(i),a(i+1))
      enddo                               !
C 0 signals end of stream in imitation. !
      a(1000) = 0                         !
      call sieve(a(1),p(1))
      end
```

Figure 95 a) Ia code using streams to define the sieve of Eratosthenes for generating primes.
b) The ia definition of a), imitated in Fortran.

```
rem_mults(int m, del int a: del b:
          ;; del int c: del d:)
{
  if (mod(a,m) == 0) {
// a is multiple of m, so throw away.
    rem_mults(m,b;;c);
  } else {
// a is not a multiple of m, so keep.
    c is a;
    rem_mults(m,b;;d);
  }
}
```

```
sieve(del int a: del b:
      ;; del int p: del q:)
{
  p is a;
  rem_mults(a,b;;c);
  sieve(c;;q);
}

next_integer(int x;; int y: del z:)
{
  y = x + 1;
  next_integer(y;;z);
}

primes(;; del int p:)
{
  next_integer(1;;a);
  sieve(a;;p);
}
```

Figure 96 As in Figure 95a), ia code using streams to define the sieve of Eratosthenes for generating primes, but using stream arguments instead of stream arrays.

```
stream_a2r(del a[1:2:];; del r: del s:)
{
  r is a[1];
  stream_a2r(a[2];; s);
}
```

```
stream_r2a(del r: del s:;; del a[1:2:])
{
  a[1] is r;
  stream_r2a(s;; a[2]);
}
```

Figure 97 Routines to convert between a stream array and a stream argument.

**a)**

```
halve(real h;; real h2);

easydiff(f(real a;;real b),
         real x, real h;; real ed);


swithin(real x0: x1:,
        real eps;; del real a)
{
  if (abs(x0-x1) < eps) {
    a is x1;
  } else {
    swithin(x1,eps;;a);
  }
}

differ(f(real a;;real b), real x, real h0
       ;; real y: del z)
{
  easydiff(f,x,h0;;y);
  halve(h0;;h1);
  differ(f,x,h1;;z);
}

sdiff1(f(real a;;real b), real x, real h0,
       real eps;; real a)
{
  differ(f,x,h0;;y);
  swithin(y,eps;;a);
}
```

**c)**

```
// Prototypes as in b).

sdiff4(f(real a;;real b), real x, real h0,
      real eps;; real a)
{
  differ(f,x,h0;;y);
  improve(y;;v);
  improve(v;;w);
  improve(w;;z);
  swithin(z,eps;;a);
}
```

**b)**

```
// Prototypes of routines from a).

halve(real h;; real h2);

easydiff(f(real a;;real b),
         real x, real h;; real ed);

swithin(real x0: x1:,
        real eps;; del real a);

differ(f(real a;;real b), real x, real h0
       ;; real y: del z);


// elimerror and order from Figure 93a).

order(real d[1:3];; int o);

elimerror(int n, real d[1:2];; real e);


improve(del real d: del e:;;
        del real i: del j:)
{
  order(d;;o);
  elimerror(o,d;;i);
  improve(e;;j);
}


sdiff2(f(real a;;real b), real x, real h0,
       real eps;; real a)
{
  differ(f,x,h0;;y);
  improve(y;;z)
  swithin(z,eps;;a);
}
```

Figure 98 a) As in Figure 92c), ia code using streams to define a numerical differentiation algorithm to differentiate a function at a given point, but using stream arguments instead of stream arrays. The routines `halve` and `easydiff` are as in Figure 78b).
b) The ia routine `sdiff2` defines a second order algorithm which improves on the first order algorithm of `sdiff1` of a).
c) The ia routine `sdiff4` defines a fourth order algorithm which improves on the second order algorithm of `sdiff2` of b).

| | |
|---|---|
| `// Prototypes as in Figure 98c).`<br><br>`super(del real y: del z:;;`<br>`      del real s: del t:)`<br>`{`<br>`  s is z;`<br>`  improve(y;;a);`<br>`  super(a;;t);`<br>`}` | `sdiffsuper(f(real a;;real b), real x,`<br>`           real h0, real eps;; real a)`<br>`{`<br>`  differ(f,x,h0;;y);`<br>`  super(y;;z)`<br>`  swithin(z,eps;;a);`<br>`}` |

Figure 99 As in Figure 93a), but using stream arguments instead of stream arrays, the ia routine `sdiffsuper` defines an increasing order algorithm.

| | |
|---|---|
| `// getchar and putchar from C language.`<br>`getchar(;;int c);`<br>`putchar(int c;;int e);`<br><br>`gs(;;int c: del_ign d:)`<br>`{`<br>`  getchar(;;c);`<br>`  if (c != EOF) gs(;;d);`<br>`}` | `ps(int c: del_ign d:;;)`<br>`{`<br>`  if (c != EOF) {`<br>`    putchar(c;;e);`<br>`    if (e != EOF) ps(d;;);`<br>`  }`<br>`}`<br><br>`main() { gs(;;c); ps(c;;); }` |

Figure 100A complete ia application using a stream to copy its input to its output.

**a)**

```
// getreal and putreal are similar
// to getchar and putchar of C.
getreal(;; real r, bool error);
putreal(real r ;;  bool error);

gsr(;; real r[0: del_ign 1:])
{
  getreal(;;r[0],error);
  if (!error) gsr(;;r[1]);
}

psr(real r[0: del_ign 1:] ;;)
{
  putreal(r[0];;error);
  if (!error) psr(;;r[1]);
}

avg(real a, real b;; real c)
{
  c = (a+b)/2.;
}

main()
{
  gsr(;; a[0:]);
  avg(a[i],a[i+1];;b[i]);
  psr(b[0] ;;);
}
```

**b)**

```
// Prototypes as in a).

avginout(real a; real b;)
{
  b = (a+b)/2.;
}

main()
{
  gsr(;; a[0:]);
  avginout(a[i+1];a[i];);
  psr(a[0] ;;);
}
```

**c)**

```
getreal(;; real r, bool error);
putreal(real r ;;  bool error);

gsr(;; real r0: del_ign r1:)
{
  getreal(;;r0,error);
  if (!error) gsr(;;r1);
}

psr(real r0: del_ign r1: ;;)
{
  putreal(r0;;error);
  if (!error) psr(;;r1);
}

avgpair(real a1: a2:;;
        real c1: del_ign c2:)
{
  c1 = (a1+a2)/2.;
  avgpair(a2;;c2);
}

main()
{
  gsr(;; a);
  avgpair(a;;b);
  psr(b;;);
}
```

Figure 101a) A complete ia application for signal processing, where each element of the output stream is the average value of two adjacent elements of the input stream.
b) As in a), but a single stream is modified in place, instead of using separate input and output streams.
c) As in a), but using stream arguments instead of stream arrays.

**a)**

```
// gsr and psr as in Figure 101a).
gsr(;; real r[0: del_ign 1:]);
psr(real r[0: del_ign 1:] ;;);

FourForThree(real D[1:4:];;
             real n[1:4: del_ign 5:])
{
  n[1] =     D[1];
  n[2] = (   D[1] + 3*D[2] ) / 4;
  n[3] = (   D[2] +   D[3] ) / 2;
  n[4] = ( 3*D[3] +   D[4] ) / 4;
  FourForThree(D[4];;n[5]);
}

main()
{
  gsr(;; a[0:]);
  FourForThree(a[0];;b[0:]);
  psr(b[0] ;;);
}
```

**b)**

```
// gsr and psr as in Figure 101a).
gsr(;; real r[0: del_ign 1:]);
psr(real r[0: del_ign 1:] ;;);

FourForThree(real D[1:3], real D4;;
             real n[1:4])
{
  n[1] =   D[1];
  n[2] = ( D[1] + 3*D[2] ) / 4;
  n[3] = ( D[2] + D[3]   ) / 2;
  n[4] = ( 3*D[3] + D4   ) / 4;
}

main()
{
  gsr(;; a[1:3,0:]);
  FourForThree(a[1:3,i],a[1,i+1];;
               b[1:4,i]);
  psr(b[1:4,0] ;;);
}
```

Figure 102a) A complete ia application for signal processing. Outputting four samples for each group of three samples in the input stream, increases the sampling rate by a factor 4/3.
b) As in a), but the transformer is unaware that it is operating on a stream.

**a)**

```
// gsr and psr as in Figure 101a).
gsr(;; real r[0: del_ign 1:]);
psr(real r[0: del_ign 1:] ;;);

TwoDimFilter(int w, real d[1:w,-1:1];;
             real n[2:w-1]);

main()
{
// assume w is defined elsewhere.
  gsr(;; d[1:w,0:]);
  TwoDimFilter(w,d[1:w,i-1:i+1];;
               n[2:w-1,i]);
  psr(b[2:w-1,1:] ;;);
}
```

**b)**

```
      subroutine TwoDimFilter(w,d,n)
      implicit none
      integer w, j,k,m
      real d(1:w,-1:1), n(2:w-1)
      real filter(-1:1,-1:1)
      data filter /1,2,1,
     +             2,3,2,
     +             1,2,1/
      do j = 2,w-1
        n(j) = 0
        do k = -1,1
          do m = -1,1
            n(j) = n(j)
     +           + d(j+k,m)*filter(k,m)
          enddo
        enddo
      enddo
      end
```

Figure 103a) A complete ia application for signal processing a stream of vectors using `TwoDimFilter` of b).
b) Fortran code for the routine `TwoDimFilter` which applies a three-by-three array `filter` to each element of a two-dimensional array.

**a)**

```
add(int a, int b;; int c) { c = a + b; }

incr(i;;j) is add(1,i;;j);

incrincr(del int k;; del int l)
{
  incr(k;;m); incr(m;;l);
}
```

**b)**

```
add(int a, int b;; int c) { c = a + b; }


incrincr(del int k;; del int l)
{
  incr(i;;j) is add(1,i;;j);
  incr(k;;m); incr(m;;l);
}
```

**c)**

```
add(int a, int b;; int c) { c = a + b; }

twice(del f(a;;b), del a;; del b)
{ f(a;;c); f(c;;b); }

incrincr(del int k;; del int l)
{
  incr(i;;j) is add(1,i;;j);
  twice(incr,k;;l);
}
```

**d)**

```
add(int a, int b;; int c) { c = a + b; }

twice(del f(a;;b), del a;; del b)
{ f(a;;c); f(c;;b); }

getincr(;;del incr(a;;b))
{
  incr(i;;j) is add(1,i;;j);
}

incrincr(del int k;; del int l)
{
  getincr(;;g);
  twice(g,k;;l);
}
```

Figure 104 The elements of the routine `incr` are defined using a variation of currying.
a) and b) `incr` may be defined at the top level or within a routine, respectively.
c) and d) `incr` may be an in or an out of another routine, respectively.

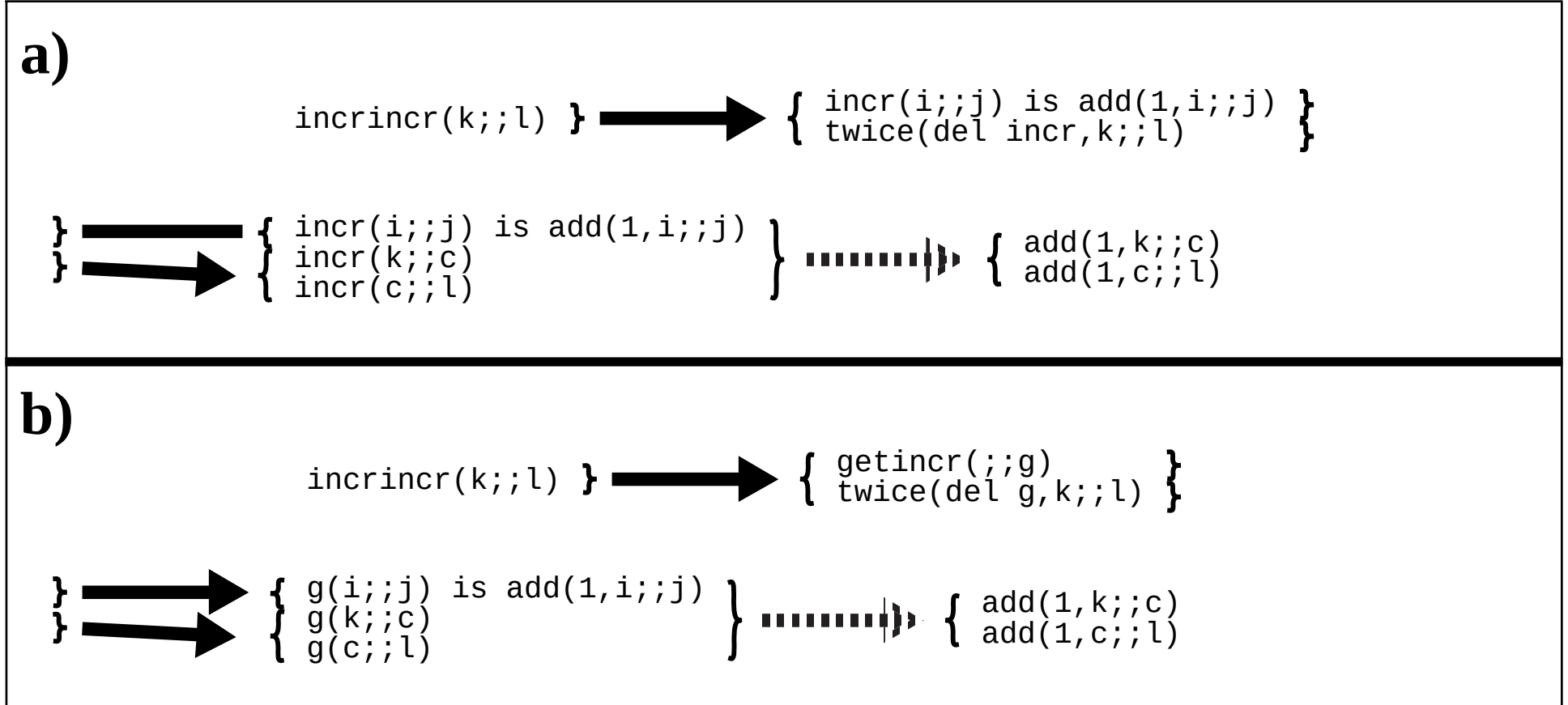

Figure 105 a) and b) The evolution in the task pool for the ia code of Figure 104c) and d), respectively. A dashed arrow indicates substitution by currying.

```
// The sort routine.
intsort(int n, cmp(int k, int e;; int c); int a[n];);

// A comparison by value.
subtract(int x, int y;; int z) { z = x - y; }

// A comparison by value/d.
divcmp(int d, int x, int y;; int z) { z = x/d - y/d; }

// A routine configuring divcmp using a constant.
f100divcmp(int x, int y;; int z) { divcmp(100,x,y;;z); }

// A routine configuring divcmp using a global item.
int globalm;
gdivcmp(int x, int y;; int z) { divcmp(globalm,x,y;;z); }

// A simple routine.
modulus(int x, int y;; int z) { z = x % y; }


main()
{
int n=11; a[n]={16, -571, 9, 4021, -16574, -123, 7, -42, 276, 8891, -3705};

// Sort by value in the usual ascending order.
intsort(n,subtract;a;);

// Sort by value/10, i.e. ignore the last digit in base 10 representation.
// Use currying to configure.
cdivcmp(x,y;;z) is divcmp(10,x,y;;z);
intsort(n,cdivcmp;a;);

// Sort by value/100. Use a constant to configure.
intsort(n,f100divcmp;a;);

// Sort by value/1000. Use a global to configure.
globalm = 1000;
intsort(n,gdivcmp;a;);

// Sort by value%10. i.e. use only the last digit in base 10 representation.
// Use currying to combine routines.
mdivcmp(x,y;;z) is { modulus(x,10;;xm); modulus(y,10;;ym); subtract(xm,ym;;z); }
intsort(n,mdivcmp;a;);

}
```

Figure 106 An application using sorting to demonstrate configuring and combining routines.

a)

```
random(;int seed; real num);

rangen(int i;;ran(;;real r))
{ int s = i;
  ran(;;r) is random(;s;r);
}

// code fragment
rangen(66;;c);
c(;;r1);
c(;;r2);
```

b)

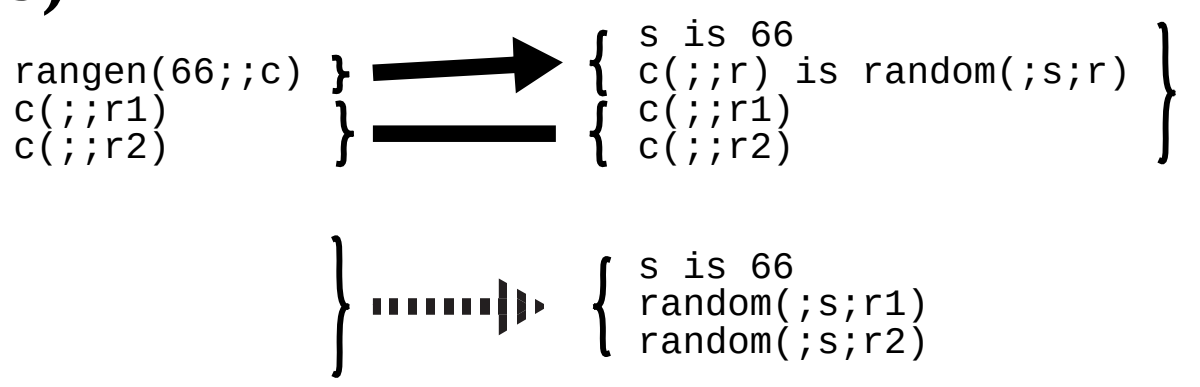

c)

```
rangen(int i;;ran(;;real r));

// code fragment
rangen(77;;a);
a(;;a1);
rangen(88;;b);
b(;;b1);
```

d)

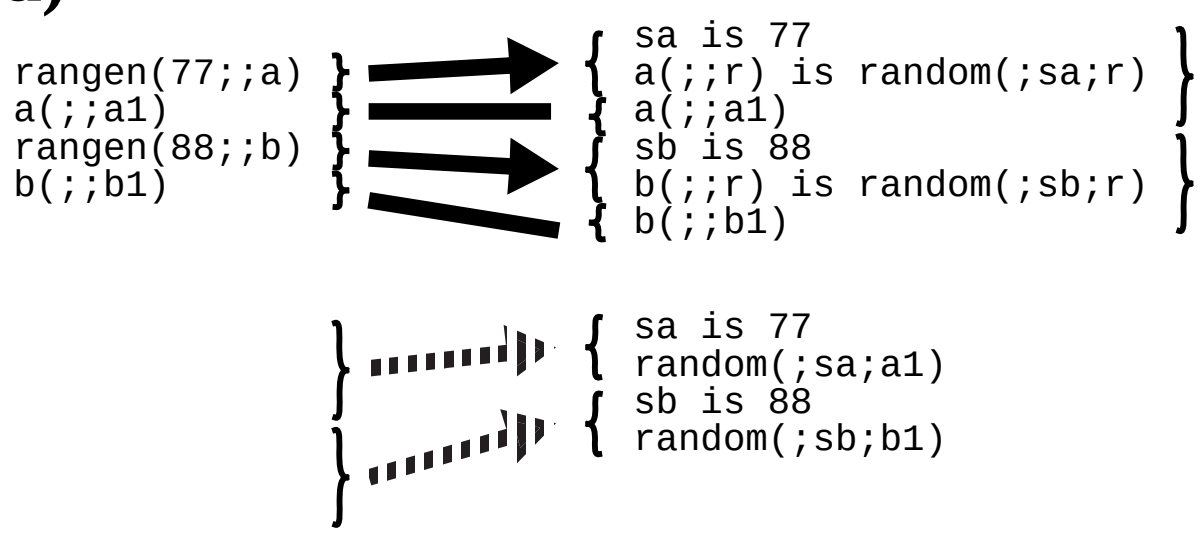

e)

```
random(;int seed; real num);
set(int x;;int y) { y = x; }

record Ran {ran(;;real), seed(;;int)};
Rangen(int i;; Ran g )
{ int s = i;
  g.ran(;;r) is random(;s;r);
  g.seed(;;e) is set(s;;e);
}

// code fragment
Rangen(42;;f);
f.seed(;;x);       // x will be 42,
f.ran(;;y);        // y will be corresponding random number.
f.seed(;;z);       // z will be sequence after 42,
f.ran(;;w);        // w will be corresponding random number.
```

Figure 107 a) A demonstration of the application defined type (ADT) `ran(;;real)`.
b) and d) Executions of the code fragments in a) and c), respectively.
c) A demonstration of two independent items of the ADT `ran(;;real)`.
e) A demonstration of the ADT `Ran`.

```
genrangen( gen(;int seed; real num) ;; rangen(int i;;ran(;;real r)) )
{
  rangen(int i;;ran(;;real r))
  { int s = i;
    ran(;;r) is gen(;s;r);
  }
}

random1(;int seed; real num); // a pseudo-random number generator.
random2(;int seed; real num); // a generator using a different algorithm.

// code fragment
genrangen(random1;;g1); // g1  is a generator of pseudo-random number generators.
genrangen(random2;;g2); // g2  Ibid., but for generators using a different algorithm.

g1(66;;c1);             // c1 is a pseudo-random number generators.
g2(66;;c2);             // c2 is different generator.

c1(;;r1);               // r1 is a pseudo-random number.
c2(;;r2);               // r2 is a pseudo-random number from a different generator.
```

Figure 108 The code of Figure 107a) is extended to show that an item of an ADT can be an ADT.

**a)**

```
// c.get(;;g) returns in g
//        value of counter.
// c.inc(;;) increments
//                 counter.

record Count { get(;;int),
               inc(;;) };

count(int i;; Count c)
{ int x = i;
  c.get(;;g) { g = x; }
  c.inc(;;) { x = x + 1; }
}
```

**b)**

```
cget(int a;;int b) {b = a;}
cinc(;int a;) { a=a+1; }

record Count { get(;;int),
               inc(;;) };

count(int i;; Count c)
{ int x = i;
  c.get(;;g) is cget(x;;g);
  c.inc(;;) is cinc(;x;);
}
```

**c)**

```
record Count { get(;;int),
               inc(;;) };

count(int i;; Count c);

// code fragment
count(10;;a);
count(50;;b);
a.get(;;a0);  // a0 is 10;
b.get(;;b0);  // b0 is 50;
a.inc(;;);
a.get(;;a1);  // a1 is 11;
b.get(;;b0);  // b0 is 50;
```

**d)**

```
// s.push(u;;e) pushes u onto the stack,
//        returning e=0 if all is well.
// s.pop(;;o,e) pops o off the stack,
//        returning e=0 if all is well.

record Stack { push(int;;int),
               pop(;;int,int) };

stack(int max;; Stack s)
{ int a[1:max], x=max;
  int p=0; // last full index.
  s.push(u;;e)
    { if (p<x) { a[++p]=u; e=0; }
      else e=1;
    }
  s.pop(;;o,e)
    { if (p>0) { o=a[p--]; e=0; }
      else e=1;
    }
}
```

**e)**

```
set(int a;;int b) { b = a; }

// s.push(u;;) pushes u onto
//        the arbitraily large stack.
// s.pop(;;o,e) pops o off the stack,
//        returning e=0 if all is well.

record Stack { push(int;;),
               pop(;;int,int) };

stack(;; Stack s)
{ int a[1:]; // arb.y large stack.
  int p=0;   // last full index.
  s.push(u;;)
    { set(u;;a[++p]); }
  s.pop(;;o,e)
    { if (p>0) { set(a[p--];;o); e=0; }
      else e=1;
    }
}
```

Figure 109 a) The routines `c.get(;;g)` and `c.inc(;;)` are examples of nested routines.
b) The nested routines of a) may be implemented by rewriting them as curries.
c) A code fragment using the routine `count` and the ADT `Count` of a).
d) and e) The routines `s.push and s.pop` are further examples of nested routines.

```
// The sort routine.
intsort(int n, cmp(int k, int e;; int c); int a[n];);

// A comparison by value/d.
divcmp(int d, int x, int y;; int z) { z = x/d - y/d; }


main()
{
int n=11; a[n]={16, -571, 9, 4021, -16574, -123, 7, -42, 276, 8891, -3705};

// Sort by value/10, i.e. ignore the digit in base 10 representation.


// Use a named currying routine.
cdivcmp(x,y;;z) is divcmp(10,x,y;;z);
intsort(n,cdivcmp;a;);

// Use an unnamed currying routine.
intsort(n, (x,y;;z) is divcmp(10,x,y;;z) ;a;);


// Use a named nested routine.
n10divcmp(int x,int y;;int z) { z = x/10 - y/10; }
intsort(n, n10divcmp ;a;);

// Use an unnamed routine.
intsort(n, (int x,int y;;int z) { z = x/10 - y/10; } ;a;);

// Ibid., but take argument types from prototype of cmp in intsort.
intsort(n, (x,y;;z) { z = x/10 - y/10; } ;a;);


// Use an unnamed curry of a nested routine.
ldivcmp(int d, int x,int y;;int z) { z = x/d - y/d; }
intsort(n, (x,y;;z) is ldivcmp(10,x,y;;z) ;a;);


}
```

Figure 110 An application using sorting to demonstrate the use of unnamed routines.

**a)**

```
whiledo(c(;;boolean l), b(;;);;)
{          c(;;l); whiledo2(l,c,b;;); }

whiledo2(boolean l, c(;;boolean),
         b(;;);;)
{ if (l) { b(;;);  whiledo(c,b;;);    }
```

**b)**

```
addcon(int n;; int r)
{ int i=1; r=0;
  while (i<=n)
        { r += i;  i++; }
}
```

**c)**

```
whiledo(c(;;boolean l), b(;;);;);

addrou(int n;; int r)
{ int i=1; r=0;
  whiledo((;;l){ l = i<=n; },
    (;;){ r += i;  i++; } ;; );
}
```

**d)**

```
whiledo(c(;;boolean l), b(;;);;);

addrou(int n;; int r)
{ int i=1; r=0;
  cond(;;boolean l) { l= i<=n; };
  body(;;) { r += i;  i++; };
  whiledo(cond,body;;);
}
```

**e)**

```
whiledo(c(;;boolean l), b(;;);;);
ccond(int i,int n;; boolean l){l = i<=n;};
cbody(;int r,int i;) { r += i;  i++; };

addrou(int n;; int r)
{ int i=1; r=0;
  cond(;;l) is ccond(i,n;;l);
  body(;;)  is cbody(;r,i;)
  whiledo(cond,body;;);
}
```

**f)**

```
      subroutine whiledo(c,b)
      external c,b
      logical l
      call c(l)
      call whiledo2(l,c,b)
      end

      subroutine whiledo2(l,c,b)
      external c,b
      logical l
      if (l) then
        call b()
        call whiledo(c,b)
      endif
      end

      subroutine ccond(i,n,l)
      integer i,n
      logical l
      l = i .le. n
      end

      subroutine cbody(r,i)
      integer r,i
      r = r + i
      i = i + 1
      end

      subroutine cond(l)
      logical l
      common /ia/ cn,cr,ci
      integer cn,cr,ci
      call ccond(ci,cn,l)
      end

      subroutine body()
      common /ia/ cn,cr,ci
      integer cn,cr,ci
      call cbody(cr,ci)
      end

      subroutine addtorial(n,r)
      external cond,body
      common /ia/ cn,cr,ci
      integer cn,cr,ci, n,r
      cn = n
      ci = 1
      cr = 0
      call whiledo(cond,body)
      r = cr
      end
```

Figure 111 a) The `whiledo` routine imitates the `while` loop construct.
b) and c) An addtorial definition using the `while` loop and the `whiledo` routine.
d) Rewriting the unnamed routines of c) as named routines.
e) Rewriting the nested routines of d) as curries of unnested routines.
f) A Fortran imitation of the ia code in e).

**a)**

```
random(;int seed; real num);
set(int x;;int y) { y = x; }

int gs = 31;

gran(;;real r) { random(;gs;r); }
gseed(;;int e) { set(gs;;e); }

// code fragment
gseed(;;x);       // x will be 31,
gran(;;y);        // y will be corresponding random number.
gseed(;;z);       // z will be sequence after 31,
gran(;;w);        // w will be corresponding random number.
```

**b)**

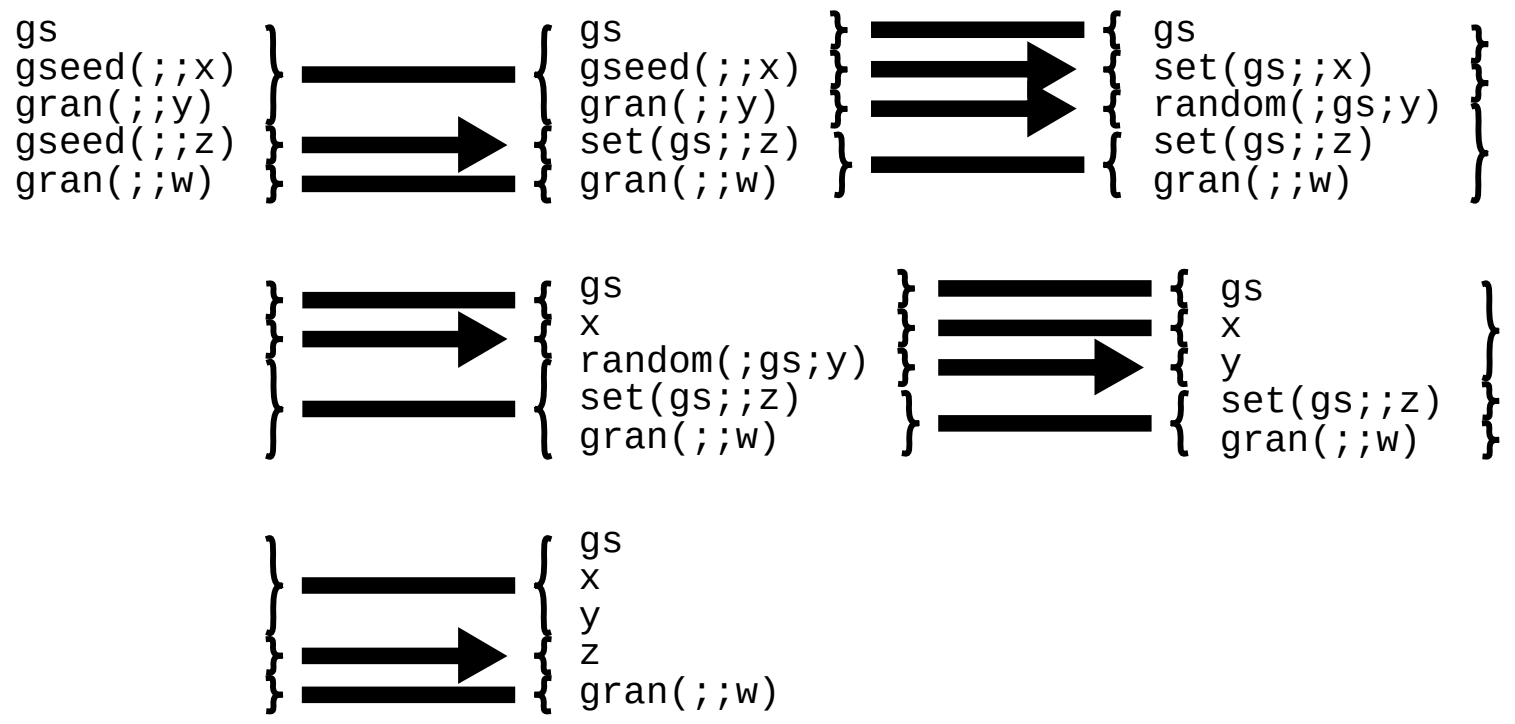

Figure 112 a) The item `gs` is an example of a global item.
b) Execution in the task pool of the code fragment in a).
The task `gseed(;;z)` is assumed to execute first.

**a)**

```
// gget(;;g) returns in g
//       value of counter.
// ginc(;;) increments
//                 counter.

int gx = 0;

gget(;;int g) { g = gx; }
ginc(;;) { gx = gx + 1; }
```

**b)**

```
cget(int a;;int b) {b = a;}
cinc(;int a;) { a=a+1; }

int gx = 0;

gget(;;g) is cget(gx;;g);
ginc(;;) is cinc(;gx;);
```

**c)**

```
gget(;;int g);
ginc(;;);

// code fragment
gget(;;d0);  // d0 is 0;
ginc(;;);
gget(;;d1);  // d1 is 1;
```

Figure 113 a) The routines `gget` and `ginc` are examples of routines which evaluate a global item.
b) The routines of a) may be implemented by rewriting them as curries.
c) A code fragment using the routines `gget` and `ginc` of a).

**a)**

```
triple(int;;int);

byadd(;;) {
  triple(a;;b) { b=a+a+a; }
}

bymult(;;) {
  triple(a;;b) { b=3*a; }
}
```

**b)**

```
triple(int;;int);
byadd(;;);
bymult(;;);

// code fragment.
// triple(5;;x); // error: triple has no value.
byadd(;;);
triple(7;;y); // y=7+7+7
bymult(;;);
triple(8;;z); // z=3*8
```

Figure 114 a) As a global and nonlocal item, the instruction `triple` is an out of the routines `byadd` and `bymult`.
b) A code fragment using the routines `triple`, `byadd` and `bymult` of a).

**a)**

```
random(;int seed; real num);

sran(;;real r) { static int sts = 31; random(;sts;r); }
```

**b)**

```
random(;int seed; real num);

int sran_sts = 31;

sran(;;real r) { random(;sran_sts;r); }
```

**c)**

```
sran(;;real r);
f(;int;);         // Some hypothetical routine.

// code fragment
f(;x;);
sran(;;a);        // a will be random number corresponding to seed 31.
sran(;;b);        // b Ibid. for seed after 31.
```

Figure 115 a) The item `sts` is an example of a static item.
b) The static item `sts` of a) is replaced by the global item `sran_sts`.
c) A code fragment demonstrating the use of the routine `sran` of a).

```
random(;int seed; real num);

record Ran {ran(;;real), seed(;;int), id(;;int)};
Rangen(int i;; Ran g )
{ int s = i;
  static int count = 0;
  int myid = count;
  count++;
  g.ran(;;r) is random(;s;r);
  g.seed(;;e) { e = s; }
  g.id(;;k) { k = myid; }
}

// code fragment
Rangen(42;;f);
Rangen(53;;h);
f.id(;;x);       // x will be 0, assuming f is the first Ran item of the application.
h.id(;;y);       // y will be 1.
```

Figure 116 The ADT `Ran` uses the item elements `s` and `myid` and the type element `count`.

**a)**

```
f(int s;;int x,real y,int z,real w)
{
  random(;int seed; real num);
  set(int x;;int y) { y = x; }

  int ns = s;

  gran(;;real r) { random(;ns;r); }
  gseed(;;int e) { set(ns;;e); }

  gseed(;;x);
  gran(;;y);
  gseed(;;z);
  gran(;;w);
}
```

**b)**

```
m(;a;);  // some hypothetical routine.

f(int s;;int x,real y,int z,real w);

// code fragment
m(;k;);
f(31;;x,y,z,w);
```

**c)**

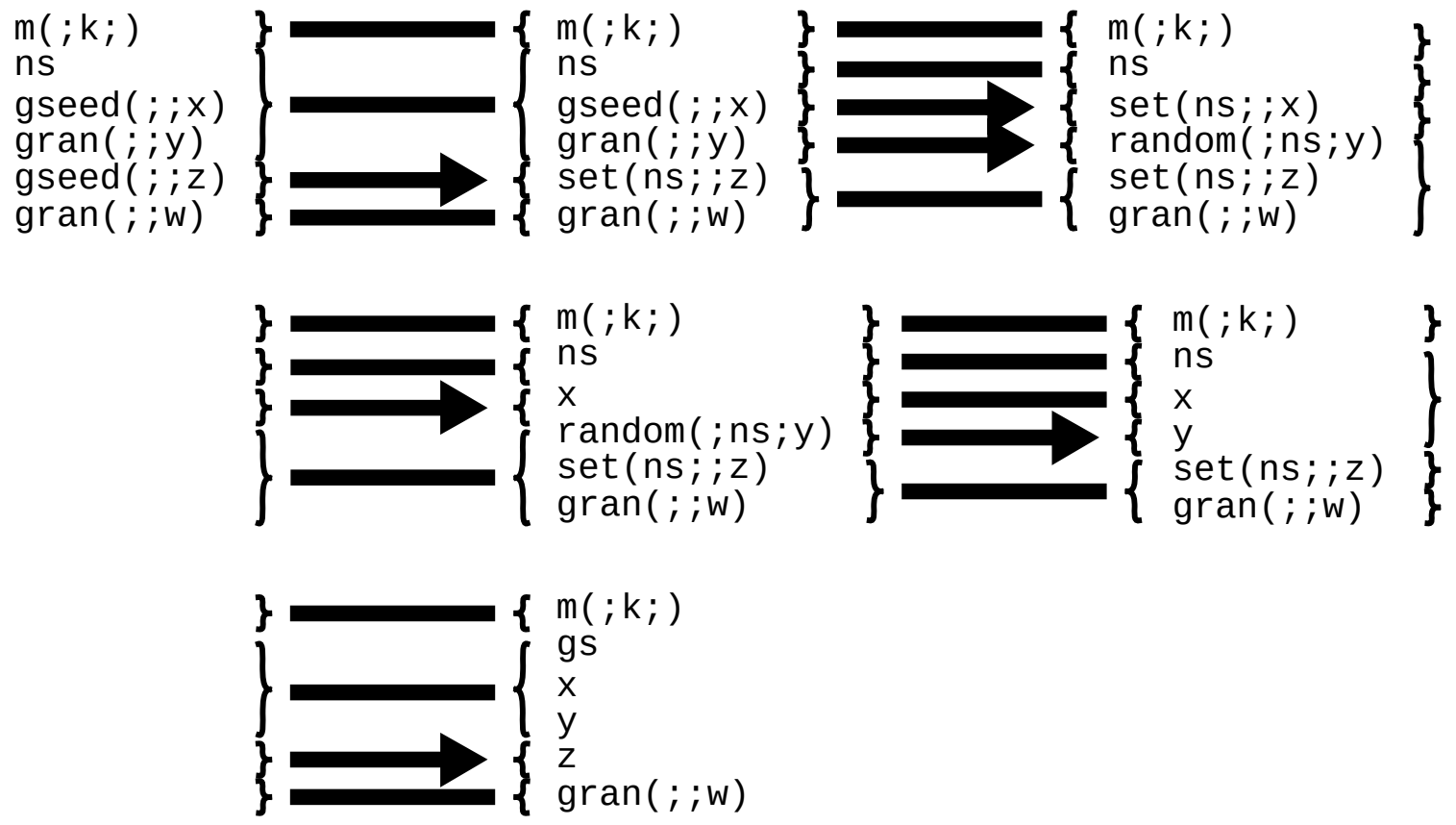

Figure 117 a) The routine `f` encapsulates the code of Figure 112a).
b) A code fragment using the routine `f` of a).
c) Execution in the task pool of the code fragment in b).
The task `gseed(;;z)` is assumed to execute first and `m(;k;)` to execute last.

**a)**

```
{
  int k = n/2;
  vseq(w  ,k  ,a     );
  vseq(w+k,n-k,a[k+1]);
}
```

**b)**

```
(int w, int n;; del int a[n]) {
  int k = n/2;
  vseq(w  ,k  ,a     );
  vseq(w+k,n-k,a[k+1]);
}
```

**c)**

```
f(int w, int n;; del int a[n]) {
  int k = n/2;
  vseq(w  ,k  ,a     );
  vseq(w+k,n-k,a[k+1]);
}

vseq(int w, int n;; del int a[n]) {
if (n == 1)
  set(w;;a);
else
  f(w,n;;a);
}
```

Figure 118 a) A block from the routine `vseq` of Figure 50c).
b) Ibid., but with a declaration of the nonlocal items.
c) The routine `vseq` of Figure 50c), but with the block of a) replaced by a routine.

```
(
  random(;int seed; real num),
  set(int x;;int y)
  ;;
  gran(;;real r),
  gseed(;;int e)
) {
  int gs = 31;

  gran(;;real r) { random(;gs;r); }
  gseed(;;int e) { set(gs;;e); }
}
```

Figure 119 The pseudo-random number generator of Figure 112a) is placed into a block.

**a)**

```
// gget(;;g) return in g value of counter.
// ginc(;;) increment counter.

int gx = 0;

gget(;;int g)(gx;;) { g = gx; }
ginc(;;)(;gx;) { gx = gx + 1; }
```

**b)**

```
    gget(;;d0)(gx;;)
    ginc(;;)(;gx;)
    gget(;;d1)(gx;;)
```

Figure 120 a) The routines `gget` and `ginc` of Figure 113a), but with nonlocal declarations.
b) The nonlocal declarations in a) make explicit in the task pool the dependencies in the code fragment of Figure 113c).

**a)**

```
// putchar from C language.
putchar(int c;; int e);

// code fragment
putchar('a';;e1);
putchar('b';;e2);
```

**b)**

```
// putchar from C language.
putchar(int c;; int e)(;stdout;);

// code fragment as in a).
putchar('a';;e1);
putchar('b';;e2);
```

**c)**

```
// putchar and puts from C language.
putchar(int c;; int e)(;stdout;);
puts(string s;; int e)(;stdout;);

// code fragment
putchar('a';;e1);
puts("bc";;e2);
putchar('d';;e3);
```

**d)**

```
// fopen, fputc and fputs from C language.
fopen(string,string;;FILE);
fputc(int,FILE f;;int)(;f;);
fputs(string,FILE f;;int)(;f;);

// code fragment
fopen("g.dat","w";;g);
fopen("h.dat","w";;h);
fputc('a',g;;e1);
fputs("bc",g;;e2);
fputs("12",h;;e3);
```

Figure 121 a) A code fragment with an indeterminate execution.
b) The code fragment with a determinate execution due to the interaction item `stdout`.
c) `putchar` and `puts` tasks are made determinate by the interaction item `stdout`.
d) Tasks with the same interaction item `f` share dependencies.

**a)**

```
// prototypes
a(;int r;)(;e;);
c(;int r;)(;e;);

// routines
pa(;int s;){a(;s;);}
pc(;int s;){c(;s;);}
b(;int t;){t=2*t;}

// code fragment
pa(;x;);
b(;y;);
pc(;z;);
```

**b)**

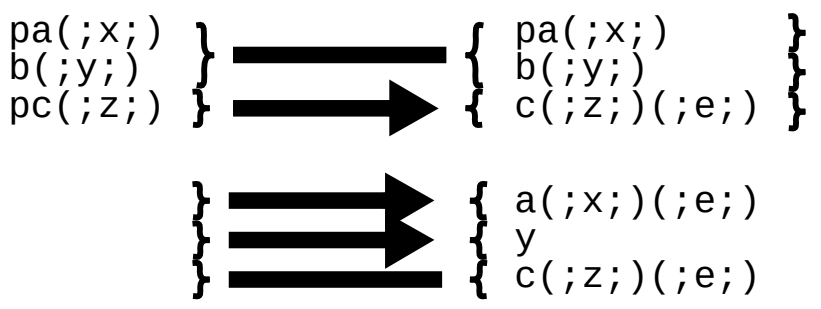

Figure 122 a) A code fragment involving interaction items.
b) An executions in the task pool of the code fragments in a).

**a)**

```
set(int x;; int y) {y=x;}

f(;;int a[1:4]) {
  set(1;;u);
  set(u;;a[1]);
  { set(u;;a[2]);
    set(2;;u);
    set(u;;a[3]);
  }
  set(u;;a[4]);
  // Out a[1:4] = 1,1,2,2.
}
```

**b)**

```
set(int x;; int y) {y=x;}

f(;;int a[1:4]) {
  set(1;; down d);
  set(d;;a[1]);
  { set(d;;a[2]);
    set(2;;d);
    set(d;;a[3]);
  }
  set(d;;a[4]);
  // Out a[1:4] = 1,1,2,1.
}
```

**c)**

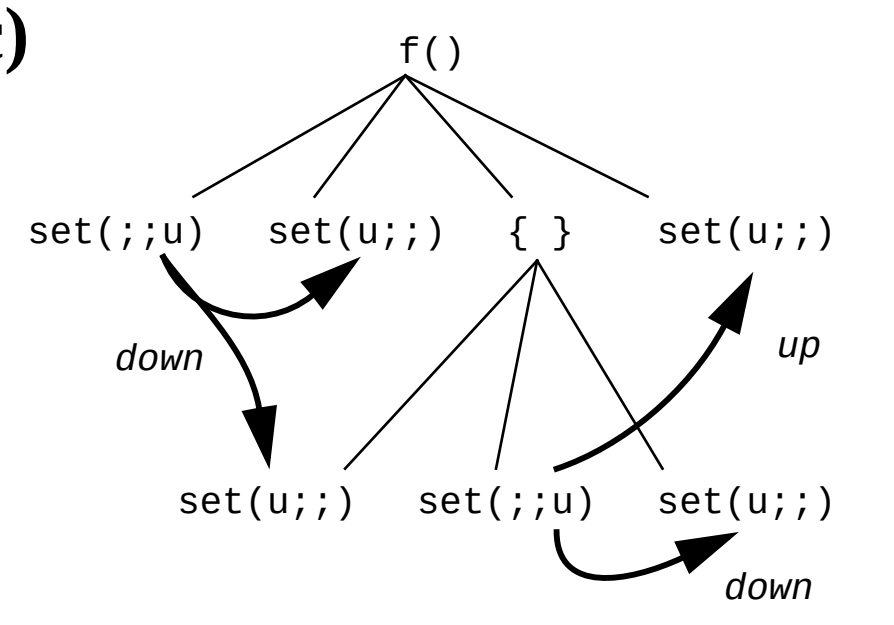

**d)**

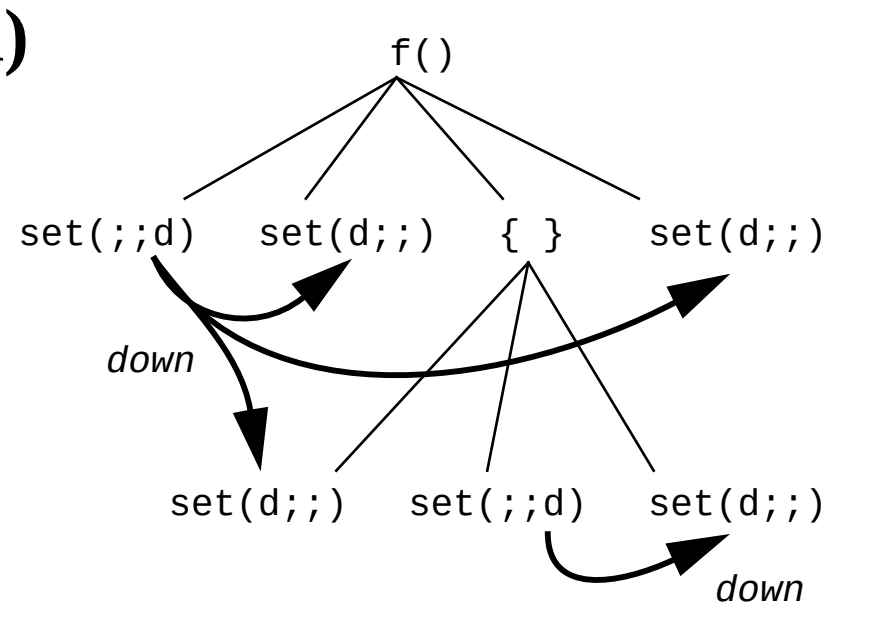

**e)**

```
f(;;int a[1:4]) {
  int u=1;
  a[1]=u;
  { a[2]=u; u=2; a[3]=u; }
  a[4]=u;
  // Out a[1:4] = 1,1,2,2.
}
```

**f)**

```
f(;;int a[1:4]) {
  down int d=1;
  a[1]=d;
  { a[2]=d; d=2; a[3]=d; }
  a[4]=d;
  // Out a[1:4] = 1,1,2,1.
}
```

**g)**

```
f(;;int a[1:4]) {
  int u=1;
  a[1]=u;
  nl(;;) {a[2]=u; u=2; a[3]=u;} // def.n
  nl(;;);                       // use
  a[4]=u;
  // Out a[1:4] = 1,1,2,2.
}
```

**h)**

```
f(;;int a[1:4]) {
  down int d=1;
  a[1]=d;
  nl(;;) {a[2]=d; d=2; a[3]=d;} // def.n
  nl(;;);                       // use
  a[4]=d;
  // Out a[1:4] = 1,1,2,1.
}
```

Figure 123 a) and b) Same code, different out `a[4]`, for `updown u` and `down d`, respectively.
c) The updown item `u` of a) propagates up and down the hierarchy of tasks.
d) The down item `d` of b) only propagates down the hierarchy of tasks.
e) and f) Assignment replaces the `set` routine of a) and b), respectively.
g) and h) The routine `nl(;;)` replaces the block of e) and f), respectively.

**a)**

```
real eps=0.01;

g(;;real eg) { eg = eps; }

h(;;real eh) {
  {  real oldeps = eps;
     eps = eps*eps;
     g(;;x);
     eps = oldeps;
  }
  eh = eps + x;
}

// code fragment
h(;;ih); // ih will be 0.0101.
g(;;ig); // ig will be 0.01.
```

**b)**

```
down real eps=0.01;

g(;;real eg) { eg = eps; }

h(;;real eh) {
  {
     eps = eps*eps;
     g(;;x);

  }
  eh = eps + x;
}

// code fragment
h(;;ih); // ih will be 0.0101.
g(;;ig); // ig will be 0.01.
```

Figure 124 a) The application definition explicitly limits the change to the updown item `eps`.
b) The change to the down item `eps` is implicitly limited.

**a)**

```
int u;

f(;; int y) { y=u; }

g(h(;;int);; int x) { u=0; h(;;x); }

// code fragment
u=1;
g(f;;z); // z will be 0.
```

**b)**

```
down int d;

f(;; int y) { y=d; }

g(h(;;int);; int x) { d=0; h(;;x); }

// code fragment
d=1;
g(f;;z); // z will be 1.
```

**c)**

```
int u;

f(;; int y) { y=u; }

k(;; m(;;int)) { u=0; m is f; }

// code fragment
k(;;p);
u=1;
p(;;w); // w will be 1.
```

**d)**

```
down int d;

f(;; int y) { y=d; }

k(;; m(;;int)) { d=0; m is f; }

// code fragment
k(;;p);
d=1;
p(;;w); // w will be 0.
```

Figure 125 a) and b) Same code, different out `z`, for `updown u` and `down d`, respectively.
c) and d) Ibid.
In a) and b) the routine `f` only is an in, but in c) and d) it first is an out.

```
// The sort routine.
intsort(int n, cmp(int k, int e;; int c); int a[n];);

// A comparison by value/d.
divcmp(int d, int x, int y;; int z) { z = x/d - y/d; }

// A routine configuring divcmp using a nonlocal item.
int nl;
nldivcmp(int x, int y;; int z) { divcmp(nl,x,y;;z); }


main()
{
int n=11; a[n]={16, -571, 9, 4021, -16574, -123, 7, -42, 276, 8891, -3705};
int       b[n]={-786, 12, 5768, -4, -89987, 619, 112, -7654, -567, -4311, 71938};

// Sort a[n] by value/10, i.e. ignore the last digit in base 10 representation.
// Use a nonlocal item to configure.
nl = 10;
intsort(n,nldivcmp;a;);

// Sort b[n] by value/100, i.e. ignore the last two digits in base 10 representation.
// Use a routine which returns a comparison routine.
getcomp(int h;;comp(int k, int e;; int c) ) { nl=h; comp is nldivcmp;}
getcomp(100;;comp);
intsort(n,comp;b;);

}
```

Figure 126 An application using sorting to demonstrate configuring routines with nonlocal items.